%% file: 21cm-physrep.tex
\documentclass{elsart}
\usepackage{amssymb,cite,epsfig,graphicx,subfigure,longtable}

\input{defns.tex}

\begin{document}

\begin{frontmatter}
\title{Cosmology at Low Frequencies:  The 21 cm Transition and the High-Redshift Universe}
\author[ad1]{Steven R. Furlanetto\thanksref{corr}},
\author[ad2]{S. Peng Oh}, \and
\author[ad3]{Frank H. Briggs}
\thanks[corr]{Corresponding author. E-mail: steven.furlanetto@yale.edu.}
\address[ad1]{Department of Physics, Yale University, PO Box 208121, \cty New Haven, CT 06520, \cny USA}
\address[ad2]{Department of Physics, University of California, \cty Santa Barbara, CA 93106, \cny USA}
\address[ad3]{Research School of Astronomy and Astrophysics, The Australian National University, Mount Stromlo Observatory, Cotter Road, \cty Weston, ACT 2611, \cny Australia}

\begin{abstract}
Observations of the high-redshift Universe with the 21 cm hyperfine line of neutral hydrogen promise to open an entirely new window onto the early phases of cosmic structure formation.  Here we review the physics of the 21 cm transition, focusing on processes relevant at high redshifts, and describe the insights to be gained from such observations.  These include measuring the matter power spectrum at $z \sim 50$, observing the formation of the cosmic web and the first luminous sources, and mapping the reionization of the intergalactic medium.  The epoch of reionization is of particular interest, because large \htwo regions will seed substantial fluctuations in the 21 cm background.  We also discuss the experimental challenges involved in detecting this signal, with an emphasis on the Galactic and extragalactic foregrounds.  These increase rapidly toward low frequencies and are especially severe for the highest redshift applications.  Assuming that these difficulties can be overcome, the redshifted 21 cm line will offer unique insight into the high-redshift Universe, complementing other probes but providing the  only direct, three-dimensional view of structure formation from $z \sim 200$ to $z \sim 6$.
\end{abstract}


\end{frontmatter}

\newpage

\tableofcontents

\newpage

\input{intro-ch1.tex}

\input{physics-ch2.tex}

\input{history-ch3.tex}

\input{powerspec-ch4.tex}

\input{dark-ch5.tex}

\input{struc-ch6.tex}

\input{obj-ch7.tex}

\input{reion-ch8.tex}

\input{inter-ch9.tex}

\input{forest-ch10.tex}

\input{synergy-ch11.tex}

\input{conc-ch12.tex}

We are grateful to our editor and referee, Marc Kamionkowski, for his patience, careful reading of the manuscript, and incisive criticism.  We also thank C. Carilli, M. Morales, J. Pritchard, and O. Zahn for their helpful comments on earlier versions.  SRF thanks the Tapir group at the California Institute of Technology for hospitality while most of this work was completed.  We are grateful to all of our collaborators on the many projects described here for sharing their insights and hard work.  We also thank M. McQuinn and K. Sigurdson for helpful discussions about various aspects of this review.  We are grateful to R. Barkana, R. Beresford, J. Bowman, C. Carilli, A. Chippendale, W. Cotton, X. Fan, G. Holder, M. Kuhlen, G. Mellema, K. Sigurdson, and R. White for permission to use their figures and/or data.  Finally, we thank J. Pritchard and M. McQuinn for assistance in generating some of the figures.  

\bibliographystyle{elsart-num}
\bibliography{Ref_21cm}

\end{document}

%% file: defns.tex
\newcommand{\cmden}{\mbox{ cm$^{-3}$}}
\newcommand{\Mpc}{\mbox{ Mpc}}

\newcommand{\kpc}{\mbox{ kpc}}

\newcommand{\Mpcinv}{\mbox{ Mpc$^{-1}$}}

\newcommand{\sqm}{\mbox{ m$^2$}}

\newcommand{\erg}{\mbox{ erg}}
\newcommand{\keV}{\mbox{ keV}}
\newcommand{\eV}{\mbox{ eV}}
\newcommand{\kel}{\mbox{ K}}
\newcommand{\mkel}{\mbox{ mK}}
\newcommand{\microkel}{\mbox{ $\mu$K}}

\newcommand{\ergsec}{\mbox{ erg s$^{-1}$}}

\newcommand{\mJy}{\mbox{ mJy}}

\newcommand{\yr}{\mbox{ yr}}
\newcommand{\hr}{\mbox{ hr}}
\newcommand{\Myr}{\mbox{ Myr}}
\newcommand{\secinv}{\mbox{ s$^{-1}$}}
\newcommand{\MHz}{\mbox{ MHz}}
\newcommand{\kHz}{\mbox{ kHz}}
\newcommand{\GHz}{\mbox{ GHz}}
\newcommand{\Hz}{\mbox{ Hz}}

\newcommand{\Msun}{\mbox{ M$_\odot$}}

\newcommand{\Zsun}{\mbox{ Z$_\odot$}}

\newcommand{\sfr}{\mbox{ M$_\odot$ yr$^{-1}$}}

\newcommand{\hunits}{\mbox{ km s$^{-1}$ Mpc$^{-1}$}}

\newcommand{\recunits}{\mbox{ cm$^{3}$ s$^{-1}$}}

\newcommand{\Junits}{\mbox{ cm$^{-2}$ s$^{-1}$ Hz$^{-1}$ sr$^{-1}$}}

\newcommand{\bxhi}{\bar{x}_{\rm HI}}
\newcommand{\xhi}{x_{\rm HI}}
\newcommand{\nhi}{n_{\rm HI}}
\newcommand{\bxion}{\bar{x}_i}
\newcommand{\xion}{x_i}
\newcommand{\dtb}{\delta T_b}
\newcommand{\bdtb}{\bar{\delta T}_b}

\newcommand{\hone}{HI }
\newcommand{\htwo}{HII }
\newcommand{\mmin}{m_{\rm min}}
\newcommand{\fcoll}{f_{\rm coll}}
\newcommand{\fesc}{f_{\rm esc}}
\newcommand{\lya}{Ly$\alpha$ }
\newcommand{\lyans}{Ly$\alpha$} % Use this version if Ly\alpha is
\newcommand{\lyn}{Ly$n$ }
\newcommand{\bhn}{{\bf \hat{n}}}
\newcommand{\bhk}{{\bf \hat{k}}}
\newcommand{\buv}{{\bf u}}
\newcommand{\bk}{{\bf k}}
\newcommand{\bC}{{\bf C}}
\newcommand{\bfmu}{\mbox{\boldmath$\mu$}}
\newcommand{\bfnu}{\mbox{\boldmath$\nu$}}
\newcommand{\bff}{{\bf f}}

\newcommand{\ga}{{\ \gtrsim \ }}
\newcommand{\la}{{\ \lesssim \ }}
\newcommand{\deriv}{{\rm d}}
\def\VEV#1{\left\langle #1\right\rangle} % This is \VEV{x} => <x>

%% file: intro-ch1.tex
%\documentclass{elsart}
%\usepackage{amssymb,cite,epsfig,longtable}

%\input{../../submission/defns.tex}

%\begin{document}

\section{Introduction} \label{intro}

\subsection{From the Dark Ages to Reionization} \label{intro-full}

Perhaps the most compelling story in all of astrophysics is the formation of structure in our Universe:  how the exceedingly complex objects that surround us today grew out of the remarkably simple and smooth medium that emerged from the Big Bang.  Recent decades have seen enormous progress in disentangling many of the threads in this story, and the basic paradigm for structure formation is now in place.  In the resulting picture, the tiny ($\sim 10^{-5}$) density fluctuations that we observe in the cosmic microwave background (CMB) grew through gravitational instability until they collapsed into the ``cosmic web" of sheets, filaments, and halos that we see around us today.

This overarching scheme has been extraordinarily successful in explaining observations of both the early Universe and local structures.  But two significant gaps remain in the story:  we have yet to directly observe the cosmic ``dark ages" between the last scattering surface of the CMB and the formation of the first luminous structures, and we are only just now beginning to explore the era of ``first light" that stretches from the formation of these sources to the full reionization of the intergalactic medium (IGM).  These epochs -- between $z \sim 1000$ and $z \sim 6$ -- constitute the next frontier of observational cosmology, where we can directly study the transition between the linear and nonlinear regimes of gravitational growth, the characteristics of the first stars and quasars, and their influence on the Universe around them.  They offer an opportunity to connect our detailed pictures of the early Universe with the galaxies around us, thus completing the narrative of structure formation.

The physics content of the dark ages\footnote{The name first became part of astrophysical parlance through W. Sargent \cite{sargent86}.} is sufficiently simple that observations of this period could directly constrain cosmology in an analogous way to the CMB.  After the hydrogen gas recombined, only a few basic processes contributed to the Universe's evolution:  its expansion, the recombination of electrons and protons, the interaction between CMB photons and the residual electrons, and gravity.  Furthermore, except at the conclusion of this era ($z \la 50$), baryonic perturbations remained linear throughout the Universe.  Thus, given a set of underlying cosmological parameters, we can straightforwardly compute the distribution of structure and its characteristics at any time during the dark ages.  As with the CMB, measurements of this era would therefore help to constrain the basic global parameters of our Universe, such as the baryon and matter densities, the shape of the underlying matter power spectrum, and the amplitude of the initial fluctuations.  Conversely, any observed deviation from the expected evolution would be a ``clean" signature of fundamentally new physics.  The dark ages even have one key advantage over the CMB:  because the IGM\footnote{Of course, the ``intergalactic medium" is a misnomer at these times, but we will nevertheless use it without reservation.} is no longer affected by photon diffusion, the baryons develop fluctuations on scales down to the Jeans mass in the neutral IGM.  In principle, this permits tests of the matter power spectrum on much smaller scales than the CMB does \cite{loeb04}.  Unfortunately, with present technology we have no way to access this treasure trove of information.  

Astrophysics became important -- and even dominant -- during the next phase of cosmic history, which stretches from the formation of the first luminous sources (most likely at $z \ga 30$) to ``reionization," when stars and quasars ionized hydrogen in the IGM.  This era obviously holds a great deal of interest for cosmologists.  Most important is the simple desire to study the first generation of stars and galaxies.  But it could also answer any number of fundamental astrophysical questions.  Although hierarchical structure formation -- in which small dark matter halos and galaxies collapse first, later merging into larger objects -- most likely describes this era as well as it does the $z \la 5$ Universe, it leaves many crucial questions unanswered.  Through which processes did the first stars form?  How massive were they, and do fossils remain in the local Universe?  When did heavy elements first form, and what processes distributed them throughout galaxies and the IGM?  Was the IGM clumpy at these early epochs, or did it remain smooth until later on? How did feedback regulate the formation of galaxies, and what types -- radiative, mechanical, or chemical -- were most important?  How and when did the first supermassive black holes form, and what role did they play in galaxy formation?  

A particularly fascinating set of questions relate to the epoch of reionization, the hallmark event of the high-redshift Universe.  It is the point at which structure formation directly affected every baryon in the IGM, even though only a small fraction of them actually resided in galaxies.  It also marked an important phase transition for galaxies:  once the IGM was ionized, it became transparent to ultraviolet (UV) photons --  a dawn (of sorts) for the young galaxies inhabiting the high-redshift Universe.

\begin{figure}[!t]
\centerline{\epsfig{file=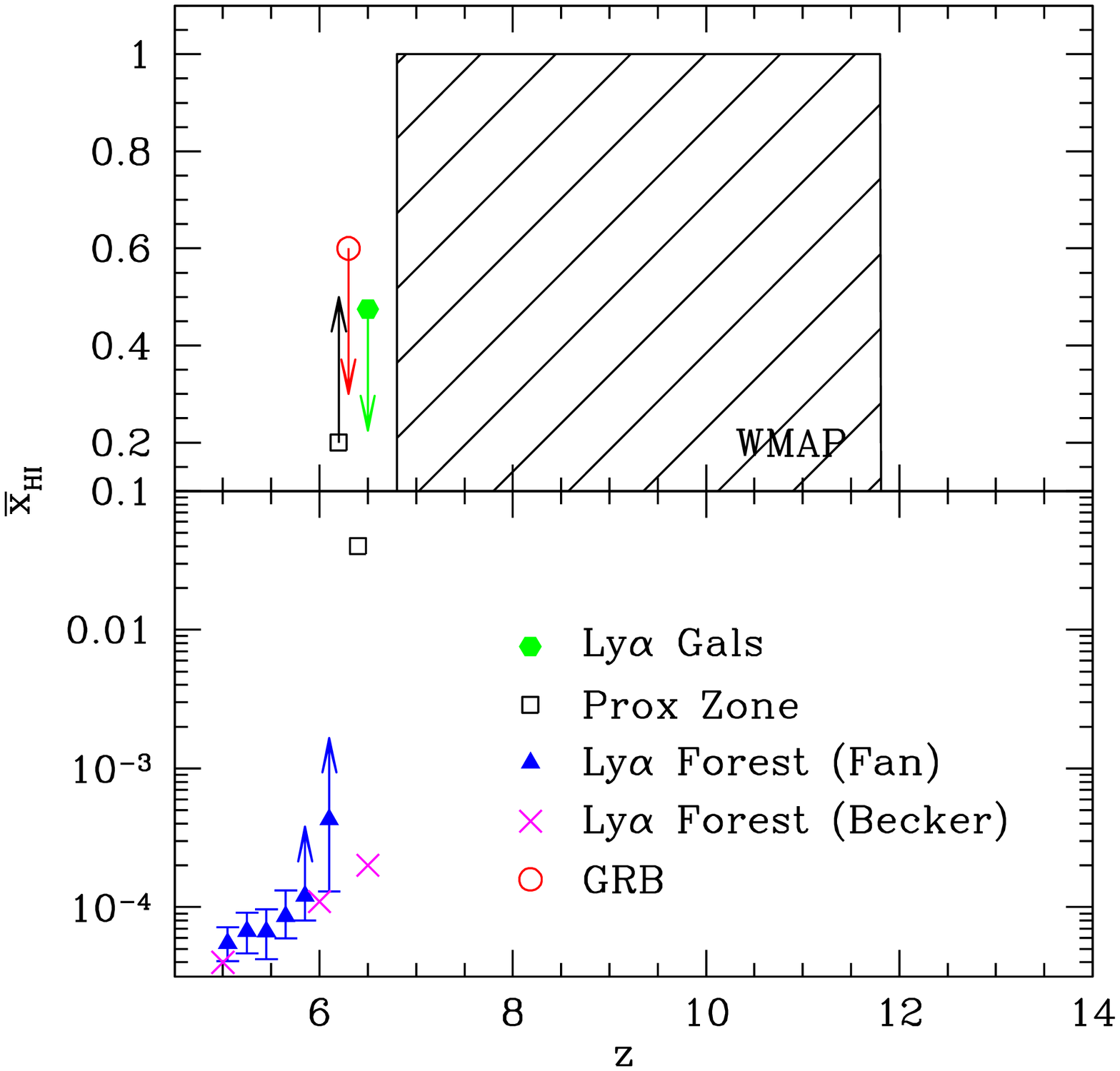,width=3.5in}}
\caption{Compilation of direct observational constraints on reionization (see also \cite{fan06-review}).  The \lya forest measurements with errors and the proximity zone point in the lower panel are taken from \cite{fan06, becker06-tau}, the other \lya forest points from \cite{becker06-tau}, the proximity zone point in the upper panel from \cite{mesinger04, wyithe05-prox}, the \lya galaxy constraint from \cite{malhotra04, furl06-lyagal, malhotra06}, and the GRB point from \cite{totani06}.  The shaded box shows the $1\sigma$ errors on the reionization redshift from the 3-year \emph{WMAP} data \cite{spergel06}, assuming that it is instantaneous.  Note that these are approximate limits (at best) and depend upon a number of theoretical assumptions (see text).}
\label{fig:obs-summ}
\end{figure}

Precisely because its astrophysics is so rich, this phase is much easier to explore than the dark ages -- although, even so, only in the last few years has it finally become accessible.  So far, nearly all of the observational attention has focused on understanding reionization itself (see \cite{fan06-review} for a recent review).  Figure~\ref{fig:obs-summ} summarizes all the existing \emph{direct} measurements of the IGM ionization state at $z > 5$.  The most straightforward come from quasar absorption spectra: as at lower redshifts, the \lya forest offers a powerful window into the IGM and specifically the ionized fraction $\xion$ (or the neutral fraction $\xhi$; we will use overbars to denote global averages of these quantities).  With its large sky coverage, the \emph{Sloan Digital Sky Survey} (SDSS) has proven particularly useful in extending this test to high redshifts and has (to date) identified 19 bright quasars at $z>5.74$ \cite{fan01, fan03, fan04, fan06}.  Unfortunately, inferring the neutral fraction from these measurements requires a theoretical model, because only the rarest voids in the IGM allow light to pass through; thus only the tail of the IGM density distribution is directly sampled, and its properties must be extrapolated to the bulk of the matter \cite{song02, oh05}.  Inserting a model based on simulations of the lower-redshift \lya forest \cite{miralda00} (shown by the solid triangles) implies that the neutral fraction evolved rapidly at $z \ga 5$  \cite{fan02, fan06}; because we expect reionization to proceed rapidly, this may indicate that it ended at about this time.  However, other models are consistent with a much more gentle evolution; for example, extrapolation from empirical fits to \lya forest measurements at $1.7 < z < 5.6$ (shown by the crosses) requires no break in the smooth evolution at $z>6$ \cite{becker06-tau} (see also \cite{song02, songaila04}).

%%%%%%%%%%%% FIGURE 1-1: Observational summary

However, because the \lya forest probes specific lines of sight and resolves features radially, it contains much more information than just $\bxion(z)$.  One example is the ``proximity zone," which is the region of the IGM directly influenced by the ionizing radiation from the background quasar itself (and thus more highly ionized than average).  Before reionization is complete, the extent of the proximity zone measures the number of photons consumed in ionizing the initially neutral gas (and hence $\bxion$) -- at least if the \htwo region has not yet reached the Str{\" o}mgren limit \cite{yu05, furl05-rec}.  This test has been used in three different ways.  First, the sizes clearly decrease with increasing redshift \cite{fan06}, which requires an increasing neutral fraction from $z \sim 5.8$ to $z \sim 6.3$ (shown by the lower open square on Figure~\ref{fig:obs-summ}; note that the errors on this extrapolation are hard to quantify and were not reported by \cite{fan06}).  Second, the inferred spatial extent of the region with non-zero transmissivity appears to indicate $\bxhi \ga 0.2$ \cite{wyithe04-prox, wyithe05-prox}.  Unfortunately, this constraint is degenerate with the quasar lifetime and luminosity and is subject to uncertainties in defining the ``edge" of the proximity zone in an inhomogeneous IGM \cite{fan06}.  Fortunately, the \emph{shape} of the edge can help disentangle these uncertainties.  The third constraint comes from such a measurement:  one quasar (at $z=6.28$) shows a significant decline in the \lya flux well before Ly$\beta$ absorption becomes strong.  This suggests a substantial damping wing, which requires $\bxhi \ga 0.2$ \cite{mesinger04}.

The shape of the IGM \lya damping wing depends on the total optical depth, so it provides another sensitive test of the neutral fraction \cite{miralda98}.  However, in practice, measuring the detailed line profile is difficult.  For example, the intrinsic \lya lines of quasars compromise its measurement in those systems.  Gamma-ray bursts (GRBs), which result from the deaths of massive stars, may provide better sources.  Their afterglow spectra are (intrinsically) featureless power laws (see \cite{piran05} for a review), providing simple templates for extracting the damping profile \cite{barkana04-grb}.  Given that they are associated with star formation, GRBs also ought to occur at high redshifts (see \cite{bromm02-grb} for one estimate), and cosmological time dilation allows the most distant bursts to be identified with realistic instruments \cite{ciardi00,lamb00}.  Indeed a GRB was recently identified at $z=6.3$ \cite{kawai06}.  Unfortunately, it now appears that most GRB spectra contain damped-\lya absorbers (DLAs) at the systemic redshifts of their hosts, as well as a rich set of other absorption features (e.g., \cite{vreeswijk04, chen04-grb}), most likely because the GRBs are embedded in rapidly star-forming regions.  Such DLAs also compromise attempts to measure the IGM absorption.  However, it is still possible to constrain the IGM damping wing, because it has a different shape than prototypical DLAs.  The $z=6.3$ GRB has no evidence for IGM absorption, setting an upper limit of $\xhi<0.6$ (at $95\%$ confidence) along this line of sight \cite{totani06}.

The sightline-to-sightline variations toward different quasars provide another approach.  The observations show surprisingly large variations on $\sim 100 \Mpc$ scales \cite{becker01, fan06}.  Some authors have argued that these require large-scale fluctuations in the ionizing background, indicative of the IGM during or soon after reionization \cite{wyithe06-var, fan06}.  However, the statistics of the \lya forest at low transmission are extremely hard to interpret because of aliasing and bias, so the evidence is not yet conclusive \cite{lidz06, liu06}.

We can also learn about reionization through galaxy surveys.  In particular, the \lya emission lines of galaxies must vanish as the IGM becomes more and more neutral, because the enormous \lya optical depth of a neutral IGM eliminates even those photons on the red side of the line \cite{miralda98-lya, madau00, haiman02-lya}.  In practice this test is difficult to make quantitative because of uncertainties in the intrinsic \lya line properties of the galaxies (and in particular winds) \cite{santos04}.  But surveys for \lyans-selected galaxies have proven useful in two ways.  First, reasonably large samples exist at both $z \approx 5.7$ and $z \approx 6.5$, bracketing the reionization epoch suggested by SDSS quasars.  The narrow range in cosmic time between these two epochs implies little intrinsic evolution between these two samples, so any observed differences between the populations may be due to evolution in $\bxion$.  However, the observed samples have no statistically significant difference in number density \cite{malhotra04}.  Unfortunately, turning this into a quantitative constraint on $\bxhi$ is difficult because of questions about the intrinsic galaxy population.  Treating each galaxy in isolation implies $\bxhi \la 0.3$ \cite{malhotra04, haiman05-lya}; including clustering weakens the constraint to $\bxhi \la 0.5$ \cite{furl06-lyagal}.  A complementary second constraint comes from requiring each observed $z \sim 6.6$  galaxy to be surrounded by a sufficiently large \htwo region to allow transmission.  This again implies $\bxhi \la 0.5$ \cite{malhotra06}, although it is also subject to uncertainties about clustering and the bubble size required for transmission.

Finally, reionization affects the CMB.  Thomson scattering of CMB photons by free electrons washes out temperature fluctuations and generates large-scale polarization anisotropies \cite{sugiyama93}.  The first-year measurements of the \emph{Wilkinson Microwave Anisotropy Probe} (\emph{WMAP}) detected a substantial large-scale correlation between their temperature and polarization maps, indicating a large optical depth for this scattering process ($\tau_{\rm es} \sim 0.17$) and correspondingly early reionization \cite{kogut03, spergel03}.  However, the three-year data now available yields a much smaller optical depth, $\tau_{\rm es} = 0.088^{+0.028}_{-0.034}$ from the \emph{WMAP} data alone or $\tau_{\rm es}=0.069^{+0.026}_{-0.029}$ when combined with a suite of other cosmological measurements \cite{page06, spergel06}.  With better data, the temperature-polarization cross-correlation is no longer detected and the constraints rely instead on the actual polarization power spectrum, which is only detected at about the $3\sigma$ level.  This illustrates the difficulty of removing strong foregrounds -- a lesson which should be noted for our later discussion.  Nevertheless, at least when taken at face value these lower measurements still imply a prolonged reionization epoch.  In Figure~\ref{fig:obs-summ}, the shaded box shows the \emph{WMAP} constraints on the reionization redshift, assuming an instantaneous transition.  Of course this is an approximation:  because the optical depth provides only an integral measurement, it is difficult to apply it directly to a plot of $\bxion(z)$.

Although these observations are beginning to constrain reionization, countless questions remain.  Most importantly, what sources drove it?  The standard assumption is that ionizing photons from massive, hot stars inside small galaxies leaked into the IGM, producing \htwo regions that grew with the galaxies until they merged and filled the Universe.  Although theoretically attractive, direct evidence for such a process is still lacking.  Could accreting black holes have played a part?  Did metal-free Population III stars contribute significantly?  Did recombinations in the IGM significantly delay reionization? Did reionization itself suppress subsequent structure formation?   The range of unanswered questions is illustrated by the proliferation of theoretical papers over the past several years; we refer the interested reader to \cite{barkana01, haiman04-rev, ciardi05-rev, loeb06} for detailed reviews of these issues.

Unfortunately, existing techniques all have intrinsic drawbacks that ultimately limit their utility for studying structure formation at high redshifts (much less the dark ages).  CMB polarization primarily provides an integrated measure of the column depth of ionized gas; information about the reionization history (much less about spatial fluctuations at any given time) is difficult to extract (see \S \ref{cmbpol}).  The patchiness of reionization also imprints small-scale temperature anisotropies on the CMB, but again the resulting constraint is an integral along the line of sight (see \S \ref{ksz}).  Quasar spectra obviously do provide redshift information, but they suffer from saturated absorption at $z \ga 6$ and can shed light only on the late stages of reionization (see \S \ref{quas}).  Finally, galaxy surveys will undoubtedly be tremendously useful for understanding high-redshift galaxies, but they are less than ideal for studying reionization.  First, these galaxies are so distant and faint that collecting substantial numbers of objects at $z \ga 8$ must await new instruments such as the \emph{James Webb Space Telescope} or a 30-meter near-infrared telescope.  Second, the objects themselves will only offer indirect information on the IGM (see \S \ref{galsurv}) and reionization, so such surveys will still leave many questions unanswered.

Thus, new probes of the high-redshift Universe -- including reionization, earlier phases of nonlinear structure formation, and the dark ages -- are urgently needed.  Our purpose is to review arguably the most promising of these techniques:  the redshifted 21 cm line of HI.  This transition, in which the electron flips its spin relative to the nucleus, has a long and successful history in astronomy, and in fact its cosmological implications were realized nearly half a century ago (see \S \ref{history}).  Put simply, the goal is to use this line to map the neutral gas before and during reionization.  The 21 cm transition has three enormous advantages.  First, as a spectral line, redshift information can be used to trace the entire three-dimensional history.  Second, it directly probes IGM gas, which contains the vast majority of baryonic matter.  (Of course, this is actually a disadvantage if one is interested in studying the galaxies themselves.)  Finally, as a forbidden transition with a mean lifetime of $\sim 3 \times 10^7 \yr$, the 21 cm line is far from saturation and is thus sensitive to the (most interesting) middle stages of reionization.

Although these advantages have long been recognized (e.g., \cite{field59-obs, sunyaev72, hogan79}), only recently has technology improved to the point where high-redshift 21 cm line observations\footnote{Note that here, and throughout this review, we will abuse the phrase ``21 cm" by applying it to this high-redshift regime without qualification, even though the observed wavelengths are of course $21 (1+z)$ cm.} are feasible.  As a result, recent years have seen a substantial effort in theoretically modeling the expected signals and in planning or building instruments to observe them.  These efforts include CoRE (Cosmic Reionization Experiment) at ATNF (Australia Telescope National Facility), designed to measure the mean background as a function of frequency, and a suite of antenna arrays designed to search for small-scale fluctuations in the 21 cm background.  The latter set includes the 21 Centimeter Array (21CMA), LOFAR (Low Frequency Array), the Precision Array to Probe Epoch of Reionization (PAPER), the Mileura Widefield Array Low Frequency Demonstrator (which we will refer to as the MWA), and the Square Kilometer Array (SKA).  All of these experiments (except for the last one) should begin gathering data within the next several years.  They hope to constrain the reionization epoch itself at $z \la 12$; observing the higher-redshift Universe at lower frequencies, where foregrounds are much stronger, will require much larger instruments.

This review will focus on theoretical predictions of the 21 cm signal over a range of cosmic time -- from $z \ga 50$, when it allows us to probe the growth of the earliest structures, to $z \sim 6$, when reionization ends.  As we will see, such predictions will be crucial to interpreting the data once it arrives, because the signals will (at least with the first generation of experiments) be buried within the huge foreground noise.  The models described here can help to isolate the cosmological signal, and they also illustrate the potential rewards of these observations -- at least those we know of today. 

This paper is organized as follows.  The remainder of this section provides a brief historical overview of 21 cm line cosmology (\S \ref{history}) and presents some basic notation and formulae (\S \ref{assump}).  The next three sections review the fundamentals necessary for understanding the 21 cm signal.  We then describe the relevant physics of the 21 cm transition in \S \ref{fund}.  From there, we forge ahead applying this knowledge to the high-redshift Universe.  First, in \S \ref{glob} we examine the mean evolution of the 21 cm background.  In \S \ref{ps} we shift gears to describe the power of statistical measurements of fluctuations in the 21 cm background; we then explore applications of this technique to the dark ages (\S \ref{dark}), the formation of the first nonlinear structures (\S \ref{struc}), the first luminous sources (\S \ref{obj}), and finally to reionization itself (\S \ref{reion}).  These four sections are (for the most part) independent of each other and need not be read sequentially.  We put these efforts in an observational context in \S \ref{int} and discuss instrumental sensitivities and the expected systematic challenges.  We then describe a complementary approach that uses bright background sources to illuminate the ``21 cm forest" of absorption features in \S \ref{forest}.  Finally, we connect 21 cm experiments to other observational probes in \S \ref{syn}, and we offer some concluding thoughts in \S \ref{conc}.

\subsection{Historical Overview} \label{history}

The 21 cm transition is a rare example (for astronomy) of a successful theoretical calculation driving observational effort:\footnote{We regard this episode as inspiration for our review and hope that the theory presented here proves even a fraction as useful!}  Indirectly inspired by Oort's hope that then-unknown radio lines would help in studying our Galaxy, the physicist van de Hulst\cite{vdhulst45,vdhulst98} first computed the transition frequency of the hyperfine transition of \hone during the Second World War and immediately realized its usefulness for tracing Galactic structure and kinematics.  Several groups of physicists, astrophysicists, and engineers subsequently detected the line in 1951\cite{ewen51,muller51,pawsey51}, followed shortly by observations of spiral structure in the Milky Way\cite{vdhulst54} and the measurement of the rotation curve of M31\cite{vdhulst57}.  At the same time, theoretical understanding of the atomic physics of the hydrogen atom itself continued to improve \cite{wild52}.

The 1960s saw a number of observational programs aiming to detect hydrogen in the nearby IGM, through both emission and absorption against bright extragalactic radio sources\cite{field59-obs,goldstein63,penziasscott68,penziaswilson69,allen69,lang76}.  It is interesting to note, in fact, that searches for IGM absorption along lines of sight to radio galaxies (what we call the ``21 cm forest" in \S \ref{forest}, although then expected to be much  more uniform) preceded the discovery of both quasars and the \lya forest.  These placed non-trivial limits on cosmology (ruling out a closed universe full of neutral hydrogen, for example) but of course were doomed to failure given the highly-ionized IGM we now know to exist.  At the same time, there was a theoretical assessment of excitation conditions  in a dilute, neutral IGM\cite{wouthuysen52,field58,field59-ts}, a topic crucial for interpreting observations and that will receive detailed treatment in this review (see \S \ref{fund}). Field reviewed the status of IGM studies (as of 1972) in \cite{field72}. 

The great leap to studying the 21 cm line at cosmological distances occurred soon after.  Zel'dovich's proposal that large-scale structure develops from the IGM through collapse of cluster-sized masses \cite{zeldovich70, sunyaev72, sunyaev75,oort84}, with subsequent fragmentation into galaxies, inspired a number of efforts to observe the 21 cm line from $z > 3$. This ``top-down scenario'' implied the existence of high-redshift``Zel'dovich pancakes'' containing ${\sim}10^{14} \Msun$ of neutral gas that would have been detectable with existing radio telescopes at $z \ga 3$--$4$; the predicted brightness temperatures were $\dtb \sim 0.1$--$1 \kel$. Table~\ref{tab:pancakesearch} summarizes the observational programs driven (at least in part) by these predictions. None were successful, though many reached limiting \hone masses comparable to predictions (and a tentative detection \cite{uson91b} did spur a brief flurry of excitement from theorists \cite{subramanian92, scott92}).  They also encountered many of the same difficulties that modern experiments will face, particularly terrestrial interference and foreground subtraction, and pioneered many of the strategies to overcome these challenges.

There was also some observational motivation for expecting an increased \hone content at $z > 3$, because the early results from surveys for DLA absorption lines against high-$z$ quasars \cite{wolfe86} appeared to indicate a factor $\sim$10 over-abundance of HI compared to the present epoch.  This has proved to be a red herring, because better measurements of the DLA statistics\cite{wolfe05} and refinements to the cosmological model now make it clear that the HI content at $z\sim 3$ is unlikely to be more than a factor two above the $z\approx 0$ value\cite{zwaan06}.

%%%%%%%%%%%% TABLE 1-1: "Hi-Z" HI emission surveys
\begin{table}
\begin{center}
\begin{tabular}{|c|c|c|c|c|}
\hline
$z$ & $\nu (\MHz)$ & Telescope & Year & Reference\\ 
\hline
8.4&151&6C Array& 1986& \cite{bebb86}\\
5.1&235&Arecibo& 1995& \cite{avruch95}\\
4.9&240&Jodrell Bank& 1978& \cite{davies78}\\
3.4&322&VLA& 1994& \cite{taramopoulos94}\\
3.4&323&Arecibo, VLA, Ooty& 1997&\cite{briggs97,kanekar97}\\
3.4&323&VLA, Arecibo, WSRT, GMRT& 1991&\cite{uson91a,briggs93,debruyn96,chandra04}\\
3.3&326&WSRT& 1992& \cite{wieringa92}\\
3.3&327&Ooty& 1990& \cite{subrahswarup90}\\
3.3&328&Jodrell Bank& 1978& \cite{davies78}\\
3.3&331&VLA& 1988& \cite{noreau88}\\
3.3&333&VLA& 1990& \cite{subrahanantha90}\\
3.3&333&VLA& 1990& \cite{uson91b}\\
3.1&349&VLA& 1995& \cite{taramopoulos95}\\
\hline 
\end{tabular} 
\end{center}
\caption{Surveys for high-redshift HI 21 cm line emission.  We list the emission redshift, the observed frequency, the telescopes used to observe the field, the date of first publication, and references.  Here the VLA is the Very Large Array, the WSRT is the Westerbork Synthesis Radio Telescope, and the GMRT is the Giant Metrewave Telescope.  \label{tab:pancakesearch} }
\end{table}

Of course, we are now convinced that structure forms ``bottom-up," with the smallest halos collapsing first and subsequently merging into larger objects.  In these kinds of scenarios, massive, neutral pancakes would not form at high redshifts, and the lack of success in the surveys listed in Table \ref{tab:pancakesearch} is not surprising.  The first predictions in the context of this kind of model came in 1979 \cite{hogan79}, but they received little attention over the next decade (largely because, in contrast to the pancake scenario, observations could not yet constrain them).  Scott \& Rees \cite{scott90} returned theoretical attention to 21 cm emission from high redshifts, examining the signals expected at $z=8.4$ in a variety of galaxy formation models.  They pointed out that, although the structures in a cold dark matter (CDM) model would be extremely faint at high redshifts, statistically measuring their scale-dependent fluctuations could constrain the matter power spectrum (remarkably similar to the strategies now under development).  

During the remainder of the 1990s, the \emph{Cosmic Background Explorer}'s measurements of the CMB, together with a wide array of other observations, cemented our cosmological paradigm.  Thus during this time, 21 cm predictions based on the CDM model began to appear \cite{subramanian93, kumar95, bagla97}, although most still focused on imaging rare, massive objects at high redshift.  The prospects for existing telescopes looked bleak in this model.  At the same time, \lya forest measurements at $z \ga 5$ showed that the Universe remained highly ionized to quite large redshifts.  Interest in these cosmological applications declined, because radio astronomers viewed the $z > 5$ Universe with trepidation, given the brightness of the radio sky and the increasing levels of terrestrially-generated radio frequency interference.  

Interest in the 21 cm signal grew again when theorists began to study reionization and the nature of the first ionizing sources.  Madau, Meiksin, \& Rees \cite{madau97} were the first to consider explicitly the effect of luminous sources on the 21 cm signal, emphasizing the role these sources played in setting the spin temperature of the IGM.  Since then, work has focused less on measuring cosmological parameters and more on understanding the interplay between luminous galaxies and the IGM.  These questions have become particularly interesting in light of the rich but confusing observational constraints on reionization described in \S \ref{intro-full}.  At the same time, technological improvements are bringing the $z>6$ Universe within our grasp.  The remainder of this review will describe the current state of theoretical predictions for the 21 cm signal within this ``modern" approach.

\subsection{Some Preliminaries} \label{assump}

We take this opportunity to summarize some useful parameters and relations.  In the numerical calculations made specifically for this review, we assume a cosmology with $\Omega_m=0.26$, $\Omega_\Lambda=0.74$, $\Omega_b=0.044$, $H_0=100 h \hunits$ (with $h=0.74$), $n=0.94$, and $\sigma_8=0.8$.  Here $\Omega_i$ is the density in component $i$ relative to the critical value, $\rho_c(z) = 3 H(z)^2/8 \pi G$, $H_0$ is the Hubble constant at the present day, $n$ is the spectral index of the matter power spectrum at inflation, and $\sigma_8$ normalizes the power spectrum; it is the amplitude of the rms fluctuations smoothed on $8 h^{-1} \Mpc$ scales.  These are consistent with the most recent \emph{WMAP} results \cite{spergel06}.  Note , however, that we have increased $\sigma_8$ to improve agreement with other observations \cite{viel06, seljak06}.  For figures taken from other sources, we refer the reader to the referenced works for their assumed parameters.  Within the existing uncertainties, the choices for these parameters do not significantly affect our predictions.  The exception is $\sigma_{8}$, which shifts the structure formation timetable but otherwise has no qualitative impact.  

In this cosmology, the hydrogen density at $z=0$ is $\nhi=2.06 \times 10^{-7} [(1-Y_p)/0.76] \cmden$, where $Y_p$ is the mass fraction of helium.  The Hubble constant evolves as $H(z) = H_0 [\Omega_m(1+z)^3+\Omega_{\Lambda}]^{1/2}$ or, at the high redshifts in which we are interested, $H(z) \approx H_0 \Omega_m^{1/2} (1+z)^{3/2}$.  In this regime, the age of the universe is 
\begin{equation}
t_0(z) = \int_0^z \frac{\deriv z'}{(1+z') \, H(z')} \approx 5.46 \times 10^8 \, \left( \frac{1+z}{10} \right)^{-3/2}
\yr.
\label{eq:tage}
\end{equation}
Unless otherwise specified, we quote distances in comoving units, $r_{\rm com}=(1+z) r_{\rm prop}$, defined by
\begin{equation}
\frac{ \deriv r_{\rm com}}{\deriv z} = \frac{c}{H(z)}.
\label{eq:rcom}
\end{equation}
The angular diameter distance is $D_A(z) = r_{\rm com}/(1+z)$, and the luminosity distance is $D_L(z) = r_{\rm com}(1+z)$.  For convenience, a useful conversion between observed angular scale $\Delta \theta$ and comoving size at $z \ga 10$ is
\begin{equation}
r_{\rm com} \approx 1.9 \, \left( \frac{\Delta \theta}{1'} \right) \, \left( \frac{1+z}{10} \right)^{0.2} \, h^{-1} \Mpc. 
\label{eq:dang}
\end{equation}
A similar approximation for the relation between observed bandwidth $\Delta \nu$ and the radial distance is
\begin{equation}
r_{\rm com} \approx 1.7 \, \left( \frac{\Delta \nu}{0.1 \MHz} \right) \left( \frac{1+z}{10} \right)^{1/2} \, \left( \frac{\Omega_m h^2}{0.15} \right)^{-1/2} \Mpc.
\label{eq:dbw}
\end{equation}

We also note that, according to common usage, we use $h$ to denote both the normalized Hubble constant and Planck's constant.  The desired meaning will normally be clear from context.

Throughout much of our theoretical discussion, we will require the dark matter halo mass function.  For simplicity, we will generally use the Press-Schechter form \cite{press74} (see \cite{cooray02} for a pedagogical review of the halo mass function and its derivation).  The comoving number density of halos with masses in the range $(m,\,m+\deriv m)$ is
\begin{equation}
n(m) \deriv m = \sqrt{\frac{2}{\pi}} \, \frac{\bar{\rho}}{m^2} \, \left| \frac{\deriv \ln \sigma}{\deriv \ln m} \right| \frac{\delta_c(z)}{\sigma(m)} \, \exp \left[ - \frac{\delta_c^2(z)}{2 \sigma^2(m)} \right]\deriv m,
\label{eq:nmps}
\end{equation}
where $\bar{\rho}$ is the mean comoving mass density and $\sigma^2(m)$ is the variance of the density field at $z=0$ when smoothed in top-hat spheres of mass $m$:
\begin{equation}
\sigma^2(m) = \int_0^\infty \frac{\deriv k}{2 \pi^2} \, k^2 P_{\rm lin}(k) \left[ \frac{3 j_1(kR)}{kR} \right]^2.
\label{eq:sigma2}
\end{equation}
Here $P_{\rm lin}(k)$ is the linear matter power spectrum (we use the transfer function of \cite{eisenstein98}), $j_1(x)=(\sin x - x \cos x)/x^2$, and $m=4\pi \bar{\rho} R^3/3$.  Finally, $\delta_c(z)\approx 1.69/D(z)$ is the linearized density threshold for collapse in the spherical top-hat model and $D(z)$ is the linear growth factor, normalized so that $D(z=0)=1$.  Note that we use the convention in which $\sigma^2$ is always evaluated at a fixed time, so that $\delta_c$ increases toward higher redshifts.  Of course this is only for notational convenience; in reality, the threshold for virialization is a fixed physical density.  

In detail, equation (\ref{eq:nmps}) is not an ideal fit to the halo mass function found in numerical simulations at low redshifts, and more accurate forms exist in the literature \cite{sheth99, jenkins01}.  The high redshift case is less clear; the mass function seems to lie in between the Press-Schechter and Sheth-Tormen forms \cite{jang01, reed03, iliev05-sim, zahn06-comp}.  Nevertheless, equation~(\ref{eq:nmps}) suffices for our purposes because the existing analytic models have many other uncertainties and approximations.  It has the enormous advantage of being easy to manipulate analytically.  For example, the collapse fraction, or fraction of the mass in halos more massive than $\mmin$, is simply
\begin{equation}
\fcoll = \int_{\mmin}^\infty \deriv m \, m\, n(m) = {\rm erfc} \left[ \frac{\delta_c(z)}{\sqrt{2} \sigma(\mmin) }
  \right].
\label{eq:fcollps}
\end{equation}
We will use $\fcoll$ in numerous calculations throughout this review.

Our Fourier transform conventions follow
\begin{equation}
f({\bf x}) = \int \frac{\deriv^3 {\bf k}}{(2 \pi)^3} \, \tilde{f}({\bf k}) \, e^{i{\bf k} \cdot {\bf x}}
\label{eq:ftx}
\end{equation}
and
\begin{equation}
\tilde{f}({\bf k}) = \int \deriv^3 {\bf x} \, f({\bf x}) \, e^{-i{\bf k} \cdot {\bf x}}.
\label{eq:ftk}
\end{equation}

Finally, for convenience we present a summary table of many of the more common symbols in this review, together with references to their definitions and brief definitions.

\begin{longtable}{ccl} 
\caption{Symbol dictionary.  Middle column shows the equation or section in which the symbol is first used or defined. \label{tab:dictionary} } \\
\hline
\hline
Symbol & Ref. & Definition \\
\hline
\hline 
\endfirsthead 
\caption{Symbol dictionary \emph{(continued).} } \\
\hline
\hline
Symbol & Ref. & Definition \\
\hline
\hline 
\endhead 
\hline
\hline
\endfoot
$A_{10}$ & (\ref{eq:sigma01}) & Spontaneous decay rate of 21 cm transition \\
$A_e$ & (\ref{eq:visnoise}) & Effective area of one antenna element \\
$A_{\rm tot}$ & \S \ref{int} & Total effective collecting area \\
$B$ & (\ref{eq:unoise}) & Total bandwidth of observation \\
$b_x$ & (\ref{eq:bias-bub}) & Bias of \htwo region \\
$C$ & (\ref{eq:bxion-evol}) & Clumping factor \\
$C_{01},\,C_{10}$ & (\ref{eq:detbal}) & Collisional spin excitation, de-excitation rates \\
$C_l^{21}$ & (\ref{eq:angps-defn}) & Angular power spectrum of $\delta_{21}$ \\
$C^N$ & (\ref{eq:cov-noise}) & Covariance matrix of noise \\
$C^{SV}$ & (\ref{eq:cov-samvar}) & Covariance matrix of sample variance \\
$D(z)$ & (\ref{eq:sigma2}) & Linear growth factor \\
$D_A$ & (\ref{eq:rcom}) & Angular diameter distance \\
$D_L$ & (\ref{eq:rcom}) & Luminosity distance \\
$D_{({\rm min,max})}$ & \S \ref{int} & Minimum/maximum baseline in interferometer \\
$\fcoll$ & (\ref{eq:fcollps}) & Collapse fraction \\
$f_{\rm esc}$ & (\ref{eq:zetadefn}) & Escape fraction for ionizing photons \\
$f_{\rm He}$ & (\ref{eq:tcomp}) & Helium fraction (by number) \\
$f_{\rm rec}(n)$ & (\ref{eq:frec}) & Recycling fraction for Ly$n$ $\rightarrow$ \lya photons \\
$f_{\rm sh}$ & (\ref{eq:fsh}) & Fraction of gas shock-heated above $T_\gamma$ \\
$f_X$ & (\ref{eq:sfrxray}) & X-ray luminosity per SFR relative to local value \\
$f_{X,h},\,f_{X,{\rm ion}}$ & (\ref{eq:fxapprox}) & Fraction of X-ray energy in heating and ionization \\
$f_\star$ & (\ref{eq:xrayemiss}) & Star formation efficiency \\
$g(z)$ & (\ref{eq:gz}) & Coefficient in approximation $\delta_T=g(z) \delta$ \\
$g_i$ & (\ref{eq:tsdefn}) & Statistical weight of level $i$ \\
$I_\nu$ & \S \ref{basic} & Specific intensity \\
$J_\alpha$ & (\ref{eq:xalpha}) & $J_\nu$ near \lya ignoring radiative transfer effects \\
$J_\nu$ & \S \ref{basic} & Angle-averaged specific intensity \\
$J_\nu^c$ & (\ref{eq:xalpha}) & Fiducial $J_\nu$ for $x_\alpha=S_\alpha$ \\
$\ell$ & (\ref{eq:cov-samvar}) & Distance to 21 cm screen \\
$n(m)$ & (\ref{eq:nmps}) & Halo mass function \\
$n_b(m)$ & (\ref{eq:nbub}) & Mass function of \htwo regions \\
$\bar{n}_b^c$ & (\ref{eq:sfemiss}) & Comoving baryon number density \\
$n(u_\perp)$ & (\ref{eq:tk}) & Baseline distribution [also $n(k,\mu)$] \\
$n_\nu$ & (\ref{eq:tcrad1}) & Photon occupation number \\
$N_a$ & \S \ref{int} & Number of elements in interferometer \\
$N_B$ & \S \ref{interferometer-patterns} & Number of baselines in interferometer \\
$N_c$ & (\ref{eq:pkerror}) & Number of measurements in $\bk$ bin \\
$N_\alpha$ & (\ref{eq:sfemiss}) & Number of $(10.2,13.6)$ eV photons per stellar baryon  \\ 
$N_{\rm ion}$ & (\ref{eq:zetadefn}) & Number of ionizing photons per stellar baryon \\ 
$P_{21}(k)$ & (\ref{eq:pkdefn}) & Power spectrum of $\delta_{21}$ (units:  volume) \\
$P_{01},\,P_{10}$ & (\ref{eq:detbal}) & Spin (de-)excitation rate from \lya absorptions \\
$P_V(\delta_{\rm nl})$ & (\ref{eq:clump}) & Volume-weighted distribution of $\delta_{\rm nl}$ \\
$P_{xx}$ & (\ref{eq:pk_iso}) & Power spectrum of ionized fraction \\
$P_{x\delta}$ & (\ref{eq:pk_iso}) & Cross-power spectrum of $\xion$, density \\
$P_\alpha$ & (\ref{eq:palpha}) & Total \lya scattering rate \\
$P_{\delta \delta}$ & \S \ref{ps} & Matter power spectrum \\
$P_{\Delta T}$ & (\ref{eq:pkerror}) & Power spectrum of $\delta_{21}$ (units:  mK$^2$ volume) \\
$S_\alpha$ & (\ref{eq:xalpha}) & Correction to $x_\alpha$ from radiative transfer \\
$S_\nu$ & \S \ref{basic} & Flux density \\
$t_{\rm int}$ & (\ref{eq:radiometer}) & Integration time \\
$t_{(\bk,{\bf u})}$ & (\ref{eq:tk}) & Integration time for the single mode $\bk$ (or ${\bf u}$) \\
$t_\gamma$ & (\ref{eq:tcomp}) & Compton cooling time \\
$T_b$ & \S \ref{basic} & Brightness temperature \\
$T_c$ & (\ref{eq:tcrad}) & Color temperature of \lya background \\
$T_K$ & (\ref{eq:xdefn}) & Gas kinetic temperature \\
$T_S$ & (\ref{eq:tsdefn}) & 21 cm spin temperature \\
$T_{\rm sky}$ & (\ref{eq:tsky}) & Sky temperature \\
$T_{\rm sys}$ & (\ref{eq:radiometer}) & System temperature \\
$T_\gamma(z)$ & \S \ref{basic} & CMB temperature at redshift $z$ \\
$T_\star$ & (\ref{eq:tsdefn}) & Energy defect of 21 cm transition ($0.068 \kel$) \\
$(u,v)$ & (\ref{eq:fduv_integral}) & Sky coordinates (in units of $\lambda$) \\
${\bf V}$ & (\ref{eq:fd_integral}) & Visibility \\
$v_\parallel$ & (\ref{eq:optdepthcosmo}) & Proper velocity along line of sight \\
$W_\nu$ & \S \ref{response-patterns} & Primary beam of antenna \\
$x_c$ & (\ref{eq:xcdefn}) & Collisional coupling coefficient for $T_S$ \\
$x_{\rm tot}$ & (\ref{eq:beta}) & Total coupling coefficient for $T_S$ \\
$x_\alpha$ & (\ref{eq:xalpha}) & Wouthuysen-Field coupling coefficient for $T_S$ \\ 
$\xhi$ & \S \ref{intro-full} & Neutral fraction \\
$\bxhi$ & \S \ref{intro-full} & Globally-averaged neutral fraction \\
$\xion$ & \S \ref{intro-full} & Ionized fraction \\
$\bxion$ & \S \ref{intro-full} & Globally-averaged ionized fraction \\
$z_c$ & \S \ref{critpt} &  Redshift at which $x_\alpha=1$ \\
$z_h$ & \S \ref{critpt} &  Redshift at which $T_K=T_\gamma$ in the IGM \\
$z_r$ & \S \ref{critpt} &  Reionization redshift \\

$\alpha_{(A,B)}$ & \S \ref{clump} & Recombination coefficient (case-A,B) \\
$\beta$ & (\ref{eq:beta}) & Expansion coefficient for $\delta$ in $\delta_{21}$ \\
$\beta_T$ & (\ref{eq:betaT}) & Expansion coefficient for $\delta_T$ in $\delta_{21}$ \\
$\beta_x$ & (\ref{eq:betax}) & Expansion coefficient for $\delta_x$ in $\delta_{21}$ \\
$\beta_\alpha$ & (\ref{eq:beta-alpha}) & Expansion coefficient for $\delta_\alpha$ in $\delta_{21}$ \\
$\gamma$ & (\ref{eq:Dkin}) & Sobolev parameter \\ 
$\gamma'$ & (\ref{eq:gammaprime}) & Sobolev parameter including spin exchange \\
$\delta$ & (\ref{eq:optdepthcosmo}) & Fractional matter overdensity \\
$\delta_{21}$ & (\ref{eq:pkdefn}) & Fractional perturbation to $\dtb$ \\
$\delta_c(z)$ & (\ref{eq:nmps}) & Linear critical density for halo collapse \\
$\delta_{\rm coll}$ & (\ref{eq:dcoll}) & Critical overdensity for collisional coupling \\
$\tilde{\delta}_{\rm iso}$ & (\ref{eq:d21ft}) & Isotropic component of $\tilde{\delta}_{21}$ \\
$\delta_{\rm nl}$ & (\ref{eq:clump}) & Nonlinear density field \\
$\delta_T$ & (\ref{eq:d21}) & Fractional perturbation to $T_K$ \\
$\delta_x$ & (\ref{eq:d21}) & Fractional perturbation to neutral fraction \\
$\delta_\alpha$ & (\ref{eq:d21}) & Fractional perturbation to $x_\alpha$ \\
$\delta A$ & (\ref{eq:visnoise}) & Physical area of one antenna element \\
$\delta P_{\Delta T}$ & (\ref{eq:pkerror}) & Error on power spectrum \\
$\dtb$ & (\ref{eq:dtbone}) & 21 cm brightness temperature relative to CMB \\
$\bdtb$ & (\ref{eq:dtbone}) & Globally-averaged $\dtb$ \\
$\delta \eta$ & (\ref{eq:unoise}) & Inverse bandwidth of measurement \\
$\Delta T^N$ & (\ref{eq:radiometer}) & Telescope noise \\
$\Delta^2_{21}$ & (\ref{eq:pkdefn}) & Variance of 21 cm brightness temperature \\
$\Delta \ell$ & (\ref{eq:cov-samvar}) & Radial width of 21 cm screen \\
$\Delta \nu$ & (\ref{eq:radiometer}) & Channel width of observation \\
$\Delta \nu_D$ & (\ref{eq:dnud}) & Doppler width of \lya transition \\
$\epsilon(\nu)$ & (\ref{eq:sfemiss}) & Comoving emissivity at frequency $\nu$ \\
$\epsilon_{\rm ap}$ & (\ref{eq:single-dish}) & Aperture efficiency \\
$\epsilon_{\rm comp}$ & (\ref{eq:tcomp}) & Heating rate from Compton scattering \\
$\epsilon_{X}$ & (\ref{eq:xrayemiss}) & Heating rate from X-rays \\
$\epsilon_{\alpha}$ & (\ref{eq:epsalpha}) & Heating rate from \lya scattering \\
$\zeta$ & (\ref{eq:zetadefn}) & Ionizing efficiency parameter \\
$\eta$ & (\ref{eq:Akin}) & Recoil parameter  \\
$\eta'$ & (\ref{eq:etaprime}) & Recoil parameter including spin exchange \\
$\eta_f$ & (\ref{eq:tb2_sensitivity}) & Array filling factor \\
$\theta_D$ & \S \ref{int-over} & Diffraction-limited angular resolution \\
$\kappa_{10}$ & (\ref{eq:xcdefn}) & Collisional spin de-excitation rate coefficient \\
$\mu$ & (\ref{eq:d21ft}) & Cosine of angle between $\bk$ and line of sight \\
$\nu_0$ & \S \ref{basic} & Rest frequency of 21 cm line \\
$\nu_\alpha$ & (\ref{eq:dnud}) & Rest frequency of \lya transition \\
$\sigma^2$ & (\ref{eq:sigma2}) & Variance of density field at $z=0$ \\
$\tau_{\rm es}$ & \S \ref{intro-full} & CMB electron scattering optical depth \\
$\tau_{\rm GP}$ & (\ref{eq:gp}) & Gunn-Peterson optical depth in \lya \\
$\tau_\nu$ & (\ref{eq:rad_trans}) & Optical depth at frequency $\nu$ \\
$\phi_\alpha$ & (\ref{eq:palpha}) & Profile of \lya line\\
$\phi_\nu$ & (\ref{eq:optdepth}) & Profile of 21 cm line\\
$\chi_\alpha$ & (\ref{eq:palpha}) & Line center \lya absorption cross section \\
\end{longtable} 

%\bibliographystyle{elsart-num}
%\bibliography{Ref_21cm}

%\end{document}

%% file: physics-ch2.tex
%\documentclass{elsart}
%\usepackage{amssymb,cite,epsfig}

%\input{../../submission/defns.tex}

%\begin{document}

\section{Fundamental Physics of the 21 cm Line} \label{fund}

\subsection{Basic Definitions} \label{basic}

We begin with some basic definitions necessary for what follows.  The fundamental quantity of radiative transfer is the {\it brightness} (or {\it  specific intensity}) $I_\nu$ of a ray emerging from a cloud  at  frequency $\nu$. This conventionally expresses the energy carried by rays traveling along a given direction, per unit area, frequency, solid angle, and time; it thus normally has dimensions ergs s$^{-1}$ cm$^{-2}$ sr$^{-1}$ Hz$^{-1}$ (see \cite{rybicki79} for a pedagogical introduction to radiative transfer).  However, for many applications of radiative transfer in an expanding universe, the units $\Junits$ are more convenient, because photon number is conserved during the expansion but energy is not; we will often work in the latter units in this review.  We will also usually slightly simplify the problem by using the angle-averaged specific intensity $J_\nu = \int I_\nu \deriv \Omega$.

For convenience, we will quantify $I_{\nu}$ by  the equivalent {\it brightness temperature}, $T_b(\nu)$, required of a blackbody radiator (with spectrum $B_{\nu}$) such that $I_{\nu}=B_{\nu}(T_b)$. Throughout the range of frequencies and temperatures relevant to the 21 cm line, the Rayleigh-Jeans formula is an excellent approximation to the Planck curve, so that $T_b(\nu)\approx I_{\nu} \, c^2/2k_B{\nu}^2$, where is $c$ is the speed of light and  $k_B$ is Boltzmann's constant.

We will be almost exclusively interested in the brightness temperature of the \hone 21 cm line, which has rest frequency $\nu_0 =$ 1420.4057~MHz.  Because of the cosmological redshift, the emergent brightness $T_b'(\nu_0)$ measured in a cloud's comoving frame at redshift $z$ creates an apparent brightness at the Earth of $T_b(\nu) = T_b'(\nu_0)/(1+z)$, where the observed frequency is $\nu=\nu_0/(1+z)$. Similarly, the brightness temperature of the CMB in a comoving frame at redshift $z$ scales from the presently observed value of $T_\gamma(0)=2.73$~K to  $T_\gamma'(z)=2.73\;(1+z)$~K.\footnote{Henceforth we will drop the prime in $T_\gamma(z)$ for notational convenience.}

An individual gas cloud produces a {\it flux} (erg~cm$^{-2}$~s$^{-1}$) at the Earth of $S=
\int_{\rm cloud} I_{\nu}\;\cos \theta \;\deriv \Omega\; d\nu$, where the integral extends over the solid angle subtended by the cloud and the frequency spread of the signal.  In most of our applications we will instead consider the frequency-dependent incident energy flux.  Radio astronomers typically measure this frequency-dependent {\it flux density}, $S_{\nu}$, in Janskys (1 Jy $=10^{-23}$ erg~s$^{-1}$~cm$^{-2}$~Hz$^{-1}$). The apparent angle $\theta$ between the cloud centroid  and the element of solid angle $d\Omega$ is generally small ($\sin\theta\approx \theta$) in the applications considered here.  Thus for  uniform brightness clouds with small apparent angular diameter, a convenient conversion is $S_{\nu}\approx I_{\nu}\Delta\Omega= 2 k_B T_b \nu^2\Delta\Omega/c^2$, where all quantities are measured in the observer's frame.  Note that, once the cosmological redshift and the variation of $\deriv \nu$ with distance are included, the flux scales like $S \propto D_L^{-2}$, while the flux density scales as $S_{\nu} \propto (1+z) D_L^{-2}$.   

In the Rayleigh-Jeans limit, the equation of radiative transfer  along a line of sight through a cloud of uniform excitation temperature $T_{\rm ex}$ states that the emergent  brightness at frequency $\nu$ is 
\begin{equation}
T_b'(\nu) = T_{\rm ex}(1-e^{-\tau_{\nu}})+T_R'(\nu)e^{-\tau_{\nu}}
\label{eq:rad_trans}
\end{equation}
where the {\it optical depth}  $\tau_\nu \equiv \int \deriv s \, \alpha_{\nu}$ is the integral of the absorption coefficient ($\alpha_{\nu}$)  along the ray through the cloud, $T_R'$ is the brightness of the background radiation field incident on the cloud along the ray, and $s$ is the proper distance. 

For the 21 cm transition, the excitation temperature $T_{\rm ex}$ is referred to as the spin temperature $T_S$.  It quantifies the relative number densities, $n_i$, of atoms in the two hyperfine levels of the electronic  ground state (we will use the subscripts 1 and 0 to denote the triplet and singlet states, respectively; these equal the total angular momentum $F$ of the atom).  It is defined via
\begin{equation}
\frac{n_1}{n_0} = \frac{g_1}{g_0}e^{-E_{10}/k_B T_S }  = 3\;e^{-T_\star/T_S}
\label{eq:tsdefn}
\end{equation}
where $g_i$ is the statistical weight (here $g_0=1$ and $g_1=3$), $E_{10}=5.9 \times 10^{-6} \eV$ is the energy splitting, and $T_\star \equiv E_{10}/k_B=0.068$~K is the equivalent temperature.  Because all astrophysical  applications have $T_S \gg T_*$, approximately three of four atoms find themselves in the excited state.  As a result, the absorption coefficient must include a correction for stimulated emission (and hence it depends on $T_S$ as well).  Note that, in detail, the assumption of a single $T_S$ applying to the entire hydrogen distribution is not necessarily correct.  Rigorously, one should solve a Boltzmann equation that couples the spin and velocity distributions \cite{hirata06}.  When the collision time is long, this introduces percent level changes to the brightness temperature.

The optical depth of a cloud of hydrogen is then:
\begin{eqnarray}
\tau_{\nu} & = & \int \deriv s \,  \sigma_{01} \, (1-e^{-E_{10}/k_B T_S}) \, \phi(\nu) \, n_0  \label{eq:optdepth} \\
& \approx & \sigma_{01} \, \left(\frac{h\nu}{k_B T_S}\right) \left(\frac{N_{\rm HI}} {4}\right) \, \phi(\nu),
\label{eq:optdepthcloud}
\end{eqnarray}
where 
\begin{equation}
\sigma_{01} \equiv \frac{3 c^2 A_{10}}{8\pi\nu^2},
\label{eq:sigma01}
\end{equation}
$A_{10}=2.85 \times 10^{-15} \secinv$ is the spontaneous emission coefficient of the 21 cm transition, $N_{\rm HI}$ is the column density of \hone (here the factor 1/4 accounts for the fraction of \hone atoms in the hyperfine singlet state), and $\phi(\nu)$ is the line profile (defined so that $\int \deriv \nu \, \phi(\nu) = 1$).  The second factor in equation~(\ref{eq:optdepth}) accounts for stimulated emission.  The approximate form in equation~(\ref{eq:optdepthcloud}) assumes uniformity throughout the cloud.  
 
In general, the line shape $\phi(\nu)$ includes natural, thermal,  and pressure broadening, as  well as bulk motion (which increases the effective Doppler spread). Our most important application is to  IGM gas expanding uniformly with the Hubble flow.  Then the velocity broadening of a region  of linear dimension $s$ will be $\Delta V\sim s H(z)$ so that $\phi(\nu)\sim c/[s H(z)\nu]$. The column density along such a segment depends  on the neutral fraction $\xhi$ of hydrogen, so $N_{\rm HI} = \xhi n_H(z)\;s$.  A more exact calculation yields, with equation~(\ref{eq:optdepth}), an expression for the 21 cm optical depth of the diffuse IGM,
 \begin{eqnarray}
 \tau_{\nu_0} & = & \frac{3}{32 \pi} \, \frac{h c^3 A_{10}}{k_B T_S \nu_0^2} \, \frac{\xhi n_{H}}{(1+z) \, (\deriv v_\parallel/\deriv r_\parallel)}  \label{eq:optdepthcosmo} \\
 & \approx  & 0.0092 \, (1+\delta) \, (1+z)^{3/2}\, \frac{\xhi}{T_S} \, \left[ \frac{H(z)/(1+z)}{\deriv v_\parallel/\deriv r_\parallel} \right],
 \label{optdepthcosmo-approx}
\end{eqnarray}
where in the second equality $T_S$ is in degrees Kelvin.  Here the factor $(1+\delta)$ is the fractional overdensity of baryons and $\deriv v_\parallel/\deriv r_\parallel$ is the gradient of the proper velocity along the line of sight, including both the Hubble expansion and the peculiar velocity.  In the second line, we have substituted the velocity $H(z)/(1+z)$ appropriate for the uniform Hubble expansion at high redshifts.  

The two applications of equation~(\ref{eq:rad_trans}) that will be most important here are:

1.  The contrast between high-redshift hydrogen clouds and the CMB.   Many of the observational strategies for the 21 cm line involve comparison of lines of sight through a cloud\footnote{Here we use ``cloud" to refer to any patch of the IGM; it need not be physically distinct from the surrounding gas.} to (sometimes hypothetical) sightlines with clear views of the CMB.  Thus we hope to measure
\begin{eqnarray}
\dtb(\nu) & = & \frac{T_S-T_\gamma(z)}{1+z} (1-e^{-\tau_{\nu_0}}) \approx \frac{T_S-T_\gamma(z)}{1+z}\;\tau_{\nu_0} 
\label{eq:dtbone} \\
& \approx & 9\;\xhi(1+\delta) \, (1+z)^{1/2}\, \left[1-\frac{T_\gamma(z)}{T_S}\right] \, \left[ \frac{H(z)/(1+z)}{\deriv v_\parallel/\deriv r_\parallel} \right] \mkel.
\label{eq:dtb}
\end{eqnarray}
Note that $\dtb$ saturates if $T_S \gg T_\gamma$, but it can become arbitrarily large (and negative) if $T_S \ll T_\gamma$.  The observability of the 21 cm transition therefore hinges on the spin temperature; we will describe below the mechanisms that drive $T_S$ either above or below $T_\gamma(z)$, which dictate whether the 21 cm signal will appear in emission, absorption, or not at all.

2. Absorption against high redshift radio sources (\S~\ref{forest}). The brightness temperatures of nonthermal radio continuum sources ($T_{\rm src}\approx 10^6-10^{10}$K) far exceed $T_S$ and $T_\gamma$, so the flux density received from the direction of  a high redshift radio source is $S_{\nu}\approx   S_{\rm src}\exp(-\tau_{\nu})$.  High-redshift radio-loud quasars or radio galaxies would make superb probes of cloud structure in the neutral or partially reionized IGM through their absorption line spectra.

Three competing processes determine $T_S$:  (1) absorption of CMB photons (as well as stimulated emission); (2) collisions with other hydrogen atoms, free electrons, and protons; and (3) scattering of UV photons.  We let $C_{10}$ and $P_{10}$ be the de-excitation rates (per atom) from collisions and UV scattering, respectively; they will be examined in detail in the following sections.  We also let $C_{01}$ and $P_{01}$ be the corresponding excitation rates.  The spin temperature is then determined in equilibrium by\footnote{Note that the relevant timescales are all much shorter than the expansion time, so equilibrium is an excellent approximation.}
\begin{equation}
n_1 \left( C_{10} + P_{10} + A_{10} + B_{10} I_{\rm CMB} \right) = n_0 \left( C_{01} + P_{01} + B_{01} I_{\rm CMB} \right),
\label{eq:detbal}
\end{equation}
where $B_{01}$ and $B_{10}$ are the appropriate Einstein coefficients and $I_{\rm CMB}$ is the energy flux of CMB photons.  With the Rayleigh-Jeans approximation, equation (\ref{eq:detbal}) can be rewritten as \cite{field58}
\begin{equation}
T_S^{-1} = \frac{T_\gamma^{-1} + x_c T_K^{-1} + x_\alpha T_c^{-1}}{1 + x_c + x_\alpha},
\label{eq:xdefn}
\end{equation}
where $x_c$ and $x_\alpha$ are coupling coefficients for collisions and UV scattering, respectively, and $T_K$ is the gas kinetic temperature.  Here we have used detailed balance through the relation
\begin{equation}
\frac{C_{01}}{C_{10}} = \frac{g_1}{g_0} e^{-T_\star/T_K} \approx 3 \left( 1 - \frac{T_\star}{T_K} \right).
\label{eq:c01db}
\end{equation}
We have then \emph{defined} the effective color temperature of the UV radiation field $T_c$ via
\begin{equation}
\frac{P_{01}}{P_{10}} \equiv 3 \left( 1 - \frac{T_\star}{T_c} \right).
\label{eq:tcolor}
\end{equation}
The goal of the next two sections will be to calculate $x_c$,  $x_\alpha$, and $T_c$.  In the limit in which $T_c \rightarrow T_K$ (a reasonable approximation in most situations of interest, as we will see in \S \ref{wf}), equation (\ref{eq:xdefn}) may be written
\begin{equation}
1 - \frac{T_\gamma}{T_S} = \frac{x_c + x_\alpha}{1 + x_c + x_\alpha} \, \left( 1 - \frac{T_\gamma}{T_K} \right).
\label{eq:xdefn-tfac}
\end{equation}

\subsection{Collisional Coupling} \label{coll}

We will first consider collisional excitation and de-excitation of the hyperfine levels, which dominate in dense gas.  The coupling coefficient for species $i$ is
\begin{equation}
x_c^i \equiv  \frac{C_{10}^i}{A_{10}} \, \frac{T_\star}{T_\gamma} = \frac{n_i \, \kappa_{10}^i}{A_{10}} \, \frac{T_\star}{T_\gamma},
\label{eq:xcdefn}
\end{equation}
where $\kappa_{10}^i$ is the rate coefficient for spin de-excitation in collisions with that species (with units of $\recunits$).  The total $x_c$ is the sum over all $i$.  We next show how these rates are calculated.  Readers who are not interested in the details can simply use the final results presented in Figure~\ref{fig:collrates} and Tables \ref{tab:hhcoll}--\ref{tab:ehcoll}.

\subsubsection{H-H Collisions} \label{hhcoll}
 
The role of hydrogen-hydrogen collisions in setting the spin temperature has received extensive attention in the literature, dating back to the first discussions of the spin temperature \cite{purcell56, field58, dalgarno61, smith63, dalgarno64, smith66, allison69, zygelman05}.  The dominant interaction is electron (and hence spin) exchange in atomic collisions.  There are several types of permissible spin-exchange collisions.  Let us label the four hyperfine states by ``a" for $F=0$ and ``b, c, d" for $F=1$, $m=(-1,\,0,\,1)$, where $m$ is the projection of the spin.  If $A$ and $B$ label the two atoms, electron exchange must obey the conservation law
\begin{equation}
m_A + m_B = m_{A'} + m_{B'}.
\label{eq:mcons}
\end{equation}
One possible set of interactions has $\Delta F=2$:  cc$\rightarrow$aa, bd$\rightarrow$aa, and db$\rightarrow$aa; we denote their cross sections with $\sigma^+$.  Another set has $\Delta F=1$:  bd$\rightarrow$ac, db$\rightarrow$ac, cd$\rightarrow$ad, dc$\rightarrow$ad, bc$\rightarrow$ab, cb$\rightarrow$ab (and their inverses), which we denote with $\sigma^-$.  Finally, we let $\Delta F=0$ transitions (bd$\rightarrow$cc, db$\rightarrow$cc, and their inverses) have cross section $\sigma^0$.  

We will now sketch how to compute $\kappa_{10}^{\rm HH}$, the effective rate coefficient for these processes; more rigorous treatments can be found in the literature \cite{dalgarno61, smith66, allison69, zygelman05}.  Although a semi-classical approach is adequate at high temperatures \cite{purcell56}, a full quantum-mechanical treatment is necessary in the general case.  The problem can be described as a typical scattering event between two identical particles that form an intermediate (virtual) hydrogen molecule before separating again.

Actually, an analogous calculation that ignores nuclear symmetry serves to illustrate the main principles \cite{dalgarno61}.  The Schr{\" o}dinger equation for the total wavefunction $X$ of both atoms is
\begin{equation}
\left( H - \frac{1}{2 M} \nabla^2_{\bf R} - E \right) X({\bf r},{\bf R}) = 0,
\label{eq:hhschrod}
\end{equation}
where ${\bf r}$ represents the positions of the two electrons, ${\bf R}$ is the vector joining the two nuclei, $M$ is the reduced mass, $E$ is the total energy, and $H$ is the Hamiltonian of the system when the nuclei are fixed in space.  The lowest energy eigenstates of $H$ are $\chi_s$ and $\chi_t$, the singlet and triplet states of the hydrogen molecule ($^1 \Sigma_g$ and $^3\Sigma_u$, respectively).  For slow collisions, the total wave function can then be written as a superposition of these two states
\begin{equation}
X({\bf r}, {\bf R}) = F_s({\bf R}) \, \chi_s({\bf r}; {\bf R}) + F_t({\bf R}) \, \chi_t({\bf r}; {\bf R}),
\label{eq:Xsimp}
\end{equation}
where $F_{s,t}$ are determined as follows.  As usual for scattering problems, in the elastic scattering limit (valid for $T_K \gg T_\star$), the asymptotic solutions must take the form (e.g., \cite{mott65})
\begin{equation}
F_{s,t}({\bf R}) \sim e^{i{\bf k} \cdot {\bf z}} + \frac{e^{ikR}}{R} \, f_{s,t}(\theta, \phi),
\label{eq:fform}
\end{equation}
where ${\bf k}$ is the relative momentum and $(R,\theta,\phi)$ are the spherical coordinate components of ${\bf R}$.  The angular dependence (and hence the functions $f_{s,t}$) can be found by substituting the form (\ref{eq:Xsimp}) into equation (\ref{eq:hhschrod}) and expanding in Legendre polynomials $P_l(\cos \theta)$.  This reduces the problem to an infinite set of radial equations, indexed by the order $l$ of the associated Legendre polynomial and known as partial wave equations.  It can be shown that the solution has the form
\begin{equation}
f_{s,t}(k,\theta) = \frac{1}{k} \sum_{l=0}^{\infty} (2 l + 1) e^{i \delta_l^{s,t}} \sin \delta_l^{s,t} P_l(\cos \theta).
\label{eq:phase-shift-defn}
\end{equation}
The phase shifts $\delta_l^{s,t}$ quantify the coherence of the scattering amplitudes over the different partial waves.  The particulars of the scattering problem enter the solution only through these phase shifts, and they are ultimately determined by the H$_2$ energy potential curves in the singlet and triplet states.  The total cross section averages over the spins of the particles; the spin exchange cross section, on the other hand, is a coherent sum $\propto |f_t-f_s|^2$.  

The only difference for identical particles is that the wavefunction $X$ must be made antisymmetric with respect to interchange of the two nuclei \cite{smith66}.  The resulting cross sections are \cite{zygelman05}
\begin{equation}
\sigma^{\pm} = \frac{\pi}{4 k^2} \, \sum_{l=0}^{\infty} (2 l + 1) \sin^2(\delta_l^s - \delta_l^t) [ 1 - (-1)^{l+1/2 \pm 1/2} ],
\label{eq:sigpm}
\end{equation}
and $\sigma^0=\sigma^+$.  The factor in square brackets results from nuclear symmetry; it would be unity for distinguishable particles (and hence there would be only one cross section).  The most recent evaluations of $\sigma^\pm$ using state-of-the-art molecular potentials and the close-coupling method can be found in \cite{zygelman05}; they are accurate to $\la 5\%$ at $T_K < 3 \kel$ (where non-adiabatic effects become important), with even smaller errors at higher energies.  Note that $\sigma^- \rightarrow 0$ as $E \rightarrow 0$ because the $S$-wave scattering term ($l=0$) vanishes in equation (\ref{eq:sigpm}); this is a result of nuclear symmetry and does not occur in collisions between distinguishable nuclei.

The cross sections $\sigma^{\pm}$ depend sensitively on $E$ because of resonances with the molecular potential.  But in real-world applications, it is the thermally averaged cross section that is relevant:
\begin{equation}
\bar{\sigma}^\pm \equiv (k_B \, T_K)^{-2} \int \deriv E \, \sigma^\pm(E) \, E \, e^{-E/k_B \, T_K},
\label{eq:thermcs}
\end{equation}
which smooths out the structure in $\sigma^\pm$.  The corresponding rate coefficients are 
\begin{equation}
k^\pm = \sqrt{ \frac{8 k_B \, T_K}{\pi M}} \, \bar{\sigma}^\pm.
\label{eq:ktherm}
\end{equation}
From detailed balance, the excitation rates are
\begin{equation}
k_x^\pm = k^\pm \, \exp(-\omega^\pm),
\label{eq:kx}
\end{equation}
where $\omega^- \equiv \omega = E_{10}/k_B \, T_K$ and $\omega^+ \equiv 2 \omega$.  

With these cross sections in hand, we can compute how the level populations evolve, assuming that they are independent of atomic velocities.  Collecting all the transitions for each state, the rate equations may be written \cite{zygelman05}
\begin{eqnarray}
\dot{n}_d = \dot{n}_b & = & n_a^2 \, k_x^+ + 2 n_a \, n_c \, k_x^- - 2 n_b \, n_d \, (k^+ + k^-) + n_c^2 \, k^+ \nonumber \\
\dot{n}_c & = & n_a^2 \, k_x^+ + 2 k_x^- \, n_a \, (n_b - n_c + n_d) - 2n_b \, n_c \, k^- 
\nonumber \\
& & + 2 n_b \, n_d \, (k^- + k^+) - 3 n_c^2 \, k^+ - 2 n_c \, n_d \, k^- \nonumber \\ 
\dot{n}_a & = & -3 n_a^2 \, k_x^+ - 2 k_x^- \, n_a \, (n_b + n_c + n_d) + 2n_b \, n_c \, k^- 
\nonumber \\
& & + 2 n_b \, n_d \, (k^- + k^+) + n_c^2 \, k^+ + 2 n_c \, n_d \, k^-,
\label{eq:level-rates}
\end{eqnarray}
where we have set $k^0 \approx k^+$ at the temperatures of interest.  In most situations, the level populations will be near thermodynamic equilibrium, so we can linearize equations (\ref{eq:level-rates}) about that point.  Writing $n_1 \equiv n_b + n_c + n_d$ and $n_0 \equiv n_a$, and assuming statistical equilibrium among these sublevels, the linearized form is
\begin{equation}
\dot{n}_1 = n_0 \kappa_{01}^{\rm HH} \nhi - n_1 \kappa_{10}^{\rm HH} \nhi,
\label{eq:rate-coeff}
\end{equation}
where we have assumed $n_a e^{-\omega} \approx n_1/3$ and defined
\begin{equation}
\kappa_{10}^{\rm HH} = (k^+ + k^-)/2 = \kappa_{01}^{\rm HH} \, e^\omega/3
\label{eq:kappaHH}
\end{equation}
to be the effective de-excitation rate.  Note that the reduction to equation~(\ref{eq:rate-coeff}) is not trivial, because equations (\ref{eq:level-rates}) are nonlinear in the level densities.  Some earlier calculations did not linearize properly and so underestimated $\kappa_{10}^{\rm HH}$ \cite{smith66, allison69}.  Fortunately, the linearization is sufficiently accurate that this single effective rate equation (as opposed to the full nonlinear system) is accurate throughout the regime of interest \cite{zygelman05}.  However, when collisions dominate but are still relatively weak, the assumption that $n_1/n_0$ is independent of atomic velocity is not always a good one at the percent level \cite{hirata06}.

%%%%%%%%%%%% TABLE 2-1: H-H collisional coupling coefficients
\begin{table}
\begin{center}
\begin{tabular}{|c|c||c|c|}
\hline
$T_K \ ({\rm K})$ & $\kappa_{10}^{\rm HH} \ ({\rm cm}^{3} \secinv)$ & $T_K \ ({\rm K})$ & $\kappa_{10}^{\rm HH} \ ({\rm cm}^{3} \secinv)$ \\ 
\hline
1 & $1.38 \times 10^{-13}$ & 80 & $1.02 \times 10^{-10}$ \\
2 & $1.43 \times 10^{-13}$ & 90 & $1.11 \times 10^{-10}$ \\
4 & $2.71 \times 10^{-13}$ & 100 & $1.19 \times 10^{-10}$ \\
6 & $6.60 \times 10^{-13}$ & 200& $1.75 \times 10^{-10}$ \\
8 & $1.47 \times 10^{-12}$ & 300 & $2.09 \times 10^{-10}$ \\
10 & $2.88 \times 10^{-12}$ & 500 & $2.56 \times 10^{-10}$ \\
15 & $9.10 \times 10^{-12}$ & 700 & $2.91 \times 10^{-10}$ \\
20 & $1.78 \times 10^{-11}$ & 1000 & $3.31 \times 10^{-10}$ \\
25 & $2.73 \times 10^{-11}$ & 2000 & $4.27 \times 10^{-10}$ \\
30 & $3.67 \times 10^{-11}$ & 3000 & $4.97 \times 10^{-10}$ \\
40 & $5.38 \times 10^{-11}$ & 5000 & $6.03 \times 10^{-10}$ \\
50 & $6.86 \times 10^{-11}$ & 7000 & $6.87 \times 10^{-10}$ \\
60 & $8.14 \times 10^{-11}$ & 10000& $7.87 \times 10^{-10}$ \\
70 & $9.25 \times 10^{-11}$ & & \\
\hline 
\end{tabular} 
\end{center}
\caption{De-excitation cross sections for H-H collisions. Data for $T_K \le 300 \kel$ is from \cite{zygelman05}; higher temperature data is courtesy K. Sigurdson (using the methods of \cite{sigurdson05-deut}).  The cross sections at $T_K \ga 5000 \kel$ are only approximate (see text).\label{tab:hhcoll} }
\end{table}

\begin{figure}[!t]
\centerline{\epsfig{file=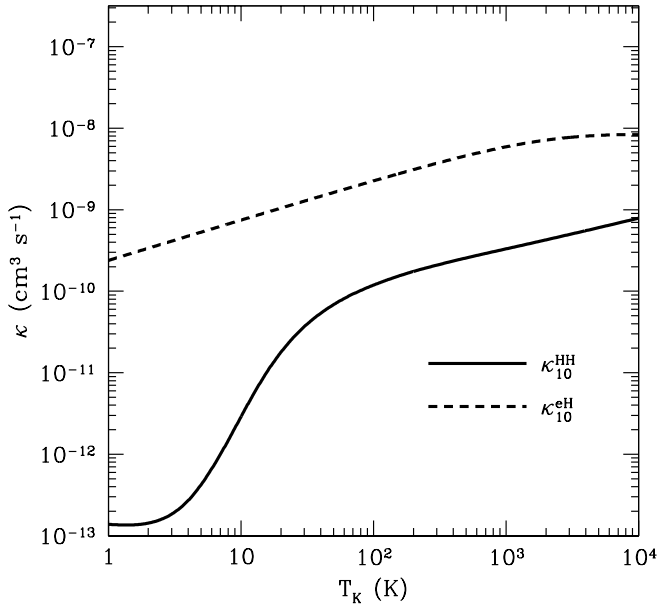,width=3.5in}}
\caption{De-excitation rate coefficients for H-H collisions (solid line) and H-e$^-$ collisions (dashed line).  Note that the net rates are also proportional to the densities, so H-H collisions still dominate in a weakly-ionized medium. }
\label{fig:collrates}
\end{figure}

Table~\ref{tab:hhcoll} lists values for $\kappa_{10}^{\rm HH}$ from \cite{zygelman05} (for $T_K \le 300 \kel$) and from \cite{sigurdson05-deut} (for higher temperatures).  For $T_K \ga 5000 \kel$, the latter uses an extrapolation for the cross section at high $l$, so the values are only approximate in this regime (however, at such high temperatures collisions with hydrogen atoms rarely dominate anyway).  At even higher temperatures, excitations to higher atomic levels begin to dominate and this calculation breaks down.  Figure~\ref{fig:collrates} shows the rate coefficient graphically.  It decreases rapidly at small temperatures but varies rather gently with temperature at $T_K \ga 50 \kel$.  Note that \cite{zygelman05} did \emph{not} assume elastic collisions and so is valid to smaller energies than previous estimates. 

We end our discussion of H--H interactions with a caveat about the use of equation (\ref{eq:rate-coeff}) \cite{zygelman05}.  Its derivation requires statistical equilibrium for the $F=1$ magnetic sublevels.  Unfortunately, collisions \emph{cannot} establish such an equilibrium on their own (for example, $\dot{n}_b - \dot{n}_d=0$ is always obeyed).  We must rely on two other mechanisms to depolarize these populations:  radiative transitions and dipolar spin-spin interactions.  The latter are long-range magnetic forces suppressed by a factor $\sim (v/c)^2$ and are ineffective in the IGM.  Fortunately, absorption of CMB photons \emph{does} efficiently mix the levels.  The weak primordial polarization of the CMB \cite{kovac02, kogut03, readhead04} may leave the atoms slightly polarized, but its effects on the spin temperature are likely to be extremely small.

%%%%%%%%%%%% FIGURE 2-1: Collision rates

\subsubsection{H--$e^-$ Collisions} \label{hecoll}

Collisions between neutral hydrogen atoms and free electrons can also induce spin exchange.  The cross section can be computed by considering the scattering of a polarized beam of electrons by hydrogen \cite{field58}.  This scattering problem can be solved with the same methods as for H-H collisions (eqs.~\ref{eq:fform}--\ref{eq:ktherm}, but with distinguishable particles).  Of course, in this case the scattering problem is a three-body one and so is more difficult to solve numerically.  But, just as before, spin exchange is driven by the differing triplet and singlet scattering amplitudes ($f_t$ and $f_s$, respectively), and the spin exchange cross section is $\propto |f_t-f_s|^2$, which reduces to the form of equation~(\ref{eq:sigpm}) -- except for the absence of the nuclear symmetry factor, of course.

The first cross section calculation \cite{field58} used the phase shifts of \cite{massey51}, while the next \cite{smith66} used the much more accurate phase shifts of \cite{schwartz61} (also see the fit of \cite{liszt01}, which is accurate to $\sim 7\%$).  These early calculations included only the $l=0$ term in equation~(\ref{eq:sigpm}), corresponding to $S$-wave scattering.  Since then, H--e$^{-}$ scattering has received a great deal of attention in the atomic physics literature, and updated spin de-excitation rate coefficients including all the $l \le 3$ partial waves have now been presented \cite{furl06-elec}.  We summarize the results in Table~\ref{tab:ehcoll} and by the dashed curve in Figure~\ref{fig:collrates}.  Note that $\kappa_{10}^{\rm eH} \gg \kappa_{10}^{\rm HH}$ because (at a fixed temperature) free electrons have much larger velocities than hydrogen atoms.  Moreover,  (unlike for H-H) the H-e$^-$ cross section does not cut off at small temperatures.  

%%%%%%%%%%%% TABLE 2-1b: H-e collisional coupling coefficients
\begin{table}
\begin{center}
\begin{tabular}{|c|c||c|c|}
\hline
$T_K \ ({\rm K})$ & $\kappa_{10}^{\rm eH} \ ({\rm cm}^{3} \secinv)$ & $T_K \ ({\rm K})$ & $\kappa_{10}^{\rm eH} \ ({\rm cm}^{3} \secinv)$ \\ 
\hline
1 & $2.39 \times 10^{-10}$ & 1000 & $5.92 \times 10^{-9}$ \\
2 & $3.37 \times 10^{-10}$ & 2000 & $7.15 \times 10^{-9}$ \\
5 & $5.30 \times 10^{-10}$ & 3000 & $7.71 \times 10^{-9}$ \\
10 & $7.46 \times 10^{-10}$ & 5000 & $8.17 \times 10^{-9}$ \\
20 & $1.05 \times 10^{-9}$ &7000 & $8.32 \times 10^{-9}$ \\
50 & $1.63 \times 10^{-9}$ & 10,000 & $8.37 \times 10^{-9}$ \\
100 & $2.26 \times 10^{-9}$ & 15,000 & $8.29 \times 10^{-9}$ \\
200 & $3.11 \times 10^{-9}$ & 20,000& $8.11 \times 10^{-9}$ \\
500 & $4.59 \times 10^{-9}$ &  & \\
\hline 
\end{tabular} 
\end{center}
\caption{De-excitation cross sections for H-e$^-$ collisions, including only collisions below the $n=2$ threshold.  Higher energy collisions become important at $T_K \ga 2 \times 10^4 \kel$.  Note that, at $T_K \ga 6200 \kel$, the \lya background generated by collisions can no longer be neglected.  From \cite{furl06-elec}.\label{tab:ehcoll} }
\end{table}

Nevertheless, H-H collisions typically dominate in the early Universe, because the relic ionized fraction from cosmological recombination is $\bxion \sim 2 \times 10^{-4}$ (see \S \ref{glob-dark}).  Only if the Universe is significantly heated and ionized can collisions with electrons become a strong coupling mechanism (see \S \ref{xray}).  Thus the high temperature behavior of $\kappa_{10}^{\rm eH}$ is relatively important.  The usual calculation (given in Table~\ref{tab:ehcoll}), which neglects scattering at energies above the $n=2$ threshold, breaks down at $T_K \ga 1.5 \times 10^4 \kel$.  Above that point, excitations, ionizations, and high-energy elastic scattering will all affect $\kappa_{10}^{\rm eH}$.  However, in practice $\kappa_{10}^{\rm eH}$ probably becomes irrelevant before that point.  H-e$^{-}$ collisions begin to excite the $2P$ level of hydrogen at $T_K \ga 6200 \kel$ \cite{furl06-elec}.\footnote{In the presence of an X-ray background, which produces a non-thermal population of high-energy photons, this threshold temperature can be much smaller \cite{chen06, chuzhoy06-first, pritchard06}.}  Because each \lya photon generated through this process scatters $\sim 10^5$ times before redshifting out of resonance, while only $\sim 10\%$ of collisions lead directly to de-excitation, the radiation background generated by collisional excitation completely dominates the spin temperature coupling at higher temperatures.  Thus the Wouthuysen-Field effect, to be described in the following section, cannot be ignored once $T_K \ga 6200 \kel$, even in the absence of luminous sources.

\subsubsection{Other Species} \label{othercoll}
 
Neutral hydrogen atoms can also collide with bare protons, deuterium atoms, and helium atoms or ions.  Proton collisions are generally subdominant:  $\kappa_{10}^{p} \approx 3.2 \kappa_{10}^{\rm HH}$ for relatively high temperatures \cite{smith66}, making them somewhat less efficient than the accompanying free electrons.  Neutral helium has a closed shell of electrons in the ground state, so the Pauli exclusion principle prevents electron exchange with it from causing spin change unless the helium atom can be excited to the triplet state (requiring significantly more energy than the cold neutral IGM can provide; see \cite{hirata06} for a detailed dicussion).  Ionized helium avoids this problem and may be significant in partially ionized gas (though the accompanying free electrons will still dominate because of their larger velocities).  To our knowledge, these rates have not yet been calculated.

Finally, we have collisions with trace elements.  Spin exchange cross sections in H-D collisions have been evaluated by \cite{sigurdson05-deut} (see \S \ref{others}).  Although they are much larger than the corresponding H-H cross sections at small temperatures, their rarity means that they still have no significant effect on $T_S$.

\subsection{The Wouthuysen-Field Effect} \label{wf}

\begin{figure}[!t]
\centerline{\epsfig{file=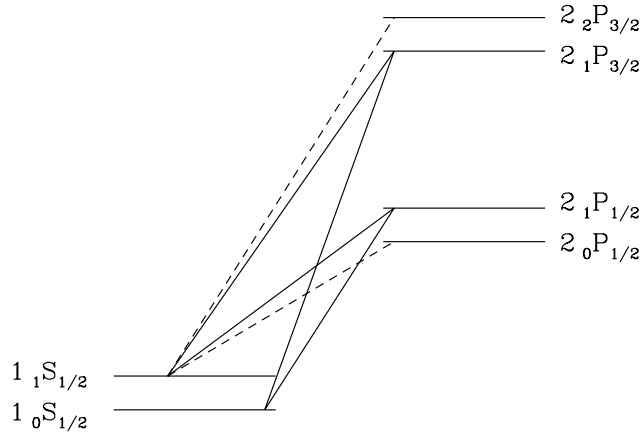,width=3.5in}}
\caption{Level diagram illustrating the Wouthuysen-Field effect.  We show the hyperfine splittings of the $1S$ and $2P$ levels.  The solid lines label transitions that mix the ground state hyperfine levels, while the dashed lines label complementary transitions that do not participate in mixing.  From \cite{pritchard05}. }
\label{fig:wf}
\end{figure}

A less obvious coupling process has become known as the Wouthuysen-Field mechanism\footnote{As a guide to the English-speaking reader, ``Wouthuysen" is pronounced as roughly ``Vowt-how-sen," although in reality the ``uy" construction is a diphthong with no precise counterpart in English.} \cite{wouthuysen52, field58}.  It is illustrated in Figure~\ref{fig:wf}, where we have drawn the hyperfine sublevels of the $1S$ and $2P$ states of HI.  Suppose a hydrogen atom in the hyperfine singlet state absorbs a \lya photon.  The electric dipole selection rules allow $\Delta F=0,1$ except that $F=0 \rightarrow 0$ is prohibited (here $F$ is the total angular momentum of the atom).  Thus the atom will jump to either of the central $2P$ states.  However, these rules allow this state to decay to the $_1S_{1/2}$ triplet level.\footnote{Here we use the notation $_F L_J$, where $L$ and $J$ are the orbital and total angular momentum of the electron.}  Thus atoms can change hyperfine states through the absorption and spontaneous re-emission of a \lya photon (or indeed any Lyman-series photon; see \S \ref{lyn} below). This is analogous to the well-known  ``Raman scattering" process, which often determines the level populations of metastable atomic states, except that in this case the atom undergoes a real (rather than virtual) transition to the $2P$ state.

%%%%%%%%%%%% FIGURE 2-2: Wouthuysen-Field effect

\subsubsection{An Approximate Treatment} \label{wf-approx}

We begin with a relatively simple and intuitive treatment of this process.  Reality is considerably more complicated; we discuss more precise calculations in \S \ref{fokker} below.  The Wouthuysen-Field coupling must depend on the total rate (per atom) at which \lya photons are scattered within the gas,
\begin{equation}
P_\alpha = 4 \pi \chi_\alpha \int \deriv \nu \, J_\nu(\nu) \phi_\alpha(\nu),
\label{eq:palpha}
\end{equation}
where $\sigma_\nu \equiv \chi_\alpha \phi_\alpha(\nu)$ is the local absorption cross section, $\chi_\alpha \equiv (\pi \, e^2/m_e \, c) f_{\alpha}$, $f_\alpha=0.4162$ is the oscillator strength of the \lya transition, $\phi_\alpha(\nu)$ is the \lya absorption profile, and $J_\nu$ is the angle-averaged specific intensity of the background radiation field (by number, not energy).  In the simplest approximation, we simply assume $J_\nu$ to be constant across the line; we will see in \S \ref{fokker} that this is often not a valid assumption.  The line often has a Voigt profile (if only thermal and natural broadening are relevant).  For concreteness, thermal broadening yields a Doppler width
\begin{equation}
\Delta \nu_D = \sqrt{\frac{2 k_B \, T_K}{m_H c^2}} \, \nu_\alpha,
\label{eq:dnud}
\end{equation}
where $\nu_\alpha = 2.47 \times 10^{15} \Hz$ is the central \lya frequency.  

Our goal here is to relate this total scattering rate $P_\alpha$ to the indirect de-excitation rate $P_{10}$ \cite{field58, field59-ts, meiksin00}.  In this section we will make the simplifying assumption that $J_\nu$ is constant across the \lya transition.  We first label the $1S$ and $2P$ hyperfine levels a--f, in order of increasing energy, and let $A_{ij}$ and $B_{ij}$ be the spontaneous emission and absorption coefficients for transitions between these levels.  We write the background flux at the frequency corresponding to the $i \rightarrow j$ transition as  $J_{ij}$.  Then
\begin{equation}
P_{01} \propto B_{\rm ad} J_{\rm ad} \frac{A_{\rm db}}{A_{\rm da} + A_{\rm db}} + B_{\rm ae} J_{\rm ae} \frac{A_{\rm eb}}{A_{\rm ea} + A_{\rm eb}}.\label{eq:psum}
\end{equation}
The first term contains the probability for an a$\rightarrow$d transition ($B_{\rm ad} J_{\rm ad}$), together with the probability for the subsequent decay to terminate in state b; the second term is the same for transitions to and from state e.  Next we need to relate the individual $A_{ij}$ to $A_\alpha = 6.25 \times 10^8 \Hz$, the total \lya spontaneous emission rate (averaged over all the hyperfine sublevels).  This can be accomplished using a sum rule stating that the sum of decay intensities ($g_i A_{ij}$) for transitions from a given $nFJ$ to all the $n' J'$ levels (summed over $F'$) is proportional to $2F+1$ (e.g., \cite{bethe57}); the relative strengths of the permitted transitions are then $(1,\,1,\,2,\,2,\,1,\,5)$, where we have ordered the lines (bc, ad, bd, ae, be, bf) and the two letters represent the initial and final states.  With our assumption that the background radiation field is constant across the individual hyperfine lines, we then find $P_{10} = (4/27) P_\alpha$ \cite{field58, field59-ts} (see \cite{meiksin00} for a detailed derivation).  

The coupling coefficient $x_\alpha$ may then be written
\begin{equation}
x_\alpha = \frac{4 P_\alpha}{27 A_{10}} \, \frac{T_\star}{T_\gamma} = S_\alpha \frac{J_\alpha}{J_\nu^c},
\label{eq:xalpha}
\end{equation}
where in the second equality we evaluate $J_\nu=J_\alpha$ neglecting radiative transfer effects 
and set $J_\nu^c \equiv 1.165 \times 10^{-10} [(1+z)/20] \Junits$.  As we will see below, the background spectrum around \lya is nontrivial, so we must include  a correction factor $S_\alpha$ that accounts for variations near the line center.  This coupling threshold for $x_\alpha = S_\alpha$ can also be written in terms of the number of \lya photons per hydrogen atom, which we denote $\tilde{J}_\nu^c = 0.0767 \, [(1+z)/20]^{-2}$.  As we will see in \S \ref{glob}, it is relatively easy to achieve in practice.

The remainder of this section will show how to calculate $T_c$ and $S_\alpha$.  For the reader who is not interested in the details, equations~(\ref{eq:tcsoln}) and (\ref{eq:salpha-approx}) provide the basic results that can be used to compute $T_S$ (see also \cite{hirata05}). Note, however, that both $T_c$ and $S_\alpha$ implicitly depend on $T_S$, so all three must be solved for simultaneously.

\subsubsection{The Color Temperature} 
\label{tc}

The \lya coupling also depends on the effective temperature $T_c$ of the UV radiation field, defined in equation~(\ref{eq:tcolor}).  This is determined by the shape of the photon spectrum at the \lya resonance.  That the effective temperature of the radiation field \emph{must} matter is easy to see:  the energy defect between the different hyperfine splittings of the \lya transition implies that the mixing process is sensitive to the gradient of the background spectrum near the \lya resonance.  More precisely, the procedure described near equation~(\ref{eq:psum}) lets us write
\begin{equation}
\frac{P_{01}}{P_{10}} = \frac{g_1}{g_0} \, \frac{n_{\rm ad} + n_{\rm ae}}{n_{\rm bd} + n_{\rm be}} \approx 3 \left( 1 + \nu_0 \frac{\deriv \ln n_\nu}{\deriv \nu} \right),
\label{eq:tcrad1}
\end{equation}
where $n_\nu = c^2 \, J_\nu/2 \nu^2$ is the photon occupation number.  Thus by comparison to equation~(\ref{eq:tcolor}), we have \cite{madau97} (neglecting stimulated emission \cite{rybicki06}) 
\begin{equation}
\frac{h}{k_B T_c} = - \frac{\deriv \ln n_\nu}{\deriv \nu}.
\label{eq:tcrad}
\end{equation}
Note that much of the literature uses a slightly different definition of the color temperature (in terms of $J_\nu$) that does not obey detailed balance (e.g., \cite{chen04, hirata05}).  Obviously the color temperature is a function of frequency; in detail, it should be harmonically averaged over the line profile \cite{meiksin06}, but this makes only a tiny difference in practice \cite{chen04}.  Note as well that slightly different definitions of $T_c$ can be useful for certain applications \cite{meiksin06}.

Simple arguments show that $T_c \approx T_K$:  all boil down to the observation that, so long as the medium is extremely optically thick, the enormous number of \lya scatterings must bring the \lya profile to a blackbody of temperature $T_K$ near the line center \cite{wouthuysen52}.  This condition is easily 
fulfilled in the high-redshift IGM, where in our cosmology the mean \lya optical depth experienced by a photon that redshifts across the entire resonance is \cite{gunn65}
\begin{equation}
\tau_{\rm GP} = \frac{\chi_\alpha \, n_{\rm HI}(z) \, c}{H(z) \nu_\alpha} \approx 3 \times 10^5 \, \bxhi \, \left( \frac{1+z}{7} \right)^{3/2}.
\label{eq:gp}
\end{equation}
Field \cite{field59-res} showed that atomic recoil during scattering, which tilts the spectrum to the red, is primarily responsible for establishing this equilibrium.  However, as we will see, spin exchange slightly modifies this expectation.  We will examine the solution in more detail below (see eq.~\ref{eq:tcsoln}).

\subsubsection{The Radiation Field} \label{fokker}

The scattering process is actually much more complicated than naively expected for two basic reasons: (1) scattering itself modifies the shape of $J_\nu$, and (2) the hyperfine sublevels must be taken into account.  Recent treatments have incorporated both these effects; the former is particularly important \cite{chen04, hirata05, chuzhoy06, rybicki06, meiksin06, furl06-lyheat}.  Intuitively, a flat input spectrum develops an absorption feature because of the increased scattering rate near the \lya resonance.  Photons everywhere continually lose energy by redshifting, but they also lose energy through recoil and spin excitation whenever they scatter.  If the drift is denoted ${\mathcal A}$, continuity would require $n_\nu {\mathcal A}$=constant (assuming photons are neither created nor destroyed in the line and neglecting diffusion from re-emission); when ${\mathcal A}$ increases near resonance, the number density must fall.  On average, the energy loss (or gain) per scattering is \cite{rybicki06}
\begin{equation}
\frac{\Delta E_{\rm recoil}}{E} = \frac{h \nu}{m_p c^2} \, \left(1 - \frac{T_K}{T_c} \right),
\label{eq:recoil-loss}
\end{equation}
where the first factor comes from recoil off an isolated atom \cite{madau97} and the second factor corrects for the distribution of initial photon energies \cite{chen04, rybicki06}; note that it vanishes as $T_c \rightarrow T_K$, which also separates the heating and cooling regimes.  The small energy defect between the hyperfine levels provides another (weak) source of energy exchange \cite{hirata05, chuzhoy06}.  The latter process can be incorporated into the scattering in nearly the same way as recoil.

Our main goal in this section is to show how to compute the photon spectrum near \lyans.  We begin with the radiative transfer equation in an expanding universe (written in comoving coordinates, neglecting stimulated emission, and using photon number rather than energy for $J_\nu$) \cite{rybicki94, meiksin06}:
\begin{equation}
\frac{1}{c n_H \chi_\alpha} \, \frac{\partial J_\nu}{\partial t} = -\phi_\alpha(\nu) \, J_\nu + H \nu_\alpha \, \frac{\partial J_\nu}{\partial \nu} + \int \deriv \nu' \, R(\nu,\nu') \, J_{\nu'} + C(t) \psi(\nu).
\label{eq:rt}
\end{equation}
Here the first term on the right-hand side describes absorption, the second the Hubble flow, and the third re-emission following absorption.  $R(\nu,\nu')$ is the ``redistribution function" that describes the frequency of an emitted photon, which depends on the relative velocities and directions of the absorbed and emitted photons as well as the absorbing atom \cite{hummer62}.  The last term describes injection of new photons:  $C$ is the rate at which they are produced and $\psi(\nu)$ is their frequency distribution.  

The redistribution function $R$ is the complicated aspect of the problem, and general solutions to equation (\ref{eq:rt}) are not available.  Fortunately, it can be simplified if the frequency change per scattering (typically of order $\Delta \nu_D$ \cite{hummer62}) is ``small" (see below).  In that case, we can expand $J_{\nu'}$ to second order in $(\nu-\nu')$ and rewrite equation~(\ref{eq:rt}) as a diffusion problem in frequency.  The steady-state version of equation (\ref{eq:rt}) becomes, in this Fokker-Planck approximation \cite{hirata05, rybicki06},\footnote{Note that we neglect stimulated emission here, which adds a term $\propto J^2$ inside the brackets \cite{rybicki06}.}
\begin{equation}
\frac{\deriv}{\deriv x} \left( - {\mathcal A} \, J + {\mathcal D} \, \frac{\deriv J}{\deriv x} \right) + C \psi(x) = 0,
\label{eq:fokker}
\end{equation}
where $x \equiv (\nu-\nu_\alpha)/\Delta \nu_D$, ${\mathcal A}$ is the frequency drift, and ${\mathcal D}$ is the diffusivity.  The problem is not, unfortunately, completely specified, because ${\mathcal D}$ is constructed from an approximation to $R$ and so depends on the physical processes incorporated.  The resulting freedom can be dealt with in a number of ways \cite{basko81, deguchi85, rybicki94, rybicki06}.  One important caveat (in this particular application) is that the approximation, just like the original problem, should obey detailed balance \cite{deguchi85}, which ensures microscopic reversibility.  This was not satisfied by the diffusivity chosen in \cite{rybicki94}, which has been used in most recent work \cite{madau97, chen04, hirata05, chuzhoy06}.  A ``corrected" form was recently presented by \cite{rybicki06}.    An alternative approach is to expand the scattering probability itself, rather than $J_\nu$ or $n_\nu$ \cite{meiksin06}.  This appears to give somewhat less accurate results but has some formal advantages.  In general, these Fokker-Planck approximations are valid when (i) the frequency change per scattering ($\sim \Delta \nu_D$) is much smaller than the width of any spectral features, and either (iia)  we are outside the line core, where $\deriv \phi_\alpha/\deriv x$ is small, or (iib) we are in equilibrium with $T_c \approx T_K$.  Fortunately, Monte Carlo simulations have verified the utility of equation~(\ref{eq:fokker}) even when these conditions are marginally violated \cite{hirata05}.

Solving for the background spectrum thus reduces to specifying ${\mathcal A}$ and ${\mathcal D}$ for each of several processes.  The first is the Hubble flow, which causes a drift ${\mathcal A}_H= - \gamma$ (without any associated diffusion), where $\gamma$ is the Sobolev parameter that parameterizes the bulk velocity of the medium.  In our case, neglecting peculiar velocities, $\gamma=\tau_{\rm GP}^{-1}$.  The remaining terms come from expanding $R$, which must incorporate all the physical processes relevant to energy exchange in scattering.  The commonly used $R_{\rm II}$, which describes frequency redistribution if the scattering is coherent in the rest frame of the atom \cite{henyey41, hummer62}, will \emph{not} suffice, because we must include recoil and spin exchange as well.  The drift from recoil is\cite{basko81, rybicki94, hirata05, chuzhoy06, rybicki06}
\begin{eqnarray}
{\mathcal D}_{\rm scatt} & = & \phi_\alpha(x)/2,
\label{eq:Dkin} \\
{\mathcal A}_{\rm scatt} & = &  -(\eta - x_0^{-1} ) \phi_\alpha(x),
\label{eq:Akin}
\end{eqnarray}
where $x_0 \equiv \nu_\alpha/\Delta \nu_D$ and $\eta \equiv (h \nu_\alpha^2)/(m_p c^2 \Delta \nu_D)$.  The latter is the recoil parameter measuring the average loss per scattering in units of the Doppler width.  The second term in equation~(\ref{eq:Akin}) follows from detailed balance and ensures the proper approach to thermal equilibrium \cite{rybicki06}.  When $T_K$ is small, we must also include energy loss during spin exchange \cite{hirata05, chuzhoy06}, which has ${\mathcal D}_{\rm se} \approx (2 \nu_0^2/9 \Delta \nu_D^2) \, {\mathcal D}_{\rm scatt}$ because the frequency change from spin exchange is $\pm \nu_0$ and only a fraction $2/9$ of absorptions actually lead to spin exchange.  When $T_K$ is extremely small, each hyperfine transition should be treated separately, with distinct \lya line profiles $\phi_{10}(\nu)$ and $\phi_{01}(\nu)$ for indirect de-excitations and excitations of the ground state hyperfine levels.

To solve equation~(\ref{eq:fokker}) we must specify the boundary conditions, which essentially correspond to the input photon spectrum (ignoring scattering) and the source function.  Because the frequency range of interest is so narrow, two cases suffice (e.g., \cite{chen04, chuzhoy06}).  These are a flat input spectrum (which describes photons that redshift through the \lya resonance) and a step function, where photons are ``injected" at line center (through cascades or recombinations; see \S \ref{lyn} below) and redshift away.  In either case, the first integral over $x$ is trivial and we can write \cite{hummer92}
\begin{equation}
\phi \frac{\deriv J}{\deriv x} + 2 (\eta' \phi + \gamma') J = 2 \gamma' K
\label{eq:fokk-simp}
\end{equation}
where \cite{chuzhoy06}
\begin{eqnarray}
\gamma' & = & \gamma \, (1 + T_{\rm se}/T_K)^{-1},
\label{eq:gammaprime} \\
\eta' & = & \eta \left( \frac{1 + T_{\rm se}/T_S}{1 + T_{\rm se}/T_K} \right) - (x + x_0)^{-1},
\label{eq:etaprime}
\end{eqnarray}
where $T_{\rm se} = (2/9) T_K \nu_{21}^2/\Delta \nu_D^2 = 0.40 \kel$.  The terms involving $T_{\rm se}$ are added by spin exchange; note that they become unimportant when the temperature is large and are often neglected. The integration constant $K$ equals $J_\infty$, the flux far from resonance, for photons that redshift into the line and for injected photons at $x<0$; it is zero for injected photons at $x>0$.  This form is \emph{not} accurate when $T_K \sim T_\star$, in which case a numerical solution must be used \cite{hirata05}.

When $\gamma$ is small (or in the limit of large optical depth, an excellent approximation for our purposes) it follows from equations~(\ref{eq:tcrad}) and (\ref{eq:fokk-simp}) that 
\begin{equation}
T_c = T_K \left( \frac{1+T_{\rm se}/T_K}{1+T_{\rm se}/T_S} \right),
\label{eq:tcsoln}
\end{equation}
agreeing with the qualitative arguments of \S \ref{tc} when spin exchange is neglected (in the limit $T_S,\, T_K \gg T_{\rm se}$).  The spin temperature affects the rate at which photons lose energy during scattering (because it determines the level populations and hence the probabilities of excitation and de-excitation) and so the background spectrum near the \lya resonance, which in turn affects the spin temperature itself.  Thus $T_S$ must actually be determined iteratively, beginning with a guess for $T_S$, calculating $T_c$ and the effective Wouthuysen-Field coupling strength, computing $T_S$, and repeating.  Fortunately the process converges rapidly \cite{hirata05}.

Several solutions to equation~(\ref{eq:fokker}) have been presented in the literature.  We begin with the formal analytic solution \cite{hummer92, furl06-lyheat}.  When $K \neq 0$, it is most compactly written in terms of $\delta_J \equiv (J_\infty - J)/J_\infty$:
\begin{equation}
\delta_J(x) = 2 \eta \int_0^\infty \deriv y \exp \left[ - 2 \eta' y - 2 \gamma' \int_{x-y}^{x} \frac{\deriv x'}{\phi_\alpha(x')} \right].
\label{eq:dj-soln}
\end{equation}
An analogous form to equation~(\ref{eq:dj-soln}) exists for photons injected at line center.  To include the full Voigt profile, equation~(\ref{eq:dj-soln}) must be solved numerically \cite{chen04, hirata05}.  But these equations can be further simplified by approximating the full line profile with the Lorentzian wings from natural broadening \cite{grachev89, chuzhoy06}; in that case the inner integral is trivial.  This assumption is quite accurate when $T_K \la 1000 \kel$; at higher temperatures, collisional broadening affects the equilibrium spectrum at the several percent level \cite{furl06-lyheat}.  

%%%%%%%%%%%% FIGURE 2-3: Lyman alpha radiation background
\begin{figure}[!t]
\centerline{\epsfig{file=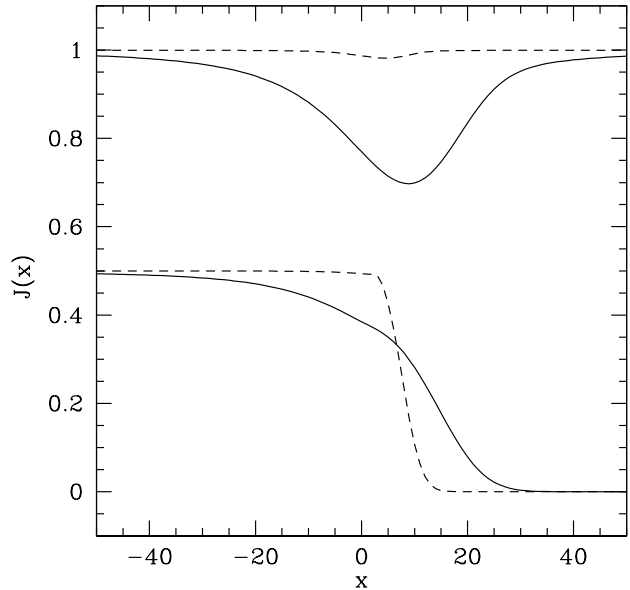,width=3.5in}}
\caption{Background radiation field near the \lya resonance at $z=10$, assuming a Voigt line profile.  The upper and lower sets are for continuous photons and photons injected at line center, respectively.  (The former are normalized to $J_\infty$; the latter have arbitrary normalization.)  The solid and dashed curves take $T_K=10$ and $1000 \kel$, respectively.  From \cite{furl06-lyheat}. }
\label{fig:lyshape}
\end{figure}

The crucial point of equation~(\ref{eq:dj-soln}) is that (as expected from the qualitative argument above) an absorption feature appears near the line center; its strength is roughly proportional to $\eta'$, our recoil parameter.  The feature is more significant when $T_K$ is small (or the average effect of recoil is large).  Figure~\ref{fig:lyshape} shows some example spectra (both for a continuous background and for photons injected at line center).  Note that the feature is asymmetric; this will be important in \S \ref{lyaheat}.  

For now, the most important result is the suppression of the radiation spectrum at line center compared to the (flat) solution without radiative transfer.  This decreases the total scattering rate of \lya photons (and hence the Wouthuysen-Field coupling) below what one naively expects.  The suppression factor is \cite{chen04}
\begin{equation}
S_\alpha = \int_{-\infty}^\infty \deriv x \, \phi_\alpha(x)\, J(x) \approx [1 - \delta_J(0)] \le 1,
\label{eq:salpha-defn}
\end{equation}
where the second equality follows from the narrowness of the line profile.  For a full Voigt profile, the integral must be performed numerically \cite{chen04, hirata05}.  But the wing approximation turns out to be an excellent one; when spin exchange can be neglected, the suppression is \cite{chuzhoy06, furl06-lyheat}
\begin{equation}
S_\alpha \approx \exp \left[ - 1.12 \eta' \left( \frac{3 a}{2 \pi \gamma'} \right)^{1/3} \right] \sim \exp \left[ -0.803 T_K^{-2/3} \left( \frac{10^{-6}}{\gamma} \right)^{1/3} \right],
\label{eq:salpha-approx}
\end{equation}
where $a=A_\alpha/(4\pi\Delta \nu_D)$ is the Voigt parameter and in the second equality we have neglected spin exchange and expressed $T_K$ in degrees Kelvin.  Note that this form applies to both photons injected at line center and those that redshift in from infinity.  As we can see in Figure~\ref{fig:lyshape}, the suppression is most significant in cool gas.  This is convenient because the Lorentzian wing approximation is least accurate at large temperatures; fortunately in that regime $S_\alpha \approx 1$ anyway, so equation~(\ref{eq:salpha-approx}) is accurate to a few percent at all $T_K \ga 1 \kel$ \cite{furl06-lyheat}.  At smaller temperatures, a numerical solution including the Voigt profile, the full integral over $\phi_\alpha$, and spin exchange, is necessary \cite{hirata05}.

\subsection{The Lyman Series} \label{lyn}

The Wouthuysen-Field mechanism describes how \lya photon scattering affects $T_S$. Of course, the neutral IGM has so much hydrogen that any photon redshifting into a Lyman series resonance will be absorbed almost immediately.  In this section, we will compute the coupling induced by photons that enter $n>1$ Lyman lines\cite{hirata05,pritchard05}.  Note that earlier work often ignored the detailed atomic physics of these transitions and incorrectly assumed their coupling to be identical to \lya photons. 

Suppose that a photon redshifts into the \lyn resonance.  After absorption, it can either scatter (through a decay directly to the ground state) or can vanish if the atom instead decays to an intermediate level.  The
scattering probability $P_{nP\rightarrow1S}$ follows directly from the Einstein coefficients,
\begin{equation}
P_{if} = \frac{A_{if}}{\sum_{f'} A_{if'}},
\label{eq:pdec}
\end{equation}
where $i$ and $f$ denote the initial and final states, respectively, and the sum is over all allowed final states.  Table~\ref{tab:recycle} shows the direct decay probabilities.  Typically, a \lyn photon will scatter $N_{\rm scatt} \approx (1-P_{nP\rightarrow1S})^{-1} \sim 5$ times before being consumed by a decay cascade.  As a result, coupling from the direct scattering of \lyn photons is unimportant relative to \lya photons \cite{pritchard05}, which vanish only when they redshift across the line.  Thus the coupling from \lyn scattering is suppressed by a factor $\sim N_{\rm scatt}/\tau_{\rm GP} \la 10^{-6}$.  

%%%%%%%%%%%% TABLE 2-2: Recycling fractions
\begin{table}
\begin{center}
\begin{tabular}{|c|c|c||c|c|c|}
\hline
$n$ & $f_{\rm rec}(n)$ & $P_{nP\rightarrow1S}$ & $n$ &
$f_{\rm rec}(n)$ & $P_{nP\rightarrow1S}$\\ 
\hline
 &  & & 16 & 0.3550 & 0.7761\\
2 & 1 & 1 & 17 & 0.3556 & 0.7754\\
3 & 0 & 0.8817 & 18 & 0.3561 & 0.7748\\
4 & 0.2609 & 0.8390 & 19 & 0.3565 & 0.7743\\
5 & 0.3078 & 0.8178 & 20 & 0.3569 & 0.7738\\
6 & 0.3259 & 0.8053 & 21 & 0.3572 & 0.7734\\
7 & 0.3353 & 0.7972 & 22 & 0.3575 & 0.7731\\
8 & 0.3410 & 0.7917 & 23 & 0.3578 & 0.7728\\
9 & 0.3448 & 0.7877 & 24 & 0.3580 & 0.7725\\
10 & 0.3476 & 0.7847 & 25 & 0.3582 & 0.7722\\
11 & 0.3496 & 0.7824 & 26 & 0.3584 & 0.7720\\
12 & 0.3512 & 0.7806 & 27 & 0.3586 & 0.7718\\
13 & 0.3524 & 0.7791 & 28 & 0.3587 & 0.7716\\
14 & 0.3535 & 0.7780 & 29 & 0.3589 & 0.7715\\
15 & 0.3543 & 0.7770 & 30 & 0.3590 & 0.7713\\
\hline 
\end{tabular} 
\end{center}
\caption{Recycling fractions $f_{\rm rec}(n)$ and direct decay
  probabilities to the ground state $P_{nP\rightarrow1S}$ for the
  first 29 Lyman-series transitions. From
  \cite{hirata05,pritchard05}. \label{tab:recycle} }
\end{table}

\begin{figure}[!t]
\centerline{\epsfig{file=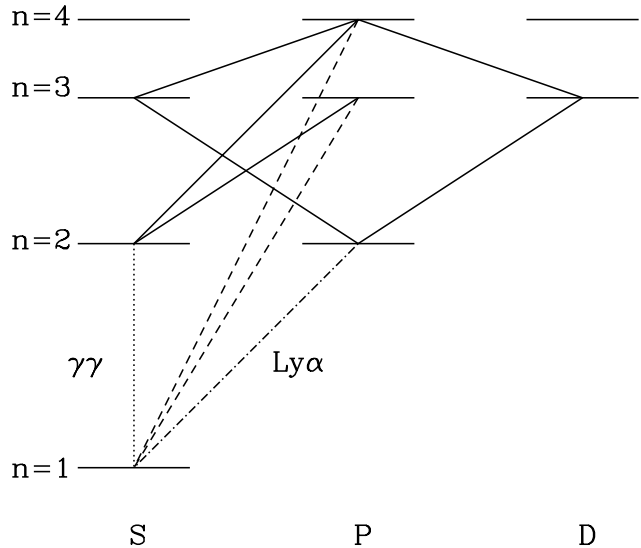,width=3.5in}}
\caption{ Decay chains for Ly$\beta$ and Ly$\gamma$.  We show
  \lyn transitions by dashed curves, \lya by the dot-dashed curve,
  cascades by solid curves, and the forbidden $2S \rightarrow 1S$
  transition by the dotted curve. From \cite{pritchard05}.}
\label{fig:lygamma}
\end{figure}

However, \lyn photons can still be important because of their cascade products.  Consider the decay chains shown in Figure~\ref{fig:lygamma}.  For Ly$\beta$, the only permitted decays are to the ground state (regenerating a Ly$\beta$ photon and starting the process again) or to the $2S$ level.  The H$\alpha$ photon produced in the $3P \rightarrow 2S$ transition (and indeed any photon produced in a decay to an excited state) escapes to infinity.  Thus the atom will eventually find itself in the $2S$ state, which decays to the ground state via a forbidden two photon process with $A_{2S\rightarrow1S}=8.2 \secinv$ (other processes, such as collisional excitation, are much slower \cite{hirata05}).  These photons too will escape to infinity.  Thus coupling from Ly$\beta$ photons can be completely neglected.

%%%%%%%%%%%% FIGURE 2-4: Decay diagram

But now consider excitation by Ly$\gamma$, also shown in Figure~\ref{fig:lygamma}.  This too can decay directly to the ground state or to $2S$.  But it can also cascade (through $3S$ or $3D$) to the $2P$ level, in which case the original \lyn photon is ``recycled'' into a \lya photon.  This photon contributes to the background responsible for the Wouthuysen-Field effect just as any other \lya photon would (it is an injected photon in the terminology of \cite{chen04}).  Thus the key quantity for determining the coupling induced by \lyn photons is the fraction $f_{\rm rec}(n)$ of cascades that terminate in \lya photons.  This can be computed iteratively from
\cite{hirata05,pritchard05} 
\begin{equation}
f_{\rm rec}(i) = \sum_f P_{if} f_{\rm rec}(f),
\label{eq:frec}
\end{equation}
where the sum is again over all allowed final states.  (Note that the sum should exclude transitions directly to the ground state, because such decays regenerate the \lyn photon and start the whole process again.)  Table~\ref{tab:recycle} lists $f_{\rm rec}(n)$ for $n \le 30$.  The recycling fractions rapidly approach $f_{\rm rec} \approx 0.36$ at large $n$.  Thus the Wouthuysen-Field coupling from \lyn photons is about one-third as efficient as that from \lya photons, with the important exception that $f_{\rm rec}(3)=0$.  We show how to compute the total coupling from all Lyman-series photons in \S \ref{tshist}.

\subsection{Polarization} \label{pol}

There are two mechanisms that could induce polarization in the 21 cm line.  The first is anisotropic excitation of the triplet state.  For example, if a single source of \lya photons dominated the Wouthuysen-Field coupling, the resulting level populations would be polarized.  However, because $\tau_{\rm GP} \ga 10^5$, the coupling is actually produced through repeated scatterings of each \lya photon.  This efficiently isotropizes the \lya radiation field, and this kind of polarization is unlikely to be important in practice \cite{babich05}.

Magnetic fields can also induce polarization.  The Zeeman effect splits the 21 cm transition into three parts \cite{nafe48,bolton57}: an unpolarized central component with frequency $\nu_{0}$ and two components with frequencies $\nu_{0} \pm \Delta \nu_Z$, where
\begin{equation}
\Delta \nu_Z = \frac{ e B_\parallel}{4 \pi m_e c} = 14 \left(
\frac{B_\parallel}{10 \, \mu{\rm G}} \right) {\rm Hz}
\label{eq:zeeman}
\end{equation}
and $B_\parallel$ is the magnetic field strength along the line of sight.  Zeeman splitting leaves the two states with momenta $\pm \hbar$, so conservation of angular momentum demands that the higher and lower frequency states be right- and left-circularly polarized, respectively.  This frequency difference is, of course, much too small to be observed directly, but a net signal will be present if the difference in the polarized intensities exceeds the detector noise.  Because the two components are proportional to $\dtb(\nu_0 + \Delta \nu)$ (for right circular) and $\dtb(\nu_0 - \Delta \nu)$ (for left circular), this can happen if there is a strong gradient in $\dtb$:  then observations corresponding to $\nu_0$ receive a strong signal in one polarization and negligible in the other.  Let $\delta T_{\rm diff}(\nu)= | \delta T_L(\nu) - \delta T_R(\nu)|$ be the difference in brightness temperature between the left- and right-handed components.  Then 
\begin{equation}
\delta T_{\rm diff}(\nu) = 28 \left( \frac{B_\parallel}{10 \, \mu{\rm G}}
\right) \left( \frac{\deriv \dtb/\deriv \nu}{{\rm mK \ Hz}^{-1}} \right),
\label{eq:dtpol}
\end{equation}
where $\deriv \dtb/\deriv \nu$ is the derivative of the brightness temperature (in this patch of the sky) with observed frequency.  This technique has been used to constrain the Milky Way magnetic field \cite{heiles04}.

Intrinsically polarized 21 cm emission or absorption is one of the few possibilities for measuring magnetic fields in the high-redshift IGM (another is Faraday rotation of radio-loud quasars).  It is an interesting prospect because the origin of intergalactic magnetic fields is controversial and essentially unconstrained \cite{gaensler04}.  There are two kinds of possible measurements.  First, targeted observations of regions with strong gradients in the underlying $\dtb$ can constrain their local magnetic fields \cite{cooray05-pol}.  For example, quasar \htwo regions typically have edges of width $\sim 1.5 \bxhi^{-1} (1+z)^{-3}$ proper Mpc, corresponding to $\Delta \nu \sim 2 \kHz$; the softer spectra of stars yield even narrower edges.  Any coherent magnetic field across these edges will yield a residual polarization.  SKA-class instruments could (in the absence of systematics) detect magnetic fields of order $\sim 100 \, \mu{\rm G}$ in these systems \cite{cooray05-pol}.

A second possibility is a statistical search for large-scale fluctuations in the polarized emission or absorption.  In this case the constraints are strongest during epochs in which the globally-averaged brightness temperature changes rapidly (see \S \ref{glob} for a discussion of the requisite conditions).  However, realistically the gradients will be modest, so the limits from SKA-class telescopes will lie well above theoretical expectations \cite{cooray05-pol}.  Moreover, such statistical techniques will likely be compromised by the much larger polarization from Thomson scattering in the post-reionization Universe (see \S \ref{sec}).

\subsection{Other Hyperfine Transitions} \label{others}

Of course, the \hone 21 cm line is only one of many hyperfine transitions.  Two others may be of particular interest for studying the IGM: the $\lambda=91.6$ cm transition of deuterium (only recently measured in the Milky Way ISM with high statistical signifiance \cite{rogers05}) and the $\lambda=3.46$ cm transition of $^3$He$^+$ (note that $^4$He has a zero nuclear spin, and neutral $^3$He has a closed shell of electrons; thus neither has any hyperfine structure).  Of course, both of these transitions are exceedingly weak because [D/H] $\approx 3 \times 10^{-5}$ and [$^3$He/H] $\approx 10^{-5}$ according to big bang nucleosynthesis \cite{kneller04}.  The brightness temperatures corresponding to equation (\ref{eq:dtb}) are then
\begin{eqnarray}
\delta T_{b,{\rm D}} & \approx & 0.079 \ \xhi \, (1+\delta) \, \left( \frac{{\rm [D/H]}}{3 \times 10^{-5}} \right) \, \left( 1 - \frac{T_\gamma}{T_{S,{\rm D}}} \right) (1+z)^{1/2} \microkel, \label{eq:dtb-D} \\
\delta T_{b,{\rm ^3He}} & \approx & 0.53 \ x_{\rm HeII} \, (1+\delta) \, \left( \frac{{\rm [^3He/H]}}{10^{-5}} \right) \, \left( 1 - \frac{T_\gamma}{T_{S,{\rm He}}} \right) (1+z)^{1/2} \microkel, 
\label{eq:dtb-He}
\end{eqnarray}
where of course $T_S$ must be defined separately for D and $^3$He and need not equal the \hone value.  For example, the de-excitation rate in D-H collisions is many orders of magnitude larger than the corresponding rate for H-H collisions at $T_K \la 10 \kel$ \cite{sigurdson05-deut}, so D has a much easier time maintaining $T_{S,{\rm D}} \ne T_\gamma$ than H; note that earlier treatments incorrectly ignored this effect \cite{field58}.  Lyman-series coupling also differs between the three species \cite{deguchi85} and can lead to some surprising effects \cite{chuzhoy06}. These brightness temperatures are so small that detecting IGM fluctuations through either of these isotopes will require a truly heroic effort.

However, detecting deuterium before reionization is not completely hopeless, because density fluctuations in $\delta T_{b,{\rm D}}$ \emph{precisely} trace those of hydrogen.  Thus a 21 cm map could be used as a ``template," and the correlation coefficient between H and D maps could be used to measure their abundance ratio \cite{sigurdson05-deut}.  By averaging over the a large fraction of the sky and a large redshift interval (and with a significantly larger instrument than any currently planned), we could, in principle, detect the deuterium signal at the expected level, although systematics may in the end prevent us from reaching the necessary precision.  On the other hand, this is the only known method to measure the truly primordial [D/H] ratio and would provide an important confirmation of big bang nucleosynthesis \cite{wagoner67, kneller04}.

$^3$He$^+$ is also potentially interesting.  HeI was likely reionized with HI, but observations indicate that HeII is not reionized until $z \sim 3$ (e.g., \cite{heap00}).  Thus it probably emits over a substantial frequency range, $\nu \sim 0.8$--$2.2 \GHz$.  In principle, the $^3$He$^+$ transition could be detected through its anti-correlation with \hone during hydrogen reionization; unfortunately, this carries much less information than the D-H correlation, because only the ionization pattern (and not small-scale density fluctuations) contribute.  On the other hand, the sky is significantly quieter at these higher frequencies, so (depending on the details of reionization) it may end up being an easier experiment \cite{sigurdson05-deut}.  In principle, it would also be possible to study HeII reionization with this transition.  Unfortunately, simple sensitivity estimates following \cite{zald04} show that such attempts are well beyond any planned telescope.  

%\bibliographystyle{elsart-num}
%\bibliography{Ref_21cm}

%\end{document}

%% file: history-ch3.tex
%\documentclass{elsart}
%\usepackage{amssymb,cite,epsfig}

%\input{../../submission/defns.tex}

%\begin{document}

\section{The Global Evolution of the IGM} \label{glob}

Now that we have reviewed the underlying physics of the 21 cm transition, we will consider how global properties of the IGM  evolve through time.  These may or may not be directly observable (see \S \ref{glob-obs}), but in any case they constitute the background against which the fluctuating signals discussed in the next five sections occur.

\subsection{The Dark Ages} \label{glob-dark}

We begin by examining the cosmic dark ages, when the physics is rather simple.  The first step is to compute how $T_K$ evolves with time.  Energy conservation in the expanding IGM demands
\begin{equation}
\frac{\deriv T_K}{\deriv t} = -2 H(z) T_K + \frac{2}{3} \sum_i
\frac{\epsilon_i}{k_B n},
\label{eq:tkevol}
\end{equation}
where the first term on the right hand side is the $p\deriv V$ work from expansion and $\epsilon_i$ is the energy injected into the gas per second per unit (physical) volume through process $i$.  Before the first nonlinear objects appear, the only relevant heating mechanism is Compton scattering between CMB photons and residual free electrons in the IGM.  The heating rate from this process can be calculated from the drag force exerted by the isotropic CMB radiation field on a thermal distribution of free electrons.  It is \cite{peebles93, seager99}
\begin{equation}
\frac{2}{3} \, \frac{\epsilon_{\rm comp}}{k_B n} = \frac{\bxion}{1 + f_{\rm He} + 
  \bxion} \, \frac{(T_\gamma - T_K)}{t_\gamma},
\label{eq:tcomp}
\end{equation}
where $t_{\gamma} \equiv (3 m_e c)/(8 \sigma_T u_\gamma)$ is the Compton cooling time, $u_\gamma \propto T_\gamma^4$ is the energy density of the CMB, $f_{\rm He}$ is the helium fraction (by number), and $\sigma_T$ is the Thomson cross section.  The first factor on the right hand side accounts for the distribution of the energy over all free particles.  Compton heating drives $T_K \rightarrow T_\gamma$ when $u_\gamma$ and $\bxion$ are large; thus at sufficiently high redshifts the gas can cool no faster than the CMB, $T_K \propto (1+z)$.

Eventually, however, the gas does thermally decouple from the CMB.  The decoupling redshift can be computed precisely from the publicly available RECFAST code \cite{seager99,seager00}, but a simple estimate provides the main result.  The recombination rate is $\dot{n}_e = - \alpha_B \bxion^2 n_b^2$, where $\alpha_B \propto T_K^{-0.7}$ is the case-B recombination coefficient (see \S \ref{clump} below).  The fractional change in $\bxion$ per Hubble time is therefore
\begin{equation}
\frac{1}{H(z) n_e} \, \frac{\deriv n_e}{\deriv t} \approx 100 \bxion
(1+z)^{0.8} \, \frac{\Omega_b h^2}{\sqrt{\Omega_0 h^2}},
\label{eq:neevol}
\end{equation}
where we have assumed that $T_K \propto (1+z)$ (i.e., coupling to the CMB is still strong).  Freeze-out occurs when this is of order unity; at later times $\bxion$ remains roughly constant because the recombination time then exceeds the expansion timescale.  Numerical calculations with RECFAST yield $\bxion = 3.1 \times 10^{-4}$ at $z=200$ (which is past freeze-out).  Inserting this into equation (\ref{eq:tcomp}), we find
\begin{equation}
\frac{1}{H(z) T_K} \, \frac{\deriv T_K}{\deriv t} \sim
\frac{10^{-7}}{\Omega_b h^2} \, (T_\gamma/T_K-1) \, (1+z)^{5/2}. 
\label{eq:thermdec}
\end{equation}
Thus thermal decoupling occurs when 
\begin{equation}
1 + z_{\rm dec} \approx 150 \, (\Omega_b h^2/0.023)^{2/5}.
\label{eq:zdec}
\end{equation}

\begin{figure}[!t]
\centerline{\epsfig{file=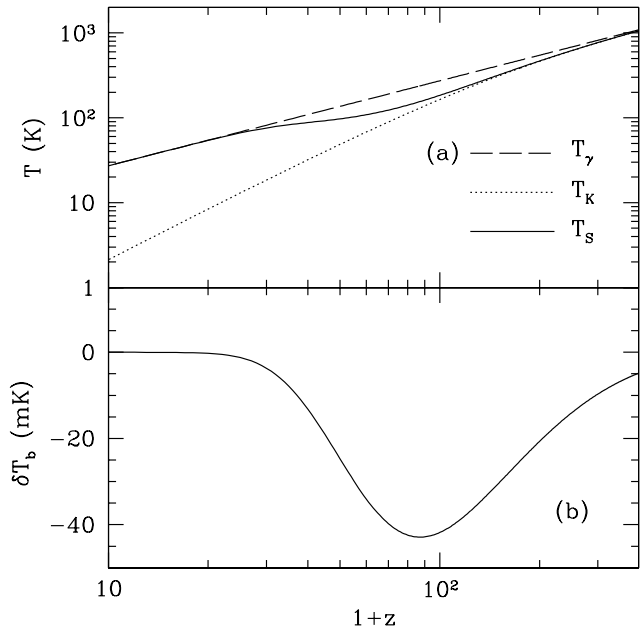,width=3.5in}}
\caption{\emph{(a)}:  IGM temperature evolution if only adiabatic cooling and Compton heating are involved.  The spin temperature $T_S$ includes only collisional coupling.  \emph{(b)}: Differential brightness temperature against the CMB for $T_S$ shown in panel \emph{a}. }
\label{fig:tevol}
\end{figure}

Figure~\ref{fig:tevol}\emph{a} shows a more detailed calculation (from RECFAST).  The dashed curve shows $T_\gamma$ and the dotted curve $T_K$.  We see that Compton heating begins to become inefficient at $z \sim 300$ and is negligible by $z \sim 150$. Past this point, $T_K \propto (1+z)^2$, as expected for an adiabatically expanding non-relativistic gas.

%%%%%%%%%%%%FIGURE 3-1: Thermal history, without sources

We must next determine the spin temperature.  Obviously $x_\alpha=0$ during this epoch (barring any exotic processes \cite{furl06-dm}).  But at sufficiently high redshifts, the Universe was dense enough for collisions with neutral atoms to be efficient in the mean density IGM.  The effectiveness of collisional coupling can be computed exactly for any given temperature history from the rate coefficients presented in \S \ref{coll}.  A convenient estimate of their importance is the critical overdensity, $\delta_{\rm coll}$, at which $x_c=1$:
\begin{equation}
1 + \delta_{\rm coll} = 1.06 \, \left[ \frac{\kappa_{10}(88
    \kel)}{\kappa_{10}(T_K)} \right] \, \left( \frac{0.023}{\Omega_b
    h^2} \right) \, \left( \frac{70}{1+z} \right)^2,
\label{eq:dcoll}
\end{equation}
where we have inserted the expected temperature at $1+z=70$.  Thus for redshifts $z \lesssim 70$, $T_S \rightarrow T_\gamma$; by $z \sim 30$ the IGM essentially becomes invisible.  It is worth emphasizing that $\kappa_{10}$ is extremely sensitive to $T_K$ in this regime (see Fig.~\ref{fig:collrates}).  If the universe is somehow heated above the fiducial value, the threshold density can remain modest:  $\delta_{\rm coll} \approx 1$ at $z=40$ if $T_K=300 \kel$.  The solid line in Figure~\ref{fig:tevol}\emph{a} shows the spin temperature $T_S$ during the dark ages, and Figure~\ref{fig:tevol}\emph{b} shows the corresponding brightness temperature.  The signal peaks (in absorption) at $z \sim 80$, where $T_K$ is small but collisional coupling still efficient.  Because of the simple physics involved in Figure~\ref{fig:tevol}, the 21 cm line offers a sensitive probe of the dark ages \cite{loeb04}, at least in principle.

\subsection{The Thermal History of the IGM} \label{therm}

Figure~\ref{fig:tevol}\emph{b} also shows that the $z \la 30$ Universe would remain invisible without luminous sources.  Our next task is to consider their effects on the 21 cm background.  We will begin by examining several processes that affect the temperature evolution.

\subsubsection{X-ray Heating} \label{xrayheat}

Because they have relatively long mean free paths, X-rays from galaxies and quasars are likely to be the most important heating agent for the low-density IGM.  In particular, photons with $E>1.5 \bxhi^{1/3} [(1+z)/10]^{1/2} \keV$ have mean free paths exceeding the Hubble length \cite{oh01}.  Given our enormous uncertainties about the nature of high-redshift objects it is of course impossible to describe the high-redshift X-ray background with any confidence (see \S \ref{xray} for a discussion of current constraints).  Our strategy (here and in the remainder of this chapter) must therefore be to construct simple, but plausible, models to examine the range of possibilities.  The most conservative assumption is that the local correlation between the star formation rate (SFR) and the X-ray luminosity (from 0.2--10 keV) can be extrapolated to high redshift \cite{grimm03, ranalli03, gilfanov04, glover03}
\begin{equation}
L_X = 3.4 \times 10^{40} f_X \left( \frac{{\rm SFR}}{1 \sfr} \right)
\ergsec,
\label{eq:sfrxray}
\end{equation}
where $f_X$ is an unknown renormalization factor at high redshift.  Note that the numerical factor depends on the photon energy range assumed to contribute to IGM heating; soft photons probably carry most of the energy, but they do not penetrate far into the IGM  (and hence induce reasonably strong fluctuations \cite{sethi05, pritchard06}).  

We can only speculate as to the accuracy of this correlation at higher redshifts.  Certainly the scaling is appropriate so long as stars dominate, but $f_X$ will likely evolve with redshift. The X-ray emission has two major sources.  The first is inverse-Compton scattering off of relativistic electrons accelerated in supernovae.  In the nearby Universe, only powerful starbursts have strong enough radiation fields for this to be significant; however, at high-redshifts it probably plays an increasingly important role because $u_\gamma \propto (1+z)^4$ \cite{oh01}.  Assuming that $\sim 5\%$ of the supernova energy is released in this form \cite{koyama95} yields $f_X \sim 0.5$ if $\sim 10^{51}$ ergs are released in supernovae per 100 $\Msun$ in star formation.  The second class of sources, which dominate in locally observed galaxies, are high-mass X-ray binaries, in which material from a massive main sequence star accretes onto a compact neighbor.  Such systems are born as  soon as the first massive stars die, only a few million years after star formation commences.  So they certainly ought to exist in high-redshift galaxies \cite{glover03}, although their abundance depends on the metallicity and stellar initial mass function, and it could be especially large if very massive Population III stars\footnote{Henceforth we will abbreviate ``Population III" with ``Pop III," and similarly for Pop II.} dominate.

X-rays heat the IGM gas by first photoionizing a hydrogen or helium atom.  The hot ``primary'' electron then distributes its energy through three main channels: (1) collisional ionizations, producing more secondary electrons, (2) collisional excitations of HeI (which produce photons capable of ionizing HI) and \hone (which produces a \lya background), and (3) Coulomb collisions with free electrons.  The relative cross sections of these processes determine what fraction of the X-ray energy goes to heating ($f_{X,h}$), ionization ($f_{X,{\rm ion}}$), and excitation ($f_{X,{\rm coll}}$); clearly it depends on both $\bxion$ and the input photon energy.  Monte Carlo simulations can be used to estimate these fractions, and an accurate fitting function exists in the high-energy limit \cite{shull85}.  However, a crude approximation often suffices \cite{chen04-decay}:
\begin{eqnarray}
f_{X,h} & \sim & (1+2 \bxion)/3 \nonumber \\
f_{X,{\rm ion}} \sim f_{X,{\rm coll}} & \sim & (1-\bxion)/3.
\label{eq:fxapprox}
\end{eqnarray}
In highly ionized gas, collisions with free electrons dominate and $f_{X,h} \rightarrow 1$; in the opposite limit, the energy is split roughly equally between these processes.

X-rays directly heat a small fraction of electrons, which must then transfer energy to the other particles \cite{madau97}.  The primary photoelectrons, with $T \sim 10^6 \kel$, rapidly cool to energies just below the \lya threshold, $\la 10 \eV$, by collisionally ionizing and exciting hydrogen atoms before equilibrating with the other electrons.  After that, the electrons and neutrals equilibrate through elastic scattering on a timescale $t_{\rm eq} \sim 5 [10/(1+z)]^3 \Myr$.  Because $t_{\rm eq} \ll H(z)^{-1}$, the assumption of a single temperature fluid is an excellent one.

Finally, to relate the X-ray emissivity to the global SFR we will assume (again for simplicity) that the SFR is proportional to the rate at which gas collapses onto virialized halos, $\deriv \fcoll/\deriv t$ (see eq. \ref{eq:fcollps}).  In that case, we can write
\begin{equation}
\frac{2}{3} \, \frac{\epsilon_{X}}{k_B n H(z)} = 10^3 \kel \ f_X \, \left(
\frac{f_\star}{0.1} \, \frac{f_{X,h}}{0.2} \, \frac{\deriv
  \fcoll/\deriv z}{0.01} \, \frac{1+z}{10} \right),
\label{eq:xrayemiss}
\end{equation}
where $f_\star$ is the star formation efficiency.  It is immediately obvious that X-ray heating is quite rapid; we will examine this in more detail in \S \ref{glob-theor}.  

Of course, even if equation (\ref{eq:sfrxray}) is accurate, there may be other contributions to the X-ray background.  Quasars are one obvious example, although their relevance is far from clear.  The known population of bright quasars, extrapolated to higher redshifts, causes negligible heating \cite{venkatesan01}.  But these bright quasars may be only the tip of the iceberg:  ``miniquasars," which are rapidly accreting intermediate-mass black holes that may form from the remnants of Pop III stars, can strongly affect the thermal history \cite{oh01, glover03, madau04, kuhlen05-sim}.  For example, if the ``Magorrian relation" between black hole and stellar mass \cite{magorrian98} (see also \cite{ferrarese00, gebhardt00}) holds at high redshifts, the equivalent normalization factor would be $f_X \sim 10$.  We review existing limits on the X-ray background in \S \ref{xray}.

\subsubsection{\lya Heating} \label{lyaheat}

In addition to setting the spin temperature, resonant scattering of \lya photons can also heat the gas through atomic recoil.  The typical energy exchange per scattering is small (see eq. \ref{eq:recoil-loss}), but the number of scatterings is large.  If the net heating rate per atom were $\approx P_\alpha \times (h \nu_\alpha)^2/m_p c^2$, the gas temperature would exceed $T_\gamma$ soon after Wouthuysen-Field coupling becomes efficient \cite{madau97}.

However, the details of radiative transfer radically change these expectations.  In a static medium, the energy exchange must vanish in equilibrium even though scattering continues at (nearly) the same rate.  Scattering induces an asymmetric absorption feature near $\nu_\alpha$ (Fig.~\ref{fig:lyshape}) whose shape depends on the combined effects of atomic recoils and the scattering diffusivity (see the first paragraph of \S \ref{fokker}).  In equilibrium, the latter exactly counterbalances the former.  Without scattering, the absorption feature would redshift away; thus the equilibrium energy exchange rate is simply that required to maintain the feature in place \cite{chen04}.  For photons redshifting into resonance, the absorption trough has total energy $\Delta u_\alpha = (4\pi/c) \int (J_\infty - J_\nu) h \nu \deriv \nu$, where $J_\infty$ is the input spectrum (thus the integration extends over the dip in Fig.~\ref{fig:lyshape}).  The radiation background loses $\epsilon_\alpha = H \Delta u_\alpha$ per unit time through redshifting; this energy goes into heating the gas.  We therefore have
\begin{equation}
\frac{2}{3} \, \frac{\epsilon_\alpha}{k_B T_K n_H H(z) } = \frac{8 \pi}{3} \, \frac{h \nu_\alpha}{k_B T_K} \, \frac{J_\infty \, \Delta \nu_D}{c n_H} \, I_c
\label{eq:epsalpha}
\end{equation}
where (using $\nu \approx \nu_\alpha$ throughout the region of interest)
\begin{equation}
I_c = \int_{-\infty}^{\infty} \deriv x \delta_J(x).
\label{eq:ic}
\end{equation}
Analogous expressions exist for photons injected at line center, except that here the corresponding integral $I_i$ measures the difference between the actual spectra shown in Figure~\ref{fig:lyshape} and a spectrum with a step function at line center.

For an arbitrary line profile these integrals must be performed numerically (see \cite{chen04}, although that does not include spin exchange \cite{hirata05, chuzhoy06} or the detailed balance correction \cite{rybicki06}, so it is only accurate to a few percent).  However, approximating the line profile with the Lorentzian wings of natural broadening turns out to be an excellent one \cite{grachev89, chuzhoy06-heat} and allows an analytic expression (in terms of Airy functions) for $I_c$ and a power series expansion for $I_i$ \cite{furl06-lyheat}.  The first few terms are
\begin{eqnarray}
I_c & \approx & 3^{1/3} \left( \frac{2 \pi}{3} \right)^{5/3} \left( \frac{a}{\gamma'} \right)^{1/3} \left[ \frac{ \beta}{\Gamma^2(2/3)} - \frac{3^{1/3} \,  \beta^2}{\Gamma(1/3) \Gamma(2/3)} + \frac{3^{2/3} \, \beta^3}{\Gamma^2(1/3)} \right]
\label{eq:icapprox} \\
I_i & \approx & \left( \frac{a}{\gamma'} \right)^{1/3} (-0.6979 + 2.4138 \beta - 2.7755 \beta^2 + 2.2724 \beta^3)
\label{eq:iiapprox} 
\end{eqnarray}
where $a=A_{21}/(4 \pi \Delta \nu_D)$, $A_{21}$ is the spontaneous emission coefficient for the \lya transition, and $\beta \equiv \eta' (4a/\pi\gamma')^{1/3}\approx 0.99 T_K^{-2/3} (\tau_{\rm GP}/10^6)^{1/3}$ (here the approximate form ignores spin exchange and the correction for detailed balance).  When $\beta$ is small and only the first terms contribute (reasonably accurate at $T_K \ga 10 \kel$ for $I_c$ but not as good for $I_i$), we expect $I_c \propto T_K^{-5/6} (1+z)$ and $I_i \propto T_K^{-1/6} (1+z)$.

Although $I_c$ is always positive (an absorption trough always forms), $I_i$ is negative at most temperatures ($T_K \ga 4 \kel$) \cite{chen04}.  Given enough time, the gas would therefore approach an equilibrium temperature determined by competition between heating from continuum photons and cooling from injected photons \cite{chuzhoy06-heat}.  The relevant parameter is the fraction of injected photons at the \lya resonance; for Pop II and very massive Pop III stars, this is $6\%$ and $12\%$, respectively \cite{pritchard05}.  However, both heating and cooling turn out to be extremely slow and this equilibrium is never reached in practice.  We find that the continuum heating rate is \cite{furl06-lyheat}
\begin{equation}
\frac{2}{3} \frac{\epsilon_\alpha}{H n_H k_B T_K} \approx \frac{0.80}{T_K^{4/3}} \, \frac{x_\alpha}{S_\alpha} \left( \frac{10}{1+z} \right),
\label{eq:lyaheat}
\end{equation}
which is usually negligibly small compared to X-ray heating.  The reason is that the scattering diffusivity acts to cancel the effects of recoil.  From Figure~\ref{fig:lyshape}, it is obvious that the background spectrum is weaker on the blue side of the line than on the red.  Scattering tends to return the photon toward line center, with the extra energy deposited in or extracted from the gas.  Because more scattering occurs on the red side, this tends to transfer energy from the gas back to the photons, canceling the recoil exchange.  We refer the interested reader to \cite{rybicki06, meiksin06} for a more formal explanation.  

Of course, so far we have assumed steady-state ($\partial J_\nu/\partial t=0$); obviously the heating \emph{rate} will be much faster before equilibrium is established.  But the \emph{total} energy transferred to the gas during this period is simply $\Delta u_\alpha$, the energy lost in producing the absorption feature of Figure~\ref{fig:lyshape}.  Thus the total energy transfer is essentially the same as in our equation~(\ref{eq:lyaheat}): this is because, although the heating rate itself is large, only a few scattering times are required to establish a steady-state \cite{meiksin06}.  This may seem a bit puzzling, because the Hubble expansion used in deriving equation (\ref{eq:epsalpha}) should not affect the approach to equilibrium.  But note that the redshift continually brings entirely new photons into the line.  Equilibrium must be re-established for these photons, so the two calculations are actually identical.

As described in \S \ref{lyn}, higher Lyman-series photons also scatter resonantly.  They rapidly cascade to lower levels, so the radiation field around these lines never assumes a distribution like \lyans.  Thus the energy exchange is more subtle.  Because the medium is so optically thick at these resonances, the few scatterings per photon that do occur take place at frequencies well blueward of resonance.  In that case the photon is generally re-emitted closer to line center; the remaining energy -- much larger than that provided through recoil -- is deposited in the gas.  This can increase the heating efficiency above what recoil would predict, but it is still much smaller than the \lya heating \cite{furl06-lyheat}.

The energy exchange through this frequency drift is maximal when scattering occurs near the line center, which requires the line to have a relatively small optical depth.\footnote{In the Doppler core, photons are preferentially absorbed by atoms moving opposite the photon; the isotropic re-emission imparts $\sim h \Delta \nu_D$ to the gas.  In the high optical depth case, absorption occurs in the Lorentz wings of the line, where the direction of the absorbing atom is much less significant.}  As a result, scattering across the Ly$\beta$ deuterium resonance (which has $\tau_{\rm GP} \ga 2$, and which also destroys photons and so produces an asymmetric background) is the most important of all the upper level transitions (including those of hydrogen itself) \cite{chuzhoy06-heat}.  However, this energy is deposited in the rare deuterium atoms and must then be shared with the hydrogen atoms through collisions, which are relatively inefficient; the resulting heating rate is usually (though not always) slower than direct \lya heating.

\subsubsection{Shock Heating} \label{shockheat}

The final heating mechanism is purely hydrodynamic: thermal energy injection through IGM shocks.  This is an integral part of our modern view of structure formation.  The initially tiny density fluctuations seen in the CMB grow through gravitational instability.  Because these fluctuations are not perfectly spherical, different directions grow at different rates \cite{zeldovich70}.  Collapse along the shortest axis produces sheets (or ``Zel'dovich pancakes,'' part of the original motivation for 21 cm experiments in the 1970s; \cite{sunyaev72,sunyaev75}), collapse along the middle filaments, and along the third virialized halos.  At each stage in this process, some of the gravitational infall energy is transformed into thermal energy through shocks, creating complex networks bounding and penetrating the sheets and filaments \cite{miniati00, kang05}. This paradigm of the ``cosmic web'' emerges naturally in cosmological simulations and nicely explains both the large-scale distribution of galaxies in redshift surveys and the distribution of IGM gas in the \lya forest.  

Simulations show that such shocks dominate the thermal energy budget of the IGM at $z \sim 0$ \cite{cen99, dave01}.  Fortunately, their characteristic properties are easy to estimate.  Each cosmic web shock is part of the ongoing (but recent) collapse of a nonlinear structure.  Thus the peculiar velocity of the shocked gas must be \cite{cen99} 
\begin{equation}
v_{\rm sh} \approx H(z) R_{\rm nl}(z),
\label{eq:shockvel}
\end{equation}
where $R_{\rm nl}^3=3 M_{\rm nl}/(4 \pi \bar{\rho})$ and $M_{\rm nl}$
is the mass scale that is just entering the nonlinear regime: in other
words, its velocity is simply the distance it has collapsed ($R_{\rm
nl}$) divided by the age of the universe.  The corresponding postshock
temperature is then (assuming strong shocks and a monatomic gas)
\begin{equation}
T_{\rm sh} = \frac{3 \mu m_p}{16 k_B} \, v_{\rm sh}^2,
\label{eq:shockT}
\end{equation}
where $\mu$ is the mean molecular weight in atomic units.  This simple argument reproduces the typical shock temperature $T_{\rm sh} \sim 10^7 \kel$ at the present day \cite{cen99} and also quantifies their importance at higher redshifts.

Of course, a shock will only form if $T_K < T_{\rm sh}$ (which depends on the typical mass of a collapsing object $M_{\rm nl}$).  At the present day, photoheating sets $T_K \approx 10^4 \kel$, and infall shocks are quite strong.  At moderate redshifts $z \ga 3$, the typical nonlinear object is much smaller while $T_K$ (again set by photoheating) is about the same.  Thus $T_{\rm sh} \sim T_K$ and shock-heating is insignificant \cite{furl04-sh}.  

However, we have already seen that \emph{before} reionization $T_K$ can be quite small.  Both analytic models \cite{furl04-sh} and simulations \cite{gnedin04, shapiro05, kuhlen06-21cm} predict that shock heating significantly affects the thermal structure of the IGM at high redshifts.  A simple analytic description of cosmic web shocks will suffice to build intuition \cite{nath01, valageas02, furl04-sh}.  The model associates these shocks with the ``turnaround" of spherical perturbations, where the flow breaks off from the Hubble expansion and begins to converge on itself; at this point, perturbations in the flow can easily induce shocks.  Turnaround occurs at a linearized overdensity $\delta_{\rm sh}=1.06$ -- just when perturbations become nonlinear.  So the model reproduces equation (\ref{eq:shockvel}) and hence the characteristic temperature of $z=0$ shocks.  We will use it to estimate the mass fraction inside cosmic web shocks in an analogous way to the collapse fraction \cite{press74, bond91}.  The fraction of gas that has been shocked to a temperature $T > T_{\gamma}$ is (cf. eq. \ref{eq:fcollps})
\begin{equation}
f_{\rm sh}(>T_{\gamma}) = {\rm erfc} \left[ \frac{\delta_{\rm
    sh}(z)}{\sqrt{2} \sigma(T_{\gamma})} \right],
\label{eq:fsh}
\end{equation}
where $T_{\gamma}$ corresponds to a mass $m_\gamma \approx 4.3 \times 10^5 \Msun$ via equations (\ref{eq:shockvel}) and (\ref{eq:shockT}).  This yields $f_{\rm sh} \sim (0.1\%,\, 3\%, \, 25\%)$ at $z=(30,\,20,\,10)$:  like halos, the cosmic web accretes gas rapidly in this regime.  Note that turnaround occurs at a physical overdensity $\delta_{\rm ta}=5.55$, so this gas is overdense and has relatively efficient collisional coupling.  In the calculations shown below, we will therefore assume for simplicity that it all emits 21 cm radiation relative to the CMB.

This model agrees reasonably well with simulations \cite{kuhlen05-sim, shapiro05}, but two caveats are necessary.  First, we have ignored X-ray heating in equation (\ref{eq:fsh}), which will smooth out these shocks.  Fortunately, when X-rays are important shocks do not affect the 21 cm signal anyway, because all the gas is hot.  Second, in the absence of X-ray heating, the IGM sound speed is so low that even tiny peculiar velocities could potentially source shocks.  In a cold IGM, such slow shocks could form even before turnaround and would therefore be missing from our model.  If so, they could dramatically affect the predictions when $T_K \ll T_\gamma$.  There is some indirect evidence for such shocks \cite{gnedin04}, but they are difficult to resolve in simulations and have not yet been studied in detail.

\subsection{The Spin Temperature} \label{tshist}

With the thermal evolution in hand, we now turn to the spin temperature.  Recall from equation (\ref{eq:dcoll}) that collisions are totally inefficient at $z \la 30$ (except inside filaments, or if $\xion$ is large enough for collisions with electrons to dominate; see \S \ref{xray}), so we must rely on the Wouthuysen-Field effect.  Of course (as for the X-ray background), we cannot yet predict the detailed evolution of $J_\alpha$, because it depends on the star formation history as well as any other radiation background (quasars, etc.).  But we can make an educated guess by assuming that it traces the star formation rate, which is again proportional to the rate at which matter collapses into galaxies.  We therefore write the comoving emissivity at frequency $\nu$ as
\begin{equation}
\epsilon(\nu,z) = f_\star \, \bar{n}_b^c \, \epsilon_b(\nu) \,
\frac{\deriv \fcoll}{\deriv t},
\label{eq:sfemiss}
\end{equation}
where $\bar{n}_b^c$ is the comoving number density of baryons and $\epsilon_b(\nu)$ is the number of photons produced in the frequency interval $\nu \pm \deriv \nu/2$ per baryon incorporated into stars.  Here we are only interested in photons between \lya and the Lyman-limit.  Although real spectra are rather complicated, a useful quantity is the total number $N_\alpha$ of photons per baryon in this interval.  For low-metallicity Pop II stars and very massive Pop III stars, this is $N_\alpha=9690$ and $N_\alpha=4800$, respectively.  More detailed spectra (and fits) appear in \cite{leitherer99, bromm01-vms, barkana05-ts}. 

Although only \lya photons efficiently couple to $T_S$, higher Lyman-series photons contribute by cascading to \lya (see \S \ref{lyn}).  The average background at $\nu_\alpha$ is\footnote{Note that we neglect absorption by the higher Lyman series resonances of deuterium here, which reduce the spectrum just blueward of the hydrogen resonances \cite{chuzhoy06}.  Some of the scattered photons are re-injected at the deuterium \lya line and then redshift into hydrogen \lyans.}
\begin{eqnarray}
J_\alpha(z) & = & \sum_{n=2}^{n_{\rm max}} J_\alpha^{(n)}(z) \nonumber
\\
& = & \sum_{n=2}^{n_{\rm max}} f_{\rm rec}(n) \int_z^{z_{\rm max}(n)} 
\deriv z' \, \frac{(1+z)^2}{4 \pi} \, \frac{c}{H(z')} \,
\epsilon(\nu_n',z'),
\label{eq:jalpha}
\end{eqnarray}
where $\nu_n'$ is the frequency at redshift $z'$ that redshifts into the \lyn resonance at redshift $z$, $z_{\rm max}(n)$ is the largest redshift from which a photon can redshift into the \lyn
resonance, and $f_{\rm rec}(n)$ is the fraction of \lyn photons that actually cascade through \lya and induce strong coupling (see \S \ref{lyn}).  The sum must be truncated at some large $n_{\rm max}$, but its precise value does not matter because the high-$n$ lines are so closely spaced.  Once we know $J_\alpha$, we can compute the coupling coefficient $x_\alpha$ from equation (\ref{eq:xalpha}).  

Of course, processes other than star formation can also create a \lya background.  These include UV photons from quasars, recombinations in a partially ionized medium, and collisional excitation from X-rays \cite{madau97}. In the latter, a fraction $f_{X,{\rm coll}} \sim 1/3$ of the energy is typically lost to excitations (see eq. \ref{eq:fxapprox}); the fraction of this that eventually winds up in \lya photons is $\approx 0.8$ \cite{pritchard06}.  The coupling coefficient induced by these line photons is \cite{chen06, chuzhoy06-first}
\begin{equation}
x_\alpha^{\rm X-ray} \sim 0.05 \, S_\alpha \, f_X  \, \left( \frac{f_{X,{\rm coll}}}{1/3} \, \frac{f_\star}{0.1} \frac{\deriv \fcoll/\deriv z}{0.01} \right) \, \left( \frac{1+z}{10} \right)^{3}.
\label{eq:wf-xray}
\end{equation}
Here we have substituted the same emissivity as in equation~(\ref{eq:xrayemiss}); thus heating is accompanied by a small, though far from negligible, \lya background.  This process is particularly important near star-forming galaxies, where most soft X-rays are absorbed \cite{chen06, chuzhoy06-first, pritchard06}.  Recombinations in a partially ionized medium can also produce a weak background \cite{madau97}.

\subsection{The Ionization History} \label{ionhist}

The next step is to compute the evolution of the ionization fraction $\bxion$.  This, of course, has been the subject of extensive study in the past three decades (see, e.g., \cite{barkana01,haiman04-rev,ciardi05-rev, loeb06} for recent reviews).  As such, we will not attempt to explore all of its aspects but only to describe its principal components.  In \S \ref{reion} we will consider the spatial fluctuations in the ionization fraction that will be the main focus of most 21 cm measurements.

The usual assumption (and the one we will make here) is that ionizing photons are produced inside of galaxies, so that their production rate can be associated with the star formation rate in a similar way to the \lya radiation background and our X-ray heating model (see eqs. \ref{eq:xrayemiss} and  \ref{eq:sfemiss}).  In the most basic approximation, we simply assign a fixed average ionizing efficiency across all galaxies, so that
\begin{equation}
\bxion = \zeta \fcoll/(1 + \bar{n}_{\rm rec}),
\label{eq:bxion}
\end{equation}
where $\bar{n}_{\rm rec}$ is the mean number of recombinations per ionized hydrogen atom and the ionizing efficiency is
\begin{equation}
\zeta = A_{\rm He} f_\star \fesc N_{\rm ion}. 
\label{eq:zetadefn}
\end{equation}
In this expression, $\fesc$ is the fraction of ionizing photons that escape their host galaxy into the IGM, $N_{\rm ion}$ is the mean number of ionizing photons produced per stellar baryon, and $A_{\rm He}=4/(4-3Y_p)=1.22$, where $Y_p$ is the mass fraction of helium, is a correction factor to convert the number of ionizing photons per baryon in stars to the fraction of ionized hydrogen.\footnote{Here we have assumed that helium is singly ionized along with hydrogen, because their ionization potentials are relatively close.}  We give some fiducial estimates for these parameters in \S \ref{ioneff}.

A more sophisticated treatment includes both ionizing sources and recombinations, so
\begin{equation}
\frac{\deriv \bxion}{\deriv t} = \zeta(z) \frac{\deriv \fcoll}{\deriv t}
- \alpha C(z,\bxion) \bxion(z) n_e(z),
\label{eq:bxion-evol}
\end{equation}
where $\alpha$ is the recombination coefficient, $C \equiv \VEV{n_e^2}/\VEV{n_e}^2$ is the clumping factor, and $n_e$ is the average electron density in ionized regions.  The precise definition of $C$ is subtle but important:  the electron density is averaged (by volume) over all regions penetrated by ionizing photons (thus excluding gas outside of ionized bubbles and gas inside self-shielded clumps).  Equation~(\ref{eq:bxion-evol}) allows both the ionizing efficiency $\zeta$ and the clumping factor $C$ to depend on redshift -- and hence implicitly on $\bxion$ or any other parameter of galaxy formation.

\subsubsection{Recombinations and the Clumping Factor} \label{clump}

Before considering $\zeta$, we first discuss some subtleties of the sink term, which is more complicated than it looks.  First of all, the recombination coefficient is uncertain by a factor of a few through both the gas temperature and an environmental factor that determines whether case-A or case-B is more appropriate.\footnote{Here ``case-A'' allows recombinations directly to the ground state while ``case-B'' does not; in the latter, ionizing photons produced from recombinations into the ground state are assumed to be immediately absorbed again, so that they do not affect the net ionization rate.}  On the one hand, consider the case in which ionizations (and hence recombinations) are distributed uniformly throughout the IGM (the usual assumption in the literature e.g., \cite{wyithe03,cen03}).  Then case-B would be appropriate, with $\alpha_B \approx 2.6 \times 10^{-13} (T_K/10^4 \kel)^{-0.7} \recunits$ \cite{osterbrock89}.  On the other hand, in the highly-ionized low-redshift universe, most recombinations actually take place in dense, partially neutral gas (so-called Lyman-limit systems) because high-energy photons can penetrate inside these high-column density systems.  However, the ionizing photons produced after recombinations to the ground state usually lie near the Lyman-limit (where the mean free path is small) so are consumed inside the systems.  Thus these photons would not help ionize the IGM, and case-A (with $\alpha_A \approx 4.2 \times 10^{-13} [T_K/10^4 \kel]^{-0.7} \recunits$\cite{osterbrock89}) would be more appropriate \cite{miralda03}.  Which of these regimes is more relevant depends on the details of small-scale clumping and radiative transfer and is not yet a solved problem.  Additional uncertainty comes from the gas temperature, which depends on the ionizing spectrum,  another large unknown \cite{abel99}.

Even more problematic is the clumping factor $C(z)$.  In principle, of course, this can be computed through numerical simulations.  But that requires overcoming two difficult problems: (1) tracing the gas distribution with sufficient precision to resolve density fluctuations on the smallest scales and (2) correctly tracing the topology of ionized and neutral gas -- because the average must be performed only over the ionized gas.  The first problem is obvious: even leaving aside the interstellar medium (ISM) of each galaxy (which is accounted for by $\fesc$ in eq. \ref{eq:zetadefn}) the Jeans mass in the cold IGM is $\la 10^5 \Msun$.  This allows the formation of a well-defined cosmic web, as well as ``minihalos,'' dense gas clouds that virialize but cannot cool or form stars.  But simulations of reionization must span $\sim 100 \Mpc$ boxes in order to adequately sample the large \htwo regions, requiring an enormous dynamic range.  Thus, even in simulations, clumping is usually accounted for through a ``subgrid" model built from semi-analytic techniques or smaller simulations \cite{sokasian03, iliev05-sim, ciardi05-mh, kohler05-sim}.  Minihalos are particularly problematic.  Fortunately, although early estimates predicted that they could increase the clumping factor by more than an order of magnitude \cite{haiman01}, more recent efforts show that they have a relatively modest effect, consuming only a few photons per baryon \cite{shapiro04, iliev05}.

The second problem is perhaps more subtle: how do the sources and absorbers relate to each other, and how does ionization affect the small-scale clumping?  For example, if low-density gas is ionized first,  $C<1$ throughout most of reionization \cite{miralda00}, because all the dense gas would remain locked up in self-shielded systems.  This is an attractive picture because (on small scales) it is certainly easier to ionize voids than filaments or dense blobs.  On the other hand, on large scales the ionizing sources actually lie inside overdense regions (sheets and filaments), where the recombination rate is relatively high.  Moreover, as the gas is ionized, the thermal pressure will increase and the clumpiness will decrease.  While widely appreciated in the context of minihalos \cite{haiman01, shapiro04, iliev05, ciardi05-mh}, the implications for the clumping of the more diffuse IGM have not been studied.

All of these issues have been considered in the literature, but unfortunately there is no single unifying model that accounts for realistic small-scale structure along with the topology of reionization and feedback.  The two most common approaches include different physics.  The first is to compute the clumping factor from high-resolution simulations, ignoring reionization and feedback.  One must still be careful to exclude gas bound to ionizing sources (which are described by $\fesc$ rather than $C$; see \cite{kohler05-sim}), or else the estimated clumping is much too large (this problem is why the widely-quoted $C \sim 30$ value of \cite{gnedin97} is much larger than more recent estimates).  One recent estimate, from a 3.5$h^{-1}$ Mpc N-body simulation resolving the Jeans mass in the IGM, is well-fit by \cite{mellema06}
\begin{equation}
C(z) = 27.466 \exp(-0.114z + 0.001328 z^2).
\label{eq:clump-mips}
\end{equation}
The other approach follows \cite{miralda00} and assumes that the lowest-density gas is ionized first; in this case recombinations are much less significant. (This model can be modified to include at least some aspects of source clustering, which increases the clumpiness by a factor of a few \cite{furl05-rec}.)  The truth lies somewhere between these extremes for two reasons:  (1) the numerical estimate neglects photoheating during reionization, while the other model of \cite{miralda00} is based exclusively on post-reionization simulations with a warm IGM, and (2) the numerical estimate averages over all gas and does not allow any to sit in self-shielded neutral systems, while assuming voids to be ionized first represents a particularly optimistic case for identifying such systems.

\subsubsection{The Ionizing Efficiency} \label{ioneff}

We now move on to the source term in equation (\ref{eq:bxion-evol}).  This has two parts:  $\deriv \fcoll/\deriv z$ and the ionizing efficiency $\zeta$.  The collapse fraction for a given cosmology depends only on $\mmin$, the mass threshold for galaxy formation.  The most common choice for $\mmin$ corresponds to a virial temperature $T_{\rm vir}=10^4 \kel$, the threshold at which hydrogen line cooling becomes efficient for primordial gas (see, e.g., \cite{barkana01}).  Above this mass, cooling and fragmentation into stars is relatively straightforward.  Other choices are, however, physically plausible in certain regimes.  For example, H$_2$ cooling could allow star formation in much smaller halos, although it is relatively weak (limiting the expected star formation efficiency) and it is subject to a variety of feedback mechanisms (see, e.g., \cite{haiman96, tegmark97, ciardi05-rev} and references therein).

We can use local measurements of $f_\star$, $\fesc$, and $N_{\rm ion}$ to guide our choices, though the extrapolation to high redshifts is always difficult. Efficiencies $f_\star \sim 10\%$ are reasonable for the local Universe, but so little gas has collapsed by $z=6$ that this does not directly constrain the high-redshift value.  Appropriate values for Pop III stars are even more uncertain.  To the extent that each halo can form only a single very massive ($\ga 100 \Msun$) star that enriches the entire halo, $f_\star \sim
(\Omega_m/\Omega_b) M_\star/M_h \la 10^{-3}$, though larger values are often taken in the literature (implicitly assuming either inefficient metal dispersal or an extremely rapid starburst).

The UV escape fraction is small in nearby star-forming galaxies (including the Milky Way), with many upper limits $\fesc < 6\%$ \cite{hurwitz97,heckman01,deharveng01, blandhawthorn99, blandhawthorn01} and only a few positive detections (at comparable levels).  An initial detection of ionizing flux from a stacked sample of blue (and hence atypical) Lyman-break galaxies at $z \sim 3$ implied $\fesc \approx 10\%$ \cite{steidel01}, but more recent observations either place upper limits $\fesc \la 5$--$10\%$ \cite{giallongo02,malkan03,fernandez03,inoue05} or claim detections at much lower levels, $\fesc \approx 2\%$ \cite{shapley06}.  Note as well that $\fesc$ shows large variance between galaxies.  Theoretically the problem is equally difficult, because it depends on the spatial distribution of hot stars and absorbing gas in the ISM.  Some studies suggest that the escape fraction in high-redshift galaxies could be much higher than the detected values (generally because higher specific star formation rates allow supernovae to blow transparent windows through the ISM) \cite{dove00,ciardi02,fujita03}, but others predict that it will remain small \cite{wood00}.  Two special cases are worth noting. First, quasars may have large escape fractions because they concentrate all of their photons in one spatial location \cite{wood00}.  Second, the shock bounding the \htwo regions of very massive Pop III stars may evacuate all the gas from their host halos if $T_{\rm vir} < 10^4 \kel$.  In that case, $\fesc \sim 1$ \cite{whalen04}.  

$N_{\rm ion}$ depends only on the stellar initial mass function and metallicity.  Convenient approximations are $N_{\rm ion} \approx 4000$ for $Z=0.05 \Zsun$ Pop II stars with a Scalo IMF \cite{kamionkowski94, barkana01,scalo98} and $N_{\rm ion} \approx 40,000$ for very massive Pop III stars \cite{bromm01-vms}.  Note that the latter assumes that \emph{all} Pop III stars are massive; metal-free stars with a normal Salpeter IMF are only $\sim 1.6$ times more efficient than their Pop II counterparts \cite{tumlinson03}.  More detailed spectra for Pop II stars can be found in \cite{leitherer99,bruzual03} and for Pop III stars in \cite{bromm01-vms,schaerer02,tumlinson03}.

\subsubsection{Feedback During Reionization} \label{feedback}

The wide range of allowed values for $\mmin$, $f_\star$, $\fesc$, and $N_{\rm ion}$ suggests that these quantities might evolve in a non-trivial way, either over time or across the galaxy population.  Such feedback effects are now considered crucial to reionization, so here we will briefly outline their effects on its global history.  In particular, we will focus on how feedback prolongs reionization.  We refer the interested reader to \cite{ciardi05-rev} for a more comprehensive discussion.

Feedback mechanisms can be conveniently divided into ``internal'' and ``external'' flavors.  The former includes  processes that affect later generations of objects within the same galaxy.  One straightforward example is self-enrichment: once Pop III stars pollute their host halo with metals, subsequent star formation will be Pop II. Supernova winds also have other internal feedback effects:  most importantly, they can eject a large fraction of the interstellar gas into the IGM, shutting off star formation (if only temporarily).  These winds are well-known in the local universe \cite{heckman02} and at moderate redshifts \cite{shapley03}, and they may be even more important at high redshifts.  This is easy to see if we compare the energy output in supernovae per baryon, $W_{\rm SN}=f_\star E_{\rm SN} \nu_{\rm SN}$, to the binding energy per baryon of a virialized halo, $E_b \sim G m_h/(\lambda_s r_{\rm vir})$:
\begin{equation}
\frac{W_{\rm SN}}{E_b} \sim 4 \, E_{51} \, \left( \frac{f_\star}{0.1} \,
\frac{\nu_{\rm SN}}{0.01 \Msun^{-1}} \, \frac{\lambda_s}{0.05} \,
\frac{10}{1+z} \right) \left( \frac{m_h}{10^8 \Msun} \right)^{-2/3}.
\label{eq:snbind}
\end{equation}
Here $r_{\rm vir}$ is the virial radius, $E_{\rm SN} = 10^{51} E_{51} \erg$ is the energy per supernova, $\nu_{\rm SN}$ is the specific frequency of supernovae (i.e., the number of supernovae per unit mass of star formation), and we have assumed that the baryons lie at a typical distance $\lambda_s r_{\rm vir}$ from the halo center (our fiducial value is appropriate for disks in normal galaxies \cite{mo98}).  Thus the supernova energy reservoir is probably comparable to the binding energy of the baryons in high-redshift dwarf galaxies, suggesting that they could have a significant dynamical effect.  Because $W_{\rm SN}/E_b \propto m_h^{-2/3}$, the feedback efficiency is a function of halo mass, and $f_\star$, $\fesc$, or any of our other parameters may also be functions of halo mass.  Indeed, low-mass galaxies in the SDSS appear to have $f_\star \propto m_h^{2/3}$ \cite{kauffmann03} that may be a result of supernova feedback \cite{dekel03}. 

Fortunately, internal feedback mechanisms are relatively easy to incorporate into reionization models.  For the most part, we can mimic them by setting $\zeta=\zeta(m_h)$.  Because these processes are local, on a global level they probably do not evolve rapidly with cosmic time and so will not introduce features into $\bxion(z)$.  For instance, self-enrichment causes each individual galaxy's star formation mode to evolve on a short timescale, but the continuous formation of galaxies throughout the Universe prevents any sharp evolution in the global initial mass function (although of course it can prolong reionization relative to a constant efficiency scenario \cite{wyithe03}).

External feedback is much more difficult to incorporate into analytic models.  This includes all those effects that galaxies can have on their neighbors.  We consider two representative mechanisms here.

{\em Metal enrichment:} As described above, supernova winds likely escape their host galaxies, spreading metals into the IGM and affecting subsequent generations of galaxies.  The effective transition between (very massive) Pop III and Pop II star formation is sudden, occurring at a critical metallicity $Z_t \sim 10^{-3.5} \Zsun$ in the gas phase \cite{bromm01-tran,bromm03} or at $Z_t \sim 10^{-5} \Zsun$ if dust is included \cite{schneider03}.  Because the transition could be accompanied by a large drop in $N_{\rm ion}$, this may have had an enormous effect on the reionization history.  

The simplest prescription for chemical feedback is to track the mean cosmic metallicity $\bar{Z}$ and to
switch the mode of star formation once $\bar{Z} > Z_t$.  Because this imposes a global drop in the ionizing emissivity, it led to several predictions of ``double reionization'' in which $\bxion(z)$ actually decreased over some finite time interval \cite{wyithe03, cen03, wyithe03-letter, cen03-letter, wyithe06-cen}.  However, this approximation is not a good one, because metal enrichment (like reionization) is highly inhomogeneous and \emph{must} be modeled physically (rather than with an arbitrary prescription) \cite{haiman03, furl05-double}.  On the one hand, internal feedback must be included:  even before $\bar{Z}=Z_t$, ongoing accretion onto existing galaxies produces Pop II stars (assuming mixing is efficient).

The other aspect is that newly-formed halos only produce Pop III stars if they collapse from pristine material.  As galactic winds expand into the IGM, more and more halos form out of pre-enriched gas, eventually choking off the supply of Pop III stars.  Wind enrichment has been studied extensively from a theoretical perspective \cite{ferrara00, madau01, scannapieco02, mori02, furl03-metals}, though firm predictions are even more difficult than for reionization because the physics is so complex.  The crucial point, however, is obvious:  winds expand relatively slowly compared to ionizing photons, broadening the transition from Pop III to Pop II \cite{scannapieco03, furl05-double}.  For the same reason, it is difficult to arrange for complete enrichment to precede reionization.

The condition for double reionization follows from equation (\ref{eq:bxion-evol}) \cite{furl05-double}
\begin{equation}
\zeta' \frac{\deriv \fcoll}{\deriv z} \, \la \, 0.1 \, C \, \bxion
\left( \frac{1+z}{10} 
\right)^{1/2},
\label{eq:doublecond}
\end{equation}
where $\zeta'$ is the ionizing efficiency of the second generation of sources (in this case, Pop II stars).  Near the end of reionization, $\zeta \deriv \fcoll/\deriv z \sim 1$, so only if $C \gg 1$ or $\zeta/\zeta' \gg 1$ can recombinations dominate.  It is important to note that, although the mean clumpiness may be large, rapidly recombining regions contain only a small fraction of the total mass.  Moreover, after reionization, equation (\ref{eq:clump-mips}) no longer applies because it neglects Jeans smoothing from photoheating -- so $C$ is likely reasonably small in the vast majority of the IGM.  Of course, double reionization is still possible if $\zeta/\zeta'$ is sufficiently large.  However, in practice the slow evolution of the Pop III fraction (thanks to self-enrichment) smoothes the transition so that, if the ratio is large, reionization takes place over a long enough period that double reionization requires $\zeta^{\rm III}/\zeta^{\rm II} \ga 200$
\cite{furl05-double}; while in principle possible, this certainly pushes parameters in uncomfortable directions.

{\em Photoheating:} A second crucial feedback mechanism is photoionization itself, which heats the gas because the liberated electrons are typically left with energies $\ga 1$ eV.  The increased thermal pressure suppresses accretion onto small halos and hence decreases the rate of star formation.  This is usually quantified by raising the Jeans mass in heated regions to a larger mass.  Unfortunately, the equivalent virial temperature $T_h$ is not entirely clear; the canonical value of $T_h \sim 2 \times 10^5 \kel$ \cite{rees86, efstathiou92, thoul96, kitayama00} may be an overestimate for halos collapsing during or soon after reionization \cite{dijkstra04-feed}.

Photoheating feedback provides a ``self-regulation" mechanism for reionization, because $\bxion$ itself (or at least the closely related ``photoheated fraction" \cite{furl05-double}) controls the transition.  Thus photoheating can significantly extend the reionization epoch (by $\Delta z \sim 4$ for the canonical $T_h$) by causing a plateau when $\bxion \sim 0.5$.  On the other hand, self-regulation makes it quite difficult for recombinations to dominate at any point in this kind of feedback mechanism \cite{furl05-double}.  Note as well that photoheating can only be important if small halos dominate the photon budget.  If $\zeta$ increases rapidly with mass, reionization is only slightly delayed.

Thus the major effect of external feedback (in either manifestation) is to prolong reionization, but it probably does not introduce sharp features into the history.  Other mechanisms, such as the suppression of molecular hydrogen cooling by soft-UV photons, likely have much smaller effects on $\bxion(z)$ \cite{haiman97a, haiman97b, haiman00, machacek01, mesinger06}, though they may affect star formation in the first galaxies.  Internal feedback affects the distribution of ionizing efficiencies within galaxies but also does not introduce strong features into the global ionization history.  The 21 cm background and its fluctuations will likely be one of our premier tools for understanding the role these various processes play.

\subsubsection{Exotic Reionization Scenarios} \label{exotic-reion}

To this point, we have implicitly assumed that stars (of one form or another) reionized the universe.  While that is clearly the most natural explanation, other sources -- such as quasars -- may also have contributed. Because quasars form in collapsed halos, they would probably produce qualitatively similar $\bxion(z)$ to star formation, although many of the ionizations would be due to X-rays (which would affect the topology of reionization; see \S \ref{xray}).  However, extrapolating the known population of quasars to higher redshifts suggests that they are unimportant \cite{wyithe03,fan04}, and the $z=0$ X-ray background prevents them from dominating the reionization era \cite{dijkstra04,salvaterra05-xray} (see \S \ref{xray}). On the other hand,  there are currently no direct observational constraints on faint, high-redshift AGN, so they may still be important.

An even more exotic possibility is reionization from some sort of decaying or annihilating particle \cite{chen04-decay,hansen04,kasuya04,avelino04, pierpaoli04, mapelli06}.  In that case, the ionization history would be entirely uncorrelated with structure formation, and there is no \emph{a priori} reason to expect sharp features.  However, the CMB does constrain such models, because long-lifetime decay leads to a protracted ionization history that affects the CMB temperature and polarization anisotropies \cite{chen04-decay, pierpaoli04}.  Several currently fashionable dark matter models do not produce significant ionization \cite{mapelli06}, though they can affect the 21 cm history \cite{shchekinov06, furl06-dm}.

\subsection{The Global History} \label{glob-theor}

We are now in a position to compute the evolution of $\bdtb$ in some representative structure formation models.  These histories illustrate the basic features of the observable signal and set the stage for the next several sections.  We will necessarily ignore many of the subtleties in constructing detailed reionization histories; we refer the interested reader to the existing literature for more examples \cite{wyithe03, cen03, wyithe03-letter, cen03-letter, haiman03, somerville03, fukugita03, onken04, furl05-double, choudhury05, sethi05, wyithe06-cen}.

\subsubsection{Some Critical Points in the 21 cm Line History}
  \label{critpt}

There are five ``critical points'' in the 21 cm history that divide the signal into several distinct epochs.  The first is $z_{\rm dec}$, when Compton heating becomes inefficient and $T_K<T_\gamma$ for the first time (eq. \ref{eq:zdec}).  This marks the earliest epoch for which 21 cm observations are possible even in principle.  The second transition is when the density falls below $\delta_{\rm coll}$ (see eq. \ref{eq:dcoll}), at which point $T_S \rightarrow T_\gamma$ and the IGM signal vanishes.  These two transitions are well-specified by atomic physics processes, at least in the absence of any exotic dark sector processes.  They define the beginning and end of ``dark ages" for the purposes of the 21 cm line.

The remaining transition points are determined by luminous sources, so their timing is much more uncertain.  These are the redshift $z_h$ at which the IGM is heated above $T_\gamma$, the
redshift $z_c$ at which $x_\alpha=1$ so that the Wouthuysen-Field mechanism couples $T_S$
and $T_K$, and the redshift of reionization $z_r$.  Their relative timing sets the observability of the 21 cm signal, so it is useful to consider them in some simple parameterized models \cite{furl06-glob}.  We first ask whether \lya coupling precedes the other two transitions.  The net X-ray heat input $\Delta T_c$ at $z_c$ is 
\begin{equation}
\frac{\Delta T_c}{T_\gamma} \sim 0.08 f_X \left( \frac{f_{X,h}}{0.2} \,
\frac{\fcoll}{\Delta \fcoll} \, \frac{9690}{N_\alpha} \,
\frac{1}{S_\alpha} \, \frac{0.023}{\Omega_b h^2} \right) \left(
  \frac{20}{1+z} \right)^3,
\label{eq:dtc}
\end{equation}
where $\Delta \fcoll \sim \fcoll$ is the effective collapse fraction appearing in the integrals of equation (\ref{eq:jalpha}).  Note that $\Delta T_c$ is independent of $f_\star$ because both the coupling and heating rates are proportional to the star formation rate.  Interestingly, for our fiducial (Pop II) parameters $z_c$ precedes $z_h$ (see also the simple models of \cite{chen04,hirata05, sethi05}).  This could create a significant absorption epoch whose properties offer a meaningful probe of the first sources.  For example, very massive Pop III stars have a smaller $N_\alpha$, and an early miniquasar population could completely eliminate the absorption epoch.

A similar estimate of the ionization fraction $\bar{x}_{i,c}$ at $z_c$ yields
\begin{equation}
\bar{x}_{i,c} \sim 0.05 \left( \frac{\fesc}{1+\bar{n}_{\rm rec}} \,
\frac{N_{\rm ion}}{N_\alpha} \, \frac{\fcoll}{\Delta \fcoll} \,
\frac{1}{S_\alpha} \, \frac{0.023}{\Omega_b h^2} \right) \left( \frac{20}{1+z} \right)^2.
\label{eq:xic}
\end{equation}
For Pop II stars, $N_{\rm ion}/N_\alpha \approx 0.4$; thus even in the worst case of $\fesc=1$ and $\bar{n}_{\rm rec}=0$ coupling would become efficient during the initial stages of reionization.  However, very massive Pop III stars have much harder spectra, with $N_{\rm ion}/N_\alpha \approx 7$. In principle, it is therefore possible for Pop III stars to reionize the universe \emph{before} $z_c$, although that would require \emph{extremely} unusual parameters (cf. \cite{ciardi03-21cm}).  Such histories cannot be ruled out at present, but we regard them as exceedingly unlikely.  Histories with $\bar{x}_{i,c} \ll 1$ are much more plausible, at least given our theoretical prejudices about high-redshift sources.

Finally, we ask whether the IGM will appear in absorption or emission during reionization.  Combining equations (\ref{eq:xrayemiss}) and (\ref{eq:bxion}), we have
\begin{equation}
\frac{\Delta T}{T_\gamma} \sim \left( \frac{\bxion}{0.025} \right) \,
\left( f_X \, \frac{f_{X,h}}{\fesc} \, \frac{4800}{N_{\rm ion}} \,
\frac{10}{1+z} \right) \, (1 + \bar{n}_{\rm rec})
\label{eq:tx-xi}
\end{equation}
for the heat input $\Delta T$ as a function of $\bxion$. Thus, provided $f_X \ga 1$, the IGM will be much
warmer than the CMB during the bulk of reionization.  This is convenient in that $\dtb$ becomes independent of $T_S$ when $T_S \gg T_\gamma$, so it is easier to isolate the effects of the ionization field.  Significant absorption during reionization becomes more plausible for very massive Pop III stars, because they have much larger ionizing efficiencies (although their remnants may also induce correspondingly large X-ray heating).

\subsubsection{Some Example Histories} \label{globreion-models}

\begin{figure}[!t]
\centerline{\epsfig{file=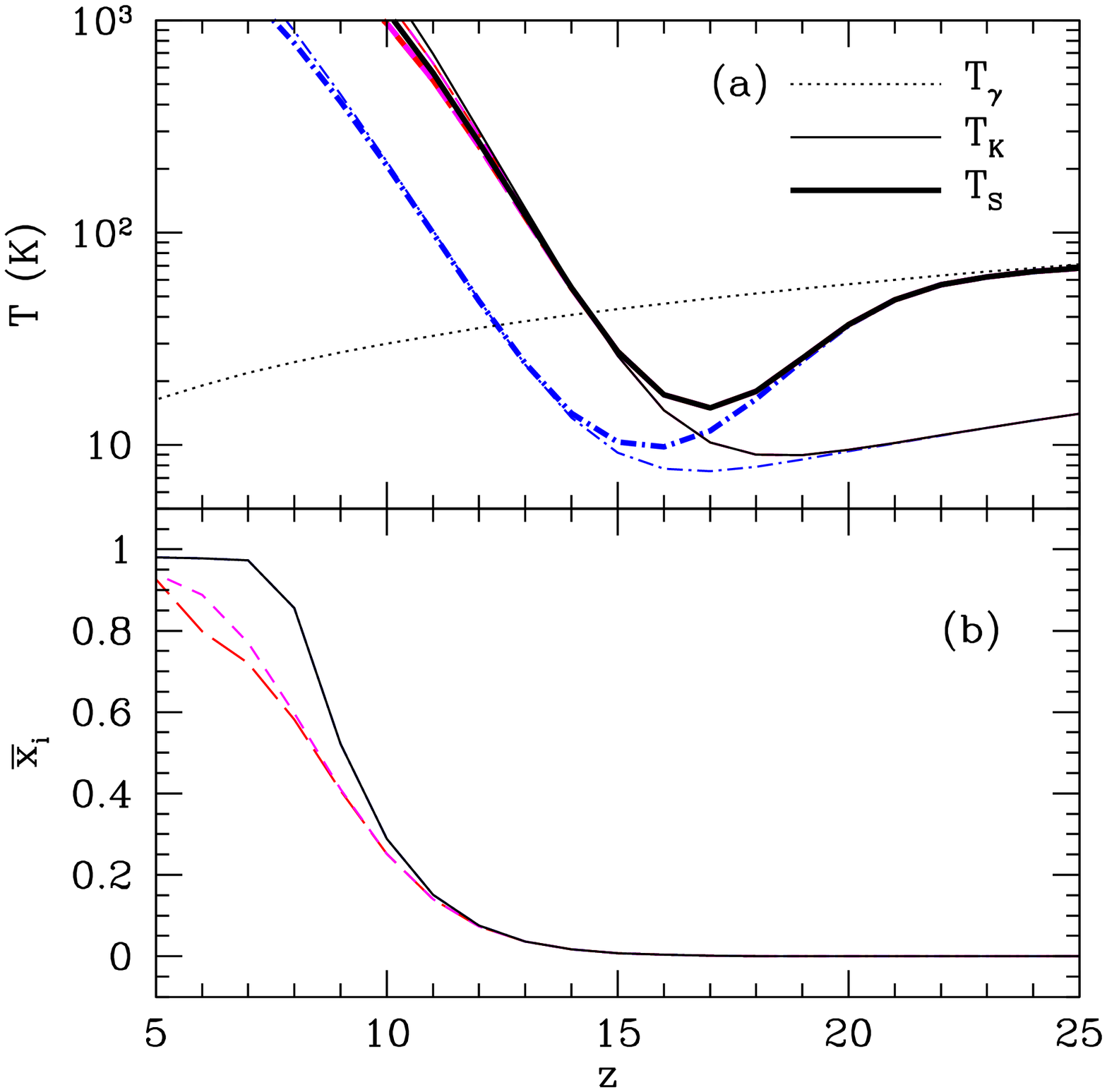,width=2.75in}
\epsfig{file=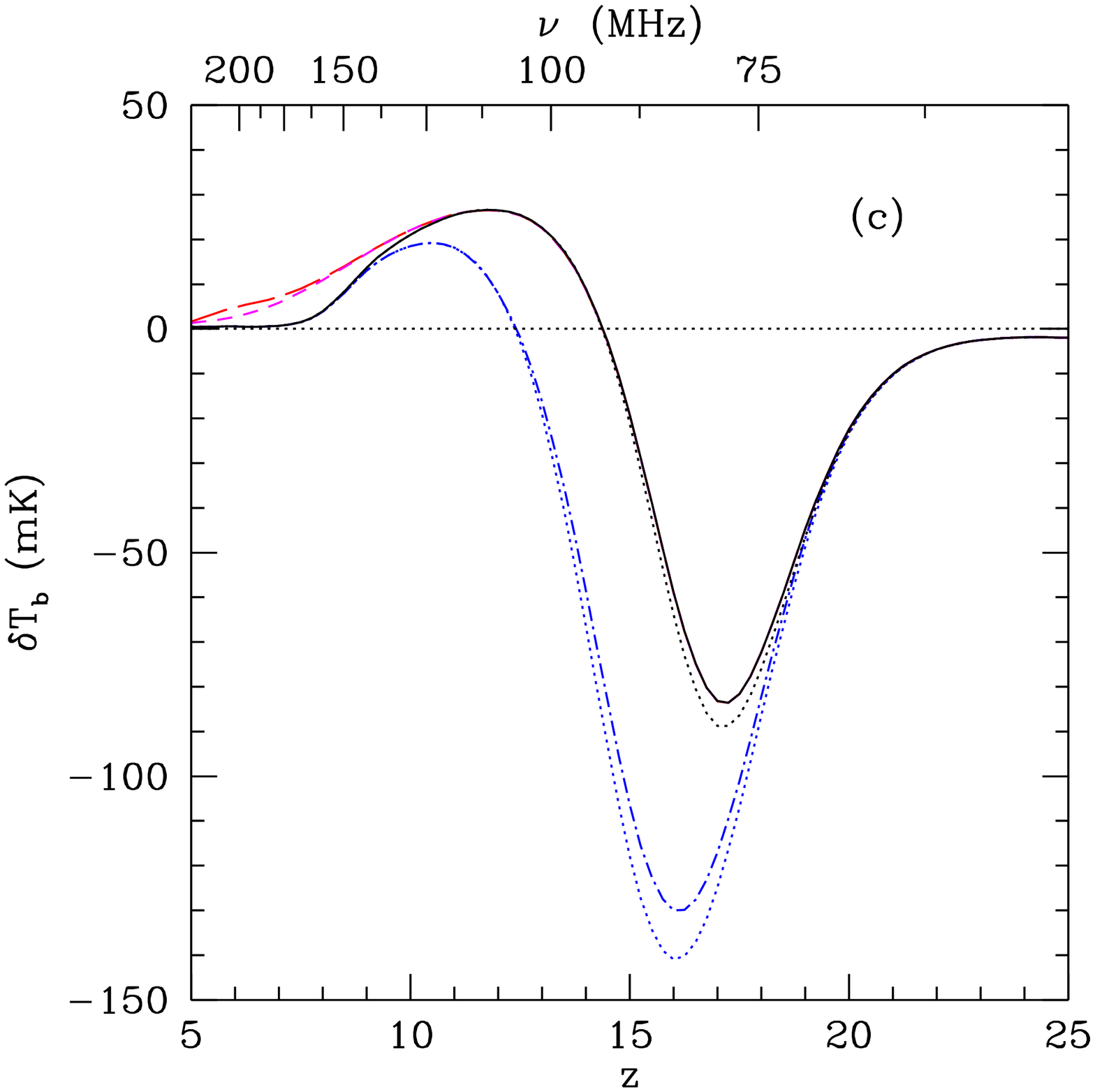,width=2.75in}}
\caption{Global IGM histories for Pop II stars.  The solid curves take our fiducial parameters without feedback.  The dot-dashed curve takes $f_X=0.2$.  The short- and long-dashed curves include strong photoheating feedback. \emph{(a)}:  Thermal properties. \emph{(b)}: Ionized fraction.  \emph{(c)}:  Differential brightness temperature against the CMB.  In this panel, the two dotted lines show $\delta T_b$ without including shock heating.  From \cite{furl06-glob}.}
\label{fig:pop2-glob}
\end{figure}

We will now use some representative models chosen from \cite{furl06-glob} to illustrate these qualitative features in a more concrete fashion (see also \cite{chen04, sethi05, hirata05}).  We begin with a fiducial set of Pop II parameters.  We ignore feedback (of all kinds) and take $\mmin$ to correspond to $T_{\rm vir}=10^4 \kel$, $f_\star=0.1$, $\fesc=0.1$, $f_X=1$, $N_{\rm ion}=4000$, and $N_\alpha=9690$.  (Thus $\zeta=40$ for this model.)  Figure~\ref{fig:pop2-glob}\emph{a} shows the resulting temperature history.  The dotted curve is $T_\gamma$, the thin solid curve is $T_K$, and the thick solid curve is $T_S$.  As expected from equation (\ref{eq:dtc}), in this case we do indeed find that $z_c>z_h$; specifically, $z_c \approx 18$ and $z_h \approx 14$.  Clearly \lya coupling is extremely efficient for normal stars.

%%%%%%%%%%%% FIGURE 3-2: Global history, pop II 

The solid curve in Figure~\ref{fig:pop2-glob}{\em b} shows the corresponding ionization history, with the clumping factor computed following \cite{miralda00} (which assumes that the lowest density regions are ionized first).  It increases smoothly and rapidly over a redshift interval of $\Delta z \sim 5$, ending at $z_r \sim 7$.  That is of course purely a function of our choice for $\zeta$, but other values do
not strongly affect the width.

Figure~\ref{fig:pop2-glob}{\em c} shows the corresponding 21 cm brightness temperature decrement $\dtb$ relative to the CMB.  Here we have also labeled the corresponding (observed) frequency $\nu$ for
convenience.  The signal clearly has interesting structure.  At the highest frequencies, reionization causes a steady decline in the signal, with $|\deriv \bdtb/\deriv \nu| \sim 1 \mkel \MHz^{-1}$.  In this model, recombinations are relatively inefficient; the only way to significantly increase the gradient during reionization would be with some positive feedback mechanism.  

However, as illustrated by the dashed curves, it is relatively easy to slow reionization.  These curves use two models for photoheating feedback \cite{furl05-double} in which the minimum virial temperature for galaxy formation increases to $T_h=2 \times 10^5 \kel$ in photoheated regions, near the upper limit of theoretical expectations \cite{dijkstra04-feed}.  It slows the evolution when $\bxion \ga 0.5$ (though it remains monotonic), decreasing $|\deriv \bdtb/\deriv \nu|$ by about a factor of two.  The ionization history can be slowed by an even larger factor by also decreasing $\zeta$ in heated regions, and in extreme circumstances this could even cause $\bxion$ to decrease slightly (see eq. \ref{eq:doublecond}).  Unfortunately, such a ``recombination epoch" would have spectral gradients no larger than reionization itself (though with opposite sign, of course).

Figure~\ref{fig:pop2-glob}\emph{c} contains an even more striking feature at higher redshifts.  At $z \sim 30$, the IGM is nearly invisible even though $T_K \ll T_\gamma$ (see Fig.~\ref{fig:tevol}).  However, as the first galaxies form, the Wouthuysen-Field effect drives $T_S \rightarrow T_K$.  Because $z_c > z_h$, this produces a relatively strong absorption signal ($\dtb \approx -80 \mkel$) over the range $z \sim 21$--$14$ (or $\nu \sim 70$--$95 \MHz$).  However, the IGM still heats up well before reionization begins in earnest, making $\bdtb$ nearly independent of $T_S$ throughout reionization.

%%%%%%%%%%%% FIGURE 3-3: Global history, pop III
\begin{figure}[!t]
\centerline{\epsfig{file=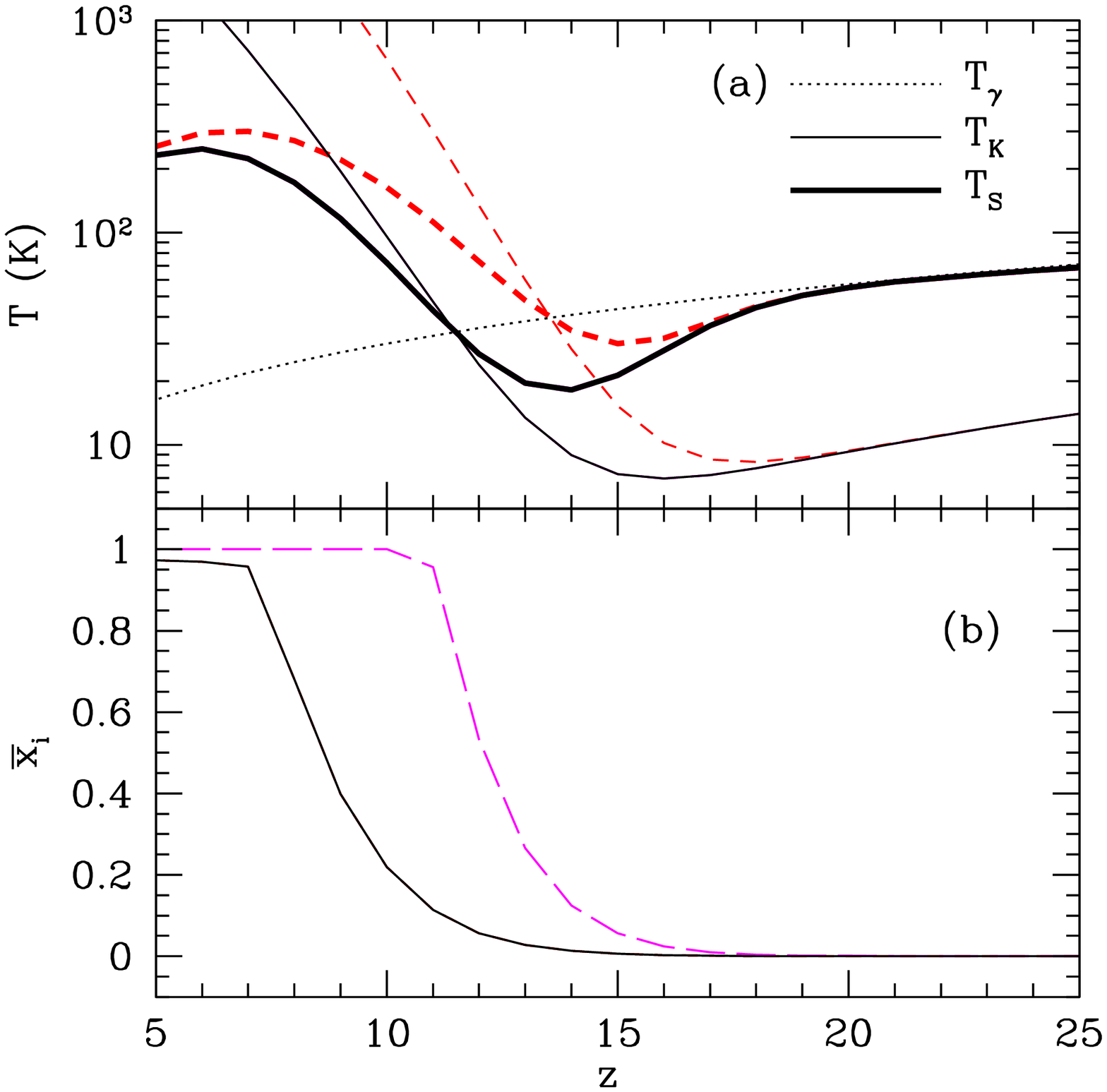,width=2.75in}
\epsfig{file=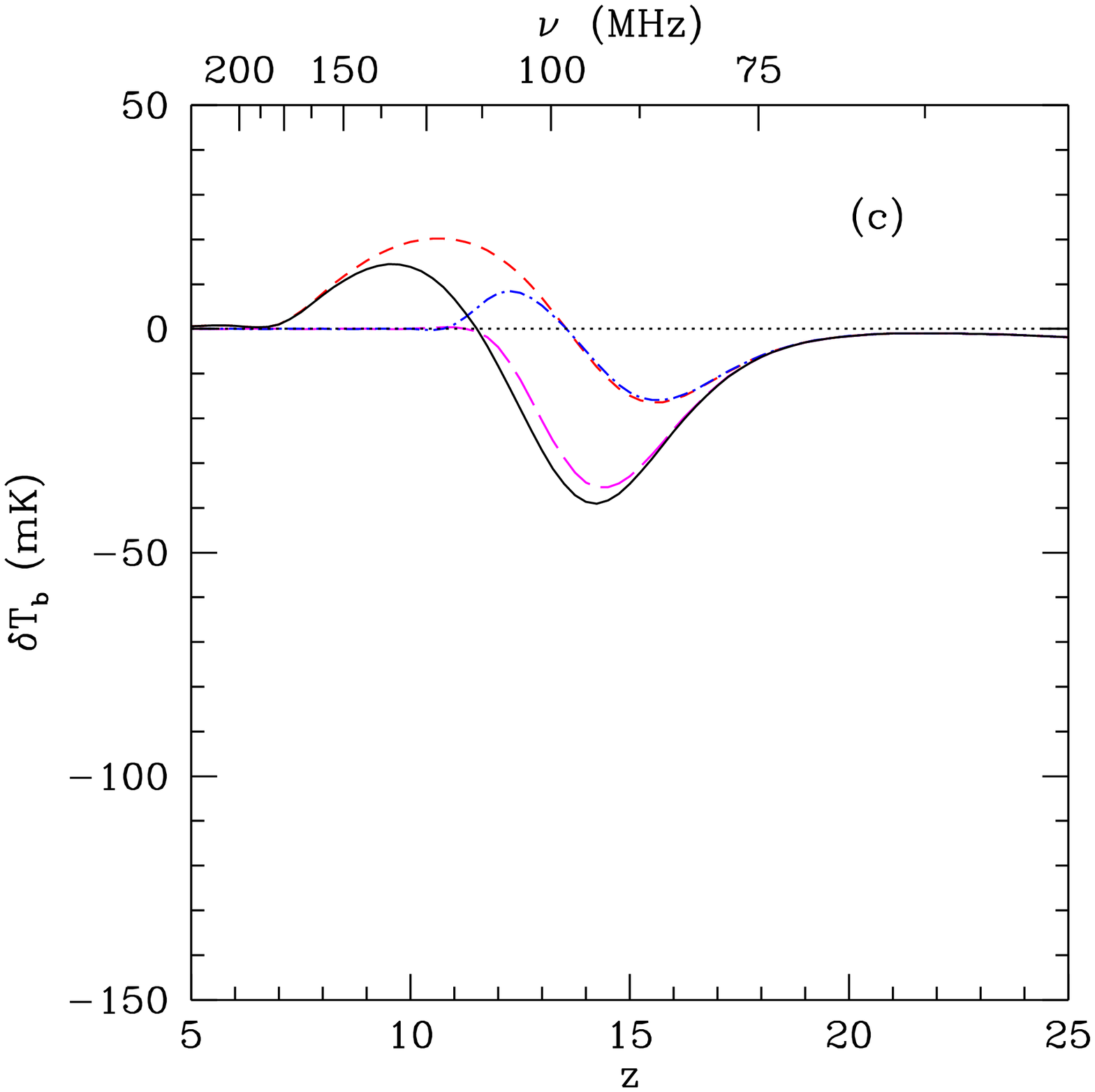,width=2.75in}}
\caption{Global IGM histories for very massive Pop III stars.  Panels are the same as in Fig.~\ref{fig:pop2-glob}.  The solid curve takes our fiducial Pop III parameters.  The long-dashed lines take $\fesc=1$, the short-dashed lines take $f_X=5$, and the dot-dashed line (shown only in {\em c}) assumes $\fesc=1$ and $f_X=5$.  From \cite{furl06-glob}.}
\label{fig:pop3-glob}
\end{figure}

Figure~\ref{fig:pop3-glob} shows similar histories for very massive Pop III stars.  The solid curves take $\mmin$ to correspond to $T_{\rm vir}=10^4 \kel$, $f_\star=0.01$, $\fesc=0.1$, $f_X=1$, $N_{\rm ion}=30,000$, and $N_\alpha=4800$, yielding $\zeta=30$.  Although the thermal history is qualitatively similar to the Pop II case, it has $z_c \sim 13$ and $z_h \sim 11$.  Thus the absorption epoch is somewhat narrower, and it is also weaker because Pop III stars produce relatively few \lya photons.  As a result, $T_S$ does not approach $T_K$ until the IGM is already hot.  Thus, if very massive Pop III stars dominate, the absorption epoch will be considerably weaker, with gradients about half as large as the Pop II case.    Moreover, $z_h$ is relatively close to $z_r$, so $T_S$ does not saturate until after reionization begins.  It may therefore be somewhat difficult to separate $T_S$ and $x_i$ at the beginning of reionization.

Obviously, measuring this background could offer strong constraints on high-redshift star formation.  The other curves in Figures~\ref{fig:pop2-glob}--\ref{fig:pop3-glob} illustrate the range of features we expect.  However, they all share one crucial property:  $z_c$ occurs long before reionization, so we can safely expect \emph{some} signal from the high-redshift IGM.

\subsection{Observational Prospects} \label{glob-obs}

One obvious application of the 21 cm signal is to measure this ``all-sky'' spectrum $\bdtb(z)$ \cite{madau97,shaver99,tozzi00,gnedin04}.  The wide range of histories shown in Figures~\ref{fig:pop2-glob}--\ref{fig:pop3-glob} illustrate how powerful such observations would be (see also \cite{sethi05}).  The models we have considered imply gradients $|\deriv \bdtb/\deriv \nu| \sim 1 \mkel \MHz^{-1}$ during reionization and possibly somewhat larger values during a preceding absorption epoch.

Because this is an all-sky signal, single-dish measurements (even with a modest-sized telescope) can easily reach the required mK sensitivity.  However, the much stronger synchrotron foregrounds (see \S \ref{int} for a detailed discussion) nevertheless make such observations extremely difficult: they have $T_{\rm sky} \ga 200$-$2000 \kel$ over the relevant frequencies.  The fundamental strategy for extracting the cosmological signal relies on the expected spectral smoothness of the foregrounds (which primarily have power law synchrotron spectra), in contrast to the non-trivial structure of the 21 cm background.  Nevertheless, these foregrounds have $|\deriv T_{\rm sky}/\deriv \nu| \ga 3 \kel \MHz^{-1}$, so extracting the high-redshift component will be a challenge that requires extremely accurate calibration over a wide frequency range \cite{shaver99, gnedin04} and, most likely, sharp localized features in $\bdtb(z)$ that can be distinguished from smoother foreground features.

The limiting factors in these measurements are likely to be systematics \cite{shaver99}.  Foregrounds vary across the sky, coupling to the frequency-dependent sidelobes (though the variation should be smooth and hence removable \cite{oh03}).  One strategy to avoid confusion with features in the foreground is to use multiple (wide) fields across which foregrounds are uncorrelated.  Another problem may be the recombination lines of hydrogen and other common elements, which give rise to spectral features throughout the relevant frequency range. Fortunately, the frequencies of these lines are known precisely, and their relative strengths are related through simple physics. Observations seeking the detection of a global signal will necessarily be made with high spectral resolution, both to separate the recombination lines and to prevent contamination from terrestrial radio frequency interference.

The instrumental calibration required to detect these faint, $\sim 10^{-5}$ inflections in the radio spectrum is an exceptionally difficult task. The usual strategy is to compare the sky spectrum with a calibration source that is known to be smooth. Two clear questions arise: first, is the comparison source truly smooth, and second, can the two signals be brought together in the receiving system through paths with identical (or at least calibrate-able) gains? One approach is to use internal resistive loads that provide  thermal power in proportion to the load's temperature. In this case, the impedance matching is critical, and there must be ways to accurately assess matching (or reflections) in the signal paths from the antenna that observes the sky and from the internal load. Alternatively, one can search for sources in the sky that could provide ``external loads" with smooth spectra; one possible source is the Moon \cite{shaver99}, which emits thermally at these freqencies (although its brightness temperature is actually cooler than the sky temperature at the low-frequency end of the relevant range). The advantage of an external load is that the calibration signal enters the receiving system through the same path as the sky signal.  However, the Moon is reflective (as well as emissive), so that a faint reflection of the Earth radio frequency interference is mirrored at the Moon, making that a likely contaminant.

The precision measurement of the shape of the CMB spectrum by the FIRAS instrument on COBE \cite{mather99} makes a sobering comparison.  FIRAS compared the CMB spectrum from the sky with an onboard black body source to determine precisely the CMB spectrum.  It found agreement with a Planck curve with an rms deviation of five parts in $10^5$ over the wavelength range 0.5 to 5~mm.  This was a space-based instrument, observing a thermal sky spectrum in comparison to a thermal load, in a wavelength range where radio interference is negligible.  The highly-redshifted 21 cm spectrum will surely be much more difficult to measure.

Fortunately, even if systematics do prove to be insurmountable, there are other ways to measure the evolution of the mean signal.  For example, Compton scattering by a massive, low-redshift galaxy cluster shifts the spectrum in frequency space in a predictable way, so that differential measurements between the cluster and its surroundings could be made \cite{cooray05-glob}.  This avoids many of the systematic difficulties, but it requires relatively high-resolution observations (and hence may not measure the truly ``global" signal).  Another problem is the association of strong radio galaxies with the cores of massive galaxy clusters, because the continuum flux density will be systematically greater than the surrounding
reference fields. Slight errors in gain calibration as a function of frequency (``passband calibration'') may leave residuals that mimic the expected global signal. The calibration error would likely be present for all clusters, although meticulous comparison with strong radio sources unassociated with X-ray clusters may be helpful.  A second possibility is to use the 21 cm fluctuations to measure the mean background through their redshift-space anisotropies \cite{barkana05-vel}, although in practice this is beyond the capabilities of the first generation of interferometers \cite{mcquinn05-param}.  Finally, if the ionized bubbles can be separated from the residual neutral gas (through imaging or one-point statistics \cite{hansen06}, for example), their contrast and the abundance of each phase provides a straightforward measurement of $\bdtb$.

%\bibliographystyle{elsart-num}
%\bibliography{Ref_21cm}

%\end{document}

%% file: powerspec-ch4.tex
%\documentclass{elsart}
%\usepackage{amssymb,cite,epsfig}

%\input{./defns.tex}

%\begin{document}

\section{The Power Spectrum} \label{ps}

While the global 21 cm background contains a great deal of information about the mean evolution of the sources, each and every component discussed in \S \ref{glob} also fluctuates significantly.  For the density field this is obvious:  the evolving cosmic web imprints growing density fluctuations on the matter distribution.  For the other aspects, the discrete nature of the luminous sources gives rise to 21 cm fluctuations.  Ionized gas is organized into discrete \htwo regions (at least in the most plausible models), and the \lya background and X-ray heating will also be concentrated around galaxies.  The single greatest advantage of the 21 cm line is that it allows us to separate this fluctuating component both on the sky and in frequency (and hence cosmic time).  Thus we can study the sources and their effects on the IGM in detail.  It is the promise of these ``tomographic" observations that makes the 21 cm line such a singularly attractive probe, and in the next several sections we will describe their expected forms in many different physical regimes.

Observing the 21 cm fluctuations has one practical advantage as well.  The difficulty of extracting the global evolution lies in its relatively slow evolution.  On the small scales relevant to the fluctuations, the gradients increase dramatically.  At the edge of an \htwo region, for example, $\dtb$ drops by $\sim 20 \mkel$ essentially instantaneously.   As a result, separating them from the smoothly varying astronomical foregrounds may be much easier.  We will discuss this and other experimental issues in \S \ref{int}.  Unfortunately, as we will also see, constructing detailed images will remain extremely difficult because of their extraordinary faintness; telescope noise is comparable to or exceeds the signal except on rather large scales.  Thus a great deal of attention has recently focused on using statistical quantities readily extractable from low signal-to-noise maps to constrain the IGM properties.  This is motivated in part by the success of CMB measurements and galaxy surveys at constraining cosmological parameters through the power spectrum.  In our case, although any number of statistical quantities may be useful (especially during reionization, when the fluctuations are highly non-gaussian), we will take the power spectrum as our primary example.  In the next several sections we will describe how both imaging and statistics can teach us about the high-redshift Universe.  The goal of this section is to develop the formalism necessary to compute the 21 cm power spectrum.  Most of our discussion will be sufficiently general that it can easily be extended to other statistics of interest.

We first define the fractional perturbation to the brightness temperature, $\delta_{21}({\bf x}) \equiv [\dtb({\bf x}) - \bdtb]/\bdtb$, a zero-mean random field (here $\dtb$ is to be evaluated at the observer).  We will be interested in its Fourier transform $\tilde{\delta}_{21}({\bk})$. Its power spectrum is defined to be
\begin{equation}
\VEV{ \tilde{\delta}_{21}(\bk_1) \, \tilde{\delta}_{21}(\bk_2) } \equiv (2 \pi)^3 \delta_D(\bk_1 + \bk_2) P_{21}(\bk_1),
\label{eq:pkdefn}
\end{equation}
where $\delta_D(x)$ is the Dirac delta function and the angular brackets denote an ensemble average.  Power spectra for other random fields (such as the fractional overdensity $\delta$, the ionized fraction, etc.), or cross-power spectra between two different fields, can be defined in an analogous fashion.  In general, a power spectrum $P(\bk)$ is the three-dimensional Fourier transform of the corresponding two-point function and thus parameterizes the correlations present in the appropriate field.  We will often use the dimensionless version $\Delta^2(\bk) = (k^3/2 \pi^2) P(\bk)$, which roughly quantifies the variance when the field is smoothed on the scale $x=2 \pi/k$.

As is obvious from equations~(\ref{eq:dtb}) and (\ref{eq:xdefn}), the brightness temperature depends on a number of input parameters.  Expanding those equations to linear order in each of the perturbations, we can write 
\begin{equation}
\delta_{21} = \beta \delta_b + \beta_x \delta_x + \beta_\alpha \delta_\alpha + \beta_T \delta_T -  \delta_{\partial v},
\label{eq:d21}
\end{equation}
where each $\delta_i$ describes the fractional variation in a particular quantity: $\delta_b$ for the baryonic density, $\delta_\alpha$ for the \lya coupling coefficient $x_\alpha$, $\delta_x$ for the neutral fraction (note that using the ionized fraction would cause a sign change), $\delta_T$ for $T_K$, and  $\delta_{\partial v}$ for the line-of-sight peculiar velocity gradient.  The expansion coefficients $\beta_i$ are
\begin{eqnarray}
\beta & = & 1 + \frac{x_c}{x_{\rm tot}(1+x_{\rm tot})},
\label{eq:beta} \\
\beta_x & = & 1 + \frac{x_c^{\rm HH} - x_c^{\rm eH}}{x_{\rm tot} (1 + x_{\rm tot})}
\label{eq:betax} \\
\beta_\alpha & = & \frac{x_\alpha}{x_{\rm tot}(1+x_{\rm tot})},
\label{eq:beta-alpha} \\
\beta_T & = & \frac{T_\gamma}{T_K - T_\gamma} + \frac{1}{x_{\rm tot}(1+x_{\rm tot})} \left( x_c^{\rm eH} \frac{\deriv \ln \kappa_{10}^{\rm eH}}{\deriv \ln T_K} + x_c^{\rm HH} \frac{\deriv \ln \kappa_{10}^{\rm HH}}{\deriv \ln T_K} \right),
\label{eq:betaT} 
\end{eqnarray}
where $x_{\rm tot} \equiv x_c + x_\alpha$ and we have split the collisional term into the dominant H-e$^-$ and H-H components ($x_c^{\rm eH}$ and $x_c^{\rm HH}$, respectively) where necessary.  Here we have assumed $T_c = T_K$ throughout; this is reasonable in most cases but, if not, the expressions become much more complicated.  Note that, by linearity, the Fourier transform $\tilde{\delta}_{21}$ can be written in a similar fashion.  Each of these expressions has a simple physical interpretation.  For $\beta$, the first term describes the increased matter content and the second describes the increased collisional coupling efficiency in dense gas.  For $\beta_x$, the two terms describe direct fluctuations in the ionized fraction and the effects of the increased electron density on $x_c$.  (The latter is only important in partially ionized regions; 21 cm emission is negligible in HII regions, of course.)  $\beta_\alpha$ simply measures the fractional contribution of the Wouthuysen-Field effect to the coupling.  The first term in $\beta_T$ parameterizes the speed at which the spin temperature responds to fluctuations in $T_K$, while the others include the explicit temperature dependence of the collision rates.  The challenge of the next several sections will be to compute these perturbation fields in the appropriate physical regimes.  Note that all of these terms, with the crucial exception of $\delta_{\partial v}$, are isotropic (see \S \ref{redshift-dis}).  

Of course, the 21 cm background directly measures the baryonic density field $\delta_b$ (or even more precisely, the hydrogen density field).  For most purposes, this is equivalent to the total matter density $\delta$ and in the following we will set $\delta_b=\delta$ throughout.  However, note that on small scales the finite pressure of the baryons introduces a cutoff absent from the dark matter \cite{naoz05}; in detail, galaxy formation processes and feedback can also work on the two separately.

\begin{figure}[!t]
\centerline{\epsfig{file=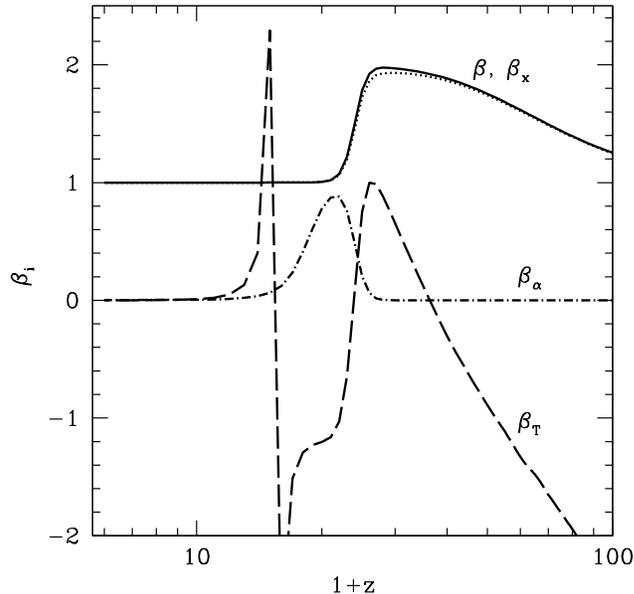,width=3.5in}}
\caption{Redshift dependence of perturbative expansion coefficients in the fiducial Pop II model of Fig.~\ref{fig:pop2-glob}.  We show $\beta$ (solid curve), $\beta_x$ (dotted curve), $\beta_\alpha$ (dot-dashed curve), and $\beta_T$ (dashed curve).  Note that the singularity in $\beta_T$ at $z=17$ is artificial in that it does not actually appear in the fluctuation amplitude.}
\label{fig:betaz}
\end{figure}

For context, Figure~\ref{fig:betaz} shows how these expansion coefficients evolve in our fiducial Pop II structure formation model (see \S \ref{globreion-models}).  The density coefficient $\beta$ increases with time until $z \sim 20$ before abruptly falling to unity.  At $z \ga 20$, collisions are only marginally important so the extra collisional coupling imparted by an increased density has a relatively large effect; at lower redshifts, collisional coupling is negligible compared to the Wouthuysen-Field effect so the second term in equation~(\ref{eq:beta}) vanishes.  $\beta_x$ behaves nearly identically, because (outside of \htwo regions) the ionized fraction remains small.  Fluctuations in the \lya background are only important over a limited redshift range (where $x_\alpha \sim 1$); at lower redshifts, all the gas is strongly coupled so fluctuations in the background are unimportant.  The temperature coefficient has the most complicated dependence because it depends on the mix of Compton heating and collisional coupling.  Note that the apparent singularity occurs where $T_K=T_\gamma$; it is not physical because $\bdtb$ also vanishes at the same point.  At lower redshifts, $T_K \gg T_\gamma$ and the emission saturates, $\beta_T \rightarrow 0$.

%%%%%%%%%%%% FIGURE 4-1: beta_i(z)

By equation~(\ref{eq:pkdefn}), the power spectrum clearly contains all possible terms of the form $P_{\delta_i \delta_j}$; some or all could be relevant in any given situation.  Of course, in most instances the various $\delta_i$ will be correlated in some way; statistical 21 cm observations ideally hope to measure these separate quantities.  We have already included some of the obvious correlations in equations~(\ref{eq:beta})--(\ref{eq:betaT}), such as the variation of the collision rate with the ionized fraction.  But we have left others implicit:  for example, as we will see later, it is likely that overdense regions are ionized first.  Or, if photoionization equilibrium applies, $\delta_x$ depends on the density field.  A more subtle example is the relation of $\delta_\alpha$ to the other quantities; as we saw in \S \ref{fokker}, it depends on the radiation spectrum and hence on density, neutral fraction, and temperature in addition to the background flux.  We have left this implicit in the interests of simplicity (and it is often ignored in the literature; \cite{barkana05-ts, pritchard05}), but it should be included in detailed calculations.

In all of these expansions, one must bear in mind that $\delta_x$ is of order unity if the ionization field is built from \htwo regions.  In that case terms such as $\delta \delta_x$ are in fact \emph{first} order and must be retained.  This leads to non-trivial four-point terms in the power spectrum (e.g., \cite{mcquinn05-param}); in practice, these terms may need to be computed along with the two-point correlations.

Because of the analogy to the CMB, some of the 21 cm literature makes use of the angular power spectrum (e.g., \cite{zald04, bharadwaj04-vel}).  This integrates out the line of sight information (which is automatically measured in any real experiment thanks to the frequency axis), so the three-dimensional version is generally preferred (especially at large angular scales, where the shape of the matter power spectrum is such that relatively small scale density fluctuations dominate \cite{barkana05-vel}).  But the angular power spectrum is still useful for some purposes -- especially in correlations with the CMB or other two-dimensional fields on the sky \cite{cooray04-lens, cooray04-cmbt, cooray04-cmbpol, mandel06, zahn05-lens, alvarez06} -- so for convenience we will define it here as well.  For the moment we neglect the velocity term.  In that case $\delta_{21}$ is isotropic, and the brightness temperature (relative to the CMB) of the sky at frequency $\nu$ can be written
\begin{equation}
\delta_{21}(\bhn,\nu) = \int \deriv r \, R(r; r_0) \, \delta_{21}(\bhn,r),
\label{eq:dtb-sky}
\end{equation}
where the integral is over the line of sight distance and $R(r;r_0)$ describes the frequency response of the experiment; it is typically sharply peaked around $r_0$, the conformal distance to the redshift of interest.  We construct the angular power spectrum from the spherical harmonic expansion of $\delta_{21}$, conventionally written
\begin{equation}
\delta_{21}(\bhn,\nu) = \sum_{lm} a_{lm}(\nu) Y_{lm}(\bhn).
\label{eq:angps}
\end{equation}
Using the identity
\begin{equation}
e^{i\bk \cdot {\bf x}} = \sum_{lm} 4 \pi i^l j_l(kr) Y_{lm}^*(\bhk) Y_{lm}(\bhn),
\label{eq:exp-ylm}
\end{equation}
where $j_l(x)$ is the spherical Bessel function of order $l$, we find
\begin{eqnarray}
a_{lm}(\nu) & = & 4 \pi i^l \int \frac{\deriv^3 k}{(2 \pi)^3} \alpha_l(k,\nu) Y_{lm}^*(\bhk) \tilde{\delta}_{21}(k,\nu),
\label{eq:alm-21cm} \\
\alpha_l(k,\nu) & = & \int \deriv r \, R(r; r_0) j_l(kr).
\end{eqnarray}
The (dimensionless) cross-frequency angular power spectrum is defined by
\begin{equation}
\VEV{ a_{l_1m_1}(\nu_1) a^*_{l_2 m_2}(\nu_2)} \equiv \delta_{l_1 l_2} \delta_{m_1 m_2} C_{l_1}(\nu_1,\nu_2).
\label{eq:angps-defn}
\end{equation}
Performing the integrals,
\begin{equation}
C_l^{21}(\nu_1,\nu_2) = 4 \pi \int \frac{k^2 \, \deriv k}{2 \pi^2} P_{21}(k) \alpha_l(k,\nu_1) \alpha_l(k,\nu_2).
\label{eq:cldefn}
\end{equation}
In temperature units, equation~(\ref{eq:cldefn}) takes a prefactor $\bdtb^2$.

To gain some intuition, consider two limiting forms of equation~(\ref{eq:cldefn}) for a pure power law input spectrum\cite{zald04}.  First suppose that the frequency response is a delta function (valid on angular scales much larger than the radial response, or small $l$); then 
\begin{equation}
\frac{l^2 C_l^{21}(\nu, \nu)}{2 \pi} \propto \bdtb^2(\nu) \Delta_{21}^2(l/r_0), \qquad \qquad l \, \delta r/r \ll 1
\label{eq:cl-lim1}
\end{equation}
where $\delta r$ characterizes the bandwidth of the observation.  Thus on large angular scales, the angular fluctuations should simply trace the corresponding density fluctuations.  (This is not, however, true for 21 cm fluctuations, because the power-law approximation is not valid \cite{barkana05-vel}.)  Small scale modes, on the other hand, suffer a cancellation from the oscillatory Bessel functions, and 
\begin{equation}
\frac{l^2 C_l^{21}(\nu, \nu)}{2 \pi} \propto \bdtb^2(\nu) \Delta_{21}^2(l/r_0) \frac{r_0}{l \, \delta r}. \qquad \qquad l \, \delta r/r \gg 1
\label{eq:cl-lim2}
\end{equation}
Here the angular fluctuations are suppressed by a factor that equals the number of wavelengths fitting across the band.  Clearly, retaining the frequency dimension will avoid this cancellation, so direct measurements of $P_{21}(k)$ are preferred.

\subsection{Redshift-Space Distortions} \label{redshift-dis}

In general, we expect fluctuations in density, ionization fraction, Ly$\alpha$ flux, and temperature, to be statistically isotropic, because the physical processes responsible for them have no preferred direction [e.g., $\delta(\bk) = \delta(k)$].\footnote{Actually, this assumption can break down on extremely large scales, because then the growth of structure with redshift becomes important.  Fortunately, the 21 cm field only contains rapidly evolving features on such large scales at the tail end of reionization.  The evolution is generally not important on the scales accessible to observations \cite{mcquinn05-param, barkana05-cone}.}  However, there are two effects that do break the isotropy of 21 cm fluctuations, both well-known from previous studies of large scale structure. The first is that transverse and light of sight distances scale differently in non-Euclidean spacetimes, which artificially distorts the appearance of any isotropic distribution. This well-known Alcock-Paczy\'{n}ski (AP) effect \cite{alcock79,nusser05-ap,saiyadali05,barkana_AP} depends on the underlying background cosmology, and we will discuss it in \S \ref{costest}.\footnote{For the moment, we shall assume that that the true underlying cosmology is known from, e.g., CMB studies; in that case the AP effect can be ignored.} Second, peculiar velocity gradients introduce redshift space distortions.  Bulk flows on large scales, and in particular infall onto massive structures, compress the signal in redshift space (the ``Kaiser" effect), enhancing the apparent clustering amplitude \cite{kaiser87,tozzi00,bharadwaj04-vel,barkana05-vel}. On small-scales, random motions in virialized regions create elongation in redshift space (the ``finger of God" effect), reducing the apparent clustering amplitude \cite{wang05}. We will now discuss each of these effects in turn. 

The anisotropy in the power spectrum induced by the Kaiser effect could allow an interesting separation of astrophysical and cosmological contributions to the 21 cm fluctuations \cite{barkana05-vel}. On large scales, where linear theory holds, the fractional perturbation in the radial peculiar velocity gradient $\deriv v_{r}/\deriv r$ has a Fourier transform proportional to that of the density field, $\tilde{\delta}_{\partial v} = - \mu^{2} f \tilde{\delta}$, where $\mu$ is the cosine of the angle between the wavevector ${\bf k}$ and the line of sight direction and $f = \deriv \ln D/\deriv \ln a \approx \Omega_m^{0.6}(z)$ \cite{kaiser87}. (Intuitively, the two powers of $\mu$ result from taking the line of sight components of the peculiar velocity and of its gradient.)  Thus in Fourier space, brightness temperature fluctuations in Fourier space have the form \cite{bharadwaj04-vel}:
\begin{equation}
\tilde{\delta}_{21} =  \mu^{2} f \tilde{\delta} + \tilde{\delta}_{\rm iso} 
\label{eq:d21ft}
\end{equation}
where we have collected  all the statistically isotropic terms in equation~(\ref{eq:d21}) into $\tilde{\delta}_{\rm iso}$. Neglecting ``second-order" terms (see below) and setting $f=1$ in the high-redshift limit, the total power spectrum can therefore be written as \cite{barkana05-vel}:
\begin{equation}
P_{21}({\bf k}) = \mu^{4} P_{{\delta} {\delta}} + 2 \mu^{2} P_{{\delta}_{\rm iso} {\delta}} + P_{{\delta}_{\rm iso} {\delta}_{\rm iso}}.
\label{eqn:p_polynomial}
\end{equation}
Equation (\ref{eqn:p_polynomial}) has some interesting features. Even in the simple case where $\tilde{\delta}_{\rm iso}=\tilde{\delta}$ (i.e., there are no spin temperature or ionization fraction fluctuations), the velocity term boosts the spherically averaged power spectrum by a factor $\langle (1+ \mu^{2})^{2} \rangle=1.87$: redshift space distortions are not a small effect. Furthermore, because of the simple form of this polynomial, measuring the power at $\ge 3$ values of $\mu$ should allow one to determine $P_{\delta \delta}, P_{{\delta}_{\rm iso} {\delta}}$, and $P_{{\delta}_{\rm iso} {\delta}_{\rm iso}}$ for each $k$. In particular, we can isolate the contribution from density fluctuations $P_{\delta \delta}$.  This would not have been possible without peculiar velocity flows:  comparison to equation~(\ref{eq:d21}) shows that, in the most general case, $P_{{\delta}_{\rm iso} {\delta}}$ and $P_{{\delta}_{\rm iso} {\delta}_{\rm iso}}$ contain several different power spectra, including those of the density, neutral fraction, and spin temperature as well as their cross power spectra \cite{saiyadali05}.  Note that $P_{\delta \delta}$ is simply the baryonic matter power spectrum at redshift $z$; assuming that we have independent measurements of that, we can also determine the astrophysically interesting prefactor $\bdtb \propto \bxhi (1 - T_\gamma/T_S)$ from the velocity field. In particular, at late times, when X-ray heating and Ly$\alpha$ coupling have driven $T_{S} \gg T_{\gamma}$, we can in principle extract the mean neutral fraction $\bxhi(z)$ \cite{barkana05-vel}. 

Disentangling the other components will be more difficult, since there are several remaining power spectra to be determined from the two measured quantities $P_{{\delta}_{\rm iso} {\delta}}(k)$ and $P_{{\delta}_{\rm iso} {\delta}_{\rm iso}}(k)$.  Anything we can learn from these components will require properly modeling the processes and the correlations between them.  This will be simplest in regimes where one or more of the terms can be neglected.  For example, during the earliest stages of reionization (when $\delta_x$ is negligible), one might be able to measure the power spectrum of spin temperature fluctuations as well as its correlations with density.  At late times (when $T_{S} \gg T_{\gamma}$ and $\dtb$ becomes independent of $T_{S}$), one might likewise ignore spin temperature fluctuations and measure the ionization fraction fluctuations $P_{\delta x}$ and $P_{xx}$. In both these cases, there is generally a mild degeneracy in parameter estimation which can likely be broken by making weak assumptions about the asymptotic behaviour of the power spectra.  
 
An additional difficulty comes from the correlations of ``second-order" terms in the perturbation expansion, such as $\delta \delta_x$, that produce four-point terms in the power spectrum.  As mentioned in the previous section, $\delta_x$ is not necessarily a small parameter, so these terms can be substantial.  Unfortunately, they also produce terms with non-trivial $\mu$ dependence that depend on the particulars of reionization \cite{mcquinn05-param}.  The presence of these terms make attempts to separate the $\mu^n$ powers during reionization more difficult; the prospects are much better before $\delta_x$ becomes important.  We will return to this question in \S \ref{anisotropy}.
 
Note that the velocity term also affects the angular power spectrum; the extra $\mu^2$ terms alter the angular dependence of the signal and hence the $l$-distribution of the power spectrum \cite{bharadwaj04-vel, barkana05-vel}.  
 
On small scales, random motions in virialized regions wash out features in redshift space \cite{wang05_FoG}. If we only study line-of-sight modes (which are the most immune to foreground contamination), this places an irreducible lower limit on the size of structures we can study. However, for all of the planned experiments, the limiting angular resolution, which dilutes the radial signal by mixing together neutral and ionized structures in the transverse direction, corresponds to much larger scales than the ``fingers of God." Thus these will not affect any observables.  Even at much higher angular resolution, virial motions should not pose a significant obstacle to detecting ionized bubbles in the 21 cm data so long as photoionized bubbles are significantly larger than the virial extent of halos (almost definitely a safe assumption; see \S \ref{reion}).  

An important caveat to recovering redshift space distortions (and the AP test below) is that it requires a high signal to noise measurement of the angular structure of the signal.  Unfortunately, the noise is anisotropic: radio foregrounds have much more power across the sky than in the line of sight direction. (Indeed, this very feature is crucial to foreground removal algorithms; see \S \ref{clean}.)  Moreover, it is much easier to probe small physical scales in the frequency direction than across the angular dimensions.  As a result, taking advantage of this ``separation of powers" will likely require second generation experiments (see \cite{mcquinn05-param} and Fig.~\ref{fig:sense-mu}).

\subsection{Cosmological Tests} \label{costest}

The preceding discussion assumed that the correct underlying cosmological model was already known (from CMB measurements, for example).  Using an incorrect cosmological model creates apparent errors in the scaling of angular sizes (which depend on the angular diameter distance $D_{A}$) compared to line of sight sizes (which depend on the Hubble parameter), introducing an artificial anisotropy even in intrinsically isotropic distributions. This Alcock-Paczy\'{n}ski (AP) effect \cite{alcock79} can be used to measure cosmological parameters, though doing so in practice has proven difficult. Galaxy redshift surveys generally do not extend to sufficiently high redshifts for it to be important (although they may become useful if supplemented by deep galaxy cluster surveys \cite{hu_haiman03}). Quasar surveys (e.g., SDSS), on the other hand, do have sufficient redshift depth but tend to be too sparse for optimal parameter estimation \cite{matsubara02}. In principle the AP effect can be measured through distortions of the Ly$\alpha$ forest power spectrum \cite{hui99}, but the practical challenges have proven substantial, and the first results are only just arriving \cite{eriksen05}. Because the 21 cm signal is all-sky and so does not suffer from sparseness problems, one might hope that it will allow a definitive detection of the AP effect \cite{scott90, nusser05-ap, barkana_AP}.

In keeping with our former discussion we consider the AP effect in terms of the power spectrum \cite{barkana_AP}, though a presentation in terms of the correlation function is equally illuminating \cite{nusser05-ap}. The AP effect distorts the shape and normalization of the 21 cm power spectrum to \cite{barkana_AP}:
\begin{equation}
P_{21}(k)= \mu^{6} P_{\mu^{6}} (k) + \mu^{4} P_{\mu^{4}}(k) + \mu^{2} P_{\mu^{2}}(k) + P_{\mu^{0}} (k).
\end{equation}
The $P_{\mu^{4}}$, $P_{\mu^{2}}$, and $P_{\mu^{0}}$ terms include the redshift space distortions due to peculiar velocity gradients (eq. \ref{eqn:p_polynomial}), but here they are further modified by the AP effect. How can we disentangle the two?  The key is the presence of the $P_{\mu^{6}}(k)$ term, which is solely due to the AP effect.\footnote{This is not precisely true, because lensing can also introduce such a term \cite{mandel06}, although it is generally small.} It therefore allows a measurement of
\begin{equation}
(1+\alpha)=\frac{H D_{A}({\rm Assumed \ Cosmology})}{H D_{A}({\rm True \ Cosmology})}.
\end{equation}
Conceptually, we constrain cosmological parameters by varying them until $\alpha=0$. Once the underlying cosmology is known, the AP contribution to the $P_{\mu^{4}},P_{\mu^{2}}$, and $P_{\mu^{0}}$ terms can be identified and removed, leaving only the intrinsic power spectrum that we have already discussed.  Note that the $k$ and $z$ dependence of the AP effect are well known and can aid in separating it from noise and foreground contributions.

In principle the peculiar velocity anisotropy is also sensitive to the background cosmological model: substituting the full expression for the linear peculiar velocity perturbation, $P_{\rm z-space} \propto (1+f \mu^{2})^{2} P_{\rm real \, space}$ \cite{kaiser87}. However, because the universe is so close to Einstein-de Sitter at high redshift, this dependence turns out to be negligible (with $D \propto a$ and $f=1$).  In contrast, the AP effect does remain sensitive to the background cosmology out to high redshifts \cite{saiyadali05}. This can be easily seen from the expression for angular diameter distance in a flat Universe:
\begin{equation}
D_{A}(a)= a \int_{a}^{1} \frac{d a^{\prime}}{a^{\prime 2} H(a^{\prime})}.
\end{equation}
The dependence on cosmology comes mostly from the contribution to the integral at low redshifts, where $\Omega_{\Lambda}(z) > 0$. Unfortunately, this also implies that the information content is essentially the same as that of the CMB, where the angular diameter distance to the surface of last scattering has been accurately measured using acoustic oscillations as a standard ruler \cite{kamionkowski94, jungman96, spergel06}. Any additional information must come from the difference between $D_A(z)$ and $H(z)$ at the redshift of the 21 cm survey and at recombination. A relative accuracy of better than $\sim 5\%$ in log$H$ and $D_{A}$ must be obtained for the 21 cm AP effect to yield improvements on the CMB \cite{barkana_AP}.  Unfortunately, real-world challenges prevent any planned experiments (or even instruments an order of magnitude larger than the SKA) from reaching this level \cite{mcquinn05-param}.

The 21 cm AP effect is best measured in the early stages of reionization, when the effect of ionized bubbles is negligible (these introduce substantial additional power in 21 cm temperature fluctuations as well as non-trivial $\mu$ dependence; see \S \ref{anisotropy}). Once reionization is underway, the large scale anisotropy pattern is strongly dominated by the distribution of \htwo regions rather than the background cosmology \cite{saiyadali05}. A clever way of disentangling these effects might still be possible (see \cite{nusser05-ap} for some suggestions) but is likely to be difficult. 

\subsection{Secondary Fluctuations} \label{sec}

Highly-redshifted 21 cm emission and absorption constitute a new source of diffuse background light.  Because it emerges from the distant Universe, it is subject to all of the same radiative transfer effects as the CMB; these are collectively termed ``secondary fluctuations'' and modify the observed
power spectrum (or specific resolved features in maps).  Secondaries allow us to use the 21 cm background to study the lower-redshift universe, yielding similar information to analogous measurements with the CMB.  However, the 21 cm background does have two advantages over the CMB.  First, because it is a spectral line, the 21 cm background is not a single full-sky map but rather a series of them at many closely-spaced redshifts \cite{pen04-lens}.  Second, it contains structure down to small physical scales (in principle the IGM Jeans mass), whereas the CMB cuts off at $\ell \sim 1200$ because of Silk damping.

These properties are particularly useful for gravitational lensing.  When considered as a single
diffuse \emph{broadband} background, the 21 cm signal is less useful than the CMB because it is difficult to unambiguously separate the lensing modifications from the intrinsic clustering of the signal
\cite{cooray04-lens}.  Most of the strategies for extracting lensing information from the CMB rely on two properties that the 21 cm background does not share.  First, they assume that any observed non-gaussianities are induced by lensing; as we shall see, this will not be true for 21 cm emission during reionization.  Second, the CMB has a well-defined acoustic peak structure from which power is transferred, whereas the 21 cm background is relatively featureless (though some structure does appear during reionization).

Fortunately, the two advantages identified above still outweigh these difficulties.  Conceptually, 21 cm maps at many different redshifts are all at sufficiently large distances that they experience nearly the same lensing potential.  This induces a cross-correlation between source screens from which the projected mass distribution can be extracted \cite{pen04-lens}.  The cross-correlation is visible both in the variance map (because lensing changes the physical scale subtended by each angular patch) and in the shear map.  Typically, CMB lensing is described through a perturbative expansion of the deflection angle $\delta \theta$.  Unfortunately, this does not work well for the 21 cm background because of the large temperature gradients on small scales, so new methods in which the deflection field is built by progressively adding smaller and smaller wavelength modes in Fourier space are required \cite{mandel06}.  Lensing typically  modifies the power spectrum by $\sim 1 \%$ \cite{pen04-lens, mandel06}.  

In practice, the cross-correlation between neighboring maps is noisy and may not be the best way to extract the lensing information.  Another way to approach it is through the aspherical perturbations induced by lensing, with the largest modifications to modes along the line of sight because they sample the smallest projected scales \cite{mandel06, zahn05-lens}.  Like redshift space distortions, this introduces $\mu^2$ and $\mu^4$ components into the power spectrum.  A quadratic estimator, generalized from those used for the CMB, can take advantage of this angular dependence to reconstruct the projected mass field from 21 cm observations \cite{zahn05-lens}.  Because such estimators extract the most information from the smallest resolved scales, they require high angular resolution.  As such, SKA-class instruments will be needed to probe lensing effectively.  However, such an instrument could prove more powerful than the CMB at reconstructing the intervening matter distribution.

One interesting (though difficult) application is to ``delense'' CMB polarization maps \cite{sigurdson05}.  CMB lensing creates both E- and B-mode (i.e., curl-free and divergence-free) polarization.  The main interest in observing the much weaker B-mode polarization is that its primordial component is generated by gravitational waves and offers a sensitive probe of inflation \cite{seljak97, kamionkowski97b}.  However, the lensing contribution must first be identified and removed; lensing aliases power from E modes to the much weaker B modes.  This can be done statistically with high-resolution, all-sky CMB maps, but 21 cm maps provide an alternative by independently measuring the projected mass distribution and hence the B-mode contamination.  In that case, the limiting systematic is the mismatch in source redshifts (and hence lensing potential) between the 21 cm screens and the CMB.  An experiment centered on $z=30$ that is sensitive to $\ell \la 5000$ could, together with a modest ground-based CMB polarization experiment, reach comparable limits to a cosmic-variance limited, all-sky CMB-only measurement after a year of (on-source) integration \cite{sigurdson05}.

Another type of secondary is generated by Thomson scattering in the post-reionization IGM, which washes out some of the intrinsic fluctuations (just as with the CMB) \cite{babich05}.  But scattering also drives polarization anisotropies sourced by the (local) quadrupole fluctuations seen by the scatterers (again just as with the CMB \cite{ng96, zalda97-reion}).  However, the local quadrupole of the 21 cm background may be much larger than that of the CMB, because fluctuations in the ionized fraction (i.e., \htwo regions) when $\bxion \sim 0.5$ have large sizes ($\ga 20 \Mpc$; see \S \ref{reion} below) and so contribute a substantial anisotropy.  A simple model with randomly distributed bubbles of a single size and instantaneous reionization predicts peak rms fluctuations in the polarized brightness temperature of $\sim 3$--$20 \microkel$ (with the amplitude increasing as reionization moves to higher redshifts) on scales $\ell \sim 100$--$500$ (depending on the typical bubble size) \cite{babich05}.  The cross-correlation with temperature is typically an order of magnitude larger and (neglecting systematics)
may be detectable with SKA-class instruments.

A third type of secondary is the Sunyaev-Zel'dovich (SZ) effect familiar from the CMB \cite{sunyaev80}, which describes the modification to the background spectrum from energy exchange between photons and electrons during Compton scattering.  For lines of sight through hot galaxy clusters at low redshifts, this modifies the sharp spectral features of the 21 cm background and introduces new, non-trivial features \cite{cooray05-glob}.  The difference between the spectra seen through the cluster and through its surroundings could allow one to measure the global 21 cm background through a differential measurement free from many of the usual systematics (see \S \ref{glob-obs}).

%\bibliographystyle{elsart-num}
%\bibliography{Ref_21cm}

%\end{document}

%% file: dark-ch5.tex
%\documentclass{elsart}
%\usepackage{amssymb,cite,epsfig}

%\input{./defns.tex}

%\begin{document}

\section{The Dark Ages} \label{dark}

As we will see in \S \ref{syn}, a number of other techniques can constrain star formation at $z \ga 6$ -- and indeed many have already had some successes.  However, thus far 21 cm observations are the {\it only} proposed means of probing the truly dark ages between recombination ($z\sim 1000$) and first light ($z\sim30$).\footnote{It has been proposed that resonant scattering by neutral Li at $z\sim 500$ might create detectable CMB anisotropies which could also probe this era \cite{loeb01,zalda02}.  Unfortunately, recent calculations show that the Ly$\alpha$ background from residual hydrogen recombinations is sufficient to keep Li ionized until low redshift \cite{switzer05}.}  By observing the era before the messy baryonic physics of galaxy formation appears, this is potentially a gold mine of information about cosmological initial conditions \cite{loeb04}. It even has two major advantages over the CMB:  (i) IGM fluctuations persist to much smaller mass scales, being unaffected by Silk damping, and (ii) cosmic variance is much smaller, because there are many more independent modes in the full three-dimensional volume than in the (single) last-scattering surface. However, the enormous difficulty of observations at such low frequencies means that they probably lie many years in the future. The ideas described in this section should therefore be regarded only as hopes for the far future; all of the observatories now being planned will focus on the ``low-redshift" regime $z \la 15$.

As discussed in \S\ref{glob}, the IGM thermally decouples from the CMB at $z\sim200$, cooling adiabatically so that $T_{K} < T_{\gamma}$. However, until $z\sim 30$ it remains sufficiently dense that collisions drive $T_{S} \rightarrow T_{K}$. Since $T_{S} < T_{\gamma}$, the IGM can be seen in 21 cm absorption against the CMB, with a peak signal at $z\sim 80$.  Figure~\ref{fig:tevol} shows the global spin temperature evolution. Because it depends only on simple, well-known physics (adiabatic cooling and Compton heating), this global temperature history is exact; any observed deviations will be an exciting signature of energy injection at these high redshifts (e.g., due to decaying particles \cite{shchekinov06, furl06-dm}). Unfortunately, the observational difficulties are immense: they are identical to those involved in detecting the ``all-sky" $\bdtb(z)$ during the reionization epoch (see \S\ref{glob-obs}), but at a much lower frequency when foregrounds are $\sim 1000$ times stronger ($T_{\rm sky} \sim 10^{5}$ K at 15 MHz), requiring telescope calibration stable to one part in $10^{7}$ over a wide frequency range. However, the fluctuating 21 cm signal -- seeded by linear density fluctuations -- might still be observable, providing a probe of the primordial matter power spectrum over a wide range of scales \cite{loeb04}.

It is useful to briefly review the evolution of these density fluctuations, which serve as initial conditions for cosmological simulations and are calculated to high accuracy in Boltzmann code solvers \cite{ma95,CMBFAST}.  Before recombination, perturbations in the baryons on sub-horizon scales are strongly suppressed by radiation pressure.  After recombination, the coupling between baryons and photons decreases sharply, and the baryon temperature and density are determined by the gravitational attraction of dark matter potential wells, in addition to thermalization with the CMB through Compton scattering off residual free electrons.  The linearized form of the thermal evolution equations~(\ref{eq:tkevol}) and (\ref{eq:tcomp}) is
\begin{equation}
\frac{\deriv \delta_T}{\deriv t} = \frac{2}{3} \frac{\deriv \delta}{\deriv t} - \frac{\xion}{1 + f_{\rm He} + 
  \bxion} \, \frac{T_\gamma}{T_K} \, \frac{\delta_T}{t_\gamma}.
\label{eq:dtcomp}
\end{equation}
Temperature perturbations are thus sourced by scale-dependent density perturbations; this leads to a spatially variable sound speed that modifies the growth of baryonic density fluctuations by up to $30 \%$ at $z=100$ and $10\%$ at $z=20$ \cite{barkana05-infall,naoz05}.  Moreover, it requires that density and temperature perturbations be tracked separately. The baryonic density perturbations gradually approach those of the dark matter, while the temperature perturbations approach those of an adiabatic gas at a somewhat later time (see below). The acoustic oscillations created as baryons fall into dark matter potential wells, along with the variable sound speed, imprint five distinct fluctuation modes on the baryons, each with an amplitude sensitive to the underlying cosmology.  Because each mode evolves differently with redshift, four of these modes may be separable with 21 cm data (the fifth is already much smaller) \cite{barkana05-infall}.  By contrast, low-redshift galaxy surveys are only sensitive to the growing mode (which we have written as $D(z)$ throughout).  

Although a complete calculation of $P_{21}(k)$ requires tracking the density and temperature modes independently, we can build some intuition by ignoring the extra scale dependence introduced by them.  Thus we write $\delta_T \equiv g(z) \delta$ \cite{bharadwaj04-vel}.  (This is a reasonable approximation on the scales most accessible to observations, especially at $z \la 100$ when the growing mode dominates.)  Equation~(\ref{eq:dtcomp}) becomes
\begin{equation}
\frac{\deriv g}{\deriv z} \approx (2/3-g) \frac{\deriv \ln \delta} {\deriv z} + \frac{\xion/t_\gamma}{1 + f_{\rm He} + \bxion} \, \frac{T_\gamma}{T_K} \, \frac{g}{(1+z)H(z)}.
\label{eq:gz}
\end{equation}
Here the first term describes adiabatic heating, which tries to drive $\delta_T \rightarrow 2\delta/3$; the second term (from Compton heating) counters this by trying to make the gas isothermal.  Obviously, during this epoch the terms $\propto \delta_x$, $\delta_\alpha$ vanish in equation~(\ref{eq:d21}).
Thus we can write
\begin{equation}
\tilde{\delta}_{21} \approx (\beta' + \mu^2 f) \tilde{\delta}
\label{eq:d21-dark}
\end{equation}
where $\beta' \equiv \beta + g(z) \beta_T$ and we have used $\tilde{\delta}_{\partial v}= - \mu^2 f \tilde{\delta}$. The crucial point is that, even in this simple approximation, fluctuations in the spin temperature (and hence $\bdtb$) do \emph{not} evolve identically to density fluctuations, because of the extra redshift-dependent factor $g(z)$.  This term describes how the modified kinetic temperature in dense gas affects the collision rate (and hence the spin temperature).

\begin{figure}[!t]
\centerline{\epsfig{file=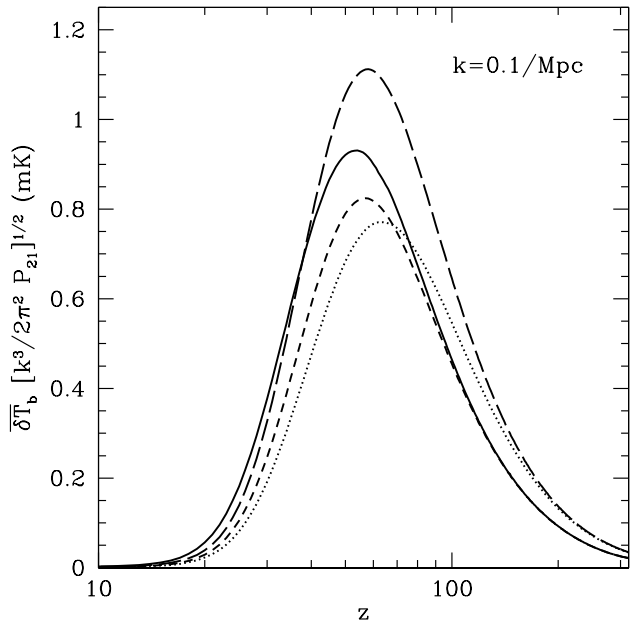,width=3.5in}}
\caption{Spherically-averaged fluctuations of the 21 cm brightness temperature relative to the CMB at $k=0.1 \Mpcinv$ as a function of redshift. The solid curve shows the true net fluctuation amplitude.  The dotted curve ignores collisions and temperature fluctuations entirely (i.e., $\beta'=1$).  The short-dashed curve ignores the temperature dependence of the collision rate, while the long-dashed curve sets $\beta_T=0$. }
\label{fig:Tfluct_density}
\end{figure}

Generically, we expect $g(z)$ to evolve from $g=0$ to $g \approx 2/3$ as the CMB energy density drops and Compton heating becomes less important.  This must compete with the evolving density perturbations and background temperature to determine the overall growth of 21 cm fluctuations.   We show the resulting spherically-averaged rms 21 cm brightness temperature in Figure~\ref{fig:Tfluct_density} (see also \cite{loeb04,bharadwaj04-vel}).\footnote{Note that the relative importance of temperature variations depends on angle because they must compete with the (temperature-independent) velocity term.}   For this illustrative figure, we have included only the growing density mode; we refer the reader to \cite{barkana05-infall, naoz05} for more detailed calculations.  We present the amplitude at a fixed wavenumber $k=0.1 \Mpcinv$ for convenience; this is chosen to be in the 
regime most accessible to observations.  The solid curve shows the net fluctuation amplitude; the other curves show the various components that contribute to this signal.  

%%%%%%%%%%%%%%%%%%%%%%%%%%%%%%%%%%%%%%%%%%%%%%%%%%%%%%%%%%%%%%%%%%%
%%%% FIG 5-1: 21cm fluctuation in response to density fluctuations. 
%%%%%%%%%%%%%%%%%%%%%%%%%%%%%%%%%%%%%%%%%%%%%%%%%%%%%%%%%%%%%%%%%%%%

First, the dotted curve sets $\beta'=1$; this is equivalent to ignoring fluctuations in both the temperature and collision rate.  Thus in this case the rms fluctuations simply trace $\bdtb(z)$ multiplied by the growth factor.  The long-dashed curve sets $\beta_T=0$, ignoring the temperature fluctuations entirely.  This always amplifies the 21 cm fluctuations because, for example, increasing the density increases the collision rate and hence drives $T_S$ closer to $T_K$.  But it overestimates the real fluctuations, because adiabatic compression in dense gas increases $T_K$ as well. The short-dashed curve includes this effect as well but ignores the temperature dependence of collisional coupling (i.e., it sets $\beta_T = T_\gamma/[T_K-T_\gamma]$).  Compared to the dotted curve, this decreases the fluctuation amplitude at $z \ga 80$ but increases it at lower redshifts.  At high redshifts, collisions efficiently couple $T_{S}$ to $T_{K}$ everywhere; thus because $T_{K}$ increases in overdense regions, so does $T_{S}$.  At lower redshifts, on the other hand, collisions become inefficient and $T_{S} \rightarrow T_{\gamma}$. Overdense regions can thus more efficiently drive $T_{S} \rightarrow T_{K} < T_{\gamma}$, causing the spin temperature to fall inside density enhancements and increasing the overall fluctuation amplitude.  Finally, the last step (included in the solid curve) is to add the temperature dependence of the collision rates.  This is unimportant at high redshifts but further boosts the fluctuations at lower redshifts, because $\kappa_{10}^{\rm HH}$ is so steep at low temperatures (see Fig.~\ref{fig:collrates}).  The true peak of the fluctuations occurs at $z\approx 55$.  

One more subtlety relevant to this regime is that, at $z \la 100$, collisional coupling is only marginally efficient.  In this case, the hyperfine level populations depend on the atomic velocity (essentially because faster atoms collide more frequently), which modifies the signal by a few percent \cite{hirata06}.

A great virtue of these observations is that they probe the power spectrum all the way to $k \la 10^{3} \, {\rm Mpc^{-1}}$ (above which Jeans smoothing effects become important). They can therefore directly constrain modifications of small scale power in the standard $\Lambda$CDM model made to account for galactic scale observations, as well as the possible running of the spectral index suggested by \emph{WMAP} \cite{spergel06}. By contrast, since Silk damping suppresses small-scale power in the CMB, that field can only probe $k \le 0.2 \, {\rm Mpc^{-1}}$.  These small scales are likewise inaccessible to Ly$\alpha$ forest observations, because the Jeans mass increases by a large factor after reionization.  Thus the 21 cm line is a truly unique cosmological probe.  In addition to Jeans smoothing, the 21 cm absorption line is also subject to thermal smoothing along the line of sight.  In principle, this could be separated out via its angular dependence, allowing a direct measure of the temperature $T_K(z)$ and Hubble parameter $H(z)$ \cite{naoz05}. In practice, this requires extremely high frequency resolution (e.g., $\sim 25$Hz at $z=50$, compared to the redshifted 21 cm frequency of 28 MHz).    
 
Another great virtue of 21 cm observations is the huge number of independent samples contained in the fully three-dimensional volume \cite{loeb04}.  The observable Universe has a volume $\sim 3000$ Gpc$^3$ between $z=30$ and $z=100$, corresponding to $\sim 6 \times 10^{11}$ patches with radii equal to 1 Mpc or $\ga 10^{18}$ independent Jeans masses.  This huge number of samples could potentially be a sensitive probe of primordial non-gaussianity, although of course instrumental systematics will be the real limiting factor.  

The main damper on all these exciting possibilities is the extraordinary difficulty of the observations.  The strong foregrounds at these low frequencies have brightness temperatures two orders of magnitude larger than those of $z\sim 10$ observations, which as we will see pose a rather difficult challenge themselves. Furthermore, the earth's ionosphere becomes opaque at $\nu \le 20 \MHz$ (or $z \ge 70$), and one must go to space. ``Dark age" experiments lie many years in the future. 

%\bibliographystyle{elsart-num}
%\bibliography{Ref_21cm}

%\end{document}

%% file: struc-ch6.tex
%\documentclass{elsart}
%\usepackage{amssymb,cite,epsfig}

%\input{../../submission/defns.tex}

%\begin{document}

\section{The First Structures} \label{struc}

By $z \sim 30$, collisions had become so rare that $T_S \approx T_\gamma$, and the 21 cm fluctuations are no longer visible.  However, at just about this time the first collapsed objects began to form.  These, and the surrounding networks of sheets and filaments, produced hot, overdense gas where collisional coupling became efficient again.  Thus the next phase in the 21 cm history is the birth of the first nonlinear structures; in this section, we will describe potential observations of this epoch.

\subsection{Minihalos} \label{mh}

As we saw in \S \ref{glob-dark}, the IGM temperature is extremely small in the absence of luminous sources.  Thus the Jeans mass is quite small ($\sim 10^5 \Msun$ \cite{gnedin98}), and tiny objects can collapse.  However, atomic line cooling in primordial gas requires virial temperatures $T_{\rm vir} \ga 10^4 \kel$ (or $M \ga 10^8 \Msun$).  Thus over a relatively wide range in mass, halos are probably unable to cool and form stars through the usual channels.  Such objects are known as ``minihalos," and they can evolve along two possible routes.  First, if H$_2$ is able to form, the gas can cool through its vibrational transitions.  It is via this process that the first minihalos are believed to cool (over relatively long time intervals) and probably produce the first stars \cite{abel02, bromm02, bromm04}.  

However, once these first stars form, subsequent generations of minihalos no longer form in isolation.  In particular, these stars flood the Universe with UV photons.  As well as initiating Wouthuysen-Field coupling, photons in the energy range $11.18$--$13.6 \eV$ (the so-called Lyman-Werner band) dissociate H$_2$ and thus suppress cooling inside minihalos (see \cite{barkana01} for a review).  As a result, minihalos virialize but do not fragment and form stars.  Instead they become dense ($\delta_{\rm mh} \sim 200$) gas clouds that float serenely through the IGM.  During reionization, minihalos become photon sinks and contribute to the gas clumpiness; thus their properties are important to measure.  Fortunately, they also become 21 cm sources \cite{iliev02}.  Given their virial temperatures, $\delta_{\rm mh}$ is much larger than the critical value of equation~(\ref{eq:dcoll}), so collisional coupling becomes extremely efficient.  Unfortunately, the typical minihalo is $\la 10 \kpc$ across, much smaller than the resolution limit (at any reasonable sensitivity) of any telescope for the foreseeable future.  Thus minihalos must be detected statistically \cite{iliev02,iliev03}.  The net signal depends on the fraction of baryons $f_{\rm mh}$ incorporated into minihalos and their average bias $b_{\rm mh}$.  

\begin{figure}[!t]
\centerline{\epsfig{file=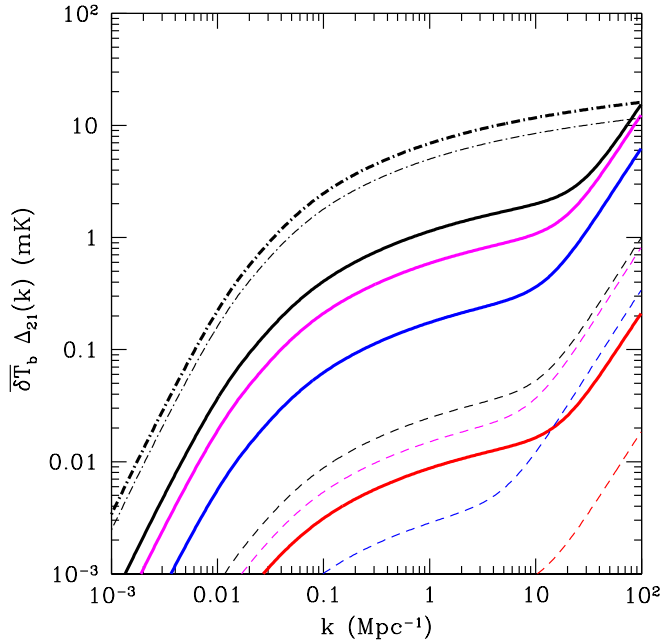,width=3.5in}}
\caption{Expected rms brightness temperature fluctuations from minihalos.  Thick solid and thin dashed curves are for $z=10$ and $20$, respectively.  From top to bottom within each set, the curves assume  $T_K=T_{\rm ad}, \, 20,\,100$, and $1000 \kel$ with $J_\alpha=0$.  The dot-dashed curves show the fluctuations from linear theory, assuming that the entire IGM has $T_S \gg T_\gamma$.  From \cite{furl06-mh}.}
\label{fig:mhbasic}
\end{figure}

Figure~\ref{fig:mhbasic} shows some example power spectra built from the halo model \cite{cooray02, furl06-mh}.  The dot-dashed curves show the rms brightness temperature fluctuations for a universe in which the entire IGM has $T_S \gg T_\gamma$; the thick and thin curves are for $z=20$ and $z=10$, respectively, assuming linear theory.  The uppermost solid and dashed curves show the rms fluctuations from minihalos at these two redshifts.  We see that the minihalo signal is $\sim 1 \mkel$ at $z=10$, and nearly an order of magnitude smaller at $z=20$.  On large scales, the net minihalo fluctuations are \cite{furl06-mh}
\begin{equation}
\bdtb \Delta_{\rm mh}(k) \approx b_{\rm mh} f_{\rm mh} \dtb^{\rm hot} \Delta_{\rm lin}(k),
\label{eq:mhfluc}
\end{equation}
where $\dtb^{\rm hot}$ is the mean brightness temperature if $T_S \gg T_\gamma$ and $\Delta_{\rm lin}$ is the linear density power spectrum. This shape only changes on scales $k \ga 10 \Mpcinv$, where individual minihalos are resolved and the power spectrum instead traces their density profiles.  Note, however, that the halo model does not include nonlinear clustering, which modifies $P(k)$ on somewhat larger scales (see below) \cite{iliev03}.

%%%%%%%%%%%%FIGURE 6-1: Minihalo signal

Of course, all the elements of equation~(\ref{eq:mhfluc}) depend on the underlying cosmological model, and in principle minihalos provide a probe of its parameters (especially the small scale matter power spectrum) \cite{iliev02}.  However, as is all too common in astronomy, there are unavoidable -- and large -- degeneracies with the astrophysics.  Most importantly, as we have seen in \S \ref{xrayheat}, luminous sources produce X-rays, which have two effects on minihalos.  First, they penetrate minihalos and increase the free electron fraction, catalyzing H$_2$ formation.  This promotes cooling and could allow minihalos to fragment and form stars \cite{haiman00, glover03,kuhlen05-sim}.  Unfortunately, the net effects of UV and X-ray feedback remain unsettled.  

But a more important aspect may be X-ray heating of the IGM, which suppresses minihalo formation by increasing the Jeans mass and preventing smaller minihalos from accreting gas \cite{oh03-entropy}.  The effects are most easily expressed through the ``entropy" $K_{\rm IGM} \equiv T_K/n^{2/3}$ injected into the IGM.  The entropy is conserved during adiabatic expansion and contraction, so the effects of feedback can be seen by comparing $K_{\rm IGM}$ to the entropy $K_{\rm mh} = T_{\rm vir}/n(r_{\rm vir})^{2/3}$ generated during halo formation.  If $K_{\rm IGM} \gg K_{\rm mh}$, the thermal pressure generated during collapse exceeds the gravitational potential, and that minihalo is unable to accrete gas \cite{oh03-entropy}.  The lower sets of curves in Figure~\ref{fig:mhbasic} show the minihalo signal if $T_K=20,\,100$, and $1000 \kel$, assuming that the energy was injected at the cosmic mean density \cite{furl06-mh}.  X-rays can decrease the signal by more than two orders of magnitude -- much larger than any cosmological uncertainty.  If the minihalo fluctuations can be seen, they will constrain astrophysical feedback rather than cosmological parameters.

\begin{figure}[!t]
\centerline{\epsfig{file=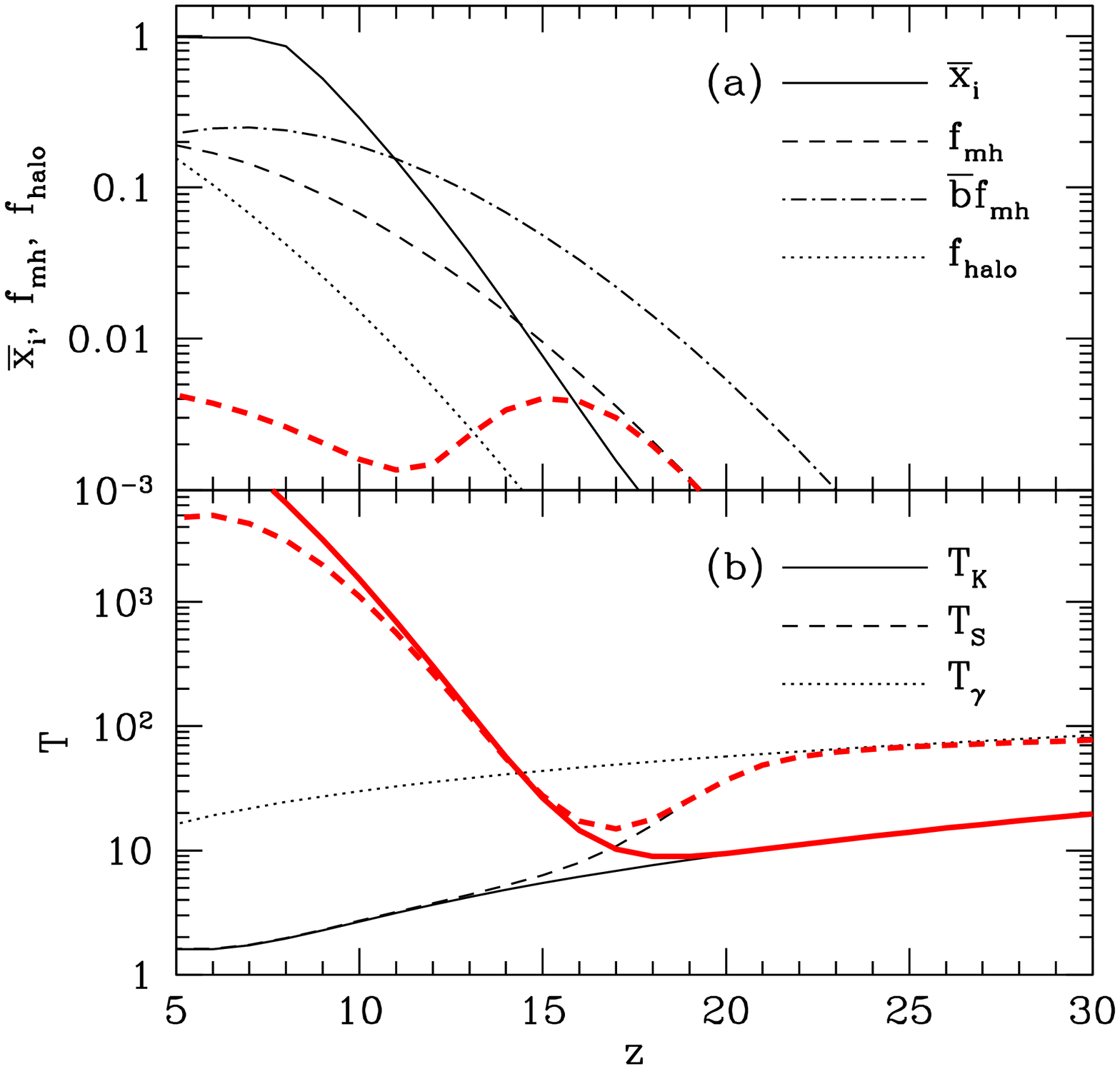,width=2.75in}
\epsfig{file=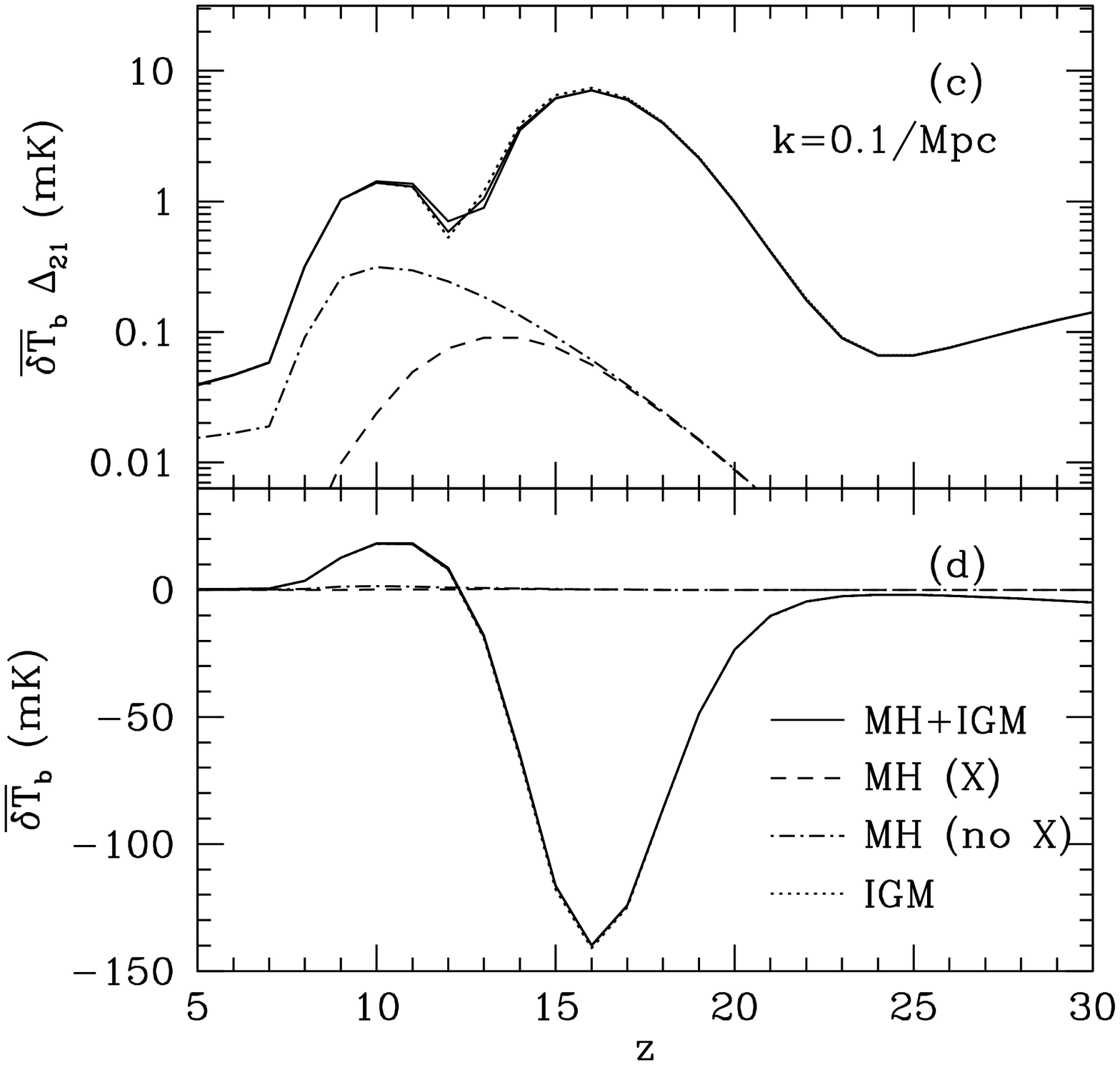,width=2.75in}}
\caption{Minihalo signal in a cosmological context.  \emph{(a)}: The mean ionized fraction $\bxion$, the collapsed fraction in star-forming halos $f_{\rm halo}$, the minihalo fraction $f_{\rm mh}$, and the bias-weighted minihalo fraction.  For $f_{\rm mh}$, the upper and lower curves respectively neglect and include X-ray heating. \emph{(b)}:  Temperature history.  \emph{(c)}:  Large-scale 21 cm fluctuation amplitude.  \emph{(d)}:  Mean brightness temperature relative to the CMB.   From \cite{furl06-mh}. }
\label{fig:mh-hist}
\end{figure}

Unfortunately, separating the minihalo signal from the IGM will be difficult, because the Wouthuysen-Field effect renders the entire IGM visible against the CMB at an early stage of structure formation.  Because the IGM contains nearly all the mass, it will outshine the minihalo component by a large factor \cite{oh03}.  Figure \ref{fig:mh-hist} illustrates how the minihalo signal evolves in our fiducial Pop II model of \S \ref{glob-theor}, except that we have set $f_X=0.2$ to minimize the importance of X-ray heating.  Panel \emph{(a)} shows $\bxion$ together with the mass fraction in minihalos.  Note that (because the effects of X-ray heating are controversial) we show two cases for $f_{\rm mh}$, one ignoring X-ray heating and one assuming suppression through the entropy floor \cite{oh03-entropy}.  

%%%%%%%%%%%% FIGURE 6-2: Histories with minihalos

Figure~\ref{fig:mh-hist}\emph{c} shows the large-scale fluctuation amplitude for the IGM, minihalo, and total signals.  Minihalos are the only objects to shine brightly at $z \ga 22$, before \lya coupling begins.  Unfortunately, at such high redshifts the minihalo component contains only a negligibly small fraction of the mass, so it is still overwhelmed by the weak IGM fluctuations.  At $z \ga 14$, the IGM is still cold (but visible), so the entropy floor is relatively unimportant.  In this regime, (hot) minihalos and the (cold) IGM actually cancel each other out in terms of the mean signal $\delta \bar{T}_{b}$.  Thus the minihalos do have observable effects -- but because they trace the linear matter power spectrum on large scales (just like the IGM), they remain difficult to separate.  In particular, the subtle change in $\bdtb$ they cause can easily be mimicked by a slightly different thermal history.  By $z \la 15$, the minihalo phase does contain a substantial fraction of the  mass, but the IGM has become hot.  In this regime, minihalos have identical emission properties to the diffuse IGM and -- even in principle -- cannot be distinguished from the IGM short of resolving them.  Note as well that this conclusion is independent of the efficacy of X-ray feedback.  

Thus our best hope of observing minihalos is to resolve them, or at least their nonlinear clustering patterns \cite{iliev03}.  The dynamics of small-scale clustering modify the \emph{shape} of the power spectrum at $k \ga 3 \Mpcinv$ and so make it distinguishable from the linear power spectrum of the IGM.  This effect is not included in the halo model.  Unfortunately, these scales are beyond the reach of any planned experiment, but if they can be measured they will provide valuable insight into the ``sinks" of reionization.  A better way to probe such small physical scales may be with the ``21 cm forest" (see \S \ref{forest}).

\subsection{The IGM} \label{first-igm}

Perhaps the best hope to observe the first non-linear structures as they form is before Wouthuysen-Field coupling becomes efficient.  But, of course, even in that case minihalos must compete with any other structures visible against the CMB.  In particular, equation (\ref{eq:dcoll}) shows that, even at $z \sim 20$, collisional coupling does not necessarily require virial overdensities if the gas is moderately warm.  Thus we would expect to see the cosmic web visible against the CMB even during the earliest phases of structure formation.  Although this is a difficult regime to study, it has been considered both analytically \cite{furl04-sh} and in numerical simulations \cite{shapiro05,kuhlen06-21cm}.  

The analytic model of cosmic web shocks described in \S \ref{shockheat} provides some intuition for the IGM signal.  It assumed that cosmic web shocks form at linearized overdensities characteristic of ``turnaround" from the IGM Hubble flow ($\delta_{\rm ta}=1.06$).  The crucial point is that this density threshold is well below virialization ($\delta_c=1.69$), so filaments form before halos.  The cosmic web contains a fraction $\sim (0.1\%,\, 3\%, \, 25\%)$ of the gas at $z=30,\,20,$ and $10$ (compare to $f_{\rm mh}$ in Fig.~\ref{fig:mh-hist}\emph{b}) so could boost the emission significantly.  Unfortunately, the emission strength also depends on the structure of the shocked gas, which the analytic model cannot accurately describe.  It is therefore best to turn to numerical simulations.

%%%%%%%%%%%% FIGURE 6-3: pre-radiation IGM and X-ray IGM
\begin{figure}[!t]
\centerline{\epsfig{file=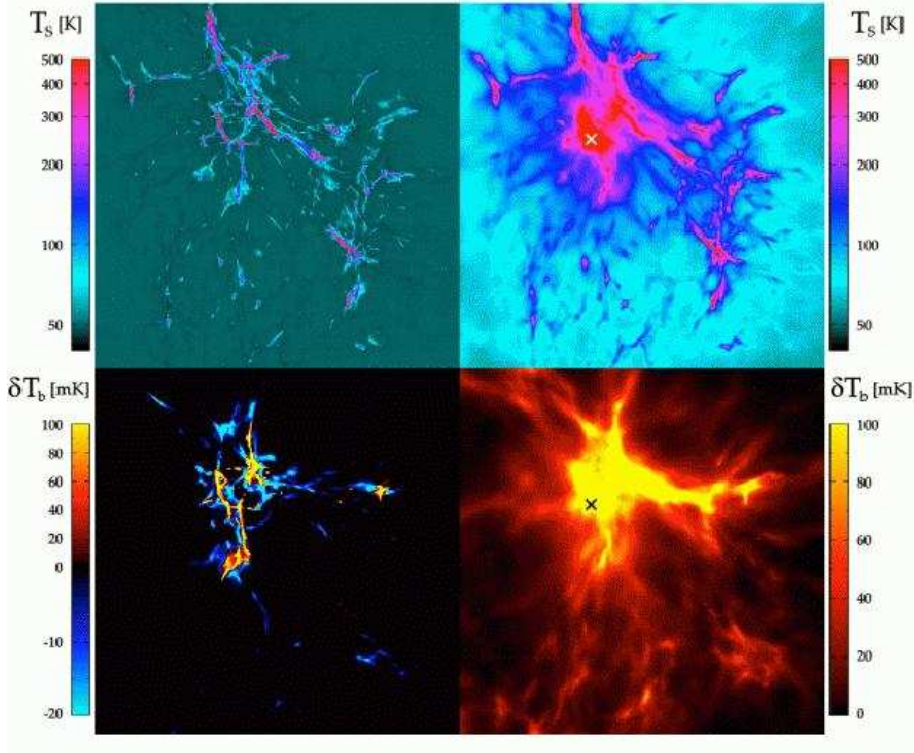,width=4.75in}}
\caption{Projected mass-weighted spin temperature (upper panels) and $\dtb$ (lower panels) from a 0.5 Mpc simulation box at $z=17.5$.  The left panels assume no radiation background, while the right panels contain a miniquasar (located at the cross) that emits X-rays (but does not induce Wouthuysen-Field coupling).  Note that $T_\gamma=50.5 \kel$ at this redshift.  From \cite{kuhlen06-21cm}. }
\label{fig:igm-pre}
\end{figure}

High-resolution simulations can compute the detailed temperature and density distributions of IGM gas (including shocks).  The left-hand panels of Figure~\ref{fig:igm-pre} (taken from \cite{kuhlen06-21cm}) illustrate the resulting signals.  As the cosmic web forms, gas collapses onto sheets and filaments.  Because the IGM has $T_K=7.2 \kel \ll T_\gamma=50.5 \kel$ at $z=17.5$, the initial contraction leaves the gas overdense but still cool.  Collisional coupling is somewhat more efficient in this gas, and the filaments are characterized by weak absorption.  As it continues to collapse, the gas shocks, heating it well above $T_\gamma$.  In these central regions, collisions are quite efficient, so filaments emit relative to the CMB.  However, the total mass locked up in this phase is still small:  only $\sim 1.7\%$ of the box has $\dtb < -10 \mkel$, with a comparable amount observable in emission.  This kind of signal therefore lies well beyond the reach of any instruments on the horizon today.

However, as the analytic model shows, the fraction of gas in the cosmic web increases rapidly at lower redshifts, because all of these structures are far off on the exponential tail of the mass function.  This is seen in much more detail in simulations that continue to lower redshifts \cite{shapiro05}.  These also show a fluctuating IGM appearing in both emission and absorption.  As the mean cosmic density decreases, the absorption phase weakens because overdensities characteristic of shocks are required for collisional coupling.  These simulations also compared the IGM emission to that of minihalos \cite{shapiro05}.   By $z \sim 9$, the nonlinear mass scale is sufficiently large that $f_{\rm mh} \ga 10\%$ (see Fig.~\ref{fig:mh-hist}\emph{b}).  At this point, the IGM emission is about half that of minihalos; at higher redshifts the IGM is more important (although its absorbing and emitting components do tend to cancel each other out).  

Of course, once a \lya background (or even a strong X-ray background) is in place, all of the IGM lights up in the 21 cm line.  Because this occurs quite early in the most plausible models,  the collisionally coupled phase will probably always be difficult to observe.  The IGM will boost the emission -- especially at higher redshifts, when both filaments and minihalos are far above the nonlinear mass scale \cite{furl04-sh} -- but not by a large enough factor to render it easily observable.

%\bibliographystyle{elsart-num}
%\bibliography{Ref_21cm}

%\end{document}

%% file: obj-ch7.tex
%\documentclass{elsart}
%\usepackage{amssymb,cite,epsfig}

%\input{../../submission/defns.tex}

%\begin{document}

\section{The First Luminous Objects} \label{obj}

Once the first sources of light appear, the character of the 21 cm sky changes completely.  These objects change the spin temperature (through Wouthuysen-Field coupling), the kinetic temperature (through X-rays), and of course the ionized fraction.  We studied the mean evolution of these quantities in \S \ref{glob}, but of course all of them also introduce fluctuations in the 21 cm signal from which we can learn about the first sources.  We have already seen that heating and Wouthuysen-Field coupling typically precede significant ionization.  To begin, we will therefore ignore \htwo regions and focus on fluctuations induced by the \lya background and X-rays.  

\subsection{Fluctuations from the Wouthuysen-Field Effect} \label{radfluc}

If star-forming galaxies dominate the UV radiation field, the average \lya background is given by equation~(\ref{eq:jalpha}):  it is simply the sum of photon fluxes redshifting into each \lyn transition, weighted by the appropriate $f_{\rm rec}(n)$.  At first sight, it might seem that this background would be extremely uniform because the volume sampled by photons that redshift directly into the \lya transition has a radius $\sim 250 \Mpc$.  However, in reality several factors combine to render the fluctuations significant \cite{barkana05-ts}:  (i) the flux is weighted by $r^{-2}$ and is hence more sensitive to nearby sources; (ii) the higher Lyman series transitions are more closely spaced in frequency and so have much smaller effective horizons; (iii) the first sources of light are highly clustered; and (iv) the finite speed of light implies that more distant sources are sampled earlier in their evolution (when they were presumably less luminous).  As a result, the UV background is bright near clumps of high-redshift sources and faint elsewhere.  Observations of this patchwork of emission or absorption would measure when these sources first appeared, their clustering, and their spectra.

The resulting fluctuation power spectrum has been calculated in the limiting case of $\bxion=0$ and uniform (or zero) X-ray heating \cite{barkana05-ts, pritchard05} and in a more general case including non-uniform X-ray heating \cite{pritchard06}.  We will first focus on the former.  The fluctuations have two parts.  The first traces the underlying density field, in a similar manner to the ionization model we will describe below:  large-scale overdensities contain extra sources and hence have large $x_\alpha$.  The resulting power spectra have the form $[P_{\alpha \alpha}(k), \, P_{\delta \alpha}(k)] \equiv [W^2(k),\, W(k)] \, P_{\delta \delta}(k)$.  Here $P_{\alpha \alpha}$ is the power spectrum of fluctuations in $x_\alpha$, $P_{\delta \alpha}$ is its cross-correlation with the density field, and $W(k)$ is a scale-dependent weighting factor.  On large scales, $W(k)$ approaches the average linear bias of the sources, as should be expected for a background produced by a population of discrete halos \cite{barkana05-ts}.  On scales much smaller than the effective horizon of Ly$n$ photons, the power vanishes since the radiation field is smooth beneath these scales.  On intermediate scales, the horizons of the various \lyn transition modulate the effective bias \cite{pritchard05}.  The stochastic distribution of galaxies provides a second set of fluctuations \cite{zuo92a, zuo92b}:  two nearby points sample nearly the same distribution of sources, inducing strong intrinsic correlations in the flux field.  As with any Poisson process, the fractional fluctuations are proportional to $N^{-1/2}$, where $N$ is the effective number of sources inside the sampled volume.  Thus this term is important on scales comparable to the mean separation of sources.  Because these two components relate to the density field (which sources the $\mu$-dependent velocity term) in different ways, they are in principle separable using redshift space distortions.

\begin{figure}[!t]
\centerline{\epsfig{file=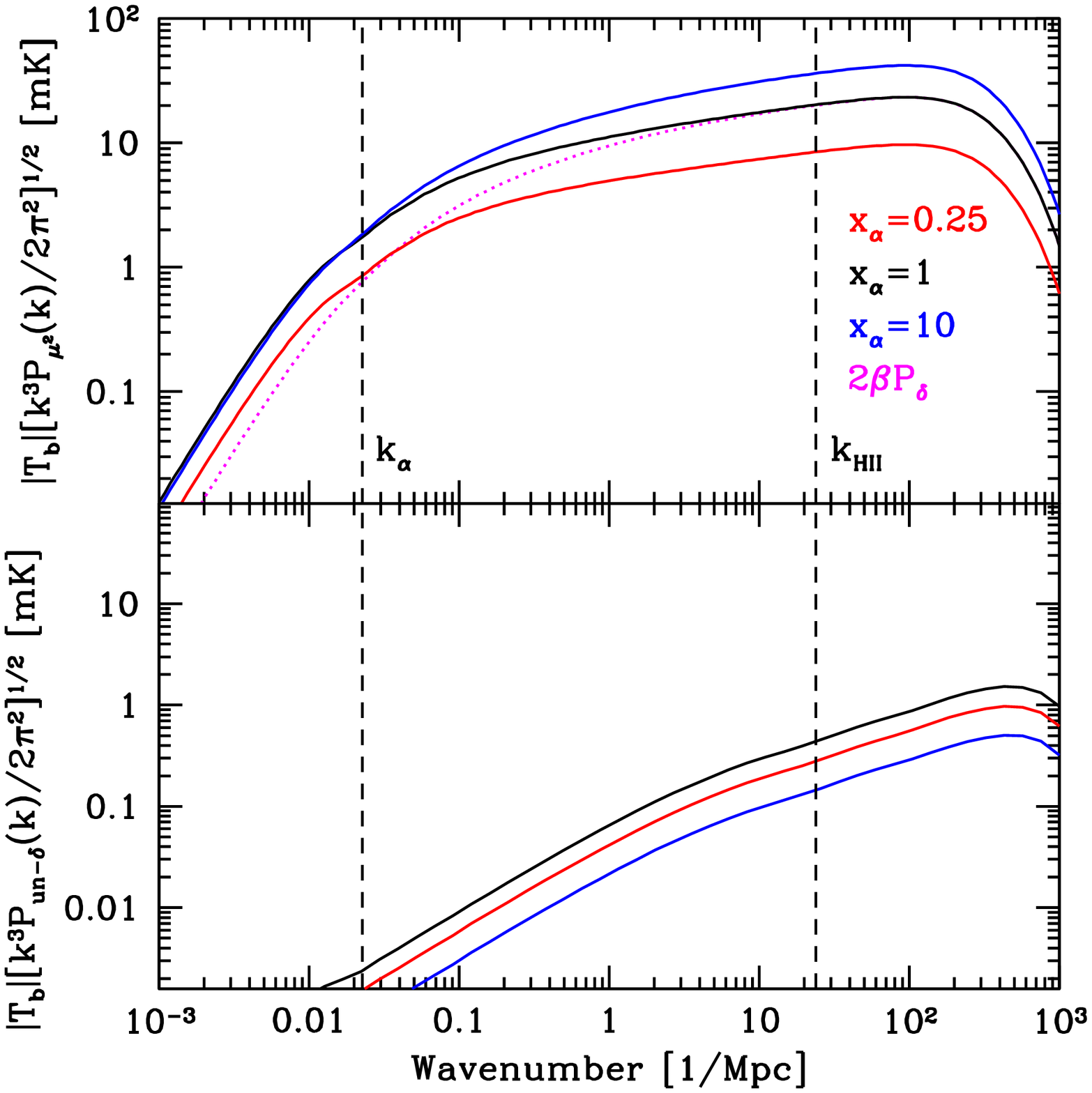,width=2.75in}
\epsfig{file=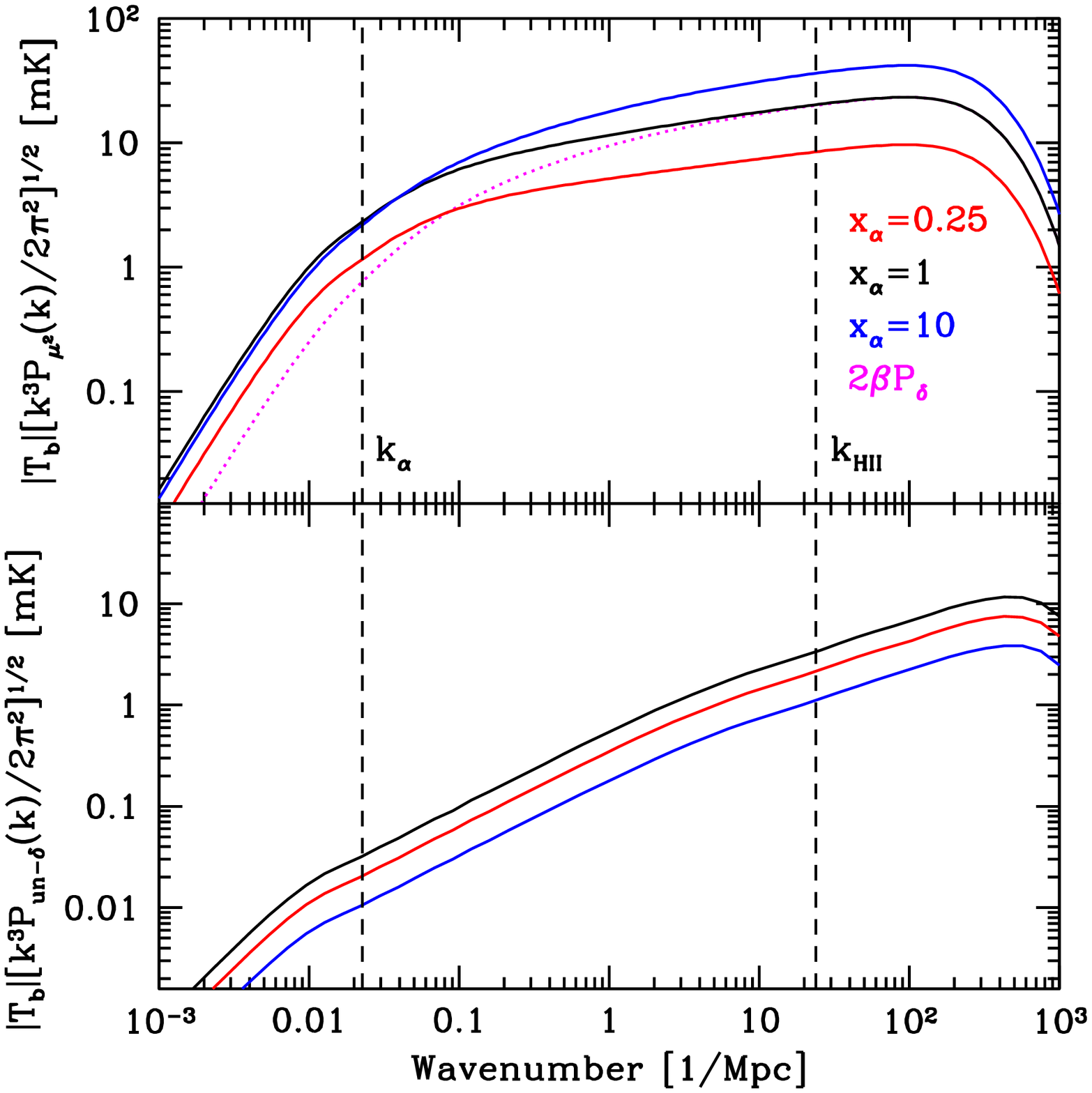,width=2.75in}}
\caption{21 cm brightness temperature fluctuations from Wouthuysen-Field coupling \cite{pritchard06}.  \emph{Upper panels}:  The $\mu^2$ component of $P_{21}(k)$.  \emph{Lower panels}:  Power sourced by stochastic variations in the galaxy density.  In each panel, we assume $z=20$ and $T_K=T_{\rm ad}$.  The left and right plots assume Pop III (Pop II) stars form in halos with $T_{\rm vir} > 500 \kel$ ($10^4 \kel$).  We show $x_\alpha=0.25,\,1$, and $10$ (solid curves, bottom to top). The dotted curves in the upper panels show the component from density fluctuations when $x_\alpha=1$.  The vertical dashed lines show the scales corresponding to the ``\lya horizon" $k_\alpha$ and the minimum \htwo region size $k_{\rm HII}$.  }
\label{fig:tsfluc}
\end{figure}

Figure~\ref{fig:tsfluc} shows some example power spectra at $z=20$ calculated following \cite{pritchard05}.  The solid curves in the upper panels show the $\mu^2$ component of the brightness temperature power spectrum \cite{barkana05-ts}, 
\begin{equation}
P_{\mu^2}(k) = 2 [\beta + g(z) \beta_T +  \beta_\alpha \, W(k) ]P_{\delta \delta}(k),
\label{eq:tsdenfluc}
\end{equation}
which includes both the density fluctuations (dotted curves, shown for $x_\alpha=1$) and those induced by Wouthuysen-Field coupling.  The lower panels show the Poisson component (uncorrelated with the density field).  The two panels span the range of plausible source parameters:  the right plot assumes that Pop II stars form in halos with $T_{\rm vir}>10^4 \kel$, while the left plot assumes that Pop III stars form in all halos with $T_{\rm vir}>500 \kel$.  In both cases, we assume a constant star formation efficiency in all halos above threshold, normalized so that $x_\alpha=0.25,\,1,$ and $10$ (bottom to top).

%%%%%%%%%%%%FIGURE 7-1:   Wouthuysen-Field fluctuations

Clearly the fluctuation amplitude depends on the source properties:  the signal increases for halos that are rarer and more biased \cite{barkana05-ts}.  The boost is much larger -- over an order of magnitude between the two cases -- for the Poisson fluctuations, because those depend directly on the source density.  The stellar spectral shapes matter only on intermediate scales where the distribution of photons between the different \lyn transitions has a weak effect on the $k$-dependence.  On large scales, the fluctuations are several times greater than the density fluctuations (because of the source bias), but the amplification vanishes at smaller scales.  In general, on observable scales the Poisson fluctuations are much weaker than the density-induced component because the corresponding volumes are rather large.

From the form of $\beta_\alpha$ in equation~(\ref{eq:beta-alpha}) and in Figure~\ref{fig:betaz}, it is obvious that the fluctuations must peak when $x_\alpha \sim 1$; this is clearly visible here as well.  The saturation when $x_\alpha \gg 1$ implies that fluctuations in the \lya background are only relevant over a limited redshift range, so identifying that epoch, and studying the power spectrum during it, constrain the parameters of the first stars. 

Figure~\ref{fig:tsfluc} assumes $T_K=T_{\rm ad}=9.25 \kel$, the temperature appropriate for an adiabatically expanding IGM with no heat sources other than Compton scattering.  In this case,  $\bdtb=-100.7 \mkel$ at $x_\alpha=1$.  The fluctuations reach $\sim 10 \mkel$ on potentially observable scales ($k \sim 0.1 \Mpc$), implying $\sim 10\%$ variations across the sky.  Because the rms fluctuations scale with $\bdtb$, these can be easily converted to other scenarios with uniform heating.  Note then that, if $T_K \gg T_S$, these fluctuations will be significantly harder to observe.

The above calculations assumed linear fluctuations.  Of course, this may not be a good approximation in all cases, especially if the sources are highly clustered \cite{pritchard05}.  For instance, for  Pop III stars with large escape fractions (which have large ratios of ionizing photons to Ly$n$ photons), recombinations near the edges of the \htwo regions would produce copious numbers of \lya photons and could create strong coupling near the edge of the region.  

To this point we have included only the coupling induced by stellar radiation; any associated X-ray component will also produce \lya photons and hence modify the coupling (see eq.~\ref{eq:wf-xray}).  As we will see below, the mean free path of X-ray photons is typically a strong function of energy, so they seed smaller-scale fluctuations and shift the observed power to larger $k$-modes \cite{pritchard06}.  \emph{Imaging} the regions around the first stars would offer even more powerful constraints on their spectra \cite{cen06, chen06, chuzhoy06-first}, although it is far beyond the capabilities of any planned instruments.  These objects would appear as \htwo regions surrounded by zones of strong emission.  Close to the star, soft X-rays heat the gas, and \lya photons (either produced directly by the the stars or through X-ray excitation) couple the spin and kinetic temperatures.  Farther out (where heating is weak but coupling still strong), absorption will dominate until the gas eventually fades into invisibility.  In either case, the Wouthuysen-Field coupling seeded by X-rays is \emph{not} a small perturbation and must be included in realistic calculations; it modifies the shape of the power spectrum on relatively small scales \cite{chuzhoy06-first, pritchard06}.  Moreover, in most circumstances these fluctuations cannot be cleanly separated from those in $T_K$, and the two must be considered together (see below).

\subsection{X-Ray Pre-ionization} \label{xray}

In order to see the high-redshift IGM in 21 cm emission, a large fraction of the gas must be heated without becoming highly ionized. An X-ray background fits the bill perfectly:  with their large mean free paths, X-ray photons can pervade the bulk of the IGM and provide a fairly uniform heating source even far away from galaxies. On the other hand, once $\bar{x}_{e} \ga 0.1$,\footnote{In this section, we will use $\bar{x}_e$ to denote the mean ionized fraction \emph{outside} of \htwo regions.} secondary ionizations become unimportant and the bulk of the primary electron's energy increasingly goes into Coulomb collisions (see eq. \ref{eq:fxapprox}); an IGM heated by X-rays will therefore remain predominantly neutral \cite{venkatesan01}. As previously mentioned in \S\ref{xrayheat}, possible X-ray sources include supernovae (which produce both free-free and inverse Compton emission), X-ray binaries, and mini-quasars. X-rays are thus an inevitable byproduct of any of the luminous sources that can source reionization. We already discussed their most important aspect, X-ray heating, in \S\ref{xrayheat}. Here we describe three other aspects relevant to 21 cm observations: the increased coupling between $T_{K}$ and $T_{S}$ in a warm, partially ionized IGM, fluctuations in the heating rate, and observational constraints on an early X-ray background. 

The importance of an X-ray background in promoting collisional coupling should be immediately obvious from Figure~\ref{fig:collrates}. First, the rate coefficient $\kappa_{10}^{\rm HH}$ (from H-H collisions alone) rises steeply at low temperatures; thus, even a small amount of X-ray heating will sharply increase the amount of collisional coupling. Second, in the range $100 \kel < T_{K} < 3000 \kel$, the H-e$^-$ collisional coupling rate is $\sim 20$ times larger than the H-H collisional coupling rate. Thus, as noted by \cite{nusser05-xray}, if  $\bar{x}_e \ga 5\%$, H-e$^-$ collisions could predominate. This is particularly important because a moderately warm IGM is typically not too far from $x_c \sim 1$ anyway.  For example, if $T_K \sim 500$ K in a purely neutral medium, an overdensity $\delta \ge 8 [(1+z)/15]^{-2}$ is required for $x_{c}^{\rm HH} \ge 1$ to unlock the spin temperature from the CMB; thus, even in such a medium, filaments will be visible in emission. However, if it is ionized, then $x_{c}^{\rm eH} \approx 2 (\bar{x}_{e}/0.1) x_{c}^{\rm HH}$, and the critical density for coupling decreases to of order the mean density $\delta \sim 1$.  

The right hand panels of Figure~\ref{fig:igm-pre} illustrate how non-uniform X-rays can affect the spin and brightness temperatures \cite{kuhlen06-21cm}.  In this simulation, a mini-quasar X-ray source has been placed at the ``X" near the center of the box.  The contrast with the left-hand panels, which lack a mini-quasar and in which only dense filaments are visible (see the discussion in \S \ref{first-igm}), is stark.  The relatively strong X-ray emission has heated the gas to $\sim 2800 \kel$ and set $\bar{x}_e \approx 0.03$ in the box; thus collisional coupling becomes quite strong in even slightly overdense filaments (especially near the mini-quasar).  This leads to a dramatic increase in the magnitude and covering fraction of 21 cm emission; indeed, because the local density determines the coupling strength (and the ionized fraction, provided that the gas is dense enough for photoionization equilibrium to apply), even a uniform X-ray background increases the 21 cm contrast between filaments and voids \cite{nusser05-xray}.  The resulting fluctuations have magnitudes of a few mK on arcminute scales.

Although illustrative, this calculation did not include the \lya background generated by X-rays (through collisional excitation; see eq.~\ref{eq:wf-xray}).  Such photons are usually not produced in sufficient quantities to drive $x_\alpha \rightarrow 1$, but they can seed strong enough coupling to substantially increase the $\dtb$ fluctuations \cite{chen06, chuzhoy06-first}, especially near sources (where most of the soft X-rays are absorbed).

Thus, even in the absence of a UV background, an early X-ray background can still drive $T_{S} \rightarrow T_{K}$, and some such scenarios are plausible and even attractive in some ways \cite{oh01, ricotti04_a, ricotti05}.  Interestingly, in the limit of uniform radiation backgrounds, adding Wouthuysen-Field coupling actually decreases the contrast of filaments because $x_{\rm tot}$ is much more uniform.  On the other hand, as discussed in \S\ref{radfluc}, the UV background varies, which seeds associated 21 cm fluctuations. By permitting $T_{S} \rightarrow T_{K}$ coupling in regions where the UV flux is sub-critical, X-ray induced collisional coupling could modulate or smooth out such fluctuations. 
 
Because of their long mean free paths, X-ray photons are often portrayed as providing a uniform background. This is of course not strictly true. The comoving mean free path of an X-ray photon with energy $E$ is:
\begin{equation}
\lambda_{\rm X} \approx 4.9 \, \bxhi^{1/3} \left( \frac{1+z}{15} \right)^{-2} \left( \frac{E}{300 \eV} \right)^{3} \Mpc;
\end{equation}
thus, the universe will be optically thick to all photons below $\sim 1.8 [(1+z)/15]^{1/2} \bxhi^{1/3}$ keV. By comparison, a photon produced just redward of Ly$\beta$ can travel a comoving distance
\begin{equation}
\lambda_{\alpha} \approx 330  \left( \frac{1+z}{15} \right)^{-1/2} \Mpc
\end{equation} 
before redshifting into the Ly$\alpha$ resonance. Thus, soft X-ray photons will fluctuate on relatively small scales but, because of the steep energy dependence, there {\it will} be a uniform component to the X-ray background (unlike in the UV).  For a spectrum with $\nu L_{\nu} \sim$const (such as the nonthermal component observed in nearby ultraluminous X-ray sources), the component uniform on $\sim 5 \Mpc$ scales provides $\sim {\rm ln(2000/300)/ln(300/13.6)} \sim 60\%$ of the heating compared to the fluctuating component.  Another consequence of $\lambda_{\rm X} \propto E^3$ is that the spectrum will harden significantly as one proceeds away from the clustered ionizing sources. Nonetheless, many of the causes of fluctuations in the Wouthuysen-Field effect ($1/r^{2}$ weighting; clustering of sources) apply with equal force to X-rays. 

Fluctuations from inhomogeneous X-ray heating can be computed using a transfer function $W_X(k)$ that parameterizes the perturbations in the heating rate relative to the density field; the only substantive difference from the \lya case is that $T_K$ depends on the entire history of heating while $x_\alpha$ depends only on the instantaneous flux at the \lya transition \cite{pritchard06}.  Fluctuations sourced by $T_K$ have one unique property:  the cross-power between temperature and density can be \emph{negative} if $T_K < T_\gamma$ (because then $\beta_T < 0$).  Physically, dense gas (which tends to sit near luminous sources) is warmer than average and has a \emph{smaller} $\dtb$ than average.  Detection of such a feature would provide a clear indication of a cold IGM with substantial temperature fluctuations.

\begin{figure}[!t]
\centerline{\epsfig{file=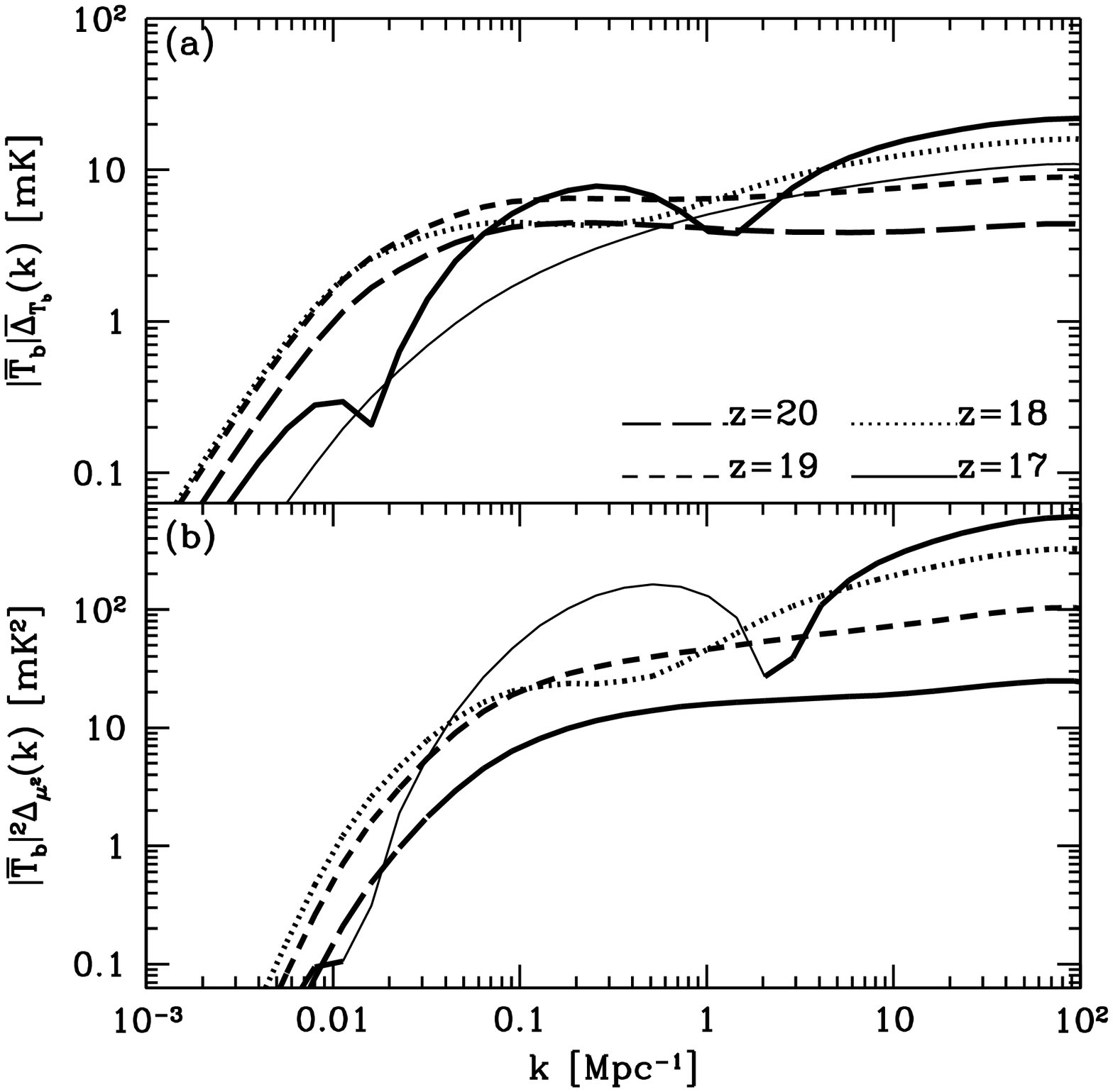,width=2.75in}
\epsfig{file=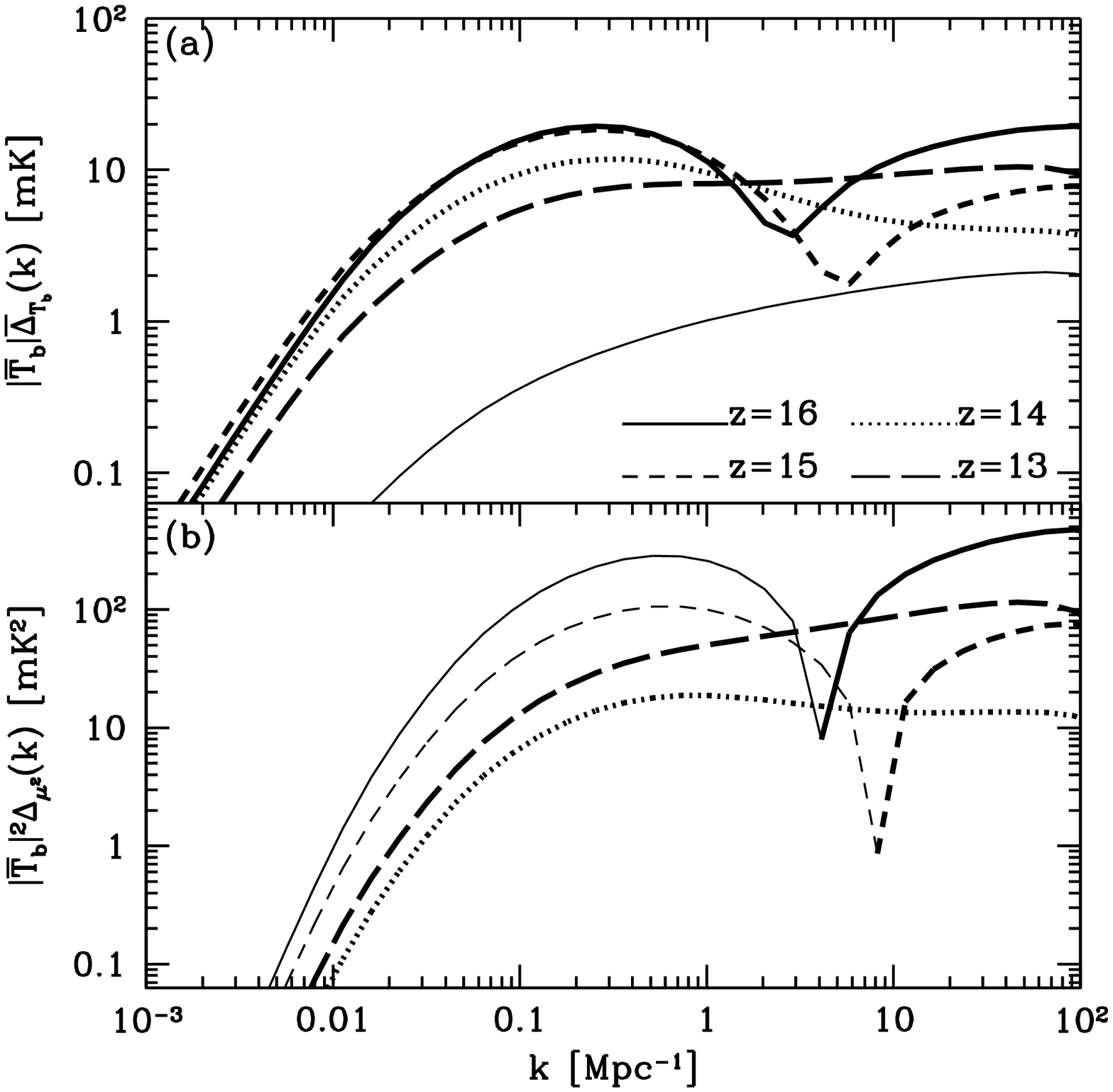,width=2.75in}}
\caption{Power spectra of $\dtb$ including fluctuations in both $T_K$ (from X-ray heating) and the Wouthuysen-Field coupling.  The upper panels show the spherically-averaged power, while the lower panels show the $\mu^2$ component.  We show several different redshifts in a reionization history similar to our fiducial Pop II model of Fig.~\ref{fig:pop2-glob}.  In the lower panels, thick curves denote a positive signal and thin curves a negative signal.  The thin curves in the top panels also show the fluctuations with uniform heating at $z=19$ and 14 (left and right, respectively) for reference.  From \cite{pritchard06}. }
\label{fig:xray}
\end{figure}

Figure~\ref{fig:xray} shows the resulting power spectra (including both the spherically-averaged component in the top panels and the $\mu^2$ component in the bottom panels) in a model similar to our fiducial Pop II history from \S \ref{globreion-models} \cite{pritchard06}.  At high redshifts, $z \ga 18$, Wouthuysen-Field fluctuations dominate and the power spectra appear similar to those in Figure~\ref{fig:tsfluc}.  X-ray heating begins to kick in at $z \sim 17$ and immediately has a dramatic effect.  $P_{\mu^2}$, which contains a term $P_{\delta_T \delta}$, becomes negative on intermediate scales, where temperature fluctuations are strong.  The spherically-averaged power remains positive (as it must), but it contains two distinctive troughs separating the regimes where density, temperature, and \lya fluctuations dominate (from small to large scales).  The peak on the largest scales eventually disappears (once $x_\alpha \gg 1$), and the troughs disappear entirely once $T_K > T_\gamma$ (at $z \sim 14$ here).  

%%%%%%%%%%%% FIGURE 7-2: X-ray fluctuations

As the mean IGM temperature increases, X-ray fluctuations become less and less important (eventually disappearing once $T_S \gg T_\gamma$ everywhere).  However, they can still be substantial even after the peak-trough component disappears, especially when $T_K$ is still relatively close to the CMB temperature:  compare the dotted $z=14$ curve to the thin solid curve in the top right panel, which shows the expected fluctuation amplitude if the heating is completely uniform.  However, at least in this scenario, these fluctuations still precede those generated by \htwo regions during reionization and can be separated cleanly.  Measuring the 21 cm power spectrum in this regime would reveal the properties of the first X-ray sources (and especially their bias and spectra) and may provide our best hope of constraining the thermal history of the IGM \cite{pritchard06}.

What observational constraints can we place on high-redshift X-ray emission? The most obvious limit is the present-day soft X-ray background (SXRB), part of which could originate from a high-redshift hard X-ray background ($\ge 10$ keV) that free streams until today \cite{ricotti04_a, dijkstra04, salvaterra05-xray}. Approximately $94^{+6}_{-7} \%$ of the SXRB has been resolved \cite{moretti03}; the mean and maximum intensity of the unresolved component is $(0.35,1.23) \times 10^{-12} \, {\rm erg \, s^{-1} \, cm^{-2} \, deg^{-2}}$ respectively in the 0.5--2 keV bands. Unfortunately, constraining X-ray reionization from the unresolved X-ray background (XRB) requires an uncertain extrapolation of the spectral energy distribution to hard energies.  For instance, in mini-quasars, most of the UV and soft X-rays photons are thought to come from an accretion disk, while  hard X-rays are part of a power-law tail produced through synchrotron/inverse-Compton emission.  The relative contribution of the two components is extremely uncertain.  

Fortunately, we can perform a simple order-of-magnitude estimate that roughly matches the more detailed constraints of \cite{dijkstra04, salvaterra05-xray}. Suppose the high-redshift XRB is emitted at a median redshift $z$ by a population of black holes with comoving mass density $\rho_{\rm BH}$ and radiative efficiency $\epsilon$, with a fraction $f_{\rm HXR}$ of the radiation emerging in the $[0.5$--$2](1+z)$keV range. The comoving energy density in hard X-rays is $\rho_{\rm HXR} \approx \epsilon \rho_{\rm BH} c^{2} f_{\rm HXR}$ and the flux received at earth (in energy units) is $J=c/(4\pi) \rho_{\rm HXR}/(1+z)$. On the other hand, the number of ionizing photons per baryon produced by these same black holes is $N_{\rm ion}=\epsilon f_{\rm UV}/\bar{n}_b (\rho_{\rm BH} c^{2}/\VEV{E})$, where $f_{\rm UV}$ is the fraction of the bolometric luminosity that emerges above 1 Rydberg and $\VEV{E}$ is the average energy per ionization. Then, the SXRB observed at the present day is
\begin{eqnarray}
J_{\rm X} & \approx & 3.2 \times 10^{-12}\left( \frac{f_{\rm HXR}/f_{\rm UV}}{0.2} \right) \left( \frac{ \VEV{E}}{ 5 \, {\rm Ry}} \right) \left( \frac{N_{\rm ion}}{10} \right) \nonumber \\ 
& & \times  \left( \frac{10}{1+z} \right)  \ {\rm erg \, s^{-1} \, cm^{-2} \, deg^{-2}},
\end{eqnarray}
where $f_{\rm HXR}/f_{\rm UV}$ and $\VEV{E}$ are appropriate for a spectrum with $L_{\nu} \propto \nu^{-1}$ ranging from 13.6 eV to 10 keV. Thus, the XRB produced if quasars or mini-quasars alone reionized the Universe probably violates observed limits, although there are considerable uncertainties. It certainly does not preclude a significant contribution ($N_{\rm ion}\sim$ few) to the observed \emph{WMAP} optical depth.\footnote{Note that this argument can be adapted for other emission mechanisms by adjusting $f_{\rm HXR}/f_{\rm UV}$ and $\langle E \rangle$.} Certainly, the amount of IGM preheating required for 21 cm emission will not violate observed constraints:
\begin{eqnarray}
J_{X} & \approx & 2.7 \times 10^{-14} \left( \frac{T_K}{1500 \kel} \right) \left( \frac{f_{\rm HXR}}{f_{\rm SXR}} \right) \left( \frac{0.13}{f_{X,h}} \right) \nonumber \\ 
& & \times \left( \frac{10}{1+z} \right)\  {\rm erg \, s^{-1} \, cm^{-2} \, deg^{-2}}, 
\end{eqnarray} 
where $f_{\rm SXR}$ is the fraction of bolometric luminosity emerging in the relevant $0.3$--$2$ keV bands and $f_{X,h}$ is the fraction of the primary electron's energy which goes into heat (this depends on the ionization fraction; see eq. \ref{eq:fxapprox}). 

Observational constraints on the X-ray luminosity associated with high-redshift star formation are less tight. In principle, the SKA will be capable of detecting synchrotron radiation from high-redshift sources \cite{oh01}. Since the same relativistic electrons producing synchrotron emission also inverse-Compton scatter CMB photons to X-ray energies, such detections could place a bound on X-ray emission from supernova remnants given reasonable assumptions about the magnetic fields in supernova remnants: $L_{\rm X}/L_{\rm sync}=u_{\gamma}/u_{\rm B}$, where $u_{\gamma}$ and $u_{\rm B}$ are the energy densities in the CMB and magnetic fields respectively. Intriguingly, the expected amount of high-redshift star formation can also produce a gamma-ray background comparable to the unresolved component from EGRET \cite{oh01}, a scenario which could be tested by GLAST.  Unfortunately, this technique would not constrain other sources of X-ray photons, especially X-ray binaries.  

%\bibliographystyle{elsart-num}
%\bibliography{Ref_21cm}

%\end{document}

%% file: reion-ch8.tex
%\documentclass{elsart}
%\usepackage{amssymb,cite,epsfig}

%\input{../../submission/defns.tex}

%\begin{document}

\section{Reionization} \label{reion}

We will now consider perhaps the most exciting aspect of the 21 cm signal:  its potential to teach us how and when the Universe was reionized.  We have already reviewed the basic physics driving $\bxion(z)$ in \S \ref{ionhist}.  Here we will focus on the ``geometry" or ``topology" of reionization (i.e., how the neutral and ionized gas was distributed at a given $\bxion$), because the 21 cm line provides by far the best avenue to study this crucial aspect.  

\subsection{Simulations} \label{reion-sims}

Because the physics governing inhomogeneous reionization is so complex, numerical simulations are the ideal way to approach it, and there has been a great deal of interest in this possibility over the past decade \cite{gnedin97, gnedin00, razoumov02, ciardi03-sim, ciardi03-sim2, sokasian03, sokasian04, kohler05-sim, iliev05-sim, zahn06-comp}.  The basic requirements are (i) code to evolve the density field, (ii) a prescription assigning ionizing source parameters, and (iii) code to compute the radiative transfer of ionizing photons through the gas density field.  

The first component solves for the growth of structure in an expanding universe; it is of course necessary for any cosmological simulation, so the relevant codes are now well-established.  The principal challenge for reionization studies is the computational expense of hydrodynamics.  Thus, often the gas dynamics are ignored and pure N-body codes, which track only the dark matter, are employed \cite{ciardi03-sim, ciardi03-sim2, iliev05-sim, zahn06-comp}.  Although this allows significantly larger volumes to be studied, it requires the gas distribution, and especially star formation rates and small-scale clumping, to be prescribed in a similar way to analytic models.

The second component is an algorithm to assign luminosities to ionizing sources.  As we have emphasized in \S \ref{ionhist}, any such procedure is fraught with uncertainties because we know so little about high-redshift galaxies.  Simulations are typically calibrated to the observed properties of $z \la 3$ galaxies, supplemented by an extrapolation to higher redshifts.  Of the parameters that determine the ionizing efficiency $\zeta$ (see eq. \ref{eq:zetadefn}), simulations can in principle most improve our estimates of $f_\star$ because they follow the gas as it collapses to high densities (though this is more difficult in pure N-body simulations, of course).  Unfortunately, the simplest prescription (the so-called Schmidt law \cite{schmidt59}), which is motivated by observed star formation rates in nearby galaxies \cite{kennicutt98}, tends to overproduce stars (hardly surprising even to its creator, given its simplified assumptions; \cite{schmidt63}).  It must therefore be supplemented by prescriptions for feedback regulation of star formation -- internal, external, or both, in the language of \S \ref{feedback}.  Examples include adding supernova winds carrying gas particles out of galaxies or introducing a multiphase interstellar medium on the subgrid level (e.g., \cite{springel03}).  The resulting models are then calibrated to specific local observations (such as the star formation rate-density relation \cite{kennicutt98, springel03} or the mean star formation rate \cite{gnedin00}).  Modern codes produce reasonable agreement with a wide range of observations at $z \la 4$ (e.g., \cite{springel03-sf, nagamine04}).  However, the star formation and feedback parameters may evolve with redshift, which could have important implications for reionization (e.g., \cite{dave05}).  Unfortunately, even with self-consistent star formation rates simulations are no better than analytic models at determining $N_{\rm ion}$ or $\fesc$, which are generally prescribed constants.  (Although the metallicity can, to some extent, be followed, that has not been factored into the ionizing efficiencies in existing simulations.)  Thus simulations suffer from the same systematic uncertainties about the source population as analytic models.  The range of source prescriptions, from purely analytic \cite{furl04-bub, furl05-charsize} to semi-analytic \cite{ciardi03-sim, iliev05-sim, benson05, zahn06-comp} to ``fully" numeric \cite{gnedin00, sokasian03}, illustrates the diversity of approaches to these problems.

Finally, radiative transfer is a cutting-edge problem that has received a great deal of attention lately.  Computing the specific intensity $I_\nu(t, \, {\bf x}, \, {\bf n}, \, \nu)$ requires solving a seven-dimensional problem:  time $t$, position ${\bf x}$, frequency $\nu$, and direction of propagation ${\bf n}$.  Furthermore, simulations can contain hundreds of thousands of sources, even excluding the diffuse light generated by IGM recombinations.  Thus the complete problem is prohibitively expensive, and approximate schemes are necessary.  Existing approaches include the ``local optical depth approximation" \cite{gnedin97, gnedin00},  the ``optically thin variable Eddington tensor" approximation \cite{gnedin01-otvet, kohler05-sim}, adaptive ray tracing \cite{abel02-ray, razoumov02, sokasian03, sokasian04, iliev05-sim}, and Monte Carlo techniques \cite{maselli03, ciardi03-sim, ciardi03-sim2}.  We will forego the details of these various approaches and refer the interested reader to the individual papers for more information.  For us, the crucial result is that, for the most part, the existing codes yield reliable and consistent results \cite{iliev06-rt}.

To further complicate matters, feedback demands that these three components interact with each other:  supernova winds and photons affect the gas distribution, which affects the halo (and hence star) formation rate, which affects the ionization rate, etc.  Prescriptions for mechanical feedback are now incorporated into many simulations, but the results can be quite sensitive to uncertainties in the observations and implementation, especially at high redshifts \cite{dave05}.  Including photoheating requires the radiative transfer to be incorporated directly into the simulation.  This is possible \cite{gnedin00}, but in most cases radiative transfer is added as a post-processing step \cite{razoumov02, sokasian03} and hence has no dynamical effect.  Of course, whenever N-body simulations are used, this kind of feedback is implicitly ignored anyway, because the photons cannot interact with dark matter particles \cite{ciardi03-sim, iliev05-sim, zahn06-comp}.  Thus, at least for now, the feedback processes described in \S \ref{feedback} must be explored by analytic or semi-analytic means.

As in most cosmological applications, the biggest computational challenge is resolving all the relevant structures while also subtending a large enough volume to be representative of the Universe as a whole.  The most obvious criterion for a ``representative volume" is that the density fluctuations on the scale of the box be small.  This is, of course, easy to check and also relatively easy to achieve:  at $z \sim 6$, the nonlinear mass scale corresponds to $\sim 10^6 \Msun$.  With this criterion in mind, most of the first generation of reionization simulations used boxes $\la 10 h^{-1} \Mpc$ on a side \cite{gnedin97, gnedin00, razoumov02, sokasian03, sokasian04} (or $20 h^{-1} \Mpc$ with pure dark matter codes \cite{ciardi03-sim, ciardi03-sim2}).  Some of these simulations could resolve all galaxies with $T_{\rm vir}>10^4 \kel$, although none included minihalos.  

These simulations focused on understanding the global evolution of reionization, tracking $\bxion(z)$ for a given source population, and on the ``breakout" phase of ionization fronts around galaxies.  But they also began to address questions about inhomogeneous reionization by observing where \htwo regions appeared and how they grew.  Unfortunately, it quickly became obvious that these volumes were insufficient for answering many questions about reionization.  One problem is that high-redshift galaxies are so highly biased that even these large boxes missed many of them \cite{barkana04}.  A second difficulty, more important for our purposes, is that (again because of clustering) reionization in each box was driven by just a few clumps of sources.  Thus all of these simulations sampled only a few \htwo regions and could not adequately model the inhomogeneity.  A third problem, relevant to 21 cm studies, is that realistic experiments will have angular resolutions of several comoving Mpc -- comparable to the total sizes of these simulations!  This motivated the development of the analytic models that we will describe in \S \ref{reion-an}.

As a result, attention has recently focused on performing reionization simulations in much larger boxes -- with sides $\ga 100 \Mpc$ \cite{kohler05-sim, iliev05-sim, zahn06-comp}.  These must use pure N-body codes, and even then the dynamic range is not good enough to resolve all of the collapsed halos, so many ionizing sources are missed (especially if reionization is tuned to occur at high redshifts, when the nonlinear mass scale is much smaller).  Moreover, the clumpiness is completely unresolved (especially without hydrodynamics).  Fortunately, both of these problems can be overcome, at least approximately, through analytic models or by ``bootstrapping" from smaller boxes where the relevant structures are better resolved.  For example, grid-based radiative transfer codes (such as ray-tracing algorithms) require the clumping factor and emissivity of each grid cell.  These can both be extracted from comparable regions in higher-resolution simulations \cite{kohler05-sim, mellema06} (note, however, that clumpiness measured in this way cannot include the back-reaction of reionization itself, which can be substantial -- see \S \ref{clump}).  The hope, of course, is that these approximations will not affect the large-scale features of reionization.  Given the resolution of 21 cm experiments, this is probably not a bad approximation.

\begin{figure}[!t]
\centerline{\epsfig{file=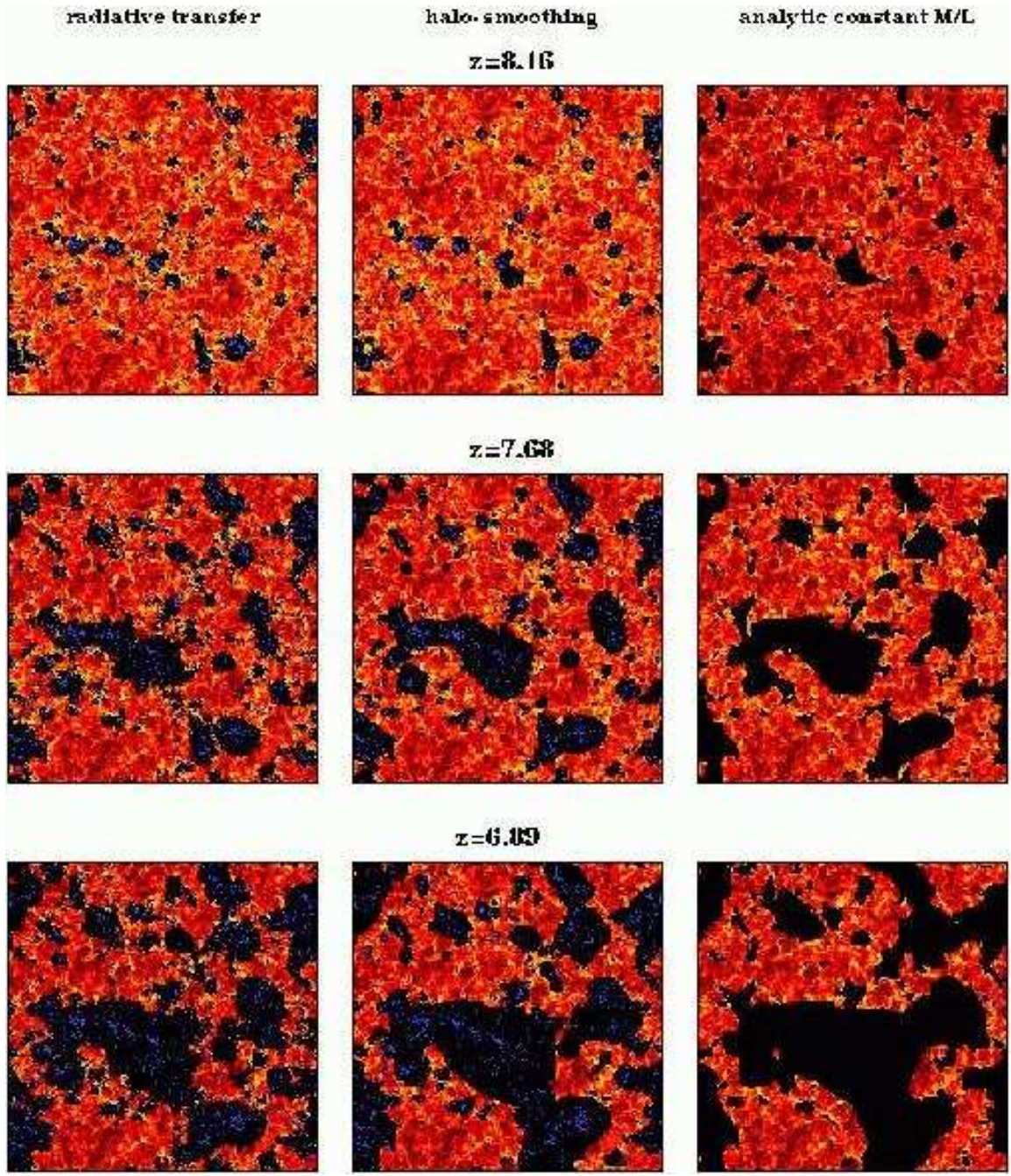,width=6.0in}}
\caption{Slices through a simulation of reionization at $z=8.16,\,7.68$, and $6.89$ (top to bottom); these have $\bxion=0.13,\,0.35$, and $0.55$, respectively.  The colorscale is proportional to the \hone density.  The slice is $65.6 h^{-1}$ comoving Mpc across and $0.25 h^{-1} \Mpc$ deep.  The left column uses a radiative transfer code; the right column applies a slightly modified version of \cite{furl04-bub} to the same density field.  The middle column applies that ionization criterion to the overdensity of bound halos.  The points inside the \htwo regions in the left and middle columns identify the ionizing sources.  From \cite{zahn06-comp}.}
\label{fig:reion-sim}
\end{figure}

The leftmost column in Figure~\ref{fig:reion-sim} shows an example of a reionization simulation in a $(65.6 h^{-1} \Mpc)^3$ N-body simulation taken from \cite{zahn06-comp} (which uses the radiative transfer code of \cite{sokasian03}).  The ionizing efficiencies were assigned so that reionization ends slightly before $z=6$.  It resolves galaxies with $M \ga 2 \times 10^9 \Msun$ and underestimates the recombination rate by ignoring small-scale clumping.  Several crucial qualitative features are apparent in Figure~\ref{fig:reion-sim}.  First, the ionized regions rapidly attain large sizes.  At $z=7.68$, when $\bxion=0.35$, the characteristic bubble size is $\sim 3 \Mpc$, with some bubbles already larger than $10 \Mpc$.  The typical size reaches $\sim 20 \Mpc$ by $z=6.89$.  As we shall see, this is a direct result of the highly clustered galaxy distribution.  These large features should make the \htwo regions much easier to observe with 21 cm surveys.  

%%%%%%%%%%%% FIGURE 8-1: Lidz/Zahn simulations

Second, reionization begins in and around the densest regions.  In Figure~\ref{fig:reion-sim}, the ionized bubbles predominantly appear around large-scale overdensities, leaving voids neutral until the end of reionization.  This ``inside-out" picture of recombination is generic to any scenario in which the sources are more biased than the recombining gas \cite{sokasian03, furl05-rec}.\footnote{This is not to say that, on a local level, reionization cannot be ``outside-in" with dense blobs ionized after their low-density surroundings \cite{miralda00, furl05-rec}.  Instead we mean that photons originate in dense environments and so \emph{must} at least ionize the more rarefied gas in their local overdensity before escaping toward large voids.  In many ways, the apparent dichotomy between these two kinds of models is a false one.}  Third, the bubbles are initially roughly spherical but take more complex shapes as they reach scales $\ga 3 h^{-1} \Mpc$.\footnote{Resolving small galaxies is important for this aspect, because they provide the small-scale structure visible in Fig.~\ref{fig:reion-sim}:  resolution much worse than used here creates a ``cookie-cutter" morphology in which merged bubbles appear to be built from nearly spherical subunits \cite{zahn06-comp,iliev05-sim}.}  Nevertheless, many galaxies contribute to each bubble even when they are still roughly spherical:  evidently, before reionization, the cosmic web plays only a secondary role in the large-scale geometry of reionization (although this is partly a product of the coarse resolution of the simulation).  

\begin{figure}[!t]
\centerline{\epsfig{file=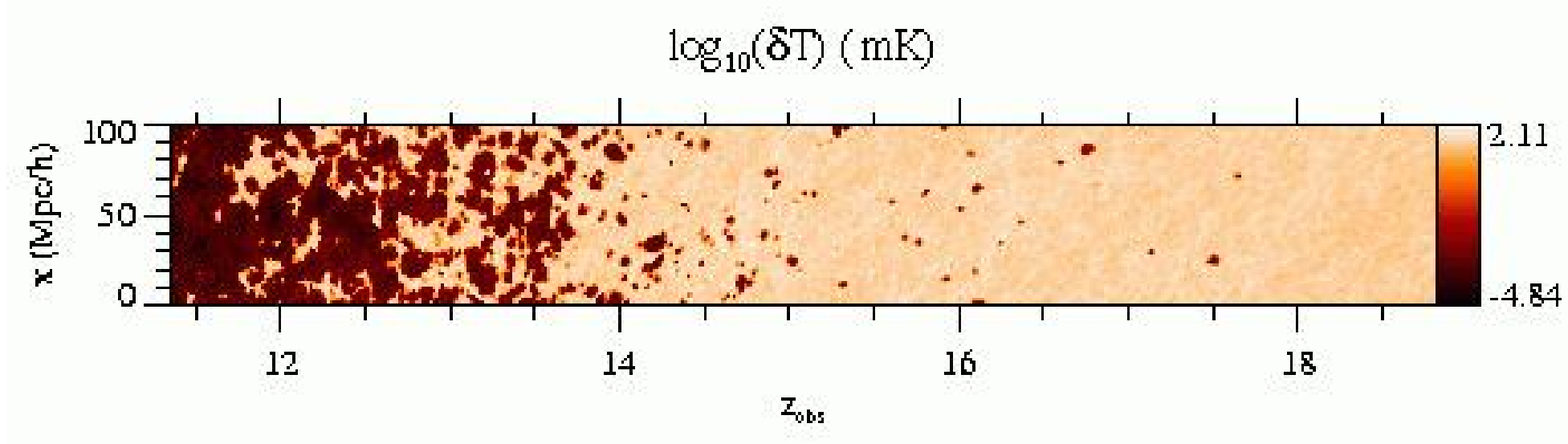,width=6.0in}}
\centerline{\epsfig{file=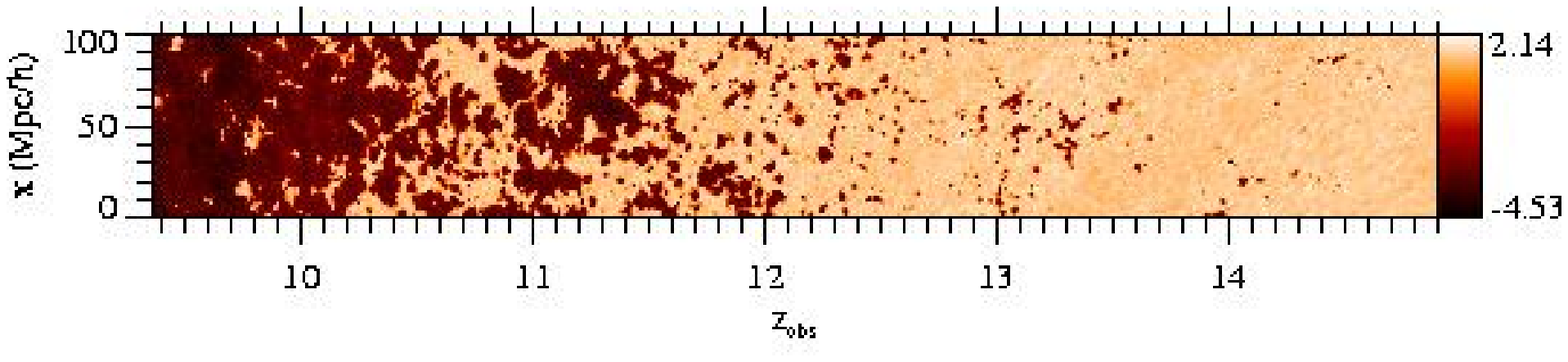,width=6.0in}}
\caption{The evolving brightness temperature of the 21 cm transition in two simulations of reionization (tuned so that reionization occurs at different times).  The box subtends $\sim 49'$ in the $x$-direction.  From \cite{mellema06}.}
\label{fig:mellema-sim}
\end{figure}

Computing the 21 cm signal from a simulation requires one additional piece of information:  the spin temperature.  Unfortunately, as we have seen in \S \ref{glob}, this depends on the UV and X-ray emissivities of the sources, which must be prescribed based on some analytic model.  (No reionization simulation has attempted radiative transfer for these components.)  A few do try to compute the evolution of all three quantities self-consistently on a global level \cite{carilli02, gnedin04, kohler05-sim}, but most simulations assume that $T_S \gg T_\gamma$ throughout the box.  Although extremely simplistic, this is most likely a reasonable assumption for $\bxion \ga 0.1$ (see \S \ref{glob} and \cite{ciardi03-21cm}).    In this limit $\dtb$ becomes independent of $T_S$, and the 21 cm signal can easily be computed from the simulation outputs.  Figure~\ref{fig:mellema-sim} shows slices through the largest calculation to date (an N-body simulation in a $100h^{-1} \Mpc$ box) \cite{mellema06}.  The slices are $\sim 49'$ across; the two panels effectively assume different ionizing efficiencies and hence have reionization end at different times.  The lower panel also imposes subgrid clumpiness that evolves with time (eq.~\ref{eq:clump-mips}), slowing down reionization.  Because the 21 cm signal essentially traces the neutral gas density, the maps look quite similar to those of Figure~\ref{fig:reion-sim}:  relatively weak fluctuations from the cosmic web in predominantly neutral gas and significantly stronger fluctuations from \htwo regions.  This is true even though the assumed reionization redshifts are quite different:  as we will see below, the qualitative features of reionization at a fixed $\bxion$ are only weakly dependent on redshift.  Note, however, that  toward the end of this simulation, the bubbles ``percolate" and the topology inverts itself,  transforming from a ``Swiss cheese" universe composed of neutral gas with ionized holes to a sea of ionized gas with islands of neutral material.  This behavior is generic to any percolation phenomenon \cite{isichenko92}. 

%%%%%%%%%%%% FIGURE 8-2: Mellema simulation

\subsection{Analytic Models} \label{reion-an}

Given its intrinsic complexity, reionization may seem a difficult problem for analytic models.  But they are useful for several reasons.  Most importantly, they give us physical insight into how the seemingly complex global ionization pattern is generated:  as we will see, to first order it is shockingly simple.  Second, they have essentially infinite dynamic range, avoiding many of the resolution problems of the simulations.  Third, they allow us to fill parameter space efficiently, which is crucial given the large uncertainties in the input parameters.  Finally, simulations themselves require analytic approximations for the sources and clumping, so it is best to understand them in detail!  Figures~\ref{fig:reion-sim} and \ref{fig:mellema-sim} give us hope that such models might be successful:  in them, reionization is driven by the large-scale clustering of ionizing sources -- which can be described by linear theory.  This allows us to construct simple models for the size distribution of \htwo regions as a function of $\bxion$.  While these models must ultimately be compared to numerical simulations, they provide a great deal of intuition about the 21 cm signal.

The key ingredients to a model of \htwo regions are their number density as a function of mass $n_b(m)$, their clustering strength, and their correlation with the underlying density field.  The only possible solution that approaches anything like ``first principles" is to examine the growth of \htwo regions around individual galaxies \cite{arons72, shapiro87}; unfortunately, this model problem is probably only relevant for the first galaxies, before clustering becomes significant.  One simple generalization is to assume that bubbles trace dark matter halos but to allow for clustering in an approximate way.  For example, specifying $\bxion(z)$ and the characteristic bubble size $R_c(z)$ determines $n_b$ uniquely, and the bubbles can then be naturally associated with dark matter halos of that number density \cite{wang05_FoG}.  (Of course, the associated halos need not provide all the ionizing photons for the bubble; their neighbors could contribute as well.)  Such a model takes advantage of the highly-developed halo-clustering machinery to describe the bubble pattern \cite{cooray02}, but it requires $R_c(z)$ to be specified ``by hand." 

A more satisfying approach is to estimate $R_c(z)$ on physical grounds.  The key ingredient is the source clustering, which implies that overdense regions (with many sources) will be ionized before underdense regions \cite{barkana04}.  We begin by assuming that $\bxion = \zeta \fcoll$, where $\zeta$ is the ionizing efficiency defined in equation (\ref{eq:zetadefn}).\footnote{This is not strictly necessary \cite{furl05-charsize}, but it makes the argument simpler.  For now we will ignore recombinations.}   Consider an isolated (large) region of the IGM with mass $m$ and mean fractional overdensity $\delta$.  It is fully ionized if the local collapse fraction $\fcoll(\delta,m)$ satisfies
\begin{equation}
\fcoll(\delta,m)>\zeta^{-1}, 
\label{eq:ioncond}
\end{equation}
where for the Press-Schechter mass function\footnote{Note that the following conclusions do not depend on the precise form of the mass function \cite{furl05-charsize}.} \cite{press74} (c.f. eq. \ref{eq:nmps})
\begin{equation}
\fcoll(\delta,m) = {\rm erfc} \left\{ \frac{\delta_c(z) - \delta/D(z)}{\sqrt{2} \, [\sigma(\mmin) - \sigma(m)]} \right\}.
\label{eq:fcoll-local}
\end{equation}
Thus, for isolated regions, we can restate the ionization condition as a criterion on the local smoothed density field.  

We would like a method to compute the statistical distribution of regions satisfying this condition, as a function of mass.  In that case we cannot treat each region in isolation:  a low-density void lying near a cluster of sources will be ionized by its neighbors.  Interestingly, stated this way the problem is similar to the ``excursion set" derivation of the Press-Schechter mass function \cite{bond91, lacey93}.  In that case, the critical overdensity $\delta_c(z)$ describes the condition for virialization:  any region with $\delta>\delta_c(z)$ is part of a halo.  But this prescription also suffers from a problem similar to our void/ionizing cluster situation (here it is known as the ``cloud-in-cloud" problem"):  a point can be part of many regions with $\delta>\delta_c(z)$ (on different mass scales).  For halos, as for \htwo regions, only the largest of these is physically relevant, because it incorporates all of the smaller ones.  The excursion set formalism treats the problem as diffusion in the $(\sigma^2, \, \delta)$ space [or equivalently $(m,\, \delta)$] with an absorbing barrier at $\delta_c(z)$;  here $\sigma^2$ plays the role of time (because the rms density fluctuation increases monotonically toward smaller scales) and $\delta$ plays the role of space.  Posed in this way, one can compute the distribution of ``crossing times" -- or masses -- from which the halo mass function (eq. \ref{eq:nmps}) follows immediately.  This procedure avoids the cloud-in-cloud problem by following trajectories from large to small mass scales (or increasing $\sigma^2$) and assigning each diffusion trajectory to the largest halo of which it is a part (this is why the barrier is absorbing).

The \htwo bubble problem has an identical structure except that the absorbing barrier $\delta_x$ comes from the condition $\fcoll(\delta,m)=\zeta^{-1}$ (instead of simply equaling $\delta_c$) and is hence a function of mass.  Fortunately, this barrier is well-approximated by a linear function in $\sigma^2$, $\delta_x(\sigma^2) \approx B(\sigma^2) \equiv B_0 + B_1 \sigma^2$,\footnote{Recently, an elegant numerical solution to the diffusion problem confirmed the accuracy of this approximation \cite{zhang05}.} which allows an analytic solution for the mass function \cite{sheth98, furl04-bub, mcquinn05}
\begin{equation}
m \, n_b(m) \, \deriv m = \sqrt{\frac{2}{\pi}} \, \frac{\bar{\rho}}{m} \, \left| \frac{ \deriv \ln \sigma}{\deriv \ln m} \right| \, \frac{B_0}{\sigma(m)} \, \exp \left[ - \frac{ B^2(m,z) }{ 2 \sigma^2(m) } \right] \, \deriv m.
\label{eq:nbub}
\end{equation}
Note the similarity to the Press-Schechter halo mass function \cite{press74, bond91}.  As a consequence of this derivation, most of the machinery used for halo mass functions, clustering, etc. can be carried over to \htwo regions.  For example, the linear bias of ionized bubbles, defined so that $n_b(m|\delta) = n_b(m) \, [1 + b_x(m) \, \delta]$ in a large region of mean overdensity $\delta$ \cite{efstathiou88, cole89, mo96}, is \cite{mcquinn05}
\begin{equation}
b_x(m) = 1 + \frac{B(m)/\sigma^2(m) - 1/B_0}{D(z)}.
\label{eq:bias-bub}
\end{equation}

\begin{figure}[!t]
\centerline{\epsfig{file=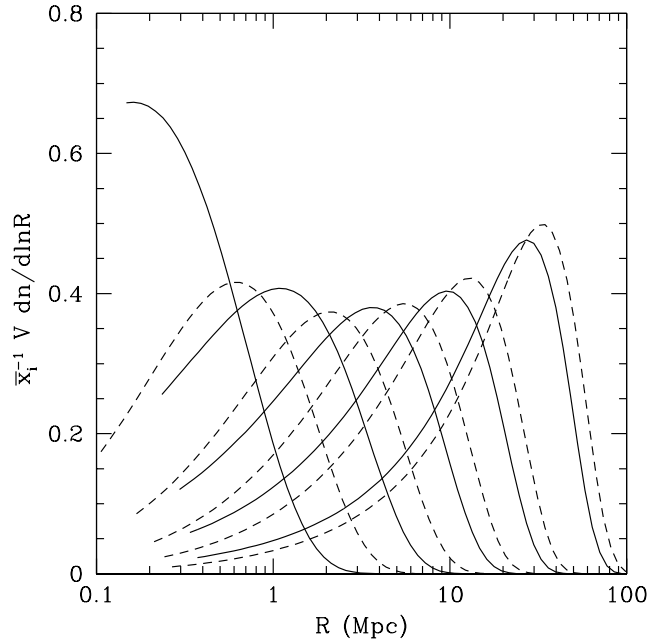,width=3.5in}}
\caption{\htwo region size distributions at $z=15$ for the model of \cite{furl04-bub, furl05-charsize}.  The solid and dashed curves assume $\zeta \propto m_h^0$ and $m_h^{2/3}$, respectively.  From left to right within each set, we assume $\bxion=0.05,\, 0.2,\, 0.4,\, 0.6$, and $0.8$.  Recombinations are assumed to be uniform throughout the IGM.}
\label{fig:nbub}
\end{figure}

The solid curves in Figure~\ref{fig:nbub} show the resulting size distributions for a range of $\bxion$ at $z=15$; the ordinate is the fraction of the ionized volume filled by bubbles of a given size.  The results are similar to those from the simulations \cite{zahn06-comp}:  most importantly, bubbles grow large during the middle stages of reionization, with characteristic sizes $R_c \sim 1,\, 4,\, 10$, and $30$ comoving Mpc when $\bxion=0.2,\, 0.4,\, 0.6$, and $0.8$.  

%%%%%%%%%%%% FIGURE 8-3: Bubble sizes

Of course it is possible to implement more sophisticated source prescriptions.  One possibility is to explicitly associate ionizing sources (either stars or black holes) with halo mergers; the results differ in detail but are qualitatively similar \cite{cohn06}.  The dashed lines in Figure~\ref{fig:nbub} show another, which assumes that $\zeta \propto m_h^{2/3}$. In this case massive galaxies provide a larger fraction of the ionizing photons, and $R_c$ increases \cite{furl05-charsize}.  To understand why, note that $R_c$ is the scale at which a ``typical" density fluctuation is able to ionize itself; mathematically, it is where $\sigma(R_c) \approx B$.  In the large bubble limit ($B \approx B_0$), our original ionization criterion becomes
\begin{equation}
\zeta \, \fcoll(\delta=B_0,\, \sigma^2=0) = 1.
\label{eq:b0}
\end{equation}
Expanding equation~(\ref{eq:fcoll-local}) to linear order, this can be written 
\begin{equation}
\sigma(R_c) \approx B_0 \approx \frac{\bxion^{-1} - 1}{D(z) b_{\rm eff}},
\label{eq:b0approx}
\end{equation}
where $b_{\rm eff}$ is the average galaxy bias \cite{mo96}.  Intuitively, a more biased galaxy population provides a larger ``boost" to the underlying dark matter fluctuations, allowing larger regions to ionize themselves.  Thus, by measuring the \htwo region sizes through the 21 cm transition, we can constrain the galaxies driving reionization.

Two more properties of equation~(\ref{eq:nbub}) deserve emphasis.  First, at a given $\bxion$, $n_b(m)$ depends only weakly on redshift.  This is because the shape of $\fcoll(\delta,m)$ evolves only slowly with redshift; quantitatively, $D(z) b_{\rm eff}$ is roughly constant for high-redshift galaxies \cite{oh99}.   Second, the width of $n_b(m)$ is ultimately determined by the shape of the underlying matter power spectrum, which steepens toward larger radii \cite{furl05-charsize}.  

Another physically motivated estimate of $R_c$ yields similar results \cite{wyithe04-var, barkana05-cone}.  It differs from this model primarily by imposing at a maximum size determined by the light crossing time at the end of reionization.  This is only one of several mechanisms that effectively limits the range of ``causal contact" between sources; the most important is likely to be recombinations.  

Incorporating inhomogeneous recombinations into this analytic model is relatively straightforward \cite{furl05-rec}.  Each \htwo region obviously contains density fluctuations.  Because the recombination rate increases like $(1+\delta_{\rm nl})^2$, where $\delta_{\rm nl}$ is the fully nonlinear fractional overdensity, dense clumps will remain neutral longer than voids will.  Thus, following \cite{miralda00}, we make the simple ansatz that there exists a threshold density $\delta_i$ below which gas is ionized and above which it is neutral.\footnote{Of course, this cannot be exactly true, because galaxies are embedded in dense filaments, so ionizing photons do not immediately reach the voids.  This model implicitly assumes that these recombinations are incorporated into $\fesc$ (see \cite{kohler05-sim} for a similar application to simulations).}  Any ionizing photons striking these dense blobs will be lost to recombinations in the neutral gas.  Thus, for an ionized bubble to continue growing, the mean separation of these dense blobs must exceed the radius of the bubble.  Given a model for the volume-averaged IGM density distribution, $P_V(\delta_{\rm nl})$, $\delta_i$ can therefore be computed by requiring the mean free path between such regions to equal the bubble radius.  Clearly it increases as the bubble grows -- so denser and denser gas is ionized.  But this will also increase the recombination rate per proton, which is
\begin{eqnarray}
A_{\rm rec} & = & \alpha(T) \bar{n}_e (1+\delta) \, \int_{-1}^{\delta_i} \deriv \delta_{\rm nl} \, P_V(\delta_{\rm nl}) \, (1 + \delta_{\rm nl})^2 \\
\label{eq:clump}
& \equiv & \alpha(T) \bar{n}_e C(\delta,R), \nonumber
\end{eqnarray}
where $C(\delta,R)$ is the \emph{local} clumping factor.  The bubble can only grow if ionizing photons are produced more rapidly than recombinations consume them; in other words if
\begin{equation}
\zeta \, \frac{\deriv \fcoll(\delta,\, R)}{\deriv t} > \alpha(T) \bar{n}_e \, C(\delta,\, R),
\label{eq:rec-cond}
\end{equation}
The crucial point is that $C$ depends on both the mean density of the bubble (recall that on large scales overdense regions are ionized first in this model) and on its size (through $\delta_i$) -- as expected from \S \ref{clump}, inhomogeneous reionization affects the clumping factor.\footnote{Note that this model is therefore both ``inside-out" on large scales and ``outside-in" on small scales.}  Recombinations become increasingly important as bubbles grow; eventually they balance ionizations and the bubbles saturate, becoming true cosmological Str{\" o}mgren spheres.

Equation (\ref{eq:rec-cond}) complements our original ionization condition, equation (\ref{eq:ioncond}), which requires that the cumulative number of ionizing photons exceeds the total number of hydrogen atoms.  Of course, in reality both conditions must be fulfilled, but in practice one of the two generally dominates \cite{miralda00, furl05-rec}.  (This is essentially because recombinations take over only when $\delta_i$ approaches the characteristic density of virialized objects, or in other words when ``Lyman-limit" systems dominate the mean free path, as in the lower-redshift Universe.)  As a consequence, it is possible to combine the two conditions in the excursion set formalism and compute the ``bubble" sizes including recombinations.  However, we must keep in mind that in this case the model describes the mean free path of ionizing photons.  When recombinations are unimportant, this equals the size of isolated bubbles.  But once the bubbles ``saturate" as Str{\" o}mgren spheres, neighboring \htwo regions can touch -- it is only that their ionizing photons will not influence each other.  The model therefore describes how the ``bubble-dominated" topology characteristic of reionization maps smoothly onto the ``web-dominated" topology of the post-reionization IGM.

The key input parameter is obviously $P_V (\delta_{\rm nl})$, which describes the IGM clumpiness.  As we have emphasized repeatedly, this is difficult to specify because it depends on reionization itself.  The distribution has been measured in simulations at $z \sim 2$--$4$ \cite{miralda00}, but they assumed the gas to be highly ionized and hence smooth on the Jeans scale of photoheated gas.  While it is still neutral, and even shortly after it is ionized, the gas could be much clumpier -- although how much depends on the unknown X-ray heating rate \cite{oh03-entropy}.  If minihalos form, they could dramatically increase the clumpiness \cite{haiman01}.  However, recent numerical simulations \cite{shapiro04, iliev05-mh, ciardi05-mh} and analytic models \cite{iliev05} show that minihalo evaporation occurs quickly enough that they do not have an enormous effect on reionization.  Thus for now the most widely-used choice is the fit to lower-redshift simulations by \cite{miralda00}.

\begin{figure}[!t]
\centerline{\epsfig{file=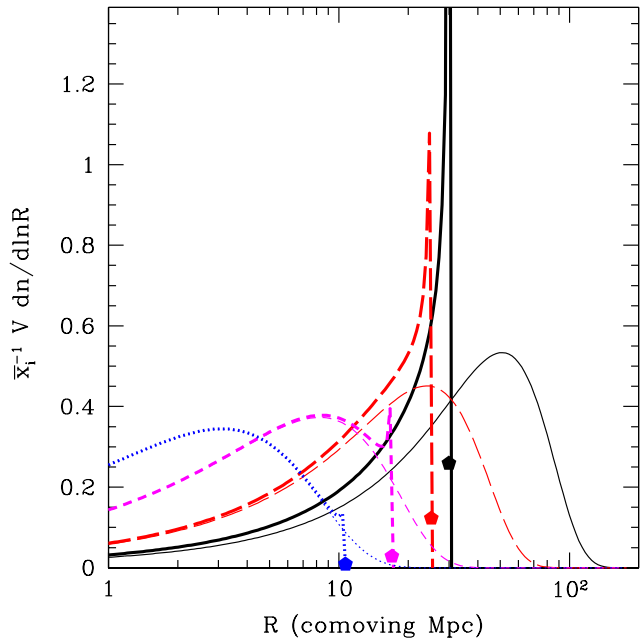,width=3.5in}}
\caption{\htwo region size distributions at $z=8$ for the models with and without inhomogeneous recombinations (thick and thin curves, respectively).  The dotted, short-dashed, long-dashed, and solid curves assume $\bxion=0.41,\,0.68,\, 0.84$, and $0.92$, respectively.  Recombinations sharply truncate the bubble size distribution at $R \approx R_{\rm max}$.  The filled hexagons show the additional fraction of the IGM volume in bubbles with $R=R_{\rm max}$.  From \cite{furl05-rec}.}
\label{fig:nbub-rec}
\end{figure}

Figure~\ref{fig:nbub-rec} shows some example bubble size distributions with this $P_V(\delta_{\rm nl})$.  The axes are identical to Figure~\ref{fig:nbub}, and the thin curves show the same model as in that figure (with $\zeta \propto m_h^0$).  The thick curves (with the filled hexagons) add recombinations as described above.  They have only a modest effect on $n_b(m)$ when $\bxion \la 0.75$, truncating the distribution's tail but not affecting most bubbles.  This is because small bubbles need not ionize any of their dense gas; it is only unusually large bubbles that ionize dense clumps deeply enough for recombinations to become significant.  However, as typical bubbles grow beyond $\sim 20 \Mpc$, recombinations completely dominate their size distribution, imposing a well-defined $R_{\rm max}$ on it.  Qualitatively, at this point the typical size of an ionized bubble equals the typical separation of Lyman-limit systems.  Thus most ionizing photons are absorbed by one of these objects before hitting the edge of a bubble; Lyman-limit systems are so dense and thick that they cannot easily be ``burned off" and so impose a sharp limit on the bubble sizes.  This saturation radius depends sensitively on the assumed IGM density distribution; measuring it would thus constrain the growth of small-scale structure.  It is also sensitive to the evolution of the ionizing emissivity \cite{furl05-charsize}.

%%%%%%%%%%%% FIGURE 8-4: Bubble sizes with recombinations

Of course, this approach must ultimately be validated against simulations.  Fortunately, because the model is constructed from the density field, it is easy to compare it to specific simulations in a detailed, and not simply statistical, sense \cite{zahn05, zahn05-lens, zahn06-comp}.  We return to our basic ionization criterion, $\zeta \fcoll(\delta,m)>1$.  This is a condition on the linear density field:  the same field that serves as the initial condition for simulations.  At any redshift, the linearly-evolved density field can be smoothed around each point on progressively smaller scales, checking whether the ionization criterion is fulfilled.  Each point satisfying it (at any scale) is tagged as ionized:  points ionized by distant neighbors pass the density threshold on large scales, while points ionized by nearby sources pass it on small scales. The rightmost column of Figure~\ref{fig:reion-sim} shows an example of this approach in the same box used for the radiative transfer simulation shown in the leftmost column \cite{zahn06-comp}.\footnote{Note that recombinations are not included in the analytic version, because small-scale clumping was also neglected in the simulation.  But they can easily be incorporated in the model by simply using both ionization criteria; in fact, given the difficulty of properly adding recombinations to simulations, this is an attractive option.}  Considering the simplicity of the analytic model, the agreement is remarkable.  Large \htwo regions are identified in a nearly one-to-one manner (although their detailed shapes differ).  Thus the basic principles of the model, and the generic predictions we have emphasized, appear to be accurate.  This scheme allows the rapid generation of large, high-resolution reionization histories, even with modest computational facilities, and it reproduces the complex geometries of ionized bubbles reasonably well.  Thus we expect it to provide a powerful tool in studying reionization and its observable implications.

Nevertheless, it is vital to explore the reasons for the disparities with the full simulations.  One interesting difference is that the simulations tend to show larger \htwo regions appearing in the later stages of reionization \cite{iliev05-sim}; this is inherent to percolation processes, where an infinitely large \htwo region containing islands of neutral gas must eventually appear.  In the simulations this manifests itself through \htwo regions much larger than expected from the naive analytic model in the later stages of reionization (beyond the stage shown in Fig.~\ref{fig:reion-sim}), although note that applying the analytic criterion to the simulation's density field does correctly describe this topology inversion, so it is not a result of new physics.  Moreover, we have already emphasized that the analytic model is better suited to describe the mean free path of ionizing photons in this regime; comparisons during the late stages must await the self-consistent incorporation of recombinations into simulations.

Another difference is that the galaxy population contains stochastic fluctuations over and above those sourced by density fluctuations.  Analytic models suggest that, in most cases, the effects of this scatter on $n_b(m)$ are modest once the bubbles become large \cite{furl05-charsize}.  But this will certainly affect the detailed distribution of ionized gas.  One way to address its importance is to apply the ionization criterion to the \emph{halo} density field rather than the dark matter \cite{zahn06-comp}.  This is shown by the center column in Figure~\ref{fig:reion-sim}, which clearly provides an even closer match to the radiative transfer results, especially in regard to the connectivity of the ionized regions.  This algorithm requires running an N-body simulation (to generate the halo field) but eliminates the radiative transfer, so it still dramatically speeds up the calculation.  Obviously there are a variety of approaches to inhomogeneous reionization, of increasing sophistication and accuracy, which can be tailored to the task at hand.  We expect the continued development of reionization models (incorporating, for example, feedback processes \cite{kramer06} and more sophisticated star formation algorithms \cite{cohn06}) to sharpen our expectations for studies of reionization.

\subsection{Statistical Predictions} \label{stat}

We now turn more specifically to predictions for 21 cm observations and what they can teach us about inhomogeneous reionization.  We will begin with statistical measurements, because they are most likely to be accomplished first.

\subsubsection{The Isotropic Power Spectrum} \label{isopower}

The simplest possible statistic is the power spectrum, neglecting peculiar velocities and other factors that introduce anisotropies into the signal (see \S \ref{redshift-dis} and \ref{anisotropy}).  This requires computing the power spectrum of ionized bubbles, which was first written down in the context of ``patchy reionization" CMB anisotropies \cite{knox98, gruzinov98}.  A number of analytic models have been presented in the literature, including simple tracers of the density field \cite{santos03} and direct association with dark matter halos \cite{wang05_FoG}.  We will first describe how to build $P_{21}(k)$ from the physically-motivated reionization model discussed above.  This provides intuition for more detailed investigations with simulations, which are necessary given the complex shapes of real \htwo regions.

We expect the two-point function to have the form \cite{furl04-bub}
\begin{equation}
\VEV{\xion({\bf r}) \, \xion({\bf r'})} = \bxion^2 + (\bxion - \bxion^2) \, f(r/R_c),
\label{eq:xxform}
\end{equation}
where $r \equiv | {\bf r} - {\bf r'} |$, $R_c$ is the characteristic bubble size, $f(r/R_c) \approx 1$ for $r \ll R_c$, and $f(r/R_c) \approx 0$ for $r \gg R_c$.  This form is necessary because the bubbles have a finite size:  thus two nearby points will either both be ionized by the same bubble (with probability $\bxion$) or both be neutral.  At large separations, on the other hand, the points must be ionized by two distinct bubbles, each with probability $\bxion$.  

In constructing $f$ we must take care because $\xion$ has a restricted range from zero to unity.  Thus $\bxion=1$ implies $\xion=1$ at \emph{every} point in the IGM, and the correlations must vanish in both limits $\bxion \rightarrow (0,\,1)$.  This is a particular concern for statistical models of the bubble distribution, because it means that bubbles \emph{cannot} overlap and we cannot simply distribute them randomly.  One solution is to treat the ionization as a Poisson process \cite{furl04-bub, wang05_FoG}, which prevents more than one bubble from covering any particular location but has the unattractive feature of failing to conserve photons.

It is therefore more illuminating to try to construct $f(r/R_c)$ directly while taking care to enforce the proper limiting behavior.  By analogy to the halo model for density fluctuations \cite{cooray02}, we separate it into two components:  the probability $p_1$ for which a single bubble ionizes both points and the probability $p_2$ for which two separate bubbles do so \cite{furl04-bub}.  Then 
\begin{equation}
p_1(r) = \int \deriv m \, n_b(m) \, V_{\rm ov}(m,\,r),
\label{eq:p1}
\end{equation}
where $V_{\rm ov}(m,\,r)$ is the volume within which the center of a bubble of mass $m$ (or radius $R$) can sit while still ionizing both points.  Note that $p_1 \sim \bxion$ for $r\ll R_c$ but $p_1 \approx 0$ for sufficiently large $r$.  The two-bubble term can be written \cite{mcquinn05}:
\begin{eqnarray}
p_2(r) & = & \int \deriv m_1 \, n_b(m_1) \int \deriv^3 {\bf r} \int \deriv m_2 \, n_b(m_2) \nonumber \\  & & \times \int \deriv^3 {\bf r'} \,  [1 + \xi_{bb}({\bf r} - {\bf r'}|m_1,\, m_2)],
\label{eq:p2}
\end{eqnarray}
where $\xi_{bb}$ is the bubble correlation function.  Clearly $P_2 \rightarrow \bxion^2$ as $r \rightarrow \infty$.  The challenge posed by the discrete bubbles is to set the spatial integration limits to simultaneously avoid overlap (to force $P_2 \rightarrow 0$ as $r \rightarrow 0$) while still allowing bubbles to reside near to each other.  This is difficult because of the assumed spherical symmetry.  (The basic problem is similar to packing a crate with oranges:  small gaps are invariably left between the bubbles.  In reality, of course, the \htwo regions can assume arbitrary sizes to fill space.)  

An approximate solution that matches simulation power spectra reasonably well (at least at $\bxion \la 0.8$) is \cite{mcquinn05} 
\begin{equation}
\VEV{\xion \, \xion}(r) = \left\{
\begin{array}{lll}
p_1(r) + p_2(r) & \ & \bxion < 0.5 \\
(1-\bxion) p_1(r) + \bxion^2 & \ & \bxion \ge 0.5,
\end{array}
\right.
\label{eq:corr}
\end{equation}
with $\xi_{bb} \approx b_x^2 \xi_{\delta \delta}$ on sufficiently large scales and $\xi_{\delta \delta}$ the dark matter correlation function.  This form is intuitively simple:  when $\bxion<0.5$, overlap is relatively unimportant and we can use the one- and two-bubble terms as is.  At larger ionized fractions, overlap becomes significant, but by this point the characteristic bubble size $R_c \ga 5 \Mpc$, and $\xi_{bb}(R_c)$ is sufficiently small that setting $p_2 \approx \bxion^2$ everywhere suffices.  Thus we only need the one-bubble term $p_1(r)$ (weighted by the neutral fraction to enforce the proper limits).  This basic form captures most of the relevant features of the correlations (see \cite{barkana05-cone} for a simplified version using bubbles of a single size).

Because $\dtb \propto \xhi n$, we also require the cross-correlation $\xi_{x\delta}$ between the ionized fraction and density.  This is implicit in the analytic model, which calculates the bubble geometry from the density field itself \cite{furl04-bub, mcquinn05}.  The resulting expressions have similar ambiguities and limiting behaviors to $\VEV{x_i x_i}$.

Armed with the correlation functions $\xi_{xx}$, $\xi_{x\delta}$, and $\xi_{\delta\delta}$ (and the corresponding power spectra),\footnote{Note that we write $P_{xx}$ for the power spectrum of $\delta_x$, the fractional perturbation in $\bxhi$ used in equation~(\ref{eq:d21}).  This differs from the definition in \cite{mcquinn05}.} the isotropic power spectrum of 21 cm brightness temperature fluctuations, $P_{21}^{\rm iso}$, is\footnote{Here we assume $T_S \gg T_\gamma$ throughout the IGM; if this is note the case, as may occur early in reionization, other terms sourced by $T_S$ will appear.} \cite{furl04-bub}
\begin{equation}
P_{21}^{\rm iso}(k) = P_{\delta \delta}(k) + P_{xx}(k) - 2 P_{x \delta}(k) + P_{x \delta x \delta}(k);
\label{eq:pk_iso}
\end{equation}
to convert to temperature fluctuations, we must multiply this by $\bdtb^2$.  The four point term $P_{x \delta x \delta}$ is in general complicated; we will take $P_{x \delta x \delta} = P_{x \delta}^2 + P_{xx} \, P_{\delta \delta}$ by dropping the connected part.\footnote{Formally, this is only valid for a gaussian field.  But comparison to numerical simulations show that it is a reasonable approximation \cite{mcquinn05}.}  

\begin{figure}[!t]
\centerline{\epsfig{file=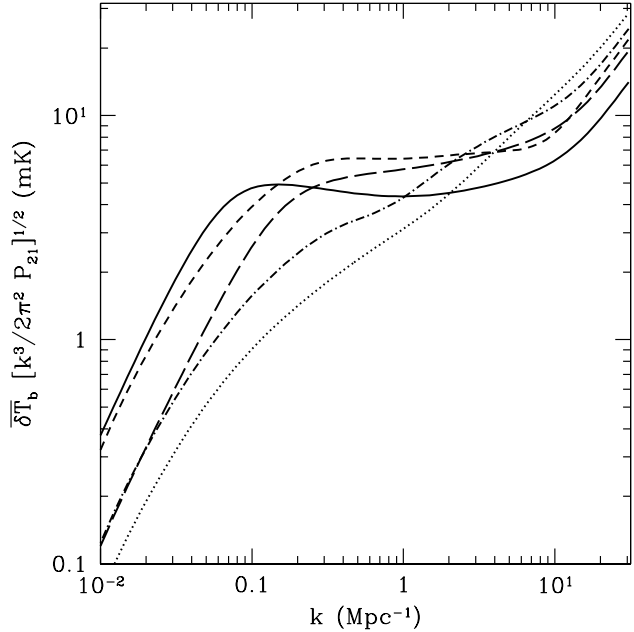,width=3.5in}}
\caption{Rms variation in the 21 cm brightness temperature as a function of wavenumber at several different stages of reionization, ignoring peculiar velocities.  All the curves assume $z=10$.  They have $\bxion=0.13$ (dotted curve), $\bxion=0.36$ (dash-dotted curve), $\bxion=0.48$ (short dashed curve), $\bxion=0.69$ (long-dashed curve), and $\bxion=0.78$ (solid curve).  Following \cite{furl04-bub, mcquinn05-param}.}
\label{fig:analytic-pk}
\end{figure}

Figure~\ref{fig:analytic-pk} shows several examples.  We set $z=10$ and vary the ionizing efficiency $\zeta$ to examine a range of $\bxion$.  When $\bxion \ll 1$, \htwo regions are unimportant and $P_{21}(k)$ essentially traces the matter power spectrum (the sharp rise at $k \ga 10 h \Mpcinv$ is due to small scale nonlinearities).  When the bubbles appear, they first suppress the power by ionizing the highest density regions first.  Soon, however, $P_{xx}$ becomes large and begins to dominate the fluctuations, creating a ``shoulder" at $k_{\rm pk}$.  The scale of this feature is directly related to the characteristic bubble size, and $k_{\rm pk}$ decreases throughout reionization; moreover, the more sharply peaked the bubble distribution, the more distinct the peak.  For $k < k_{\rm pk}$, $P_{21} \propto P_{\delta \delta}$ when $\bxion$ is small (it is simply amplified by the effective bias of the bubbles), but once the bubbles become large their Poisson fluctuations dominate and $P_{21}$ approaches white noise.  This peak is probably the single most important feature for 21 cm experiments, because it directly and clearly constrains reionization and the sources responsible for it.  Its detection is one of the major goals of 21 cm observatories.

%%%%%%%%%%%% FIGURE 8-5: Analytic, isotropic power spectrum

\begin{figure}[!t]
\centerline{\epsfig{file=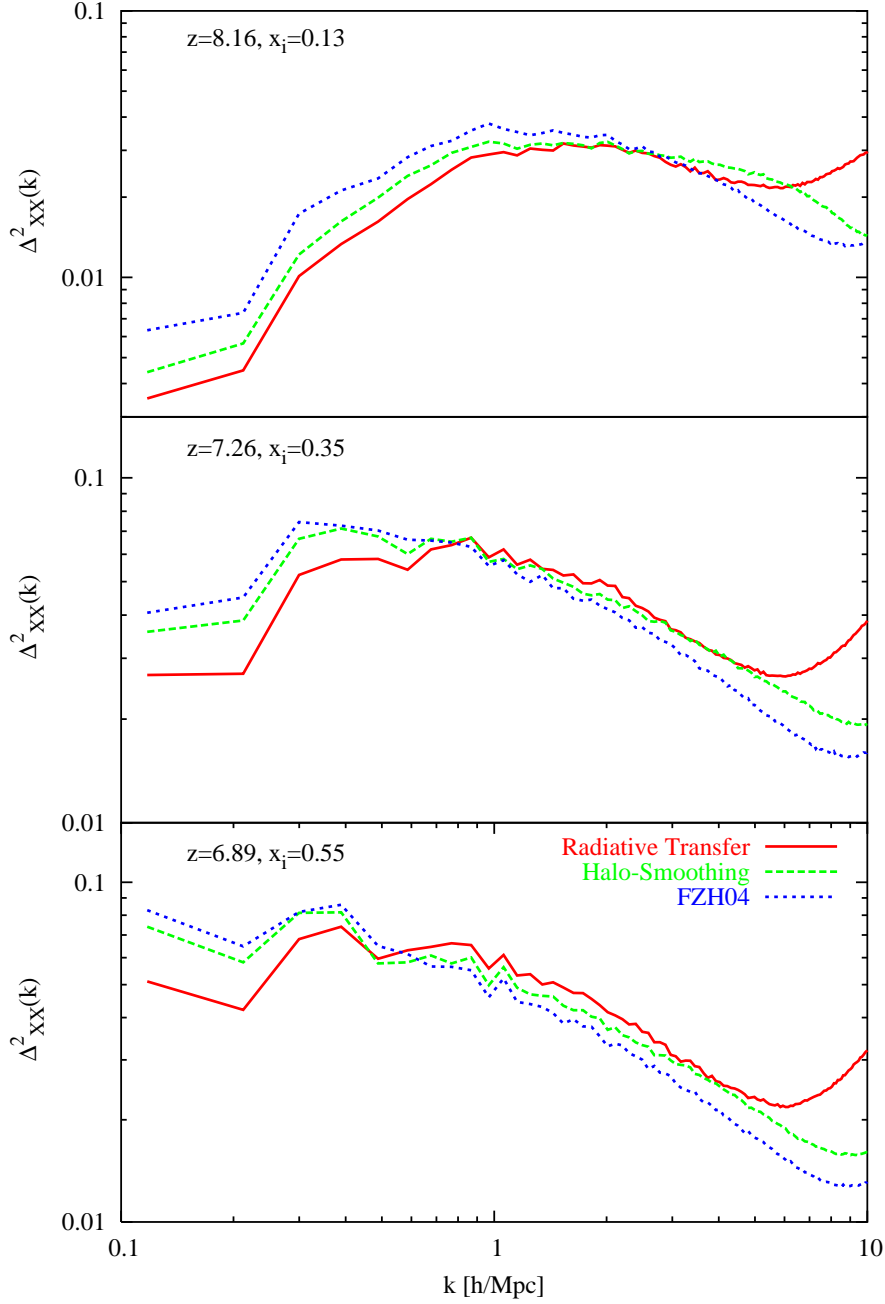,width=4.75in}}
\caption{Comparison of dimensionless power in the ionization fraction between a simulation and the analytic model.  The solid, dashed, and dotted curves correspond to the three different columns in Fig.~\ref{fig:reion-sim}. From \cite{zahn06-comp}.}
\label{fig:sim-pk}
\end{figure}

Of course, the power spectrum is best measured in simulations of reionization, which can capture its complex geometry -- although large volumes are obviously  necessary (cf. \cite{ciardi03-21cm, furl04-21cmsim}).  Figure~\ref{fig:sim-pk} compares the power spectra of the ionized fraction in full radiative transfer simulations (solid curves) to implementations of the analytic model in the same box (using the linear density field in the dotted curves and the halo distribution in the dashed curves); these are the same fields shown in Figure~\ref{fig:reion-sim} \cite{zahn06-comp}.  Qualitatively the results are similar in the three cases:  the power spectrum has a well-defined peak that moves to larger scales and eventually fades into Poisson noise as $\bxion$ increases.  We see $k_{\rm pk} \sim 0.01$--$0.1 h \Mpcinv$ over this range, as expected from \cite{furl04-bub}.  However, the radiative transfer simulations do have somewhat less power on large scales and somewhat more on small scales; the latter is due to fluctuations in the recombination rate and in the halo distribution.  The agreement in $P_{21}$ is similarly good \cite{zahn06-comp}, and other simulations show qualitatively similar behavior, with a peak developing in both $\Delta_{xx}^2$ \cite{iliev05-sim} and the angular power spectrum of 21 cm fluctuations \cite{mellema06} that moves to larger scales throughout reionization.  

%%%%%%%%%%%% FIGURE 8-6: Comparison of power spectra

Thus, although numerical implementations are clearly required to produce detailed power spectra (and by extension other statistical measures), the analytic model, especially as implemented in realistic density fields, can provide a great deal of intuition about the signal.  For example, measuring the shape of $P_{21}$ will teach us about the sources of reionization (through their clustering) \cite{furl05-charsize}, the role of recombinations \cite{furl05-rec}, feedback during reionization, and even whether X-rays have uniformly ionized the IGM \cite{furl04-21cmtop}.

\subsubsection{Anisotropies} \label{anisotropy}

The velocity field makes $P_{21}$ anisotropic (see \S \ref{redshift-dis}) and hence considerably more complicated.  Using the linear theory approximation $\tilde{\delta}_{\partial v} = -\mu^2 f \tilde{\delta}_L$, where $\delta_L$ is the linear density field, we can write (with $f=1$)\footnote{In practice $P_{\delta_L \delta_L}$ and $P_{\delta \delta}$ are identical on the scales accessible to experiments, but we separate them in order to label those terms arising from velocities.} \cite{mcquinn05-param}
\begin{eqnarray}
P_{21}({\bf k}) &= & P_{21}^{\rm iso}(k) + 2 \mu^2 [P_{\delta_L \delta}(k) - P_{x \delta_L}(k)] \nonumber \\
& & + \mu^4 [P_{\delta_L \delta_L}(k)] + 2 P_{x \delta \delta_L x}({\bf k}) + P_{x \delta_L \delta_L x}({\bf k}).
\label{eq:pk_vel}
\end{eqnarray}
The first three terms in equation (\ref{eq:pk_vel}) show the $\mu^0$, $\mu^2$, and $\mu^4$ dependencies originally derived by \cite{barkana05-vel}, but the last two terms are considerably more complicated.  Although they appear to be higher order, they must be included because $\delta_x$ is \emph{not} small during reionization, so these four-point terms are still only second order in $\delta$.  Unfortunately,  they have nontrivial $\mu$ dependence; for example, 
\begin{eqnarray}
P_{x \delta \delta_L x}({\bf k}) & = & \int \frac{\deriv^3 {\bf k'}}{(2 \pi)^3} \, ({\bf \hat{n}} \cdot {\bf \hat{k}'} )^2 \, [P_{x \delta_L}(k') \, P_{x \delta}(|{\bf k} - {\bf k'}|) \nonumber \\
& & + P_{\delta  \delta_L}(k') \, P_{x x}(|{\bf k} - {\bf k'}|) ].
\label{eq:4pt-terms}
\end{eqnarray}
Thus, when they are large, they will ruin the decomposition in powers of $\mu$ proposed by \cite{barkana05-vel}.

\begin{figure}[!t]
\centerline{\epsfig{file=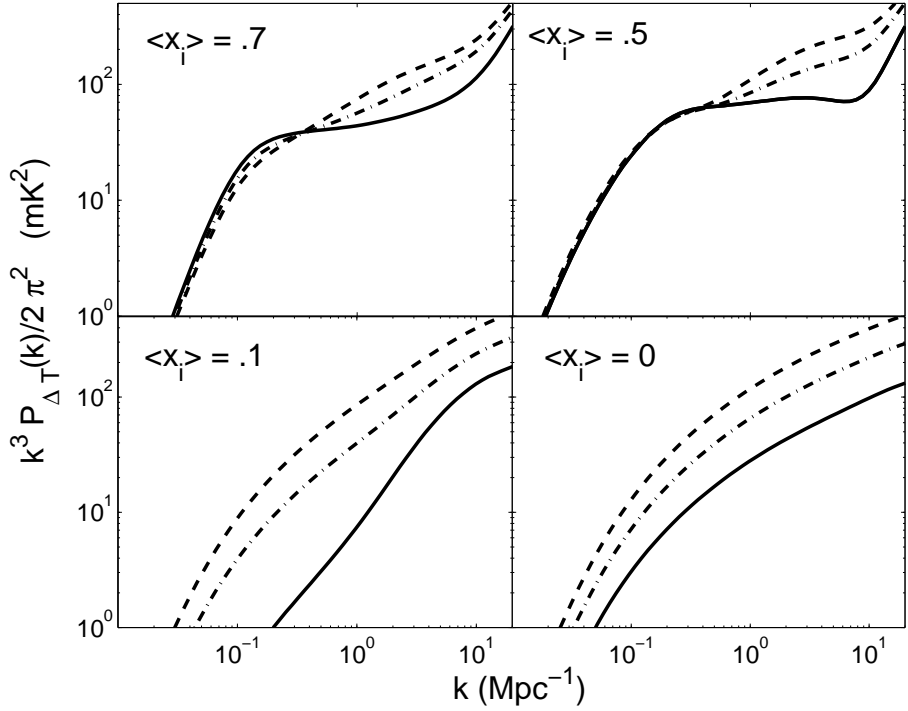,width=4.75in}}
\caption{21 cm brightness temperature power spectra at four different stages of reionization.  In each panel, the solid, dash-dotted, and dashed curves take $\mu=0,\,0.5$, and $1$, respectively.  The model follows \cite{furl04-bub} with $\zeta=12$.  From \cite{mcquinn05-param}.}
\label{fig:pk-ang}
\end{figure}

Figure~\ref{fig:pk-ang} shows the 21 cm power spectra at several stages of reionization for $\mu=0,\,0.5$, and $1$ (solid, dash-dotted, and dashed curves, respectively).  When $\bxion=0$, only the $P_{\delta \delta}$ and $P_{\delta \delta_L}$ terms remain and redshift space distortions significantly enhance the power in line of sight modes, as usual.  This remains true when $\bxion$ is small, because the enhancement from the velocity field is of course independent of reionization.  However, the $P_{xx}$ term is isotropic, so once it  begins to dominate, the total power becomes nearly independent of the underlying velocity field (and hence of $\mu$).  Only on scales smaller than the characteristic bubble size does the power spectrum remain anisotropic, partly because the four-point terms can be large in this regime \cite{mcquinn05-param}.  Thus, once reionization gathers steam, redshift space distortions will be difficult to separate and to interpret cleanly; they can probably only be used to measure cosmological parameters \emph{before} reionization.

%%%%%%%%%%%% FIGURE 8-7: Anisotropic power spectrum

Another source of anisotropy is evolution in the density field and neutral gas distribution across the observational bandwidth.  When $P_{21} \propto P_{\delta \delta}$ (i.e., when $\bxion$ is small or on scales much larger than the bubbles), this only affects the overall amplitude of the power spectrum and can be modeled reasonably well.  Only in the late stages of reionization, and only if the remaining voids are ionized rapidly, do these ``light-cone" effects introduce measurable anisotropies that dominate those of the velocity field \cite{mcquinn05-param, barkana05-cone}.

\subsubsection{Nongaussianity} \label{ng}

To this point, we have focused exclusively on the power spectrum.  To many readers familiar with the CMB and galaxy surveys, this will seem natural:  in those cases, the power spectrum provides a nearly perfect statistical description, because the relevant fields are (to an excellent approximation) gaussian.  Of course, for the 21 cm signal during reionization, this gaussian assumption breaks down.  A glance at Figures~\ref{fig:reion-sim} or \ref{fig:mellema-sim} shows immediately that the fluctuations are dominated by the striking contrast between the neutral IGM and the \htwo regions; the (gaussian) density fluctuations in the residual neutral gas play only a secondary role.  Only on scales so large that $P_{21} \propto P_{\delta \delta}$ does the gaussian approximation hold \cite{cooray04-ng}.  Exploring these signatures of nongaussianity is crucial for extracting the maximal information from upcoming experiments, especially because it helps to distinguish the cosmological signal from the foregrounds.  Unfortunately, it is also difficult and so has received relatively little attention.  

\begin{figure}[!t]
\centerline{\epsfig{file=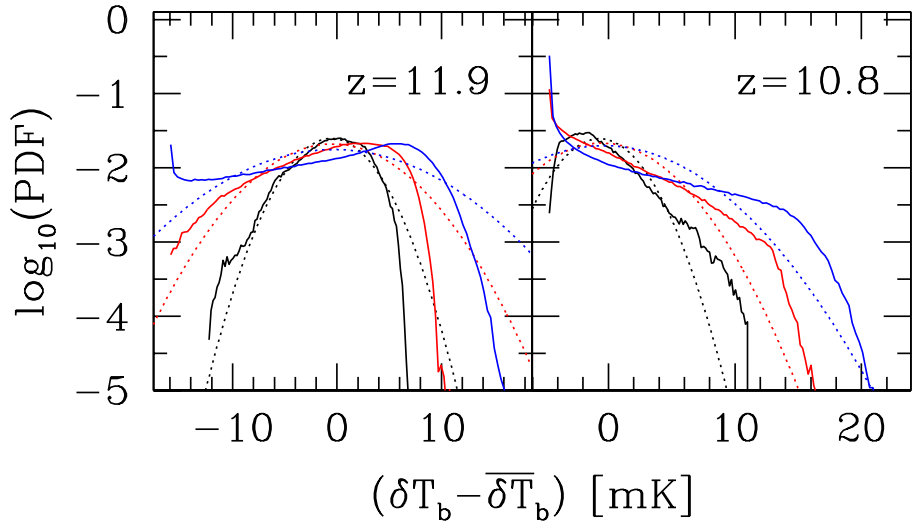,width=5in}}
\caption{Pixel flux PDFs in a simulation of reionization.  The left and right panels have $\bxion=0.5$ and $0.75$, respectively.  Within each panel, the maps are smoothed on scales $20h^{-1},\,10h^{-1}$, and $5h^{-1} \Mpc$ (solid curves; narrowest to widest).  The dotted curves show the corresponding best fit gaussians.  From \cite{mellema06}. }
\label{fig:ng}
\end{figure}

The obvious place to begin is with the probability distribution function (PDF) of pixel fluxes considered as a function of smoothing scale.  Analytic models show that, because of the correlation between $\xion$ and $\delta$, this will have interesting and nontrivial structure during reionization \cite{furl04-21cmtop}.  Figure~\ref{fig:ng} shows several examples taken from a simulation of reionization \cite{mellema06}.  The maps have been smoothed over $20,\,10$, and $5 h^{-1} \Mpc$; the Figure also shows the best fit gaussians.  When $\bxion=0.5$, the PDF has a cutoff at large $\dtb$ and an excess at negative $\dtb$ (from \htwo regions).  The cutoff is caused by ``inside-out" reionization, in which sources ionize their (dense) surroundings first.  It is a direct result of the ionization criterion $\zeta \fcoll>1$; if voids were instead ionized first, the PDF would look much different, even if the power spectra were comparable \cite{furl04-21cmtop}.  Note also that the distribution narrows as the smoothing scale increases, because bubbles are no longer resolved.  Later in reionization, at $\bxion=0.75$, the PDF has a strong peak at negative $\dtb$ and a long tail toward higher fluxes.  This occurs because, by this point, the simulation contains large, connected \htwo regions with discrete islands of neutral gas.  Because these islands still have high contrast, they may be the easiest features to detect at the end of reionization \cite{mellema06}.

%%%%%%%%%%%% FIGURE 8-8: Nongaussianity

The simplest way to approach nongaussianity is to try to measure this PDF directly.  In particular, if the bubbles are resolved, the distribution will have two peaks:  one centered on $\dtb=0$ for pixels entirely inside \htwo regions and one centered on $\dtb \approx 20 \mkel$.  Thus, even factoring in noise near to or larger than this separation, the underlying distribution will be bimodal.  Sensitive tests for bimodality in large datasets are available and relatively easy to implement \cite{hansen06}.  These techniques essentially compare the goodness of fit of single and multiple gaussians to the data; they succeed because of the enormous volume subtended by the fields of view of even the first generation of telescopes.  Numerical simulations show that this simple idea can work in realistic datasets (although foreground contamination may be a problem).  

Of course, the PDF throws away all the information contained in spatial correlations so it is certainly not optimal.  A complementary approach is to study ``global" properties of the ionization pattern, such as the genus number \cite{gleser06}.  This is particularly useful in identifying the topological transition from discrete \htwo regions to discrete \hone clouds.

We expect the continued development of strategies to detect the signatures of \htwo regions to be a major step toward reliable 21 cm data analysis algorithms.  It is imperative that we identify robust and powerful statistics to extract as much information as possible from the data.  A particularly important application is to develop tests for ``bubble" structure or sharp edges in the data; statistical detection of such features, through methods like the bimodality test of \cite{hansen06}, will vastly increase the believability of the results, because foreground contamination and removal are unlikely to introduce sharp edges like these (at least in the frequency direction, where all sources have smooth spectra).

\subsection{The Imaging Regime} \label{image}

The statistical measures described in the previous section will shed a great deal of light on the reionization epoch. However, even more exciting is direct three-dimensional tomography of the high-redshift IGM (e.g., \cite{tozzi00,wyithe05-qso,kohler05}). The 21 cm line is by far the best way to perform such tomography:  the CMB is a snapshot of a two-dimensional surface, and quasar absorption spectra only sample the IGM in sparse one-dimensional skewers. The low S/N of upcoming 21 cm experiments precludes direct imaging of the cosmic web in the near term (see \S \ref{sens-basic}). However, the large HII regions blown by quasars or clustered galaxies will appear as holes in 21 cm emission and will have sufficiently high S/N to be detected individually. They may very well be the first unambiguous features of the epoch of reionization detected by these experiments. We therefore focus here on the rich science returns from direct imaging of high-redshift HII regions. 

Detecting HII regions in the face of telescope noise requires that they be sufficiently large and that the contrast between neutral and ionized regions be sufficiently high. As we will see in \S \ref{sens-basic}, the brightness temperature sensitivity decreases dramatically as the physical scale to be probed decreases (eq. \ref{eq:if-sens}).  It should be compared to the temperature contrast $\delta T_{b} \approx 22 x_{\rm HI} (1+\delta) [(1+z)/7.5)]^{1/2}$ mK (assuming $T_{\rm S} \gg T_{\gamma}$) between neutral and ionized regions (note that this contrast falls if the IGM is partially ionized from X-rays or incomplete recombination in a fossil HII region). The angular and frequency scales required for a $\sim 5 \sigma$ detection correspond to $\sim 20 h^{-1}$ Mpc with the first generation of experiments.  To prevent beam dilution across the bubble, the beam size should be somewhat smaller than the bubble radius; as with power spectrum measurements \cite{zald04}, the optimal S/N generally results from matching the frequency resolution to the beam size:  reducing the channel width when the bubble is unresolved yields no gains (because of beam dilution) but increases the noise level. 

The need for such large bubbles pushes us toward either those blown by bright quasars \cite{tozzi00, wyithe04-qso, wyithe05-qso, kohler05} or those characteristic of clustered galaxies late in the reionization process \cite{furl04-bub}. The former are particularly attractive, since a number of $z>6$ SDSS quasars with large proximity zones are already known \cite{fan06}; if the IGM is significantly neutral $\bxhi \ge 0.2$ at $z \sim 6.0$--$6.5$, their HII regions should be detectable and will likely be the the first imaging targets of the new generation of 21 cm experiments. In this context, the 21 cm fluctuations discussed in the last several sections (including the linear density field, the spin temperature, and unresolved bubbles) become noise \cite{zald04,kohler05,wyithe05-qso}. This is analogous to confusion noise in ordinary imaging and provides an irreducible background in that additional integration does not reduce the effective noise level (except the contribution from bubbles, in so far as smaller bubbles can be identified). Thus, the optimal strategy is to integrate until the telescope noise reaches the same level as that due to the cosmic web, which is typically a few mK on the scales of interest. This limit will probably only be reached with SKA-class instruments.\footnote{See \S \ref{int} below for a detailed discussion of the instruments mentioned in this section.}

Given that we are looking for bright quasars, one might expect features to be extremely rare. However, 21 cm experiments have broad bandpasses and wide fields of view encompassing huge volumes. For instance, a simple extrapolation of the empirical quasar luminosity function at lower redshifts yields at least one quasar with $L_{\rm B} > 2 \times 10^{10} L_{\odot}$ within the $\sim 500,000 (h^{-1} \, {\rm Mpc})^{3}$ volume subtended by a single $10^{\prime}$ synthesized beam from $z=6$--$12$ \cite{kohler05}; such an object will blow an HII region of sufficient size to overcome the cosmic web variance. Similar estimates based on a semi-analytic model for the quasar luminosity function \cite{wyithe04-qso, wyithe05-qso, WL-qso-model} (and also from a different extrapolation of the empirical luminosity function \cite{fan04}), predict that HII regions should appear in each MWA-5000\footnote{The MWA-5000 is a proposed larger version of the MWA described in \S \ref{int}, increasing the collecting area by about an order of magnitude.  See \cite{wyithe05-qso, mcquinn05-param} for details on its capabilities.} pointing with a field of view of 400 deg$^{2}$ in a bandpass of 16 MHz. They emphasize that the short duty-cycle of quasars ($\sim 0.01$ at these redshifts) implies that the abundance of fossil HII regions -- where the quasar has turned off but where (due to the long recombination time) the HII cavity remains -- may be up to two orders of magnitude larger. They estimate $\sim 1$ active/fossil quasar HII region with $R > (24,40) \Mpc$, $R > (18,36) \Mpc$, and $R > (11,22) \Mpc$ (in comoving units) at $z=7,8$, and $10$, respectively, in a single MWA-5000 field of view \cite{wyithe05-qso}.  

\begin{figure}[!t]
\centerline{\epsfig{file=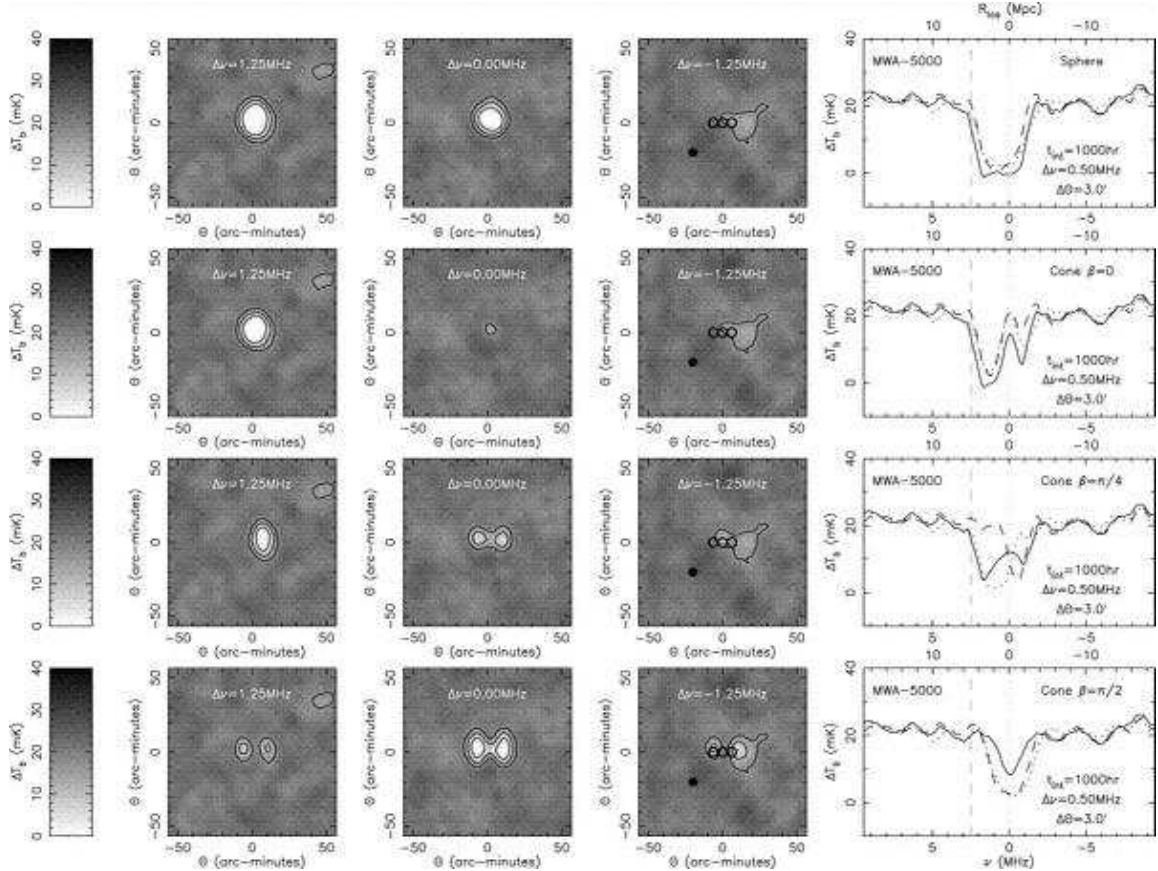,width=6.0in}}
\caption{Simulated maps of HII regions around $z=6.5$ quasars with the MWA-5000, assuming a 3$^\prime$ Gaussian beam top-hat smoothed at 0.5 MHz in a 1000 hr integration. The first row shows an intrinsically spherical HII region. The second, third and fourth rows show intrinsically bi-conical HII regions oriented along, at $\pi/4$ to, and perpendicular to the line of sight, respectively (with opening angles of $\pi/3$). The three maps show different slices of the frequency cube. The contours correspond to 5, 11, and 17 mK.  The solid black dot at (-20,-20) shows the beam size.  The dashed, solid, and dotted spectra in the right-most panel correspond to the left, center, and right lines of sight marked  in the third frequency slice, while the dotted and dashed vertical lines show the locations of the quasar and of the front of the HII region. (Note that distances are in physical units here.)  From \cite{wyithe05-qso}. }
\label{fig:WL-imaging}
\end{figure}

What is the best observational strategy to pick out these HII regions? Statistical detection of boundaries in a noisy field is a classic problem with many applications from oceanography to medical imaging. The optimal strategy for 21 cm surveys is still unclear. One possibility is to throw away spatial information and simply examine the PDF of pixel temperatures \cite{hansen06}. However, if the bubbles are large and the S/N per bubble is high, more straightforward techniques are applicable. For experiments operating solely in the frequency domain, both the equivalent width and depth of spectral dips sourced by quasar \htwo regions are possible indicators \cite{kohler05}. The equivalent width depends on the lifetime of the quasar; for short-lived objects, it cannot be distinguished from the tail of Gaussian fluctuations.  However, the depth is more robust since the gas is always highly ionized regardless of the HII region size.  The optimal strategy for detecting dips depends on their distribution along a line of sight, $[\deriv N/\deriv({\rm LOS})](> d_{\rm min}) \propto d_{\rm min}^{-\alpha}$ (where $d_{\rm min}$ is the limiting detectable depth in a given observation). Since $d_{\rm min} \propto t^{-1/2}$, the total number of detected dips $N \propto n t^{\alpha/2} \propto t^{\alpha/2-1}$ (where $n$ is the number of sightlines, $t$ is the time each is observed, and we have assumed $nt=$constant). If $\alpha > 2$, it is more efficient to integrate longer than to increase the number of lines of sight.  Figure~\ref{fig:WL-imaging} shows some simulated three-dimensional cubes illustrating how quasar HII regions can be detected (from \cite{wyithe05-qso}). The MWA should be able to detect HII regions around the most luminous quasars; the follow-up MWA-5000 should be capable of mapping the detailed geometry of HII regions (which can be complex if quasar emission is beamed), while the SKA can detect narrow spectral features and may measure the sharpness of the HII region boundary. 
 
%%%%%%%%%%%% FIGURE 8-9: Wyithe-Loeb imaging sims

What could we learn by imaging quasar HII regions?  Some of the exciting possibilities are summarized in \cite{wyithe05-qso}.  Most importantly, the contrast between HII regions and the IGM yields $\xhi$ in a more reliable way than trying to fit for $\bdtb(z)$ since the HII regions (within which only the radio foregrounds appear, provided the mass fraction of Lyman limit systems inside the HII regions is small) calibrate the foregrounds as a function of frequency.  Inferences about high-redshift quasars are also possible, since full 3D data will be available (unlike with the \lya forest).  Measurement of the shape of the HII region could reveal the emission profile of the quasar (as in the conical profiles of Fig.~\ref{fig:WL-imaging}). The HII region sizes jointly constrain the quasar lifetime $t_{\rm QSO}$ and $\bxhi$. Asymmetry in the transverse and line of sight sizes (with additional constraints on the latter from the Ly$\alpha$ forest) can arise due to finite time travel effects. 21 cm and Ly$\alpha$ forest measurements sample different epochs in the evolution of the HII regions; photons along the transverse direction were emitted earlier, when the HII region was smaller.  The asymmetry depends on the expansion speed of the HII region, which for a given ionizing flux is proportional to $x_{\rm HI}^{-1/3}$.  Thus it could potentially break the degeneracy between $\xhi$ and $t_{\rm QSO}$, directly constraining $x_{\rm HI}$  (with only a small dependence on quasar beaming) \cite{wyithe04-qso}. These possibilities are of course subject to caveats about radiative transfer through a filamentary IGM -- which also breaks spherical symmetry -- but at the least a statistical detection might be possible. Finally, near-IR followup of large ($R \ga 20 \Mpc$) HII regions, which should be dominated by quasars, may be an efficient means of discovering high-redshift quasars, since only $\sim 100$ pointings would be needed to find an active quasar -- a vast improvement over blind searches.  Such a survey would also measure the duty cycle (and lifetime) of quasars, provided that the largest bubbles are indeed sourced by bright quasars rather than clustered galaxies. These followup surveys should not be too difficult; quasars driving HII regions with $R \ga 20 \Mpc$ should be detectable with 8-meter class telescopes in 200-300 s. The next generation of large telescopes will allow us to probe further down the luminosity function and to search for bright galaxies. 

One final possibility for high resolution instruments such as the SKA is to measure the structure of the ionization front and emission/absorption shells around a quasar.  The thickness of the ionization front (which depends on the hardness of the ionizing source spectrum) could be used to discriminate between quasars and stars as ionizing sources \cite{zaroubi05}. This requires that the HII region sit in a neutral IGM; partial ionization by X-rays or relic HII regions significantly reduce the power of this discriminant. In addition to the ionization front, isolated quasars also have non-trivial temperature structure \cite{tozzi00}:  the HII region would then be surrounded by a ring of 21 cm emission (where X-rays from the quasar heat the IGM) fading into 21 cm absorption (where the X-rays peter out and $T_K < T_{\gamma}$). However, this structure will disappear if X-rays from other sources have heated the IGM on a more global level (as seems likely by $z \sim 6.5$; see \S \ref{glob}).  If it does exist, the detection of absorption will constrain the background UV flux of Ly$\alpha$ photons, because the Ly$\alpha$ photons from the quasar itself are unable to couple the spin and kinetic temperatures (largely because the HII region expands so fast) \cite{wyithe04-qso}. 

%\bibliographystyle{elsart-num}
%\bibliography{Ref_21cm}

%\end{document}

%% file: inter-ch9.tex
%\documentclass{elsart}
%\usepackage{amssymb,cite,epsfig,graphicx,subfigure}

%\input{../../submission/defns.tex}

%\begin{document}

\section{Low-Frequency Radio Observations} \label{int}

In the past several chapters, we have described the power of 21 cm tomography for studying basic cosmology, high-redshift structure formation, and reionization.  A number of experiments are currently being built to explore this signal; we list some of the largest in Table~\ref{tab:lfexps}.  Observing fluctuations during and before reionization requires low-frequency telescopes (the observed frequency is $\nu=1420/[1+z] \MHz$) with baselines of order a kilometer (in order to achieve sufficiently good angular resolution).  Thus all the experiments we will discuss are interferometer arrays.  The 21 Centimeter Array (21CMA, formerly known as PAST) is located in the Xinjiang province of western China and is currently being commissioned \cite{pen04}.  The Low Frequency Array\footnote{See www.lofar.org for more information.} (LOFAR) is under construction in the Netherlands, with plans for a possible extension into the rest of Europe.  LOFAR is a general purpose low-frequency radio telescope that will have both long and short baselines; it is the largest (both in terms of baselines and collecting area) of the first generation instruments.  The compact core, which is most relevant to 21 cm studies, is scheduled for completion in 2008.  The Mileura Widefield Array Low Frequency Demonstrator\footnote{See web.haystack.mit.edu/arrays/MWA/ for more information.} (which we will refer to as simply the MWA) is a smaller telescope to be built in western Australia, specializing in observations of the 21 cm background, radio transients, and the heliosphere.  It should also begin observations in two to three years.    Finally, the Square Kilometer Array\footnote{See www.skatelescope.org for more information.} (SKA) is the next-generation multi-purpose radio telescope, aiming for completion toward the end of the next decade.  It is sufficiently early in the design process that its parameters in Table~\ref{tab:lfexps} should be considered educated guesses at best.  Not included in Table~\ref{tab:lfexps} is the Precision Array to Probe Epoch of Reionization (PAPER), which aims to study the observational challenges, and strategies to overcome them, in detail as sensitivity is slowly increased.   Thus its parameters are sufficiently dynamic that a summary as in Table~\ref{tab:lfexps} is not useful.  An 8-dipole array has already been deployed in Green Bank, West Virginia, and plans call for later expansion to Western Australia (D. Backer, private communication).

%%%%%%%%%%%% TABLE 9-1: Telescopes
\begin{table}
\begin{center}
\begin{tabular}{|c|c|c|c|c|c|c|c|}
\hline
Array & $N_a$ & $A_{\rm tot}$ ($10^3$ m$^2$) & $\nu$ (MHz) & FoV (deg$^2$) & $D_{\rm min}$ (m) & $D_{\rm max}$ (km) \\
\hline
21CMA & ~20 & 8.0 & 70--200 & $\pi 15^2$ & 100 & 10 \\ 
MWA & 500 & 7.0 & 80--300 & $\pi 16^2$ & $4$  & 1.5 \\
LOFAR & 64 & 42 & 115--240 & $4 \times \pi 2^2$ & $100$ & 2c \\
SKA & 5000 & 600 & 100--200 & $\pi 5.6^2$ & $10$ & 5c \\
\hline 
\end{tabular} 
\end{center}
\caption{Existing and planned low-frequency radio telescopes and their approximate parameters.  The second column is the number of antenna elements.  The third shows the total collecting area $A_{\rm tot}$ at $z=8$.  We also quote the field of view (FoV) at $z=8$; it scales approximately with the diffraction limit $\theta_D^2$ (or with $[1+z]^2$).  $D_{\rm min}$ is the minimum baseline (in most cases determined by the area of a single element).  $D_{\rm max}$ is the maximum baseline; for experiments labeled with ``c" this is actually the extent of the compact core (which is the only part included in this table).  Parameters are taken from \cite{bowman05, carilli04} and the experiments' websites; all are approximate.  \label{tab:lfexps}}
\end{table}

The purpose of this section is to review these challenges and to ground the great promise of 21 cm studies in the practical considerations of the real world.  We begin with a brief discussion of low-frequency radio astronomy in \S \ref{int-over}.  We then describe how to estimate the sensitivity for statistical measurements in \S \ref{sens}.  Finally, we end with discussions of foreground cleaning (\S \ref{clean}) and some other systematic difficulties (\S \ref{other-fg}).

\subsection{Overview} \label{int-over}

Low frequency radio telescopes operate as phase coherent apertures.  The ``diffraction limit'' dictates that the finest achievable angular resolution $\theta_{\rm D}$ depends on the largest dimension  $D_{\rm max}$  of the telescope, $\theta_{\rm D} \sim \lambda/D_{\rm max}$; in our case, we have $\theta_D \sim \lambda_0 (1+z)/D_{\rm max}$, where $\lambda_0=21.1$ cm.  The exact value of $\theta_{\rm D}$ depends on the details of the aperture illumination, whether the telescope is a single dish or an array of sub-apertures, and the procedures invoked during calibration and image processing; a useful estimate for the redshifted 21 cm line is $\theta_D \sim 1.2^\circ \, [(1+z)/10] \, (D_{\rm max}/100 \,  {\rm m})^{-1}$.

Telescope mirrors must be smooth to a fraction of a wavelength in order to achieve diffraction-limited operation.  In the $\nu< 200$~MHz regime relevant for us, $\lambda >1.5$~m, and this smoothness requirement is lax.  This permits the use of coarse mesh as a reflecting surface.  Furthermore, simple, 
inexpensive technologies, such as yagi antennae\footnote{Yagi or Yagi-Uda antennae (named after their inventors) are  the most common type of roof-mounted antenna for receiving television signals. 
They consist of a driven dipole mounted between parallel elements (rods) that form a reflector behind
the dipole and a series of directors, which serve to increase the antenna gain in rough proportion
to the number of director elements. } and dipole arrays, can be used in telescope designs, making it possible to build the required large collecting areas at affordable costs.

A further consequence of diffraction-limited operation is that the sensitivity to a distant radio source whose angular size is less than $\theta_{\rm D}$ results from the phase coherent sum of the source's incident electric field over the entire telescope collecting area.  In this regime, the detectability of compact radio sources is nearly independent of the telescope aperture distribution and can be estimated quite simply from the total telescope area.  But the surface brightness sensitivity to radio sources resolved by the telescope ($\theta_{\rm S}>\theta_{\rm D}$) is highly dependent on the design of the telescope, as we discuss in \S~\ref{sens-basic}.
 
 The sensitivity of a telescope system depends on the competition between the strength of the
 celestial signal collected by the antenna and the noise, which can have different origins.
 In many radio astronomy applications, the dominant noise contribution arises in the first
 amplifiers that are connected to the output of the antenna.  The electrical junction between the antenna and this first amplifier of the receiving system is a convenient place to compare the celestial signal strength to the receiver noise level.  The signal output of the antenna can be specified as an {\it antenna temperature}, $T_a$, which is the temperature of a matched resistive load that would produce the same power level  ($P_a=k_{B}\,T_a\,\Delta\nu$ for the resistor) as the signal power $P=A_e\,S_{\nu}\,\Delta\nu/2$ received in one of two orthogonal antenna polarizations \cite{ThompsonMoranSwenson2001}.  Here $S_\nu$ is the source flux density (assuming an unpolarized source), $\Delta\nu$ is the observed bandwidth, and $A_e$ is the effective collecting area of the antenna.  These equations define the antenna sensitivity factor $K_a=T_a/S_{\nu}=A_e/2k_B$ in units of K~Jy$^{-1}$.\footnote{Some confusion surrounds the definition of $K_{a}$, because many early radio astronomy systems received only a single polarization. Then, a $K_{a}'$ was defined as $A_e/k_B$, with the implicit assumption of unpolarized sources. This definition persists in many applications, and observers  compensate by averaging (rather than summing) the measurements from dual polarized systems to obtain the total flux density.} 

The signal-to-noise ratio is assessed by comparing $T_a$ and $T_{\rm sys}$, the {\it system temperature},  similarly defined as the temperature of a matched resistor input to an ideal noise-free receiver that produces the same noise power level as measured at the output of the actual receiver.  Noise fluctuations $\Delta T^N$ decline with increased bandwidth and integration time 
$ t_{\rm int}$ according to the  well-known radiometer equation  (e.g. \cite{rohlfs04}):
\begin{equation}
\Delta T^N = \kappa_c\frac{T_{\rm sys}}{\sqrt{ \Delta \nu \, t_{\rm int}}} \approx\frac{T_{\rm sys}}{\sqrt{ \Delta \nu \, t_{\rm int}}},
\label{eq:radiometer}
\end{equation}
where $\kappa_c\ge 1$ is a loss factor accounting for the details of the signal detection scheme.
For radio spectrometers, $\kappa_c$ depends on the number of bits of precision used to quantify the signals when they are digitized. For interferometers, there may be  loss of signal when compensation is made for rapid fringe rotation. For our purposes, this instrumental parameter will be close to unity, and we will set $\kappa_c=1$ throughout the remainder of this discussion.

%%%%%%%%%%%% FIGURE 9-1: Radio Sky
\begin{figure}[!t]
%\centerline{\epsfig{file=f9-1.eps,width=3.0in,angle=-90}} % Used for Frank's original version
\centerline{\epsfig{file=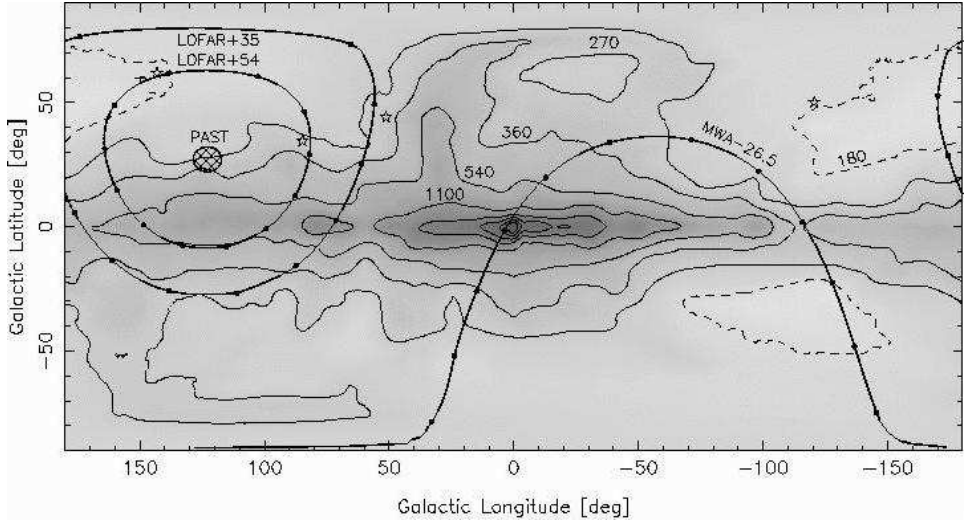,width=5.0in,angle=0}} % Used for compressed version
\caption{Brightness temperature of the radio sky at 150 MHz (from \cite{landecker69}) in Galactic coordinates.  Contours are drawn at 180 (dashed), 270, 360, 540, 1100, 2200, 3300, 4400, and 5500 K.  The 21CMA/PAST \cite{pen04} survey field at the North celestial pole is cross-hatched. Heavy lines indicate constant declinations:$-26.5^{\circ}$, $+35^{\circ}$, and $+54^{\circ}$
with dots to mark 2 hour intervals of time. Star symbols indicate the coordinates of the 4 highest redshift 
($z>6.2$) SDSS quasars (found with the NASA Extragalactic Database, nedwww.ipac.caltech.edu).
}
\label{fig:radiosky}
\end{figure}

\subsubsection{The Radio Sky}\label{radiosky}

Pointing the beam of the radio telescope into a region of sky whose brightness temperature is $T_{\rm sky}$ has the effect of adding noise power to that already present in the receiver, so that $T_{\rm sys} \ge T_{\rm sky}$.  In fact, at the low radio frequencies relevant to the redshifted 21 cm line, the sky is so bright that $T_{\rm sys} \approx T_{\rm sky}$.  Figure~\ref{fig:radiosky} maps the radio sky at 150 MHz, corresponding to $z=8.5$ for the 21 cm line.  Everywhere on the sky, $T_{\rm sky}$ is dominated by synchrotron radiation from fast electrons in the Milky Way; we will discuss the characteristics of this emission in more detail in \S \ref{clean}.  At these frequencies the Galactic Center is the strongest feature, and the brightness declines rapidly with Galactic latitude, making the Galactic poles attractive survey targets.  Also of significance are two minima in the sky brightness located in the general direction of the anti-center, where the sky brightness reaches $\sim$150~K. A rule of thumb for typical high-latitude, ``quiet" portions of the sky is
\begin{equation}
T_{\rm sky} \sim 180 \ \left( \frac{\nu}{180 \MHz} \right)^{-2.6} \kel.
\label{eq:tsky}
\end{equation}
The challenge at the heart of these observations is immediately apparent:  the foregrounds are four to five orders of magnitude larger than the expected signal, so substantial collecting areas and/or long integrations will be required to separate the cosmological component, even ignoring systematic effects!  Also note the rapid increase toward lower frequencies; for this reason, we expect the lowest redshifts (corresponding to reionization) to be by far the most accessible to observations.

To minimize $T_{\rm sys}$, experiments must therefore choose fields where the sky brightness is small. Figure~\ref{fig:radiosky} indicates the location of the 21CMA/PAST\cite{pen04} field, fixed by the telescope design to point at the north celestial pole.  Both LOFAR and MWA have steerable beams, which allows the selection of low sky brightness survey fields. Solid lines in Figure~\ref{fig:radiosky} trace the paths of the zenith through the course of a single day (at declinations $+54^{\circ}$ for LOFAR and at $-26.5^{\circ}$ for MWA).  LOFAR fields  chosen at declinations around $+35^{\circ}$ would lie in the heart of the northern cool spot.  MWA will be able to observe fields in both the northern and southern cool zones.

At low radio frequencies, non-thermal radiation from the Sun becomes extremely bright and variable during periods of high solar activity. To minimize solar contamination, redshifted 21 cm observations must be made at night.  Together with the restriction that cold regions of sky be accessible, 21 cm studies will have limited operating ``seasons."  For example, over a calendar year, any single field will only be accessible for $\sim 1000$ hours with MWA.

\subsubsection{Telescope Sensitivity: Imaging} \label{sens-basic}

Computing the sensitivity of compact sources unresolved by the diffraction-limited beam $\theta_{\rm D}$ of the array is straightforward. The noise level (in flux density units) can be written
\begin{equation}
\sigma_S = \frac{T_{\rm sys}/K_a}{\sqrt{ \Delta \nu \, t_{\rm int}}},
\label{eq:flux_noise}
\end{equation}
where $K_a$ now includes the total effective collecting area of the telescope.  As a concrete example, the flux density corresponding to $\dtb \sim 10 \mkel$ across a 25$'$ \hone cloud (or hole) at $z=8.5$ 
(spanning a comoving diameter $\approx 65 \Mpc$) is $\sim$270~$\mu$Jy.  In theory, a telescope like Arecibo ($K_{a}\sim 10$~${\rm K\,Jy}^{-1}$) observing a  quiet patch of sky with 1~MHz bandwidth and dual polarizations could detect such a hole in $t_{\rm int}\sim 35$ hours.

There is often confusion about the sensitivity of radio telescopes to low surface brightness features, especially when the telescope is a sparse ``array'' of small apertures.  Such a configuration is called an ``unfilled aperture.''  One approach to computing this important quantity uses the relation between brightness temperature and flux density from \S \ref{basic}.  We obtain an equivalent brightness temperature uncertainty,
\begin{equation}
\Delta T^N = \frac{\sigma_S c^2}{2k_B \nu^2\Omega_B},
\label{eq:tb_sensitivity}
\end{equation}
where the telescope beam subtends a solid angle $\Omega_B\approx \theta_{\rm D}^2$ if the telescope collecting area is distributed over an area of diameter $D$.  Combined with the  definitions for $\sigma_S$, $K_{a}$ and $\theta_{\rm D}$, equation~(\ref{eq:tb_sensitivity}) reduces to
\begin{equation}
\Delta T^N =\left(\frac{D_{\rm max}^2}{A_{\rm tot}}\right)\frac{T_{\rm sys}}{\sqrt{ \Delta \nu \, t_{\rm int}}}
\equiv \frac{T_{\rm sys}}{\eta_{f}\sqrt{ \Delta \nu \, t_{\rm int}}},
\label{eq:tb2_sensitivity}
\end{equation}
where $A_{\rm tot}$ is total effective area and $\eta_{f} \equiv A_{\rm tot}/D_{\rm max}^2$ is the {\it array filling factor}.\footnote{Note that, because it is the ratio of the effective area to the \emph{physical} size, $\eta_f$ need not be unity even for a single dish.}  An appreciation of this dependence on $\eta_{f}$ is crucial:  the integration time required to detect a given surface brightness grows as $t_{\rm int}\propto D_{\rm max}^4$ if the (fixed) total collecting area is spread over larger areas in order to achieve better angular resolution.

We can develop more intuition about the telescope response through a thought experiment in which a radio telescope is encased in a blackbody of temperature $T$.   Regardless of its size, and with proper impedance matching, the telescope would produce an antenna temperature $T_{a}=T$ at its output.  For this reason, attempts to observe the global 21 cm background are more concerned with issues of matching and gain calibration than with antenna size (see \S \ref{glob-obs}).

On the other hand, a telescope constructed with a beam of solid angle $\Omega_{B}$ will still deliver  $T_{a}=T$  at its output if  (i) it is embedded in a black body radiation field {\it or}  (ii) an emitter of $T_{B}=T$ entirely fills its beam. Unfortunately, real radio telescopes do not form perfectly defined beams, and all suffer from sidelobes whose shapes and responses are dictated by diffraction and scattering of
the incident radiation through the telescope. This is especially true of arrays, where a fraction $(1-\eta_{\rm f})$ of the total response lies outside the beam defined by  $\theta_{\rm D} \sim \lambda/D_{\rm max}$.  

Using equation (\ref{eq:tsky}) with $T_{\rm sys} \approx T_{\rm sky}$ to estimate the telescope noise $\Delta T^N$ for a single-dish measurement of an unresolved source, we find
\begin{equation}
\Delta T^N|_{\rm sd} \sim 0.6 \epsilon_{\rm ap}^{-1} \ \mkel \ \left( \frac{1+z}{10} \right)^{2.6} \,  \left( \frac{{\rm MHz}}{\Delta \nu} \, \frac{100 \hr}{t_{\rm int}} \right)^{1/2},
\label{eq:single-dish}
\end{equation}
where we have replaced $\eta_f$ with the aperture efficiency $\epsilon_{\rm ap}$, the ratio of effective and physical areas for a single dish.  Recall that the mean 21 cm signal has $\bdtb \sim 10 \mkel$; thus single dish telescopes can easily reach the sensitivity necessary to detect the global 21 cm background.  Instead, the challenge is to separate the (relatively slowly varying) cosmological signal from the foregrounds through careful gain calibration (see \S \ref{glob-obs}).  This may be possible if the 21 cm background has strong features (such as an absorption dip from early Wouthuysen-Field coupling or a sharp break from rapid reionization).

Of course, given the limited resolution of single-dish experiments at these low frequencies, interferometry is required to make maps with even relatively coarse resolution; for realistic collecting areas, the array dilution factor $\eta_f$ dramatically decreases the sensitivity.  Again using equation (\ref{eq:tsky}) for the system temperature, we find
\begin{equation}
\Delta T^N|_{\rm int} \sim 2  \mkel \ \left( \frac{A_{\rm tot}}{10^5 \sqm} \right) \, \left( \frac{10'}{\Delta \theta} \right)^2 \, \left( \frac{1+z}{10} \right)^{4.6} \,  \left( \frac{{\rm MHz}}{\Delta \nu} \, \frac{100 \hr}{t_{\rm int}} \right)^{1/2}.
\label{eq:if-sens}
\end{equation}
From equations (\ref{eq:dang}) and (\ref{eq:dbw}), these angular and frequency scales correspond to $\sim 20 h^{-1} \Mpc$:  clearly, at least for the first generation of interferometers, imaging will only be possible on coarse scales that exceed the typical sizes of bubbles during most of reionization.  It is for this reason that near-term imaging experiments focus primarily on large quasar \htwo regions.

To this point, we have used the conventional radio astronomy approach and explicitly separated the angular and frequency dimensions.  In most applications, these \emph{are} physically different:  the former correspond to physical distances while the latter provides spectral information from each source.  Our application is different in that all three are equivalent:  because of the cosmological redshift, the frequency dimension also corresponds to a distance (albeit in redshift space).  Thus it is usually better to think in terms of three-dimensional volumetric cells rather than pixels on the sky, or (for the power spectrum) the Fourier transform of this representation. 

\begin{figure}[!t]
\centerline{\epsfig{file=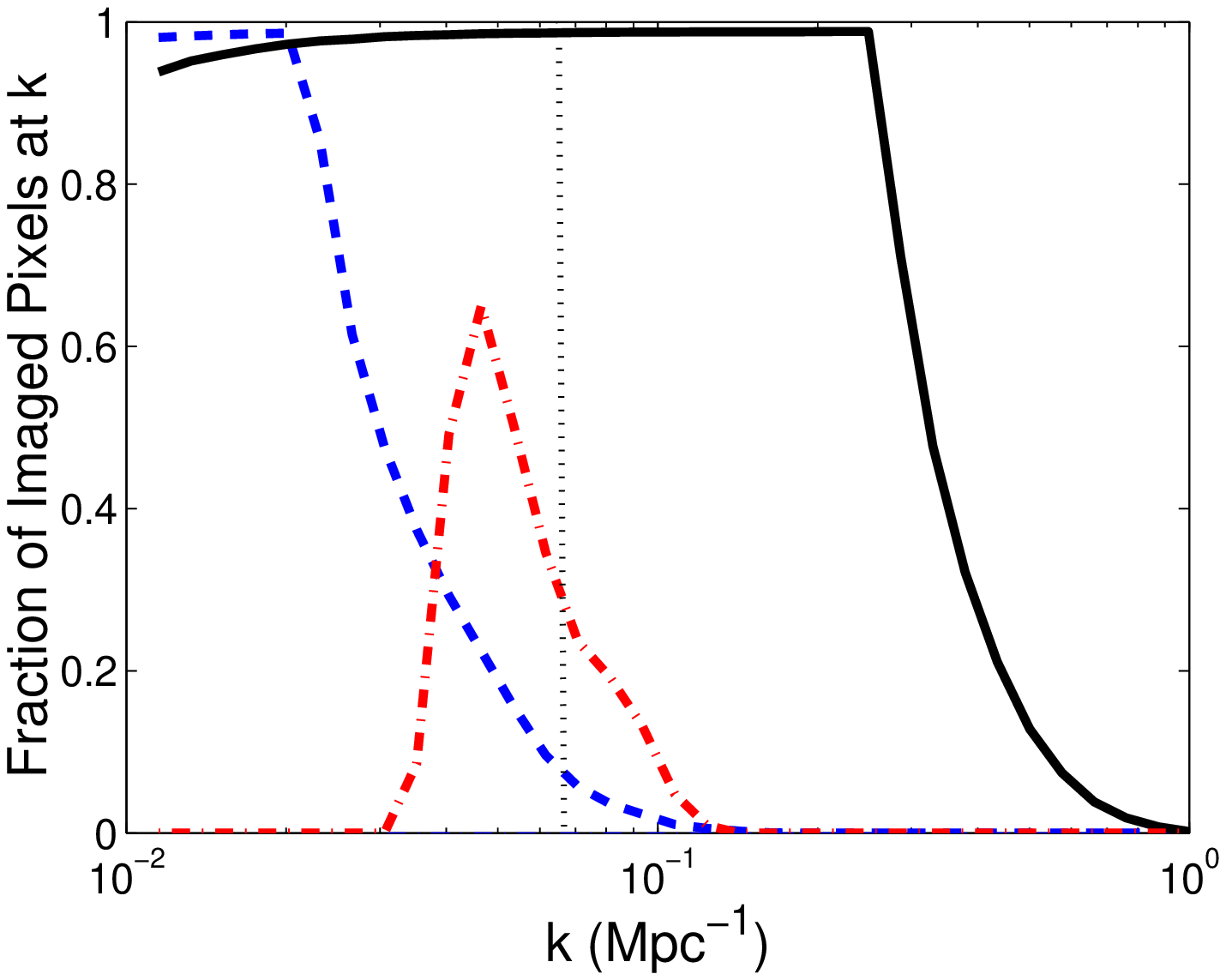,width=2.75in}
\epsfig{file=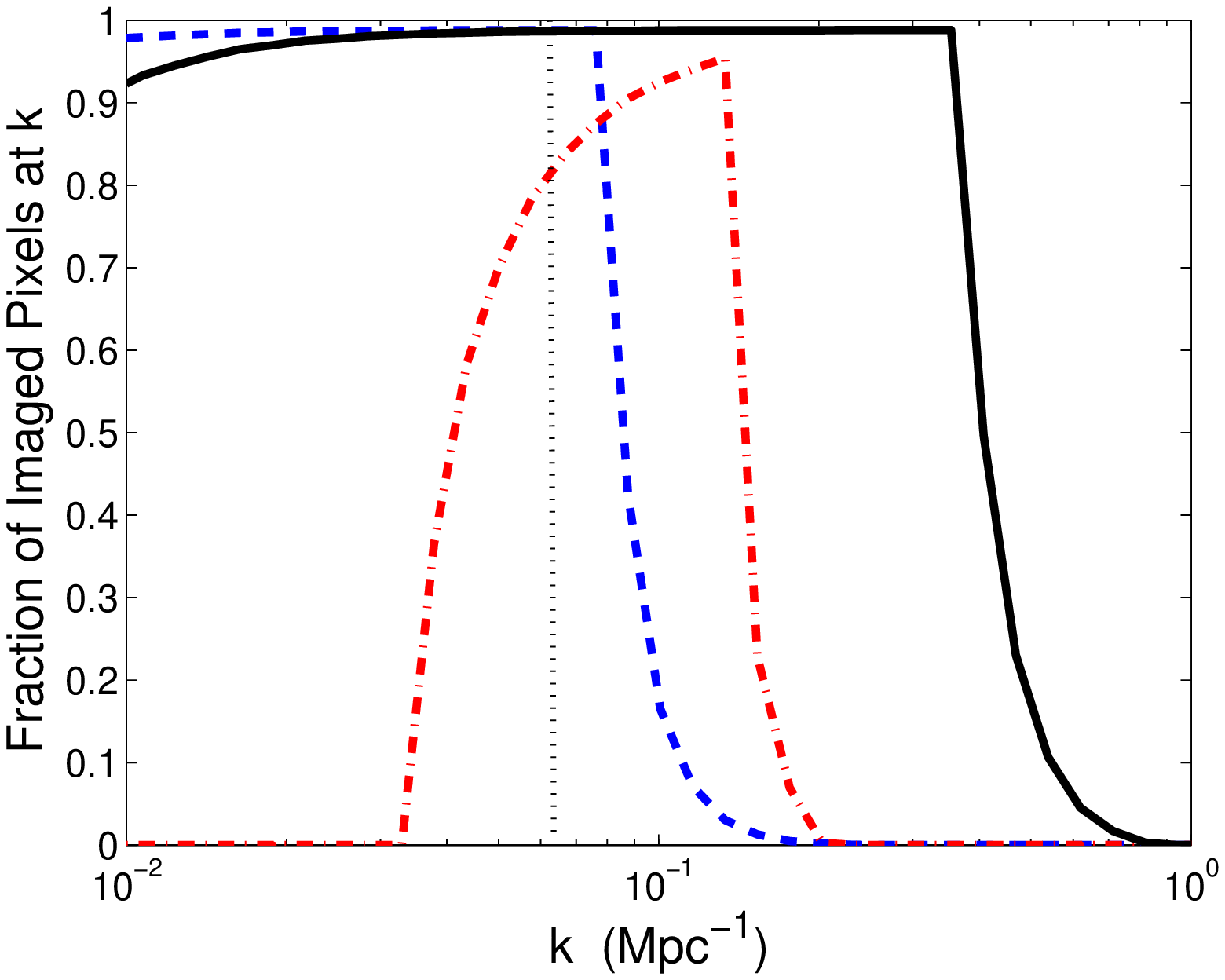,width=2.75in}}
\caption{Fraction of three-dimensional Fourier-space ``pixels" with signal-to-noise greater than unity in several upcoming 21 cm experiments:  the MWA (dashed), LOFAR (dash-dotted), and the SKA (solid).  \emph{Left:}  Scale-dependent signal originating from pure density fluctuations.  \emph{Right:} Scale-independent 12 mK signal (half the contrast between fully neutral and fully ionized gas).  We have assumed a 1000 hour integration at $z=8$.  In each panel, the vertical dotted line marks the scale corresponding to $6 \MHz$; smaller-$k$ modes will likely be overwhelmed by residual foregrounds.  From \cite{mcquinn05-param}.}
\label{fig:sense-image}
\end{figure}

In detail, the effective collecting area at a given angular resolution actually depends on the distribution of baselines in the array, so careful attention must be paid to antenna placement during array design.  Figure~\ref{fig:sense-image} illustrates this for some realistic array configurations as specified in Table~\ref{tab:lfexps}.  It shows the fraction of Fourier-space ``pixels" (each corresponding to a particular baseline and frequency pair; see below for a more precise definition) for which the rms signal exceeds the rms noise in a 1000 hour observation at $z=8$ with the MWA (dashed curve), LOFAR (dot-dashed curve), and the SKA (solid curve); note that these pixels can be identified a priori because they correspond to baselines with long effective integration times (and hence low noise).  The vertical dotted line shows the scale corresponding to a $6 \MHz$ bandwidth observation; as we will see in \S \ref{clean}, larger scales are essentially removed by foreground-cleaning.  Thus only wavenumbers to the right of this line are really relevant to observations.  

%%%%%%%%%%%% FIGURE 9-2: Imaging Power

The two panels make different estimates for the signal.  In the left plot, we let it equal the (scale-dependent) rms brightness temperature fluctuation if \emph{only} density fluctuations contribute.  Because reionization tends to enhance the fluctuations, this is somewhat pessimistic.  In the other plot, we let the signal equal 12 mK regardless of scale:  this is one-half of the contrast between completely neutral and completely ionized gas.  This plot therefore represents the ability to image isolated \htwo regions at each scale, essentially providing a ``best case" scenario for imaging during reionization.  Clearly, the prospects for imaging with LOFAR or the MWA are dim at best:  only a small fraction of modes will be well-resolved.  LOFAR fares considerably better because of its larger collecting area \cite{valdes06} -- and is even reasonably efficient at detecting \htwo regions with $k < 0.1 \Mpcinv$ -- but note that it also has a significantly smaller field of view.  Thus the total \emph{number} of imaged modes per field is actually comparable in the two experiments.  The SKA, on the other hand, has a large enough collecting area to image all modes up to $k=0.3$--$0.5 \Mpcinv$, with a tail extending to higher wavenumbers.  Still, these scales correspond to radii of $\sim 5 \Mpc$, so detecting small \htwo regions will always be difficult.

\subsubsection{Telescope Response Patterns}\label{response-patterns}

We now briefly discuss antenna response patterns in order to prepare for our later assessment of systematic errors and astrophysical foregrounds.

\begin{figure}[!t]
\epsfig{file=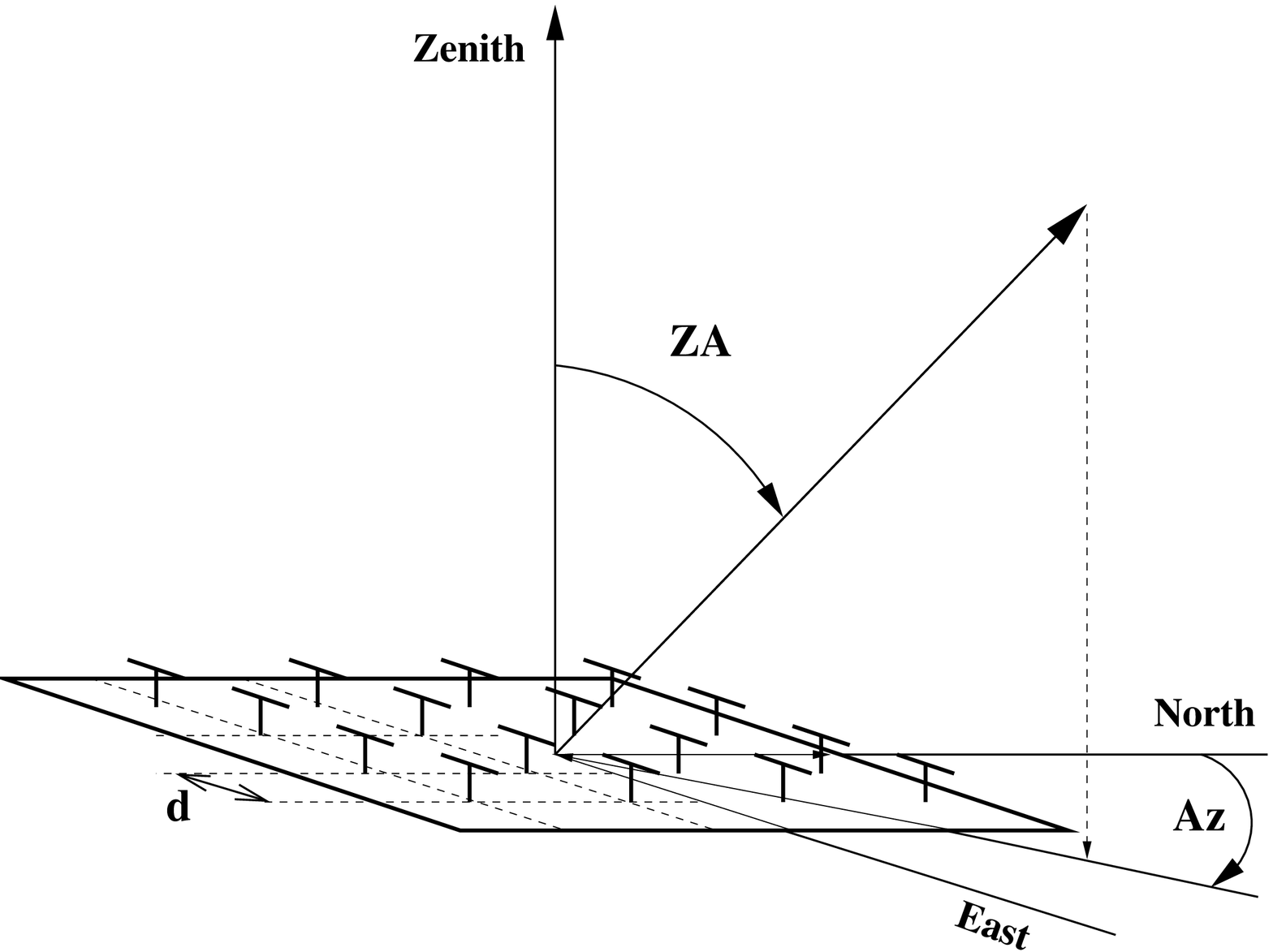,width=3in}
\vglue -2.5in\hglue 3.1in
\epsfig{file=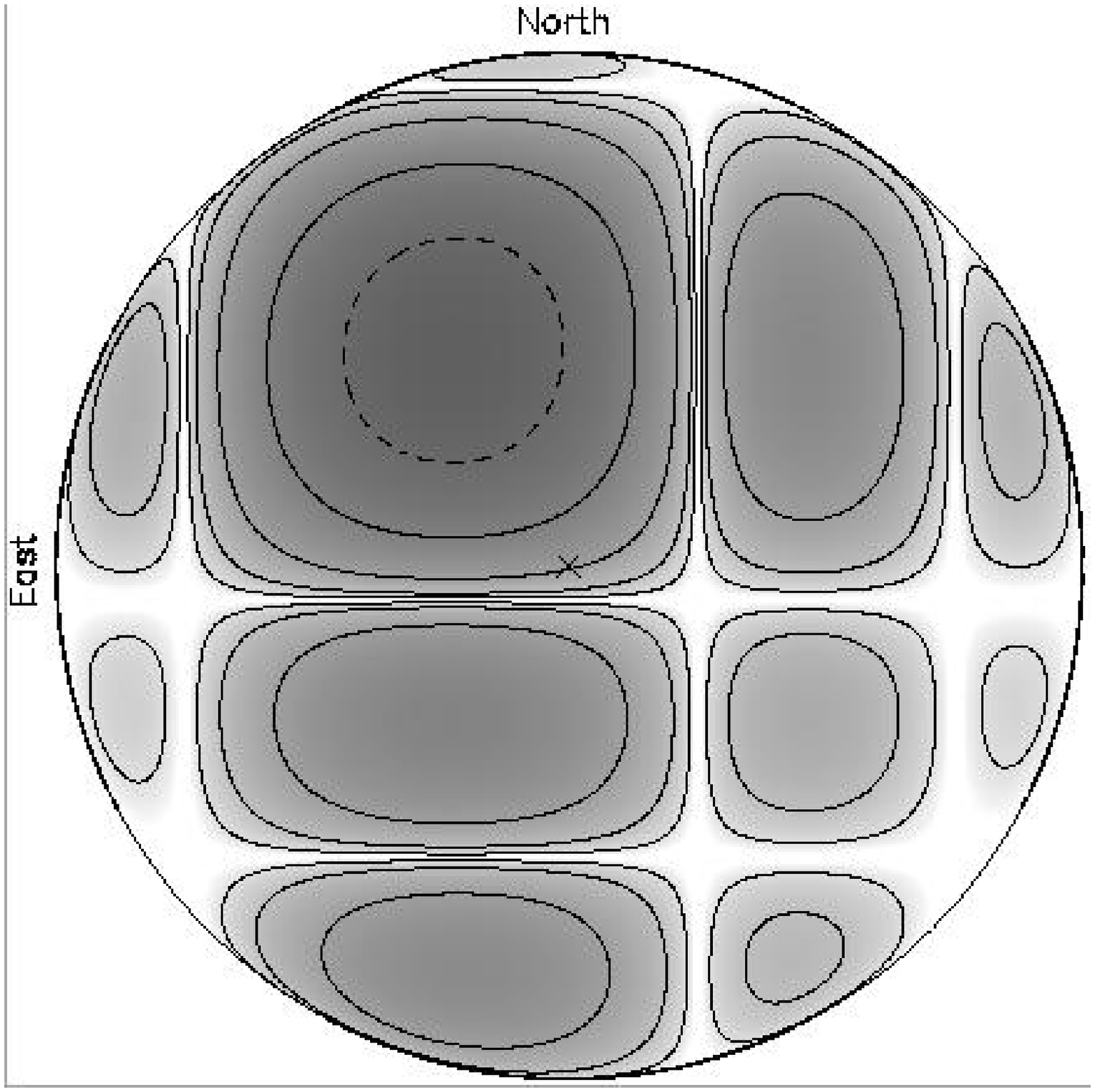,width=2.35in,angle=0.} % Used for compressed version
\caption{``4${\times}4$ tile'' and its response pattern.  {\it Left:} Sketch of  a ``tile'' with 16 horizontal dipoles arrayed in a   4${\times}$4 grid, spaced at $d\approx \lambda/2$. For optimum gain at zenith, the dipoles must be positioned at $\lambda/4$ above the conducting ground plane. {\it Right:} Response  of 
the tile, projected on the sky,  for phasing that points toward an azimuth angle Az=$30^{\circ}$ and zenith angle ZA=$30^{\circ}$. Radial distance is $\propto \sin\, $(ZA). Contours lie at $-3$ (dashed), $-10$, $-20$, $-30$, and $-40$ decibels relative to maximum. }
\label{fig:tile_response}
\end{figure}

The left panel of Figure~\ref{fig:tile_response} sketches the basic antenna station that the MWA and LOFAR designs use as a fundamental element of the interferometer arrays. This ``tile'' is composed of
16 dipoles positioned in a $4{\times}4$ grid on a conducting ground plane. The dipoles shown in
Figure~\ref{fig:tile_response} are aligned parallel to the ground plane.  For optimum gain near zenith at observed wavelength $\lambda$, they should be positioned on a grid with $\lambda/2$ spacing at $\lambda/4$ above the ground plane.  

The effective collecting area of a dipole  scales as $\lambda^2$ in applications where the linear size of the dipole is adjusted in proportion to wavelength.   However, in the tile design, a fixed  size for the elements and fixed spacing must serve a range of wavelengths in order to observe a range of redshifts. Under these circumstances, the elements are operated as ``active short dipoles'' \cite{Bregman99} in which preamplifiers are built into the hubs of the dipoles to assist in impedance matching, and the shortness of the dipoles causes their effective collecting areas to drop  toward long wavelengths.  Despite this apparent loss of effectiveness, the noise power from the sky brightness continues to dominate $T_{\rm sys}$, and some benefits accompany the reduced effective collecting area toward long wavelengths, since it carries with it a weaker coupling of the fields between the elements and a reduced cross section for mutual shadowing. 

The prototype designs  have a second set of orthogonal, north-south dipoles co-located with the east-west ones in order to receive both polarizations simultaneously.  The optimized designs from extensive
computer modeling are  generally complex structures that look little like the conventional
picture of the linear dipoles sketched in Figure~\ref{fig:tile_response}.

%%%%%%%%%%%% FIGURE 9-3: Tile Response

The right panel of Figure~\ref{fig:tile_response} shows the simulated response pattern of a simple $4{\times}4$ tile, assuming that it is phased to receive radiation from azimuth angle $30^{\circ}$ and zenith angle $30^{\circ}$ (gray scale image with contours).\footnote{Although dipole tiles lie flat on the ground and are not physically steerable, properly phasing the elements allows us to point the combined beam by introducing electrical delays in the signal paths from the dipoles.}  This response pattern is known as the \emph{primary beam} of the antenna and will be denoted $W_\nu(\bhn_0,\bhn)$. Here $\bhn$ is a unit vector denoting an arbitrary direction in the sky, and $\bhn_0$ is the direction for which the tile is phased to produce maximum response. Note the presence of ``sidelobes" far from the nominal pointing direction; these pose a significant problem for experiments.  The sine projection onto the plane tangent to the celestial sphere at zenith has the useful property that nulls between the main beam and sidelobes retain a rectangular grid pattern. The depth of the nulls in real telescopes is unlikely to equal those indicated in this figure, because slight variations in gain among the dipoles and quantization in the delays made to steer the tile beam imply that adding the 16 signals to form the tile beam will probably introduce errors greater than the $10^{-4}$ level depicted in Figure~\ref{fig:tile_response}.  

An important characteristic of the tile design (true, in fact, for all diffraction-limited telescopes) is that the sidelobe pattern scales with $\lambda$.  Thus the nulls systematically move across the sky as a function of frequency, even if the center of the beam remains phased to point at  specific coordinates independently of frequency. This chromatic behavior is a serious concern in the discussion of astrophysical foregrounds, because the interferometer array detects sources all across the sky through the frequency-dependent sidelobes.  Thus as the earth turns and the maximum of the response pattern tracks the chosen celestial coordinates, the reception pattern projected on the sky rotates and distorts, causing the signals received from sources lying in the sidelobes to vary in strength.

%%%%%%%%%%%% FIGURE 9-4: Telescope Polarization Calibration
\begin{figure}[!t]
\centerline{\epsfig{file=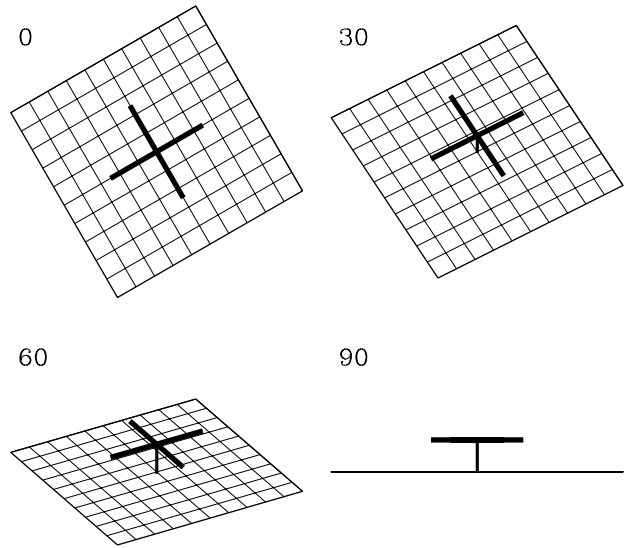,width=4.0in}}
\vglue -0.8in
\caption{Schematic view of crossed dipole antennae above a ground plane for four viewing angles, varying from face-on at  zenith angle ${\rm ZA}=0^{\circ}$ to ${\rm ZA} =90^{\circ}$. }
\label{fig:polarization_response}
\end{figure}

\subsubsection{Polarization Properties of Crossed Dipole Antennae}\label{crossed-dipole}

These 21 cm line observations will require extraordinarily precise instrumental calibration because of the high dynamic range necessary to identify mK structures in the bright radio sky (see eq.~\ref{eq:tsky}), with the additional complication of bright discrete radio sources.  While they are relatively inexpensive, dipole antennae require unconventional calibration methods. As we will see, a particular problem is instrumental coupling between polarized flux and total intensity.

The calibration problem is made difficult by the varying polarization response of crossed dipoles with angle on the sky,  as illustrated by Figure~\ref{fig:polarization_response}.  Dipoles that receive orthogonal polarizations from sources near zenith become progressively more correlated in their reception patterns as the source's zenith angle increases.  For sources close to the horizon, the crossed dipole antenna can receive only a single polarization; fortunately, this limit is not a severe problem, because dipoles above ground planes project nulls toward the horizon.  

Although antenna tiles will not be used to track sources at zenith angles greater than 60$^{\circ}$, their large sidelobes (Fig.~\ref{fig:tile_response}) imply that flux from bright sources and Galactic continuum emission enter the telescope from nearly the entire hemisphere above the horizon. For applications such as ours where precise intensity measurements or polarization information is desired, the antenna response must be carefully calibrated to decode the incident signals into Stokes parameters.  The calibration of antenna tiles will take advantage of the well-developed ``measurement equation''  formalism \cite{hamaker96,sault96,hamaker00} to describe this instrumental response.  A single dipole above a ground plane has a readily calculable response with this formalism.  But an array of  crossed dipoles on a tile is complicated by shading and electrical coupling between the dipoles, and these effects are more difficult to quantify with the measurement equation.  Developmental studies are now in progress to implement the necessary degree of precision into instrumental calibration.

\subsubsection{Interferometer Response Patterns}\label{interferometer-patterns}

\begin{figure}[!t]
%\centerline{\epsfig{file=f9-5.eps,width=2.0in,angle=-90}} % Used for Frank's original version
\centerline{\epsfig{file=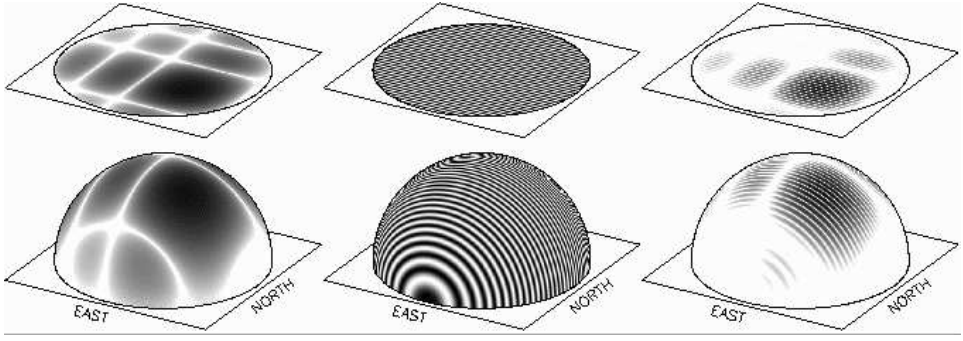,width=5.0in,angle=0}} % Used for compressed version
\caption{Telescope response patterns. {\it Upper row:} The sine projection of the response pattern on the plane tangent to the sky at zenith. {\it Lower row:} The response pattern projected on the celestial hemisphere above the horizon. {\it Left:} Response of a single interferometer element (a  ``tile'' in the case of MWA), also known as the primary beam.  {\it Center:} Response of a two element, east-west interferometer of omni-directional elements with a baseline of 20$\lambda$. {\it Right:} Response  of an interferometer with two MWA tiles as receiving elements.}
\label{fig:telescope_response}
\end{figure}

When two antennae are coupled together electronically to form an interferometer, the combined response projected on the sky resembles the characteristic diffraction pattern from a double slit (see Fig.~\ref{fig:telescope_response}).  In general, the interferometer response to the sky brightness distribution $I_{\nu}(\bhn)$ is 
\begin{equation}
\int \deriv \Omega\, I_{\nu}(\bhn\,) {\bf E}_1(\bhn_0,\bhn,\nu)\,{\bf E}_2^*(\bhn_0,\bhn,\nu)\, e^{2\pi i \,\bhn\cdot {\bf B}/\lambda}, 
\label{eq:fdgen_integral}
\end{equation}
where ${\bf E}_i(\bhn_0,\bhn,\nu)$
is the complex electric field response pattern of the $i$th element. The term $\exp({2\pi i \,\bhn\cdot {\bf B}/\lambda})$, where  ${\bf B}$ is the baseline vector from one interferometer element to the other, is purely geometric.  With the assumption
of identical beam patterns for all the elements, the product of the electric field patterns becomes the primary beam power pattern $W_\nu(\bhn_0,\bhn)$ introduced in the previous section.  Radio astronomers conventionally write the response for a particular ``visibility" ${\bf V}$, corresponding to a particular baseline and frequency pair, in units of flux density
\begin{equation}
{\bf V}_{\rm Jy}({\bf B}, \bhn_0, \nu) =  \frac{2k_B}{\lambda^2}\int \deriv \Omega\, \dtb(\bhn,\,\nu) W_\nu(\bhn_0,\bhn)\, e^{2\pi i \,\bhn\cdot {\bf B}/\lambda}.
\label{eq:fd_integral}
\end{equation}

%%%%%%%%%%%% FIGURE 9-5: Telescope Response

Usually, an approximate form of equation~(\ref{eq:fd_integral}) can be adopted:
\begin{equation}
{\bf V}_{\rm Jy}(\bhn_0,u,v,\nu) \approx  \frac{2k_B}{\lambda^2}\int \deriv x\, \deriv y\, \dtb(x,y,\nu) W_\nu(\bhn_0,\bhn)
\, e^{2\pi i \,(ux+vy)},
\label{eq:fduv_integral}
\end{equation}
where ${\bf B} = \lambda(u{\bf \hat{i}_0},v{\bf \hat{j}_0},w\bhn_0)$.  In the orthogonal $(u,v,w)$ coordinate system, the $w$ axis aligns with the direction toward the sky at the center of the tile beam, and the $u$-$v$ axes are oriented so that the $v$ axis projects onto the local meridian. The coordinates $x$ and $y$ are angles measured in the ``sky plane'' relative to the intersection of $\bhn_0$ with the celestial sphere.  In this Fourier transform of the sky, $u$ and $v$ represent spatial frequencies and the $w$ axis produces a phase offset in the interferometer fringe that can be calibrated.  This is the standard equation for describing aperture synthesis techniques, because radio interferometer arrays are designed to sample ${\bf V}_{\rm Jy}(u,v,\nu)$ at many $u$-$v$ coordinates and then reconstruct images of the sky through the inverse Fourier integral (e.g., \cite{ThompsonMoranSwenson2001}).  For our purposes, it is convenient to define the equivalent response in temperature units, 
\begin{equation}
{\bf V}(\bhn_0,u,v,\nu) \equiv \lambda^2 {\bf V}_{\rm Jy}(\bhn_0,u,v,\nu)/2 k_B.
\label{eq:visib_temp_units}
\end{equation}

For most of the following discussion, we adopt the assumptions and nomenclature of equation~(\ref{eq:fduv_integral}). However, its approximations are not always valid, in which case more elaborate solutions than a simple Fourier inversion are required. Complications arise because we must observe large areas of sky to build sufficient statistics and to identify interesting objects for followup at other wavelengths. Thus in order to keep the primary beam patterns large, the individual interferometer elements are intentionally kept small. There are several consequences.

First, the assumption of a single spatial frequency ($=\sqrt{u^2+v^2}$) across the entire primary beam does not hold for beams as broad as our example tile's (Fig.~\ref{fig:tile_response}).  Figure~\ref{fig:telescope_response} shows that the fringe spacing (or ``resolution'' $\theta$) of the interferometer varies with direction according to $\theta\propto \lambda/B\sin\alpha$, where $\alpha$ is the angle between ${\bf B}$ and $(\bhn - \bhn_0)$.  Thus the spatial frequency of structures in the sky sensed by the interferometer depends on their directions on the sky, and the interferometer actually senses a band of spatial frequencies rather than a single one.  More sophisticated imaging techniques will be required than those commonly used in aperture synthesis applications. Extracting the power spectrum will of necessity be a multi-stage procedure, involving an interplay between image plane and $u$-$v$ constraints during calibration and foreground cleaning.

Second, for  telescope designs relying on electronically steered tiles, slight differences in primary beam  patterns ${\bf E}_i(\bhn_0,\bhn,\nu)$ will result from small imbalances in the gains and the delays used to sum the signals from the individual elements. Especially with regard to removing sources outside the primary beam, these gain patterns will be difficult to calibrate, because every pointing direction suffers from slightly different gain imperfections. Thus the gain variations from element to element void the assumptions of the simple Fourier inversion approach. In the course of a day,
the rotation of the reception pattern, with its lobes and nulls, also defeats the assumptions adopted in
conventional Earth-rotation aperture synthesis.

Third, the Fourier integral does not properly account for sources far outside the primary beam; in effect, these add a noise-like contribution entering through the sidelobes.  Were it not for the variability of the primary reception pattern, the signals from these outlying sources would be suppressed by the systematic cancellation of the aperture synthesis method. But as it is, the residuals can remain problematic because of the extreme brightness of the foregrounds relative to the redshifted 21 cm signal. 

The problem is that a source with a flat continuum couples to the field of interest in a frequency-dependent way.  Each baseline in the array (phased to point in the direction $\bhn_0$) senses an unresolved source $S$ in the direction ${\bhn}$ with response
\begin{eqnarray}
{\bf V}_{\rm Jy}({\bf B}, \bhn_0, \nu) & = & W_\nu(\bhn_0,\bhn)S \exp{[2\pi i\,({\bhn}-{\bhn_0})\cdot{\bf B} \nu/c]} \nonumber \\
& \equiv & W_\nu(\bhn_0,\bhn)S\exp{(2\pi i\,\nu/\nu_{c})},
\label{eq:sideground}
\end{eqnarray}
where $\nu_c$ is the characteristic frequency scale of the chromatic contamination; for a field observed on the meridian by an east-west interferometer, a source rising in the east would yield $\nu_c=c/|{\bf B}|$, which for a 1~km baseline inserts a ripple into the spectrum with period 0.3~MHz.  As the source rises higher, the period of the ripple in the spectrum lengthens. The superposition of signals, with pseudo-random phases, from all over the sky, creates a noise-like background that is unlikely to be gaussian if it is dominated by several bright sources.  Because methods for foreground suppression rely on spectral smoothness (see \S \ref{clean}), this chromatic behavior will escape the conventional foreground removal. For this reason, reduction of 21 cm observations will need to begin with all-sky catalogs of the bright sources and perform iterative calibration and source subtraction.

The strength of this ``sideground'' contamination can be estimated by integrating over the differential
extragalactic source counts \cite{subrahmanyan02},
\begin{equation}
n(S)\,\deriv S=10^2S_{\mu{\rm Jy}}^{-2.2}\nu_{\rm GHz}^{-0.8} \, \deriv S_{\mu{\rm Jy}}\,{\rm arcmin}^{-2},
\label{eq:source-counts}
\end{equation}
where $S_{\mu{\rm Jy}}$ is the flux density in microJanskys and $\nu_{\rm GHz}$ is the frequency in GHz.  For a simple estimate, assume that $4\pi\,f_{\rm sky}$  steradians of sky (with $f_{\rm sky}=1/8$) is viewed through $-20$~dB sidelobes of the tile pattern (corresponding to a gain $g_{sl}=0.01$).  The vector responses of the sources can be combined as a random walk,  $\sigma_{sg}=g_{sl}\,[f_{\rm sky}4\pi\int n(S)\,S^2\,\deriv S]^{1/2}$ . We consider $z=8.5$ ($\nu =$ 150~MHz) and ignore the phase smearing caused by rotation (i.e., we integrate over a short interval of time). Since the integral diverges mildly at the bright end, we truncate it at 10~Jy, which implicitly assumes that all $\sim$800 sources of 10~Jy and brighter have been dealt with individually through self-calibration and subtraction. The cumulative effect of the remaining weaker sources on a single baseline is $\sigma_{sg}$ $\sim$1.6~Jy. An array such as the MWA has $N_{a}=500$  tiles, yielding $N_B=N_{a}(N_{a}-1)/2 = 1.2{\times}10^5$ baselines. If this sideground noise from each baseline were to add incoherently for all baselines, then the rms fluctuation would fall to $\sim 1.6\,{\rm Jy}/\sqrt{N_B}= 4$~mJy.  However, the baselines of an interferometer are of course calibrated to preserve systematic phases in a coherent sum, so the reduction in sideground fluctuations will actually be much larger.  Moreover, the Earth's rotation also smears the signals from outlying sources, and the total suppression should bring the residual
sideground contamination below the tens of $\mu$Jy signal expected from the the $z \sim 10$ Universe.  Similar estimates at lower frequencies show that observing the dark ages themselves will be extremely difficult.  However, this is an image processing regime that has not yet been fully explored and conquered, so for the moment these expectations should be viewed with some caution.

\subsection{Sensitivity: Statistical Measures} \label{sens}

Given the difficulty of high signal-to-noise imaging, attention has recently focused on statistical measurements.  We will now turn to estimating the sensitivity of 21 cm experiments to the power spectrum.  Error estimates for other statistical measures must still be developed, but the basic principles are the same (see \cite{bharadwaj05-ng, saiyad06}).  In this section, we will only consider the effects of thermal noise and cosmic variance, which provide a fundamental limit.  Systematics (especially foregrounds) present equally large difficulties, and the community is hard at work developing strategies to mitigate them (see \S \ref{clean}).  Over the past several years, the CMB community has developed expertise at both imaging and power spectrum estimation with interferometers, which provides an important launching point for describing the 21 cm signal (e.g. \cite{white99}).  The theory has now been further developed and applied directly to the 21 cm case (different from the CMB primarily because it is three-dimensional) \cite{bharadwaj01, bharadwaj03, bharadwaj04-gmrt, bharadwaj05, zald04, morales04, morales05, bowman05, bowman05-param, mcquinn05-param}.  Note that we will work exclusively in terms of the three-dimensional power spectrum.  Similar considerations apply to the angular power spectrum \cite{zald04}.

We begin with the complex visibility of equation~(\ref{eq:fduv_integral}).  The detector noise for a single visibility measurement is closely related to equation~(\ref{eq:radiometer}) \cite{rohlfs04}.  In the limit in which $T_{\rm sys} \approx T_{\rm sky}$,
\begin{equation}
\Delta T^N(\nu) = \frac{\lambda^2 \, T_{\rm sky}}{A_e \, \sqrt{\Delta \nu t_{\bf u}}},
\label{eq:visnoise}
\end{equation}
where here $t_{\bf u}$ is the integration time of this particular baseline\footnote{Typically, with these large interferometers, the $uv$ coverage changes with the earth's rotation, so this is \emph{not} the same as the total integration time.  Note that this is qualitatively different from CMB experiments, which work at high enough frequencies that the instruments can be rotated mechanically to achieve the desired $uv$ coverage.} and $A_e=\epsilon_{\rm ap}\delta A$ is the effective collecting area of each antenna element, which is equal to its physical area $\delta A$ multiplied by the aperture efficiency $\epsilon_{\rm ap}$. For simplicity, we will set $\epsilon_{\rm ap}=1$ in the following; its propagation through the error estimate depends on the particulars of each experiment (especially whether the beam is tapered).  To mark the difference, we will use $\delta A$ exclusively below.  This expression follows naturally from equation (\ref{eq:flux_noise}) for the noise level measured in flux density, combined with the definition for $K_a$, and the conversion from flux density to temperature in equation (\ref{eq:visib_temp_units}).

The  observed ``visibility data cube" is actually a hybrid of Fourier space ($u,\,v$) and redshift-space ($\nu$) coordinates and is thus inconvenient for comparing to theoretical models.  One can either transform the visibility data to the sky plane to obtain the  ``image cube" or transform the frequency (redshift) coordinate to its Fourier-space equivalent in order to obtain a representation with spatial frequency for all three dimensions,
\begin{equation}
\dtb(\buv) = \int_B \deriv \nu \, {\bf V}(u,v,\nu) \, e^{2 \pi i \eta \nu},
\label{eq:etadefn}
\end{equation}
where the integration extends over the full bandwidth $B$ of the observation, $\buv \equiv u {\bf \hat{i}} + v {\bf \hat{j}} + \eta {\bf \hat{z}}$, and $\eta$ has dimensions of time.  In this representation, the effective noise can be obtained by Fourier transforming the signal across the frequency axis \cite{morales05}, yielding
\begin{equation}
\Delta T^N(\buv) = \frac{\lambda^2 \, T_{\rm sys} \, \sqrt{B}}{\delta A \, \sqrt{t_{\bf u}}} \approx 
\frac{T_{\rm sys}}{\sqrt{B \, t_{\bf u}}} \times \frac{\lambda^2}{\delta A \, \delta \eta}.
\label{eq:unoise}
\end{equation}
In the second equality, we have set $\delta \eta$ equal to the inverse bandwidth.  The factor $\delta A/\lambda^2 \times \delta \eta$ then represents the Fourier space resolution of the observation (or the inverse volume sampled by the primary beam, in the appropriate units); note the similarity to equation (\ref{eq:radiometer}) when written in this form.  Here $\Delta T^N(\buv)$ has units of temperature divided by time, because of the Fourier transform in the frequency direction.

To estimate the statistical errors, we need the covariance matrix of the noise for antenna pairs at baselines $\buv_i$ and $\buv_j$.  Because the thermal noise errors are uncorrelated between measurements, this is simply a diagonal matrix with each element the square of equation (\ref{eq:unoise}).  In transforming to the physical wavevector $\bk$, we distinguish between the component $\buv_\perp$ oriented along the sky (corresponding to $\bk_\perp = 2 \pi \buv_\perp/\ell$, where $\ell$ is the comoving distance to the observed 21 cm screen) and the component $\bk_\parallel$ along the line of sight.  This is useful because interferometers can have arbitrarily good frequency resolution while the $\buv_\perp$ coverage is fixed by the baseline distribution.  

We define the number density of baselines that observe a given $\buv_\perp$ as $n(\buv_\perp)$; this is normalized so that its integral over the half-plane is $N_B = N_a \, (N_a-1)/2$, the total number of baselines in the array.  Two properties of $n(\buv_\perp)$ are worth emphasizing.  First, because of the earth's rotation, it is azimuthally symmetric and only a function of $u_\perp = | \buv_\perp |$.  Second, for a smooth antenna distribution, $n(u_\perp)$ is virtually always a decreasing function of $u_\perp$.  This is simple geometry:  it is difficult to arrange the antenna distribution to have many more long baselines than short ones.  In practice, a ring of antennae provides the flattest possible distribution (though that is rarely optimal; see below).  We can write:
\begin{equation}
t_{\bk} \approx n(u_\perp) \, \left( \frac{\delta A}{\lambda^2} \right) \, t_{\rm int}.
\label{eq:tk}
\end{equation}
As before $\delta A/\lambda^2 \approx \delta u \, \delta v$ is the angular component of the Fourier-space resolution.  Thus \cite{morales05, mcquinn05-param}
\begin{eqnarray}
C^N(\bk_i,\, \bk_j) & \equiv & \VEV{\Delta T^N(\buv_i)^* \ \Delta T^N(\buv_j)} \nonumber \\
& = & \left( \frac{\lambda^2 \, B \, T_{\rm sys}}{\delta A} \right)^2 \, \frac{\delta_{ij}}{B \, t_{\bk}}.
\label{eq:cov-noise}
\end{eqnarray}

Equation (\ref{eq:cov-noise}) represents the thermal noise contribution to the covariance matrix; even in an ideal experiment with no systematics from foregrounds, we must also include errors from sample variance.  This component is \cite{mcquinn05-param}
\begin{eqnarray}
C^{SV}(\bk_i,\, \bk_j) & = & \VEV{\dtb^*(\bk_i) \, \dtb(\bk_j)} \nonumber \\
& \approx & \delta_{ij} \bdtb^2 \, \int \deriv^3 \buv \, |\tilde{W}(\buv_i-\buv)|^2 \, P_{21}(\buv) \nonumber \\
& \approx & \bdtb^2 P_{21}(\bk_i) \, \frac{\lambda^2 \, B^2}{\delta A \, \ell^2 \, \Delta \ell} \, \delta_{ij},
\label{eq:cov-samvar}
\end{eqnarray}
where $\ell$ is the distance to the 21 cm field and $\Delta \ell \propto B$ is the line-of-sight depth of the observed volume in comoving units (see eq.~\ref{eq:dbw}).  In the first line, the average is over baseline and frequency pairs indexed by $\bk_i$ and $\bk_j$ (or equivalently $\buv_i$ and $\buv_j$).  In the second line, $\tilde{W}$ is the Fourier-transform of the primary beam response function, including the finite bandwidth, and is most naturally expressed in the ``observed" units $\buv$.   It typically differs from zero in an area $\delta u \, \delta v \, \delta \eta \approx \delta A/(\lambda^2 B)$ and (ignoring efficiencies) integrates to unity over the beam.  For the last line, we have assumed that $\buv$ is much larger than the width of this response function.  Then $P_{21}(\buv)$ is constant across the beam and can be pulled out of the integral, which becomes simply $(\delta u \, \delta v \, \delta \eta)^{-1}$.  We have also transformed to the more physically relevant wavenumber $\bk$, which introduces a factor $B/(\ell^2 \Delta \ell)$.

Equation~(\ref{eq:cov-samvar}) has a simple physical interpretation:  it is essentially a normalization factor ($\bdtb^2 B^2$) multiplied by $P_{21}/V_\ell$, where $V_\ell \approx \ell^2 \Delta \ell (\lambda^2/\delta A)$ is the total volume observed by the telescope.  The second factor counts the number of estimates available for the measurement; thus the cosmic variance decreases as the volume increases.

The Fisher information matrix gives an estimate of the errors on a power spectrum measurement from the total covariance matrix $\bC = \bC^N + \bC^{SV}$.  Given a vector of parameters ${\bf \Psi}$, the $(i,\,j)$ element of the Fisher matrix is defined as (see, e.g., \cite{dodelson03})
\begin{eqnarray}
F_{ij} & \equiv & \VEV{ - \frac{\partial^2 \ln {\mathcal L}}{\partial \Psi_i \, \partial \Psi_j}} \\
& = & {\rm Tr} \left[ \bC^{-1} \, \frac{\partial \bC}{\partial \Psi_i} \, \bC^{-1} \frac{\partial \bC}{\partial \Psi_j}  \right],
\label{eq:fisherdefn}
\end{eqnarray}
where ${\mathcal L}$ is the log-likelihood function.  For the simple case of measuring the binned power spectrum from the datapoints, the ``parameters" are the power spectrum amplitudes in each of the bins, $\Psi_i = P_{\Delta T} \equiv \bdtb^2 P_{21}(\bk_i)$; in more general cases they are the parameters of a theoretical model meant to describe the data.  The Cramer-Rao inequality states that the errors on any unbiased estimator of the power spectrum must satisfy 
\begin{equation}
\delta P_{\Delta T}(\bk_i) \geq \frac{1}{\sqrt{N_c(\bk_i)}} \, \sqrt{ ({\bf F}^{-1})_{ii}},
\label{eq:crao}
\end{equation}
where $N_c$ is the number of measurements in the appropriate bin and ${\bf F}^{-1}$ is the inverse of the Fisher matrix. 

In our case, the Fisher matrix is particularly simple to use because the covariance matrix is diagonal.  (This will not be true for real data, because foreground cleaning and other systematic effects induce correlated residual errors, but it provides a rough estimate of the noise limits.)  Also, $\bC^N$ is of course independent of the underlying power spectrum, so the errors become \cite{mcquinn05-param}
\begin{equation}
\delta P_{\Delta T}(\bk_i) \approx \frac{1}{\sqrt{N_c(\bk_i)}} \, \frac{\delta A \, x^2 \, y}{\lambda^2 \, B^2} [C^N(\bk_i,\bk_i) + C^{SV}(\bk_i,\bk_i)].
\label{eq:pkerror}
\end{equation}

The last step is to count the number of Fourier cells in each bin, which depends on the Fourier-space resolution of the instrument.  Recall that $P_{21}$ is not truly isotropic, but it is azimuthally symmetric.  
Thus we use Fourier cells grouped into annuli of constant $(k,\,\mu)$.  Then, in the limit that $k \mu$ is much larger than the effective $k_\parallel$ resolution,
\begin{equation}
N_c(k) \approx 2 \pi \, k^2 \, \deriv k \, \deriv \mu \times \left[ \frac{V_\ell}{(2 \pi)^3} \right],
\label{eq:nc-annulus}
\end{equation}
where the last term represents the Fourier space resolution.  One can also average the measurements spherically \cite{morales05, bowman05, bowman05-param}; although intuitively simpler, this eliminates information whenever redshift-space distortions are significant.

\subsubsection{Implications} \label{sens-imply}

Equations (\ref{eq:cov-noise}), (\ref{eq:cov-samvar}), and (\ref{eq:pkerror}) fully specify the effects of noise in the absence of systematic effects (which we will discuss in the next two sections).  But to make estimates we must determine the effective observing time $t_{\bk}$ for each mode -- and hence the baseline distribution $n(u_\perp)$ by equation (\ref{eq:tk}) -- as well as the sampling density (eq. \ref{eq:nc-annulus} for a measurement in annuli).  These two quantities are obviously highly dependent on the instrumental configuration.  Thus it is useful to consider the simple thermal noise-dominated case in order to develop some intuition for array design.  Substituting for $N_c$ in equation (\ref{eq:pkerror}) and assuming $C^N \gg C^{SV}$, we find
\begin{equation}
\delta P_{\Delta T} \propto \delta A^{-3/2} \, B^{-1/2} \, \left[ \frac{1}{k^{3/2} \, n(k,\, \mu)} \right] \, \left( \frac{T_{\rm sys}^2}{t_{\rm int}} \right),
\label{eq:sens-scale}
\end{equation}
where of course $u_\perp \propto k\sqrt{1-\mu^2}$.  Here we have assumed that the power spectrum is measured in bins with constant logarithmic width in $k$ but constant linear width in $\mu$.  From equation (\ref{eq:sens-scale}), we can deduce a number of fundamental considerations driving array design \cite{morales05}.  For instance, we see that $\delta P_{\Delta T} \propto t_{\rm int}^{-1}$; this is because the power spectrum depends on the square of the intensity.  

A more subtle question is the improvement possible by increasing the collecting area, which could be accomplished in either of two ways.  First, we can add antennae while holding the dish size $\delta A$ constant.  Recall that $n(k,\, \mu)$ is normalized to the total number of baselines $N_B \propto N_a^2$:  thus, adding antennae of a fixed size decreases the errors by the total collecting area squared.  (Of course, the number of correlations needed also increases by the same factor, so this strategy has other costs.)  Second, we can make each antenna larger but hold their total number fixed.  In this case, the total number of baselines, and hence $n(k,\, \mu)$, remains constant, but $\delta P_{\Delta T} \propto \delta A^{-3/2}$.  Increasing the collecting area in this way is not as efficient because it decreases the total field of view of the instrument.

Adding bandwidth increases the sensitivity relatively slowly: $\delta P_{\Delta T} \propto B^{1/2}$, because it adds new volume along the line of sight without affecting the noise on any given measurement.  Of course, one must be wary of adding too much bandwidth because of systematics (especially foregrounds) and possibly intrinsic evolution.  

Finally, as a function of scale $k$,  $\delta P_{\Delta T} \propto k^{-3/2} \,n(k,\, \mu)^{-1}$.  The first factor comes from the increasing (logarithmic) volume of each annulus as $k$ increases.  But in realistic circumstances the sensitivity actually decreases toward smaller scales because of $n$.  This is most obvious if we consider a map at a single frequency.  In that case, high-$k$ modes correspond to small angular separations or large baselines; for a fixed collecting area the array must therefore be more dilute and the sensitivity per pixel decreases as in equation (\ref{eq:if-sens}).  In the (simple but unrealistic) case of uniform $uv$ coverage, the error on a measurement of the angular power spectrum increases like $\theta_D^{-2}$ for a fixed collecting area \cite{zald04}.  

Fortunately, the three-dimensional nature of the true 21 cm signal moderates this rapid decline toward smaller scales:  even a single dish can measure structure along the line of sight on small physical scales.  Mathematically, because $n(k,\, \mu)=n(k_\perp)$ (neglecting the slow variation of $uv$ coverage with frequency across the band), each baseline can image arbitrarily large $k_\parallel$, at least in principle.  For an interferometer, this implies that short baselines still contribute to measuring large-$k$ modes.  Thus, provided that they have good frequency resolution (which is anyway required for RFI mitigation), compact arrays like the MWA are surprisingly effective at measuring small-scale power \cite{morales05}.  There is one important caveat to this trick:  if short wavelength modes are only sampled along the frequency axis, we can only measure modes with $\mu^2 \approx 1$.  Thus we recover little, if any, information on the $\mu$ dependence of the redshift-space distortions.  Studying this aspect of the signal \emph{does} require baselines able to measure the short transverse modes with $\mu^2 \approx 0$.

The preceding discussion shows that understanding the expected errors on statistical measurements is crucial to both the array design and the observing strategy.  To that end, we next review some of the main results of such estimates \cite{bharadwaj01, bharadwaj03, bharadwaj04-gmrt, bharadwaj05, zald04, morales04, morales05, bowman05, bowman05-param, mcquinn05-param}.  We emphasize again that we neglect systematic errors in this section.  Table \ref{tab:lfexps} shows the basic parameters for several upcoming experiments (although many of these are only educated guesses).  The major element missing is the baseline distribution between $D_{\rm min}$ and $D_{\rm max}$.  The figures below assume that the baselines are distributed as $r^{-2}$ with a compact, filled core.  This is reasonable if $N_a$ is large; if not, the Fourier space coverage may be patchier.

\begin{figure}[!t]
\centerline{\epsfig{file=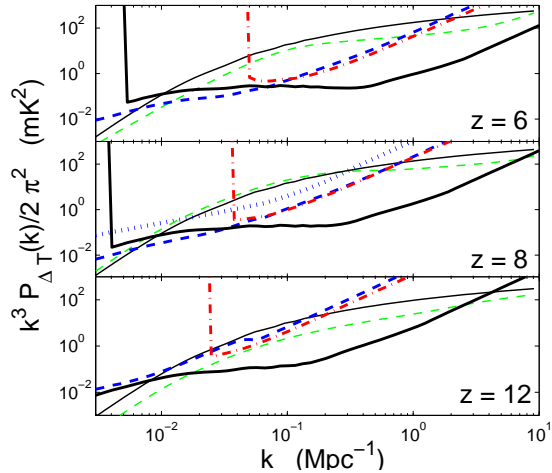,width=3.5in}}
\caption{Isotropic power spectrum sensitivity, in logarithmic bins with $\Delta k = k/2$, for several experimental configurations.  In each panel, the thin solid and dashed curves show estimates of the signal with and without reionization.  The thick solid, dashed, and dot-dashed curves show error estimates for 1000 hour observations over $6 \MHz$ with the SKA, MWA, and LOFAR, respectively.  Each assumes perfect foreground removal.  The dotted curve in the middle panel assumes a flat antenna distribution for the MWA.  From \cite{mcquinn05-param}.}
\label{fig:sense}
\end{figure}

Figure~\ref{fig:sense} summarizes the expected errors for the MWA (thick dashed curves), LOFAR (thick dot-dashed curves), and the SKA (thick solid curves) on the isotropic power spectrum \cite{mcquinn05-param}.  In each panel, estimates of the signal (with and without reionization) are shown by the thin dashed and solid curves.  The setup assumes 1000 hours of total integration on a single field (about a one-year observing campaign),\footnote{Unfortunately, $t_{\rm int}$ cannot be increased to arbitrarily large values because systematics will eventually intervene.  Current estimates suggest that integrating for several hundred hours should be reasonable, but going too much beyond that is likely to be difficult.} a $6 \MHz$ band (corresponding to $\Delta z \sim 0.5$), bins of width $\Delta k=k/2$, and $T_{\rm sys}=(250,\,440,\,1000) \kel$ at $z=(6,\,8,\,12)$.  We must also guess how $A_{\rm tot}$ scales with $\nu$ (see details in \cite{mcquinn05-param}; for now note that we optimistically assume that the SKA is optimized for each of these frequencies).

%%%%%%%%%%%% FIGURE 9-6: Overall sensitivities

Several key points are apparent in Figure~\ref{fig:sense}.  First and foremost, thermal noise and cosmic variance do not pose an insurmountable problem.  Both the MWA and LOFAR can make reasonably precise  measurements at $z \la 10$ and $k \la 1 \Mpcinv$, and the SKA can extend this to $z \ga 10$ and $k \la 10 \Mpcinv$ (although foregrounds probably prevent measurements at $k \la 0.01 \Mpcinv$).  The differences with redshift are mostly because of the rapidly increasing sky background toward lower frequencies (see eq. \ref{eq:tsky}):  clearly, our prospects for signal detection rapidly worsen toward higher redshift.  It is for this reason that exploring the first galaxies and bound structures, as well as the dark ages, will be much more difficult than reionization, even though the intrinsic signals are comparable. 

The most surprising aspect of Figure~\ref{fig:sense} is that the MWA and LOFAR have comparable performances, despite the nearly order of magnitude difference in collecting area.  There are several reasons for this.  First, the MWA has a much larger field of view as a consequence of its significantly smaller antennae, allowing it to beat down statistical errors.  Second, the smaller antennae of the MWA allow it to sample modes with $k_\perp \la 0.03 \Mpcinv$, which LOFAR cannot do (hence the cutoff near this value in the LOFAR curves); on the other hand, its extremely compact configuration does not allow it to observe \emph{any} modes with $k_\perp \ga 0.5 \Mpcinv$.  For many radio observations, this poor angular resolution would be a crippling disadvantage -- and that is why LOFAR, meant as a more general-purpose radio telescope, has both long and short baselines.  But for this particular application, high-$k$ modes can be sampled when they are oriented along the line of sight.

Finally, the thick dotted curve in the middle panel of Figure~\ref{fig:sense} shows the errors if the MWA has a flat baseline distribution (rather than $r^{-2}$).  This configuration reduces the sensitivity on all scales, which may seem surprising if one is used to measuring angular power spectra:  in that case, flat distributions are more sensitive to high-$k$ modes (e.g., \cite{zald04}).  But, thanks to the frequency dimension, even compact arrays can measure small-scale structure.

%%%%%%%%%%%% FIGURE 9-7: MWA sensitivity
\begin{figure}[!t]
\centerline{\epsfig{file=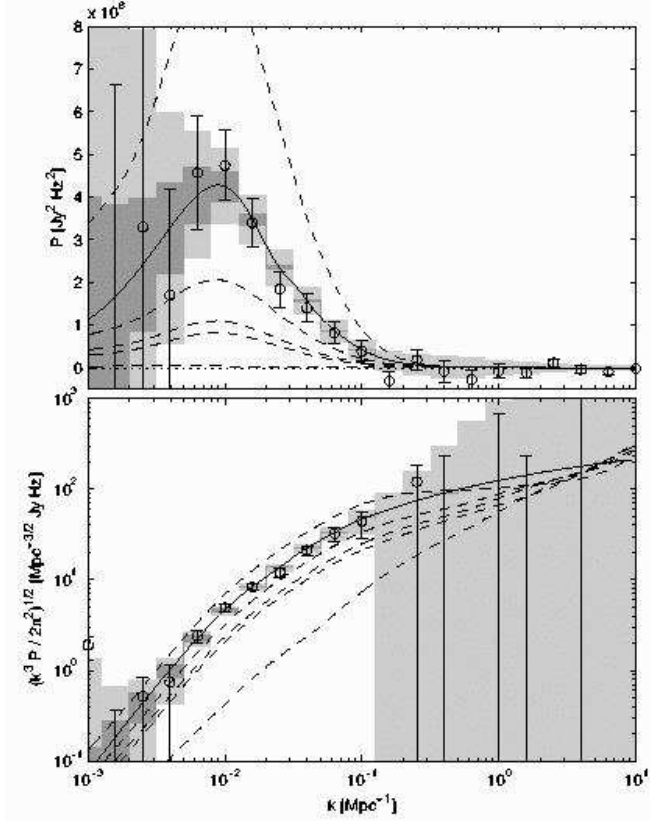,width=3.5in}}
\caption{Isotropic power spectrum sensitivity of the MWA at $z=8$.  The dark gray regions show errors from cosmic variance; the light gray regions include thermal noise as well.  The data points show a simulated realization of the measured power spectrum with $1\sigma$ errors.  The solid line assumes $\bxion=0$, while the dashed lines assume $\bxion=0.51$, $0.43$, $0.38$, $0.25$, and $0.13$, from top to bottom \cite{furl04-bub}.  The setup assumes an $8 \MHz$ band observed for $360$ hours.  From \cite{bowman05}.}
\label{fig:mwa-sense}
\end{figure}

Figure~\ref{fig:mwa-sense} shows another view of the MWA sensitivity \cite{bowman05}, this time with an explicit comparison to some analytic models of reionization \cite{furl04-bub}.  The Figure assumes a similar experimental setup as before, except with a 360 hour integration over $8 \MHz$ at $z=8$.  The points show simulated data (again neglecting systematics) with $1\sigma$ errors;\footnote{Note that some of the simulated datapoints have negative power here; this is a result of the simplified error propagation used by \cite{bowman05}.  In a realistic experiment, the estimator will be constructed to preserve the positive-definite nature of the signal.} the dark gray regions show the error envelope from cosmic variance (binned in the same manner as the data), and the light gray regions include thermal noise as well.  The solid curve is the power spectrum if $\bxion=0$; the other curves show predictions for the signal at various stages of reionization from \cite{furl04-bub}.  The purpose of this figure is to show explicitly that, over the range $0.01 \Mpcinv \la k \la 0.2 \Mpcinv$, the next generation of radio telescopes will be able to distinguish models at high confidence levels, provided that the thermal noise limit can be reached (the low-$k$ cutoff comes from foreground removal; see below).

\begin{figure}[!t]
\centerline{\epsfig{file=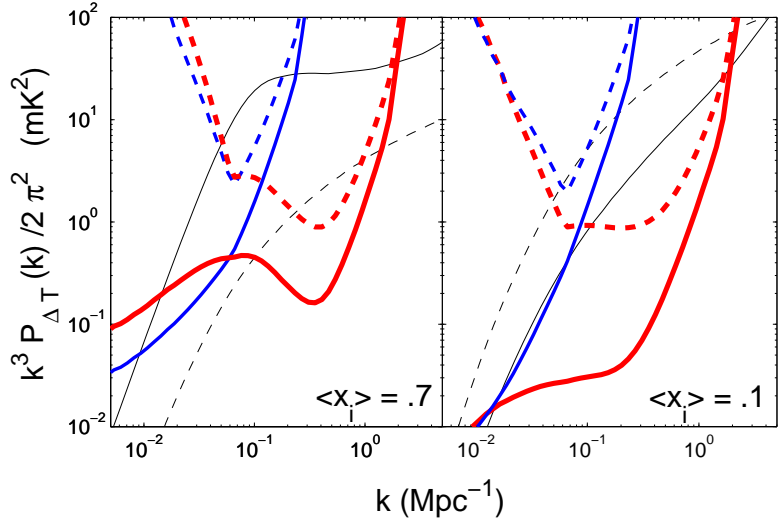,width=4.5in}}
\caption{Power spectrum sensitivity to the fully anisotropic signal.  The thin solid and dashed lines show $P_{\mu^0}$ and $P_{\mu^4}$ for $\bxion=0.1$ (right panel) and $\bxion=0.7$ (left panel).  The medium width and thick curves show the expected errors in bins of width $\Delta k=k/2$ for the MWA and SKA, respectively (solid and dashed correspond to the $\mu^0$ and $\mu^4$ components).  We assume 2000 hours of integration split equally between two fields at $z=8$.  From \cite{mcquinn05-param}.}
\label{fig:sense-mu}
\end{figure}

To this point, we have only considered the spherically-averaged signal, which ignores the cosmological information available from redshift-space distortions.  Figure~\ref{fig:sense-mu} shows the sensitivity of the MWA (medium width curves) and the SKA (thick curves) to anisotropies in the 21 cm signal \cite{mcquinn05-param}.  The thin solid and dashed curves show $P_{\mu^0}$ and $P_{\mu^4}$; the others show the errors in bins of width $\Delta k=k/2$.  Here we have assumed 2000 hours of integration time split between two fields at $z=8$.  The two panels assume different stages of reionization; at $\bxion=0.1$ the anisotropic component is strong because density fluctuations dominate, but it is weak at $\bxion=0.7$ when the bubble signature (which is isotropic) dominates the power.  The Figure shows how difficult it will be to extract detailed information on the angular dependence of the 21 cm power spectrum, especially for the MWA.  The MWA sensitivity to the $\mu^4$ component is sharply peaked because it relies on the frequency axis to provide all of its high-$k$ information.  Thus it cannot constrain the angular dependence of the power.  The larger collecting area and longer baselines of the SKA allow it to measure transverse modes to $k \sim 1 \Mpcinv$.  However, even it will have difficulty seeing the $\mu^4$ component during reionization.  

%%%%%%%%%%%% FIGURE 9-8: Anisotropic Power

Because the redshift-space distortions (and the AP effect) contain the most robust cosmological information, firm measurements of fundamental parameters will be difficult to extract from 21 cm experiments.  In the absence of the $\mu^n$ terms, these surveys can only measure the matter power spectrum over a relatively limited range of scales.  Thus, only when combined with other datasets (such as the CMB) will they be able to effectively constrain cosmological parameters \cite{bowman05-param, mcquinn05-param, santos06}.  The first generation of experiments will likely be restricted to studying features from reionization and to offering marginal improvements on cosmological parameter determinations from other methods (see \S \ref{costest}).  SKA-class instruments can significantly improve constraints on parameters that depend on the small-scale power spectrum, such as the primordial scalar spectral index $n_s$ and the neutrino mass.  But, of course, any such attempt to extract fundamental parameters relies on understanding the astrophysical factors as well.  Cosmological studies will be most straightforward during an epoch in which $\bxion$ and fluctuations in $T_S$ are both small.  Such an era may or may not exist (see \S \ref{glob}); if not, parameter estimation will likely be degenerate with these astrophysical processes \cite{santos06}.

\subsection{Foreground Removal Strategies} \label{clean}

As mentioned in \S\ref{radiosky}, foregrounds are formidable at such low radio frequencies, where the mean brightness temperature of our Galaxy, the primary foreground, exceeds that of the expected 21 cm signal by at least four orders of magnitude.  Fortunately, while this large \emph{mean} brightness resets the zero point and adds a great deal of noise, it does not introduce systematic difficulties:  instead, we are interested only in its fluctuations on the sky.  These are not yet well-studied at the frequencies and angular scales relevant to 21 cm studies.  Extrapolation from CMB foreground studies (e.g., \cite{giardino01, giardino02}) suggests that Galactic fluctuations will be relatively gentle on arcminute scales \cite{shaver99}.  An observational program at the Westerbork Synthesis Radio Telescope using the Low Frequency Front-End (LFFE) receiver system,\footnote{See http://www.ursi.org/Proceedings/ProcGA05/pdf/J03-P.14(0817).pdf.}  covering 120--180 MHz, is underway to measure the fluctuation spectrum of the Galactic emission.  However, even if these turn out to be small, the 21 cm signal is still swamped by temperature fluctuations from extragalactic sources: radio galaxies \cite{dimatteo02} and free-free emission from the reionizing sources themselves \cite{oh99,oh03,cooray04-ff} both create fluctuations that exceed the signal by one or two orders of magnitude (for a unified treatment of extragalactic foregrounds, see \cite{dimatteo04}).  

Fortunately, these difficulties are not insuperable:  it should be possible to recover the 21 cm signal through its structure in frequency space, because all the foreground contaminants mentioned above are spectrally smooth \cite{zald04,morales04,santos05,wang05}.  In other words, while the 21 cm signal is expected to be isotropic in 3D space (neglecting redshift space distortions and evolution), foregrounds have strong fluctuations in the transverse direction across the sky but weak ones in the radial direction. There {\it are} other foregrounds that are \emph{not} spectrally smooth, including radio recombination lines (\S\ref{RRL}) and terrestrial radio frequency interference, both atmospheric and man-made (\S\ref{RFI}). The telescope response can also introduce frequency structure:  sidelobes change with frequency (\S\ref{response-patterns}), implying that different parts of the sky (with different brightness temperatures) are surveyed at different frequencies \cite{oh03}.  Perhaps most worrying, in large part because so little is known about it, is whether polarized features in the Galactic foreground will suffer significant Faraday  rotation.  The polarized component can have brightness temperatures of several Kelvins and \emph{does} have substantial structure on arcminute scales \cite{wieringa93Gal, haverkorn03, haverkorn04}.  If it is Faraday rotated on frequency scales of $\sim 1 \MHz$ -- not at all unreasonable at these frequencies, given the known Galactic magnetic fields -- the polarized Galactic signal would appear as a chromatic fluctuation in the measured total intensity if the polarization response of the instrument is not perfectly calibrated (see \S \ref{crossed-dipole}).  This is another question being addressed by the LFFE system at Westerbork.  If these conditions occur, then the onus falls on the precision of the calibration algorithms to properly apportion the polarized component between the Stokes $Q$ and $U$ parameters without falsely coupling to variations in total intensity that could be mistaken for the redshifted 21 cm line signal. It is not yet clear how stringent these calibration requirements must be and whether they can be realized in practice.  In general, these frequency-dependent foregrounds could be much more difficult to remove and require specialized techniques.  Here we focus instead on continuum foreground removal. 

As mentioned above, the essential property employed by cleaning techniques is the strong frequency coherence of foregrounds. Indeed, if foregrounds were perfect power laws, then they could be trivially removed. However, (i) the Galactic foreground spectral index fluctuates as a function of both frequency (steepening toward lower frequencies) and position, and (ii) extragalactic foregrounds are a sum of power law spectra with different spectral indices; the result in general will not be a power law.\footnote{Note as well that continuum foregrounds have not been probed with such high frequency resolution before; one might worry for instance that the Galactic foreground could exhibit tiny $\sim 10^{-4}$ temperature blips at these small frequency scales due to irregularities in the electron momentum distribution in synchrotron knots.  Fortunately, this is highly unlikely because (i) one averages over a huge number of electrons along each line of sight, and (ii) the range of frequencies over which a single electron emits either synchrotron or free-free radiation is broad. The convolution of this broad emission profile with the electron momentum distribution washes out any sharp features in the latter \cite{chang06}.}The stochastic decoherence of foregrounds as a function of frequency sets a fundamental limit on the efficacy of foreground removal. These properties are well-understood in the CMB literature \cite{tegmark98,tegmark00} and have recently been applied to the 21 cm problem \cite{zald04,santos05}.

It is simplest to begin by neglecting practical difficulties of the telescopes.  We imagine an ideal instrument and ask whether foregrounds impose an absolute limit on the sensitivity.  Although we have emphasized the three-dimensional nature of the 21 cm signal, foreground removal is perhaps most intuitively understood with the angular power spectrum, because that explicitly separates the angular power spectrum (which is contaminated) from the (much more pristine) spectral fluctuations used for the cleaning.  To quantify the frequency decoherence between maps at two different frequencies, consider the correlation coefficient \cite{zald04,santos05}:
\begin{equation}
I_{l}(\nu_{1},\nu_{2}) \equiv \frac{C_{l}(\nu_{1},\nu_{2})}{\sqrt{C_{l}(\nu_{1},\nu_{1}) C_{l}(\nu_{2},\nu_{2})}} \approx 1 - \frac{{\rm log}^{2}(\nu_{1}/\nu_{2})}{2 \xi^{2}}
\label{eqn:decorrelation}
\end{equation}
where the second step represents a Taylor expansion about a power law, and (since all odd terms must vanish) is the lowest order deviation from complete correlation. The parameter $\xi$ is the ``correlation length;" smaller values of $\xi$ imply more rapid decorrelation over a fixed frequency separation. Note that $I_{l}$ can be $l$-dependent (e.g., if sources with different spectral indices $\alpha$ have different power spectra $C_{l}$ on the sky), so in general decorrelation will also be scale-dependent. 

Obviously, the crucial parameter is $\xi$. What is its value for realistic foregrounds?  For Poisson fluctuations, $\xi=1/\Delta \alpha$, where $\Delta \alpha = \langle (\alpha - \bar{\alpha})^{2} \rangle^{1/2}$ is the dispersion in spectral indices among sources \cite{tegmark98}.  If, on the other hand, $C_{l}$ is dominated by clustering, and in the simple case $\deriv (\ln C_{l})/\deriv(\ln \alpha)=0$, then $\xi=\sigma_{\alpha}/\Delta \alpha^{2}$, assuming that angular correlations between sources with spectral indices $\alpha_{1}$ and $\alpha_{2}$ fall off as ${\rm exp}[-(\alpha_{1}-\alpha_{2})^{2}/(2 \sigma_{\alpha}^{2})]$ \cite{zald04}. To the extent that all sources trace the underlying dark matter distribution, we expect them to be well-correlated on the sky. Even if they have different linear bias factors, we would have perfect correlation $\sigma_{\alpha}=\infty$ so long as they all still trace the same dark matter distribution; only the stochastic part of the bias contributes to the decorrelation. Thus, we expect $\sigma_{\alpha} \gg \Delta \alpha$; a firm limit might be $\sigma_{\alpha} = \Delta \alpha$. In that case, even though source clustering is expected to be the dominant source of angular fluctuations \cite{dimatteo02,oh03}, the Poisson component likely provides an upper bound to frequency decorrelation. For extragalactic radio sources, $\Delta \alpha \approx 0.2$ \cite{cohen04}, and for the Galactic foreground, $\Delta \alpha \approx 0.15$ \cite{tegmark00}.\footnote{Note that the dominant contribution to both of these spectral index fluctuations is synchrotron emission; the flat spectrum of free-free emission -- which varies only slightly due to the temperature dependence of the Gaunt factor and the exponential cutoff -- provides $\Delta \alpha_{\rm ff} \sim 0.03$ \cite{santos05}.}

Thus, assuming $\xi \approx 1/\Delta \alpha \approx 5$, $\nu_{1}=150 \MHz$, and $\Delta \nu=1 \MHz$, equation (\ref{eqn:decorrelation}) yields $(1-I) \approx (1$--$9) \times 10^{-7}$. It can be shown from a simple $\chi^{2}$ or Fisher matrix calculation that fitting for this frequency coherence will suppress foreground power spectra by a factor $(1-I)$; thus, as long as $(1-I)C_{l}^{\rm fg} \ll C_{l}^{\rm 21 cm}$, foregrounds can be effectively removed.  Even for the most pessimistic estimates, the raw foreground spectrum has $C_{l}^{\rm fg}/C_{l}^{\rm 21 cm} < 10^{5}$\cite{dimatteo04}, so suppression by $(1-I)$ suggests that residual foreground contamination should be small.  In fact, cleaning can be even more effective, because this estimate used only one pair of maps, instead of the  $N(N-1)/2$ pairs generated with $N$ frequency channels.  However, their errors are correlated, and difference maps from larger frequency separations have greater frequency decorrelation, so the net errors on $N$ maps will improve more slowly than $\sim [(N-1)/2]^{1/2}$. More precise error forecasts can be achieved via a Fisher matrix calculation \cite{santos05}. 

Over large bandwidths, the increasing decorrelation with larger frequency separation implies that the Taylor expansion in equation (\ref{eqn:decorrelation}) is no longer valid, so the (unknown) exact form of the coherence function, $I(\nu_{1},\nu_{2})=f({\rm log}[\nu_{1}/\nu_{2}]/\xi)$ becomes important.  The results can then be counterintuitive and careful study is required \cite{santos05}.
One could attempt to fit a parametric function to the coherence function of the data, which could improve the accuracy of the analysis. Two general points are worth noting. Accuracy diminishes toward the ends of the frequency interval, since there are fewer neighboring channels to correlate against. Also, these estimates assume that the 21 cm signal has no frequency coherence, with each channel probing a statistically independent slice of the universe. Obviously this assumption fails with narrower channel widths.  This (together with the increasing receiver noise) imposes a limit on how narrow a bandwidth one should use for foreground cleaning, even though in principle $I \rightarrow 1$ as $\Delta \nu \rightarrow 0$. The relative distribution of large- and small-scale power in foregrounds and the 21 cm signal also affects the appropriate set of basis functions to use for foreground cleaning, as described below. 

Thus the astrophysics itself does not present an insurmountable problem.  Much more worrisome is how this blanket of sources couples with practical telescope design issues.  The most important aspect is that the diffraction limit of the telescope (and hence the $uv$ vector corresponding to each baseline, the field of view, and the sidelobes) scales with wavelength.  Thus the beamsize increases steadily toward smaller frequencies, and new sources are added to each frequency channel.  This provides another -- and probably more important -- source of de-correlation than the spectral index fluctuations \cite{oh03}.  For example, in the simplest approximation we would have $I \approx 1 - (\nu_1/\nu_2)^2$ from adding new sources (with a constant density on the sky) to the beam.  With $\nu_1=150 \MHz$ and $\Delta \nu =1 \MHz$, we would therefore have $1-I \approx 4 \times 10^{-5}$, which could introduce fluctuations comparable to the cosmological signal.  Clearly, careful attention must be paid to the beam pattern.  Moreover, the changing $uv$ coverage implies that simply binning the data in a frequency-independent manner will introduce artifacts into the measurements that must be modeled properly, especially when the $uv$ coverage is not dense.  Monte Carlo runs with simulated data will allow us to understand the amplitude of such effects, and develop optimal binning strategies.

How will foreground removal be done in practice? The first step is to excise all bright point sources which exceed some detection threshold.  Ideally, this should be done directly from the $uv$ visibilities to account for the instrumental response and sidelobes, though reality may require more complicated algorithms (see \S \ref{interferometer-patterns}).  The remaining unresolved sources (and fitting errors from the subtraction of bright sources) are then the target of foreground cleaning procedures. 

The next step excises the spectrally smooth component of the observed sky.  Of course this will be done on the three-dimensional datacube, so the algorithms differ in detail from the ones we have described above.  In fact, there are many possible variants with no one optimal procedure for foreground cleaning; the best solution will likely depend on what statistic one is trying to extract from the data.  The most widely discussed approach -- which we focus on -- is to fit and subtract a smooth function to the spectral data, generally on a pixel-by-pixel basis (e.g., \cite{furl04-skarev, morales05_foregrounds,wang05,mcquinn05-param,chang06}). This process has a rigorous statistical justication as a means of extracting a tiny fluctuating signal from a huge, slowly varying background \cite{press92} (in statistical parlance, it is often called trend removal). Similar problems crop up in time series analysis, and trend removal has also been used in analysis of quasar absorption spectra to estimate the underlying continuum before extracting the Ly$\alpha$ forest \cite{hui01}. We summarize the present understanding for 21 cm foreground cleaning below.

Consider the foreground amplitude $f_{i}=f({\bf k}_{\perp},\nu_{i})$ in a pixel with angular index $\bk_{\perp}$ and frequency $\nu_{i}$.  We let the vector ${\bf f}$ denote $f$ measured at the frequencies ${\bf \nu}=(\nu_{1},.....\nu_{\rm N})$ with resolution $\Delta \nu$.  The fundamental idea is to fit $\bff$ with a  set of smooth basis functions.  This corresponds to projecting out or marginalizing over the corresponding modes in the data \cite{mcquinn05-param}; thus to remove the slowly varying foregrounds, we only fit low-order, slowly-varying basis functions.  Such techniques are commonly used in CMB analyses to marginalize over the temperature monopole and dipole, which suffer strong contamination. If the set of basis functions is complete and orthonormal (such as the Legendre polynomials), then the error analysis can be performed analytically.\footnote{Note that the most intuitive set of basis functions, polynomials in log($\nu$) -- corresponding to a Taylor series around a power law foreground -- is not orthogonal, so the error analysis is somewhat more complicated.} The residual foreground contamination after cleaning with $n$ basis functions is:
\begin{equation}
\tilde{\bf f}= \left( 1- \sum_{l=0}^{n} {\bf P}_{l} {\bf P}_{l}^{T} \right) {\bf f} = \sum_{l=n+1}^{\infty} {\bf P}_{l} {\bf P}_{l}^{T} {\bf f}.  
\end{equation}
The power spectrum of residual foregrounds is then given by \cite{mcquinn05-param}:
\begin{equation}
Q_{{\rm k}_{\perp}}(k,n) \approx \frac{\bfmu_k^\dag \, \tilde{\bff} \, \tilde{\bff}^\dag \, \bfmu_k}{ w \,\left(\tilde{\bfmu}_k^\dag \, \tilde{\bfmu}_k \right)^2},
\label{eqn:Q_residual}
\end{equation}
where the Fourier vector $\bfmu_k \propto \exp[i y k \bfnu]$.  Here $w \approx \lambda^2 B^2/(A_e \, \ell^2 \, \Delta \ell)$; see \S \ref{sens} for definitions of its components. $Q_{{\rm k}_\perp}$ is simply the square of the Fourier coefficients of $\tilde{\bf f}$, appropriately normalized. Foreground cleaning is effective provided that $Q_{{\rm k}_{\perp}}(k,n) \ll  P_{\Delta T}$.

Of course, during this cleaning process, the signal ${\bf s}$ will be attenuated as well: ${\bf s} \rightarrow \tilde{\bf s} = \sum_{l=n+1}^{\infty} {\bf P}_{l} {\bf P}_{l}^{T} {\bf s}$. The fundamental assumption of this procedure is that the scales on which the signal and foregrounds have significant fluctuation power are sufficiently well-separated that the signal is unaffected by foreground removal (or, at least, one is willing to sacrifice the smoothly varying modes in the signal).  One must then determine the optimal cleaning basis $P_{l}$, which will maximize $\langle \tilde{\bf s} \tilde{\bf s}^{\dag} \rangle$ while minimizing $\langle \tilde{\bf f} \tilde{\bf f}^{\dag} \rangle$. For instance, the order $n$ of a fitting polynomial plays an analogous role to the smoothing parameter in any regularization problem, yielding a trade-off between smoothness and fidelity to the data. If $n$ is too low, there are insufficient degrees of freedom to remove the foregrounds efficiently; if $n$ is too high, some of the cosmological signal is removed. The choice of expansion basis (Chebyshev polynomial, Legendre polynomial, broken power law, smoothing spline, etc) likewise affects the relative amounts of foregrounds and cosmological signal removed \cite{chang06}.  

%%%%%%%%%%%% FIGURE 9-9: Mcquinn foregrounds
\begin{figure}[!t]
\centerline{\epsfig{file=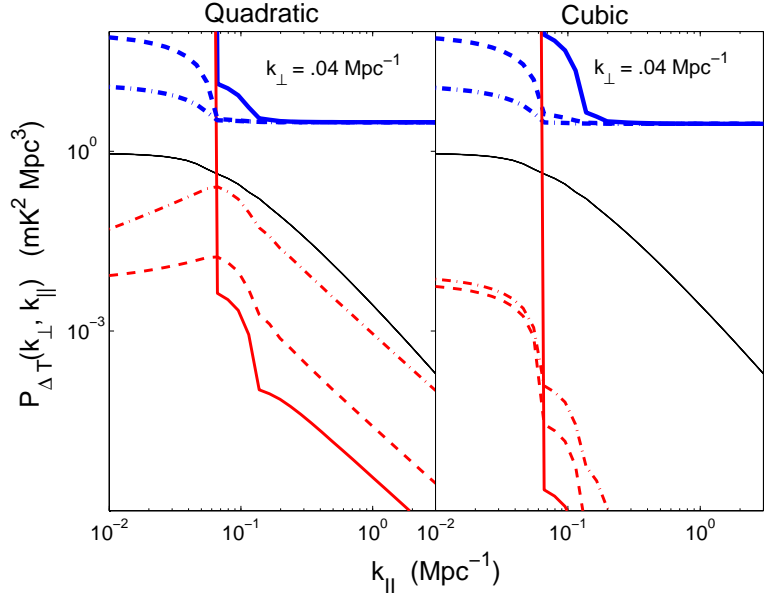,width=4.0in}}
\caption{Foreground removal for a 1000 hour observation with the MWA over a bandwidth $B=6 \MHz$ at $z=8$, using a quadratic (left panel) or cubic (right panel) Legendre polynomial. The dark thin solid curve is the 21 cm signal for a neutral universe. The thick top curves show the sensitivity for a single Fourier pixel, while the medium width bottom curves are the foreground residual power spectra (eq.~\ref{eqn:Q_residual}). In each set, the solid, dashed, and dot-dashed curves assume that 6, 12, and 24 MHz spectral chunks are used to fit the foregrounds.  Note how foreground cleaning reduces sensitivity on large scales, especially for a high-order polynomial fit to a small spectral length.  On the other hand, the same procedure also more effectively removes foregrounds. From \cite{mcquinn05-param}.}
\label{fig:mcquinn_fg}
\end{figure}

Fortunately, the effects of a particular basis choice on a particular power spectrum can be calculated a priori (thus quantifying the amount of downward bias in the recovered large scale 21 cm power spectrum). Figure~\ref{fig:mcquinn_fg} (from \cite{mcquinn05-param}) shows an example. As mentioned above, foreground cleaning reduces the sensitivity to the signal at large scales, and increasing $n$ removes more of both the cosmological signal and the foregrounds. The scales over which foreground cleaning is performed also plays an important role: over a smaller bandwidth, more of the small-scale cosmological signal is removed (conversely, over a larger bandwidth, a higher order polynomial is required to fit the foregrounds effectively). A simple rule of thumb is that, for a given bandwidth, one should choose the minimum $n$ such that $Q_{{\rm k}_{\perp}}(k,n) \ll  P_{\Delta T}(k_{\perp},\mu)$ \cite{mcquinn05-param}.
 
Of course, each stage of foreground removal (including point source removal and spectral fitting) is imperfect.  Thus the final stage, performed simultaneously with the signal extraction, is to fit for these residuals statistically using prior knowledge of their systematic shapes \cite{morales05_foregrounds}.  The details of this phase will depend sensitively on the instrument and algorithms chosen, so its effectiveness has not yet been explored quantitatively.
 
\begin{figure} %%%[!t]
\vspace{+0.2cm}
\centering
\subfigure[Input box] % caption for subfigure a
{
    \label{fig:sub:a}
    \includegraphics[width=6cm]{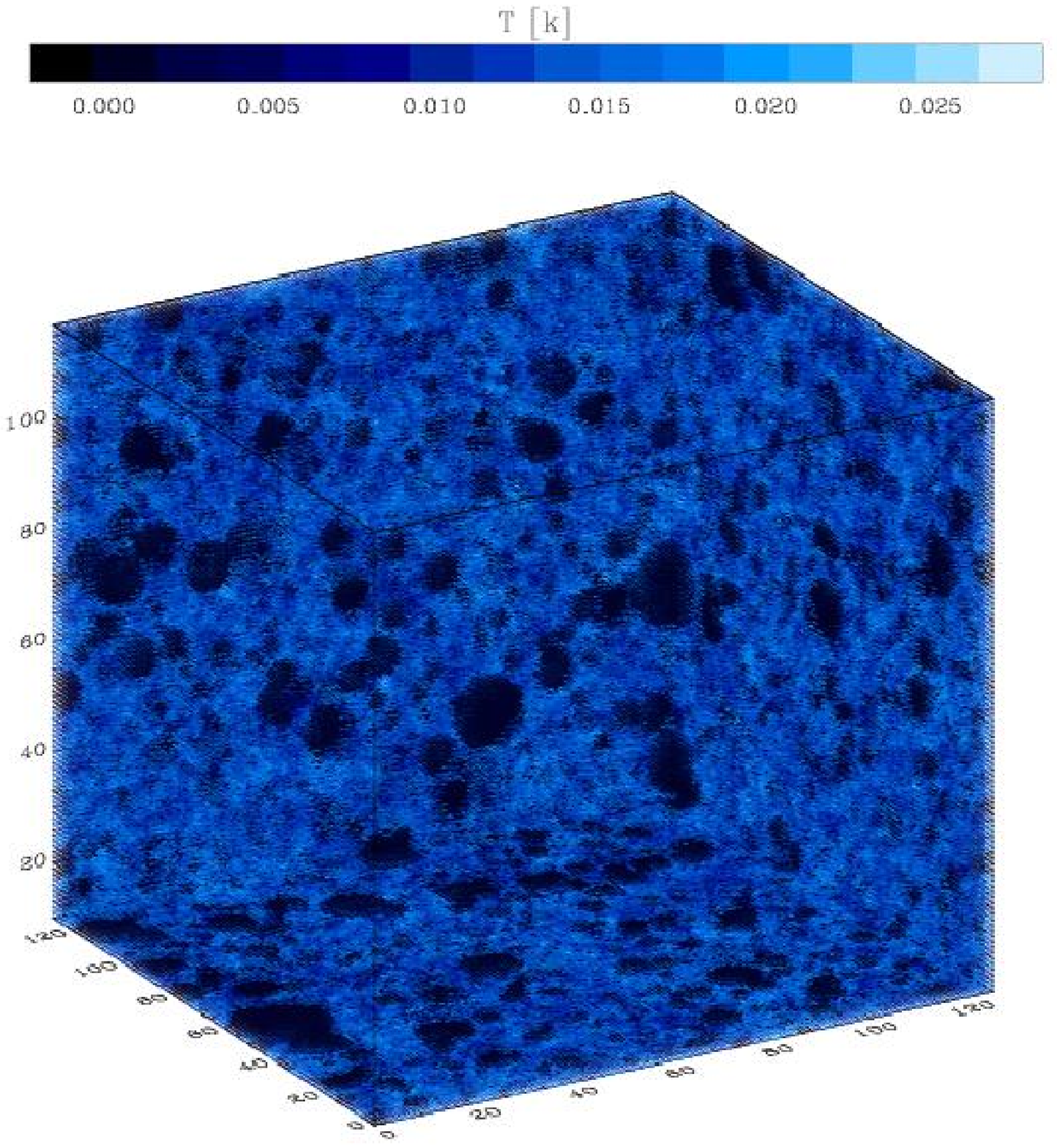}
}
\hspace{0.6cm}
\subfigure[Recovered box] % caption for subfigure b
{
    \label{fig:sub:b}
    \includegraphics[width=6cm]{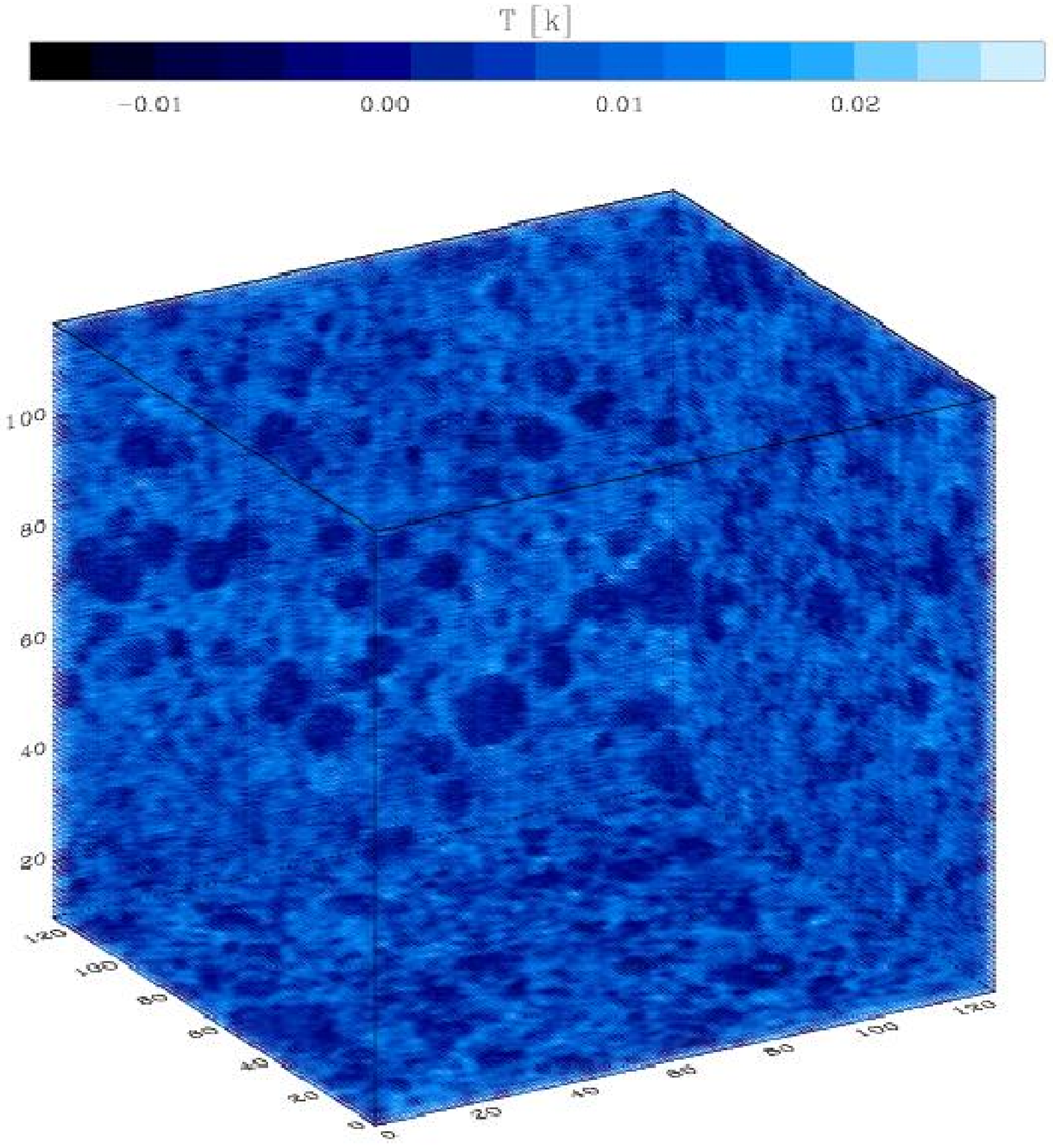}
}
\caption{A simulation of foreground removal for the SKA, when telescope noise is low and 21 cm tomography might be possible. The main features of the box, especially the largest ionized bubbles (shown in black), are robustly recovered. This simulation utilizes a fifth-order Chebyshev polynomial basis for foreground removal, as well as the additional step described in the text of zeroing the large bubbles to calibrate the foregrounds. The box is $\sim 400$ comoving Mpc across at $z\sim 9$ and contains $128^{3}$ pixels. From \cite{chang06}.}
\label{fig:sub} % caption for the whole figure
\end{figure}

The requirement in these foreground removal algorithms that the signal and foregrounds fluctuate on highly disparate scales appears to impose a fundamental limit on our ability to probe the large-scale 21 cm signal. However, at least in the high S/N regime where individual HII bubbles can be imaged (\S\ref{image}), this limitation can be broken: since HII regions should contain \emph{pure} foreground emission, they provide firm calibration points that can be renormalized to $\delta T_{b}=0$ K in the foreground-cleaned maps.  Calibration using Monte Carlo simulations of an analytic bubble size distribution (from \cite{furl04-bub}) show that, without such a correction, any measurement of the pixel PDF or of $\bdtb(z)$, as well as the recovered map itself, will have severe artifacts from large scale unphysical features introduced by foreground removal \cite{chang06}. However, once the recalibration is done, fidelity to the input maps -- as well as detailed statistics, including the PDF and large-scale power spectrum -- is remarkable, except possibly near the box edges. Figure~\ref{fig:sub} shows an example of the reconstruction.  The main limiting factor becomes the number density of sufficiently large bubbles that can be unambiguously identified as such.  Their number density must be of order $n_{\rm bub}^{1/3} \sim k$ to perform reliable foreground recalibration on wavenumber $k$.   

%%%%%%%%%%%% FIGURE 9-10: Peng foregrounds

As with the optimal set of basis functions, many of the details of foreground cleaning -- such as the appropriate space to work in (real, Fourier, or some admixture) -- depend on the details of one's application (e.g., foreground recalibration is only possible in real space), as well as the particulars of array design. They are likely best addressed in end-to-end simulations where full error probability distributions (and not just the Fisher matrix approximation, because systematic errors introduced by cleaning may not be gaussian) and biases (such as the reduction of large scale power) can be carefully quantified.

\subsection{Other Systematics} \label{other-fg}

The Galactic and extragalactic backgrounds (including the polarized component; see \S \ref{crossed-dipole} and \ref{clean}) are only two of many systematic concerns for these experiments.  Below we will briefly describe three additional concerns: foreground lines, terrestrial interference, and ionospheric distortion.

\subsubsection{Radio Recombination Lines} \label{RRL}

Unlike free-free and synchrotron emission, foreground radio recombination lines (RRLs) can introduce significant structure in frequency space. These lines, which are generated by recombination cascades through high-$n$ electronic levels in \htwo regions, could therefore be serious contaminants. Little is known about the RRL background at the low radio frequencies accessed by redshifted 21 cm instruments; indeed, these experiments could turn out to be a major source of new information about both galactic and extragalactic RRLs. Here we briefly summarize our existing knowledge of the RRL background (for a comprehensive review, see \cite{gordon_rrl_book}; see also the discussions in \cite{shaver99,oh03,OF_RRL}).

The most likely source of contamination is our Galaxy. The frequency of a hydrogen RRL between levels $n$ and $n-\Delta n$ is
\begin{equation}
\nu \approx 153 \Delta n \left( \frac{n}{350} \right)^{-3} {\rm MHz}. 
\end{equation}
Observationally, the lines tend to occur every 1--2 MHz over the frequency range of interest. Galactic ridge RRLs make the transition from emission to absorption in the range 100--200 MHz and so are at a minimum at the relevant frequencies. They can reach peak brightness temperatures of $T_L=1 \kel$, but their narrow line widths ($\Delta \nu \sim 3 \kHz$ at 100 MHz) imply that their spectral occupancy is quite small.  The first line of defense is therefore to simply excise contaminated regions of the spectrum (which can be identified a priori); all of the experiments must already work at high spectral resolution to excise terrestrial interference (see below).

The degree to which this will be necessary depends on whether RRLs will provide a significant source of contamination along lines of sight outside the Galactic plane (see Fig.~\ref{fig:radiosky}). Unfortunately, there have been only a few RRL line searches outside the Galactic plane (except toward known bright sources), and in any case RRL surveys generally have low-frequency detection limits well above the strength of the 21 cm features we seek. We can, however, place a useful limit using the fact that the observed brightness of both RRL and H$\alpha$ emission depends on the emission measure $EM=\int \deriv s n_{e}^{2}$, where $n_e$ is the local electron density and $s$ is the path length along the line of sight, and that optical H$\alpha$ surveys have much lower limiting sensitivities. Fabry-Perot surveys have detected H$\alpha$ emission from every Galactic latitude, with minimal values $0.25$--$0.8$ Rayleighs\footnote{Note $1 \, {\rm Rayleigh} = 10^6/4\pi$ photons cm$^{-2}$ s$^{-1}$ sr$^{-1}$.} toward the Galactic poles \cite{reynolds90}. The emission measure corresponding to an observed H$\alpha$ intensity $I_{\alpha}$ (in Rayleighs) is: $EM({\rm H} \alpha)=2.75 \, T_{4}^{0.9} I_{\alpha} \, {\rm cm^{-6} \, pc}$, where $T_4=T_e/(10^4 \kel)$ and $T_e$ is the temperature of the emitting region. Assuming local thermodynamic equilibrium (LTE) and that $T_L \approx \tau_{\rm RRL} T_e$ (see \S \ref{basic}), we find \cite{oh03}: 
\begin{equation} 
T_{L} = 1.3 \left( \frac{T_{e}}{8000 \kel} \right)^{-3/2}
\left( \frac{EM}{0.5 \,  {\rm cm^{-6} \, pc}} \right) \left( \frac{\Delta
  \nu}{1 \MHz} \right)^{-1} \microkel.  
\end{equation}
Thus, unless stimulated emission becomes important at low frequencies, the Galactic hydrogen RRL background should be negligible away from the Galactic plane. 

Carbon radio recombination lines are another possible source of contamination. They have been detected at $\nu=34$--$325 \MHz$ toward Cass A in the Galactic plane \cite{payne89} and also in a 327 MHz survey centered at Galactic longitude $l=14^{\circ}$ \cite{roshi02}.  Optical depths are generally $\tau \sim 10^{-3}$, and the lines are thought to arise in dense, cold ($T_{e} \sim 20$--$200 \kel$), partially ionized regions where non-LTE effects are important. Carbon RRL emission is generally confined to latitudes $b < 3^{\circ}$, and its intensity out of the plane is probably small \cite{roshi02}.   

Because Galactic lines can probably be excised anyway, a much more worrisome contaminant could be emission from unresolved extragalactic sources: since these would be randomly distributed in redshift and hence frequency, the line structure would be much harder to remove. There are no observations of extragalactic RRLs at these low frequencies and hence any estimate has considerable uncertainty, but at present such contamination is not expected to pose a significant threat \cite{OF_RRL}. Neglecting stimulated emission, a star forming galaxy producing a brightness temperature perturbation $T_L \sim 0.01\kel$ would have to lie within a distance
\begin{equation}
D < 11 \, {\rm Mpc} \left( \frac{T_L}{0.01 \kel} \right)^{-1/2} \left( \frac{\rm SFR}{100 \, {\rm M_{\odot} \, yr^{-1}}} \right)^{1/2} \left( \frac{\Delta \nu}{1 \, {\rm MHz}} \right)^{-1/2} \left( \frac{T_{e}}{10^{4} \kel} \right)^{-3/4} \left( \frac{\theta}{5^{\prime}} \right)^{-1},
\end{equation}
where $\theta$ is the angular resolution of the beam.  Such bright nearby sources can easily be identified and excised. The integrated RRL background from unresolved sources is similarly small \cite{OF_RRL}.  However, one large uncertainty is the role that non-LTE effects and stimulated emission could play in boosting the RRL flux; these may be important at low frequencies because $\tau_{\rm RRL} \propto \nu^{-1}$. It was previously thought that emission stimulated by the continua of radio galaxies and quasars could allow much more distant sources to be seen \cite{shaver78}. But existing observations at higher frequencies (e.g., at 1.4, 8.1, 84, 96, and 207 GHz \cite{ana00}) indicate that emission stimulated by luminous sources outside \htwo regions is unimportant, probably because the volume filling factor of HII regions around radio quasars is small \cite{ana93} (stimulated emission from internal sources may be important, but it does not significantly boost the flux). Furthermore, the observed line flux falls off unexpectedly rapidly towards lower frequencies. Deviations from these expectations would be interesting in their own right, useful in probing the ISM of galaxies. 
 
Extragalactic RRLs could also be a foreground in the search for 21 cm absorption against high-redshift radio-loud sources (see \S \ref{forest}) \cite{oh03}. Since high-redshift objects are potentially much denser, $n_{e}^{2} \propto (1+z)^{6}$, their emission measures could be much higher. Thus, for instance, a high-redshift disk galaxy could have $\tau_{\rm RRL} \sim 10^{-2}$, comparable to that of the IGM or a minihalo in 21 cm absorption. If such disks are abundant, this could make the task of picking out the 21 cm forest lines much tougher, though the RRL features themselves would be a marvelous probe of gas clumping. However, the existence of such RRL absorbers is extremely speculative, and only a small fraction of lines of sight are expected to intersect disks \cite{furl02-21 cm}.

\subsubsection{Terrestrial Radio Frequency Interference} \label{RFI}

Modern society's needs for high speed communication and precise navigation rely on intensive use of the radio spectrum. Strong signals from these services are present in any location in which people live in significant numbers and form one of the strongest ``foregrounds''  affecting radio astronomy.
In addition to the strong signals broadcast within specifically allocated frequency bands (through national and international coordination of the radio spectrum), there is a forest of spurious emission permeating areas of high population density, because most electrical equipment emits low levels of radio emission that falls below legal thresholds set by governmental agencies but whose cumulative effect  is substantial given the high sensitivity of radio telescopes.

Radio astronomy's first line of defense against this radio frequency interference (RFI) is to situate telescopes in remote locations, far from dense human habitation.  Unfortunately, as the world's population and the demand for broadband communication services increase, it is becoming difficult to escape RFI. For example, satellites used to obtain global access to instantaneous communications cause some radio bands to be occupied at every place on Earth at all times.  

Through the national and international agencies that coordinate spectrum use, radio astronomers have successfully negotiated legislation that provides for a number of narrow ``protected'' bands centered on selected bands of astrophysical interest, such as the rest frequency of the 21 cm line. Of course, it is impossible to reserve the sort of large bandwidths necessary for fundamentally broadband scientific applications such as ours.

Preparation for new observatories, such as the SKA, includes intensive political discussion  to define ``radio quiet zones'' in areas of low population density that are unlikely to see future growth, in order to provide some local protection for the considerable investment needed to build and operate the telescopes. These zones will enforce limits on nearby transmissions, but, due to the high sensitivity of radio telescopes, they are still vulnerable to distant high-power broadcasts and satellites.  In fact, distant TV broadcasts and other ambient RFI problems have already ruined an early attempt to use the VLA to observe the putative \htwo regions around the SDSS quasars (L. Greenhill, private communication).

%%%%%%%%%%%% FIGURE 9-11:  Radio Frequency Spectra - comparison
\begin{figure}[!t]
\centerline{\epsfig{file=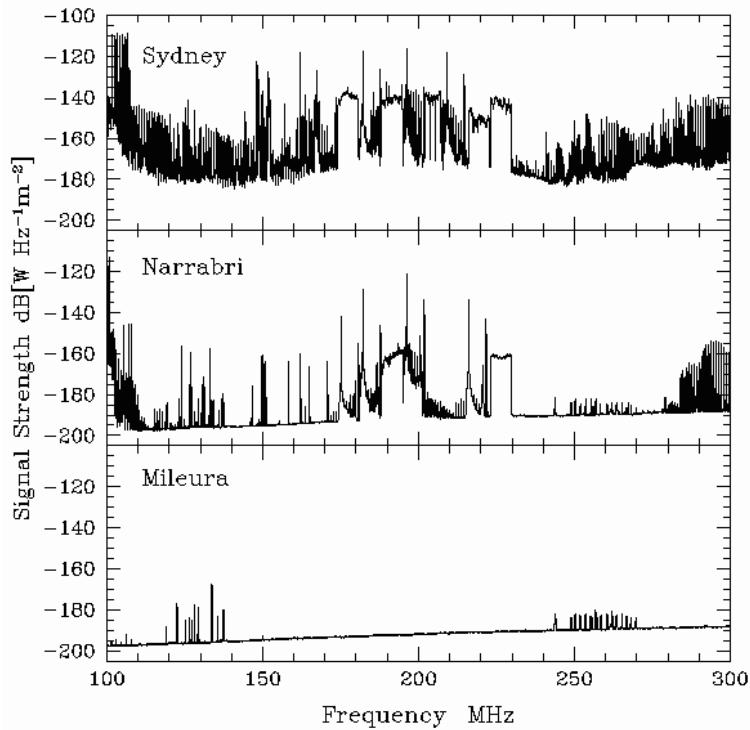,width=4.0in}}
\caption{Radio environment at three Australian locations: a suburb of Sydney, NSW; the Australia Telescope Compact Array site near Narrabri, NSW; and the Mileura, WA site (proposed by Australia for the SKA). The FM radio band (87.5 to 108 MHz throughout most of the world) enters the spectrum at left. The strong features in the center of the plot (174 to 230 MHz) are Australian high-VHF television channels 6, 7, 8, 9, 9A, 10, 11 and 12.  Shown courtesy of  A. Chippendale and R. Beresford (taken as part of the ATNF SKA Site Monitoring Program).}
\label{fig:rfispectra}
\end{figure}

Figure~\ref{fig:rfispectra} compares the radio environment at three sites in Australia  with a range of population densities (Sydney with population $\sim$4 million, Narrabri with $\sim 4,\,000$ and Mileura Station with $\sim$4). The plot spans 100--300~MHz, corresponding to $z =13.2$--$3.7$ for the 21 cm line. The tail end of reionization (at $z \sim 6.2$) falls at 197~MHz.  This is within the high-VHF television band, which spans 174--220~MHz throughout most of the world. The band at 87.5 to 108 MHz finds heavy use around the globe for FM radio.  

The frequency range from 120 to 140~MHz includes aircraft communications and a low earth orbit satellite band at 137~MHz; these are hard to avoid since air traffic and satellite transmitters are likely to pass within the reception patterns of nearly all Earth-based telescopes from time to time. Fortunately, these signals have extremely narrow bandwidths, and, in the case of aircraft communications, have only a tiny temporal duty factor. This means that they can be efficiently edited (or ``blanked") from the data stream.

Not surprisingly, Figure~\ref{fig:rfispectra}  shows  that the radio spectrum in densely populated places is saturated, and nowhere in the Sydney spectrum does the ambient power level drop down to the baseline set by the instrumental sensitivity.  Fortunately, remote locations such as Mileura are able to
escape (for the most part) the radio and TV contamination of modern civilization. However, it must be noted that the sensitivity of the measurements shown in the figure was limited by the equipment used in this monitoring exercise; here -200~db (W~m$^{-2}$~Hz$^{-1}$) corresponds to radio astronomical flux densities of  10$^6$~Jy, while observations of reionization will need to be sensitive to $\sim 10 \, \mu$Jy signals. Fortunately, more sensitive measurements \cite{barnes06} at the Mileura site show that few signals appear in the blank areas of the spectrum in  Figure~\ref{fig:rfispectra}, even after long integrations sensitive to a small fraction of the sky brightness.

Technological advances in high speed digital electronics are now permitting the development of a range of RFI ``mitigation'' techniques whose goal is to permit observations in bands contaminated with
human-generated signals.  This is necessary in studies of spectral lines, such as the  redshifted 21 cm line, that require measurements in frequency bands already in use.  The algorithms and hardware now being developed are the subject of several recent reviews \cite{boonstra05,ellingson05,fridman01}.

RFI mitigation methods include a number of different approaches.  Some focus on efficient automated editing and blanking of signals that have rapid or distinctive temporal variability or that are confined to narrow frequency ranges. Since 21 cm emission from $z>6$ is relatively smooth on scales $\la 1 \MHz$,  narrow signals with widths $\la 50$ kHz can be discarded and the broader band of interest constructed from the clean bands between the RFI spikes. 

Other techniques are being developed to identify, characterize, and ultimately subtract (or cancel) interfering signals \cite{boonstra05,ellingson05,fridman01}, permitting astronomical observations to be performed in actively used communication bands. So far, these experimental methods work best for broadcast transmitters that are fixed to the Earth, and there are limits to the achievable precision of the cancellation. Thus the first line of defense remains locating radio telescopes in the most benign environments available.

Telescope design also plays a role in mitigating interfering signals. For example, the antenna tiles
illustrated in Figure~\ref{fig:tile_response} project nulls in the their response patterns toward the horizon.
This discriminates against terrestrial transmitters. Furthermore, tiles on the ground gain additional protection from hills and foliage.  But, like all radio telescopes, they remain vulnerable to satellite emission.

\subsubsection{Ionospheric Distortions} \label{distort}

Because of its temporally and spatially variable index of refraction, the Earth's ionosphere distorts  low frequency radio signals as they propagate through it.  This creates significant calibration and imaging problems that must be solved in order to clean the strong foreground contamination reliably.

\begin{figure}[!t]
\centerline{\epsfig{file=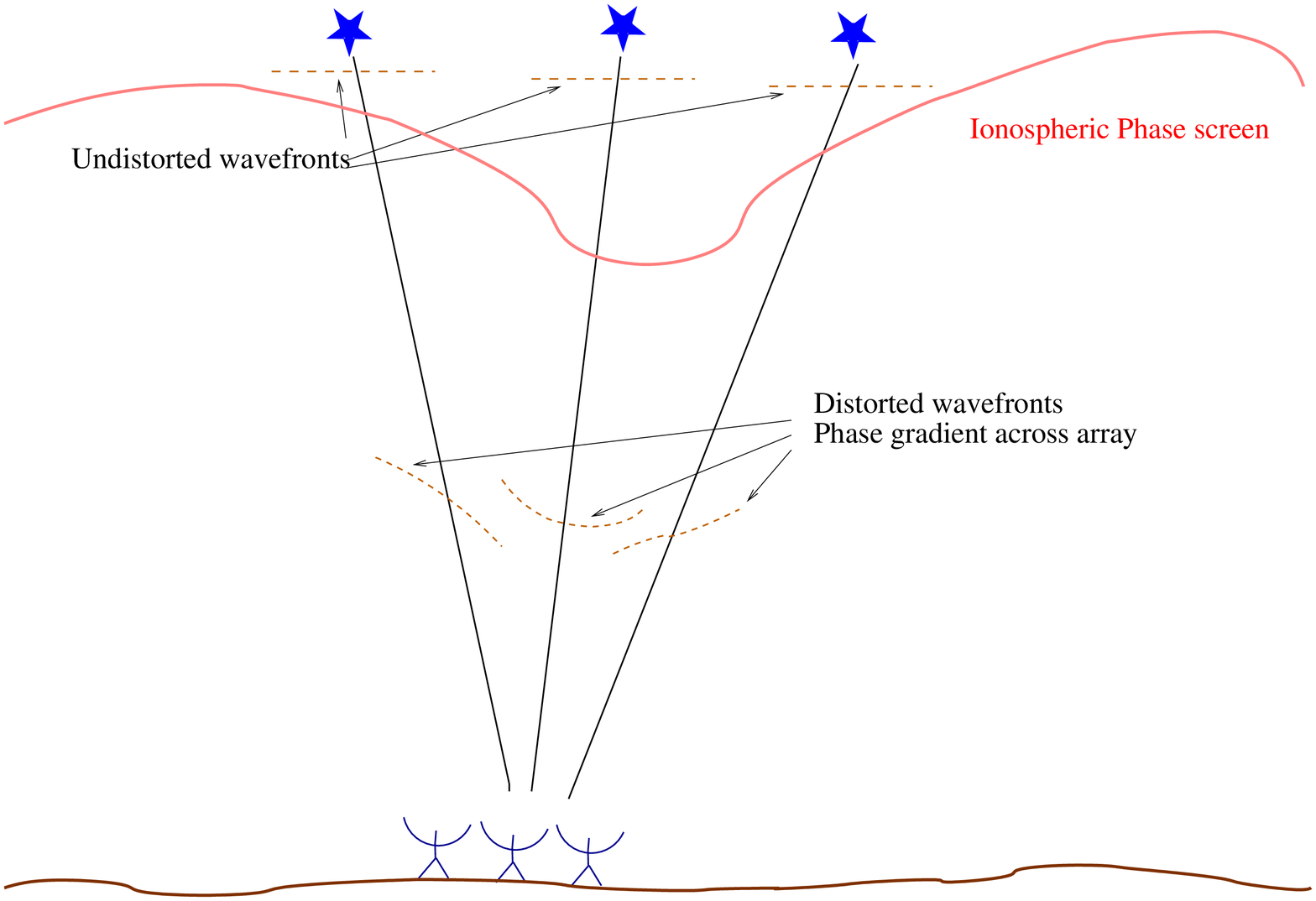,width=3.0in,angle=0}} % Used for Frank's original version
%\centerline{\epsfig{file=f9-1.eps,width=5.0in,angle=0}} % Used for compressed version
\caption{Cartoon illustration of ionospheric distortions. Wavefronts (represented by dashed lines) are
distorted as they pass through the phase screen of the ionosphere (solid wavy line) and develop nearly
linear phase gradients, which then project onto the array. The gradient introduced by the screen varies
with position across the sky.  From \cite{Cotton04}.}
\label{fig:ionosphere}
\end{figure}

Figure~\ref{fig:ionosphere} is a cartoon sketch of the wavefront distortion experienced by a compact array \cite{Cotton04}.  Lines of sight from all the antennae in the array observe a given celestial radio source through nearly the same ``refractive wedge," which to first order causes a linear gradient to develop in the phase of a wavefront as it passes through the ionosphere. This causes the array to perceive the wave to originate from a different direction on the sky than the true one.  Wavefronts arriving from sources at different locations will pass through independent wedges, so their apparent positions will be displaced by different amounts. The net effect is that the radio sources will remain coherent to the array, but they will appear to wobble on time scales of tens of minutes (the characteristic scale of fluctuations in the ionosphere).  Calibration will require the inclusion of algorithms to track and update an ionospheric  distortion model using an all-sky grid of bright radio sources.  The problem and its solutions are qualitatively similar to adaptive optics in optical astronomy, although here all the corrections will be done on the software (rather than hardware) level and the spatial and temporal scales are significantly larger.  Foreground subtraction and imaging will  rely on the corrected catalog of source positions.

In addition to the variable index of refraction, the ionosphere also induces variable Faraday rotation, which rotates the position angle of the electric polarization vector so that polarized foregrounds artificially appear to vary in time at the telescope.  This too must be evaluated and calibrated on time scales of minutes as part of the observational procedures.

Considerable experience with ionospheric distortion has been obtained during the VLA Low Frequency Sky Survey at 74~MHz \cite{Cohen06}. Among the principal findings has been the significant increase in complexity for interferometer arrays extending over more than a few tens of kilometers, because then the array samples many independent refractive wedges in the ionosphere.  This further complicates both calibration and imaging.  Fortunately, the 21 cm line observations that concern this review need only short baselines of a few kilometers in length to be sensitive to the low surface brightness features with characteristic scales of arcminutes expected during reionization.

%\bibliographystyle{elsart-num}
%\bibliography{Ref_21cm}

%\end{document}

%% file: forest-ch10.tex
%\documentclass{elsart}
%\usepackage{amssymb,cite,epsfig}

%\input{../../submission/defns.tex}

%\begin{document}

\section{The 21 cm Forest} \label{forest}

To this point we have focused on constructing large-scale three-dimensional 21 cm maps of the IGM (or at least characterizing their statistical properties).  Such observations promise to provide powerful constraints on reionization and early structure formation, but they pose three major difficulties:  (1) they require $T_S \neq T_\gamma$ and may therefore be impossible during certain epochs; (2) they cannot realistically resolve structures smaller than $\sim 1 \Mpc$; and (3) they present a number of observational challenges (see \S \ref{int}).

\begin{figure}[!t]
\centerline{\epsfig{file=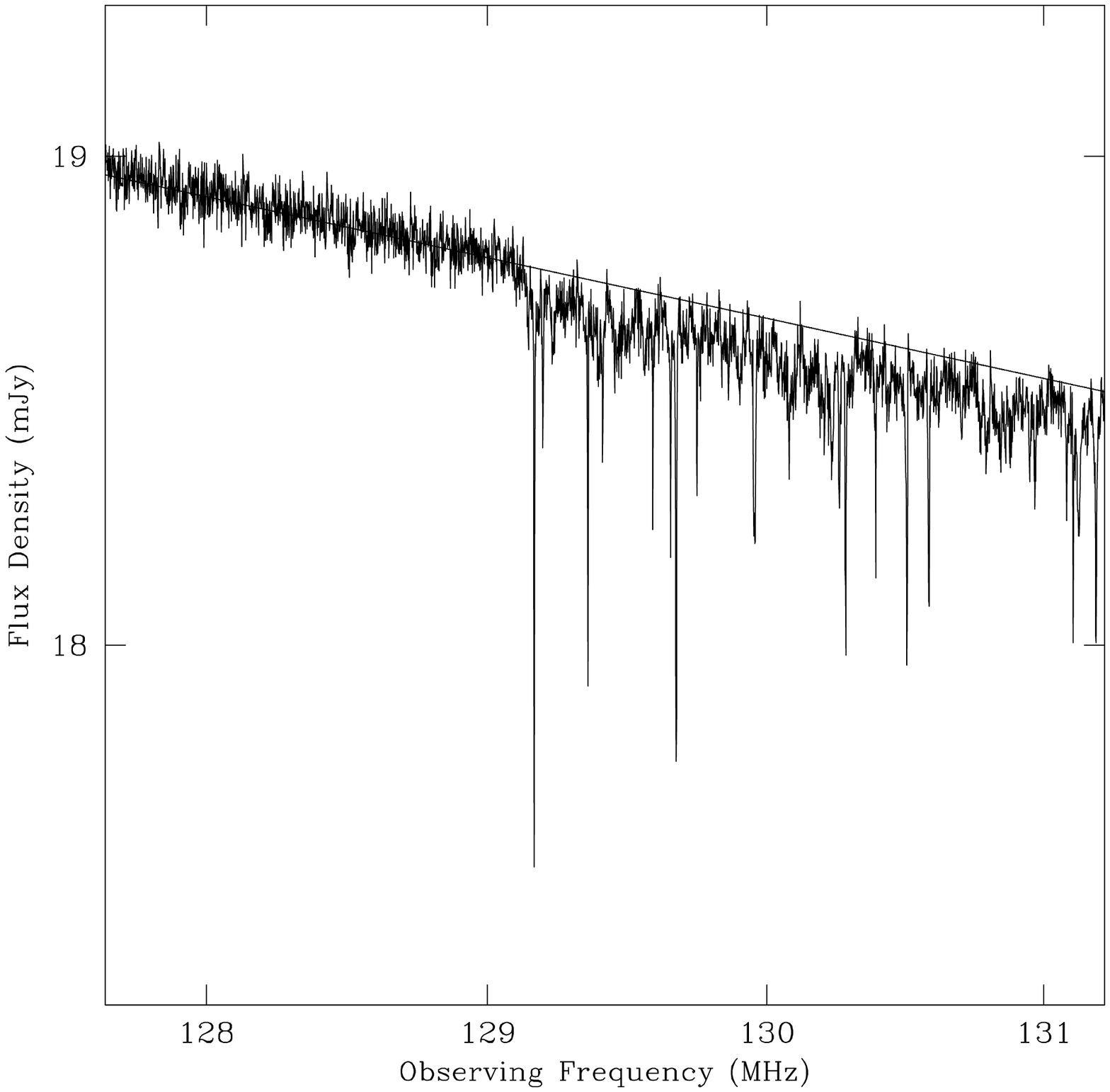,width=2.75in}
\epsfig{file=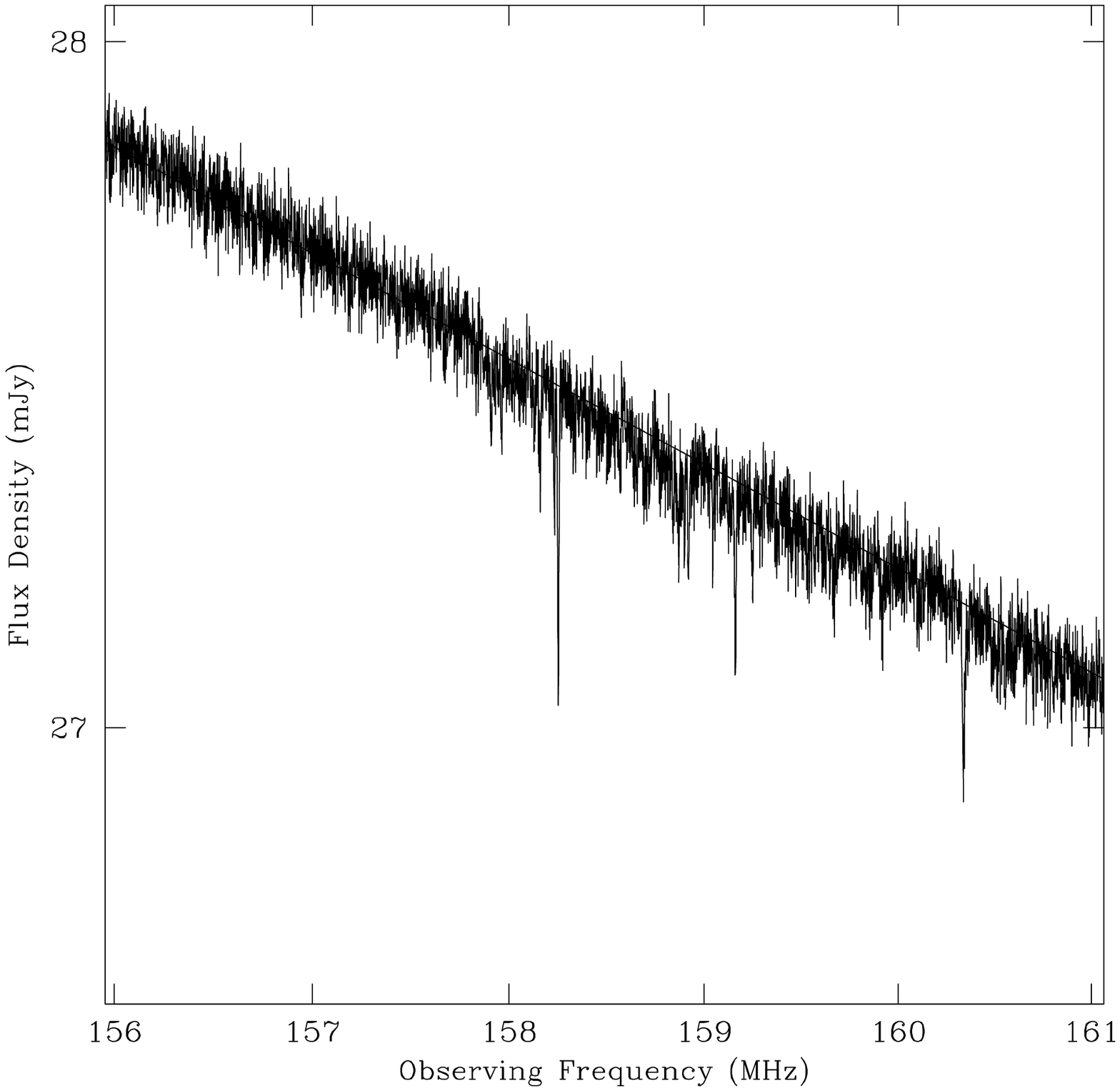,width=2.75in}}
\caption{Mock spectra of high-redshift radio sources at $z_s=10$ (left panel) and $z_s=8$ (right panel) produced from cosmological simulations.  In each case a source (with intrinsic luminosity equal to Cygnus A) is placed at the appropriate redshift and ``observed" in a week-long integration with an SKA-class instrument.  A ``forest" of absorption features appears blueward of $21(1+z_s)$ cm, caused by the cosmic web.  Note that the level of absorption depends sensitively on the assumed thermal and ionization history of the IGM.  From \cite{carilli02}.}
\label{fig:forest-sim}
\end{figure}

A complementary probe immune to most of these problems is the ``21 cm forest" \cite{carilli02, furl02-21cm, furl06-forest}.  The technique, illustrated in Figure~\ref{fig:forest-sim}, is the exact analog of the \lya forest that has proved so useful for studying the $z \la 6$ IGM.  Neutral hydrogen along the line of sight to a distant radio source will resonantly absorb photons that redshift into the 21 cm transition, creating a forest of absorption features in the spectrum due to the diffuse IGM, sheets and filaments in the cosmic web, minihalos, and \htwo regions.  The major advantage of this method is that, with a sufficiently bright source, arbitrarily high frequency -- and hence spatial -- resolution can be achieved, allowing us to probe individual filaments and minihalos.  Moreover, because it requires only a high signal-to-noise spectrum of a relatively bright radio source, the 21 cm forest is a much simpler observation than tomography.

%%%%%%%%%%%% FIGURE 10-1: Simulation figure

\subsection{Spectral Features} \label{forest-test}

The fundamental quantity of interest for the 21 cm forest is the optical depth, given by equation (\ref{eq:optdepth}) for an isolated cloud or equation (\ref{eq:optdepthcosmo}) for the IGM.  At high redshifts, the latter becomes
\begin{equation}
\tau \approx  0.011\, \xhi \, (1+\delta) \, \left( \frac{1+z}{10} \right)^{3/2} \, \left[ \frac{T_\gamma(z)}{T_S} \right] \, \left[ \frac{H(z)/(1+z)}{\deriv v_\parallel/\deriv r_\parallel} \right].
\label{eq:tauforest}
\end{equation}
As usual, the observable quantity is the brightness temperature decrement $\dtb$.  However, in this case the background radiation field comes from a source with brightness temperature $T_b \gg T_S$; thus $\dtb \propto T_S^{-1}$.  Unlike tomography, the forest's brightness always depends on $T_S$.

\begin{figure}[!t]
\centerline{\epsfig{file=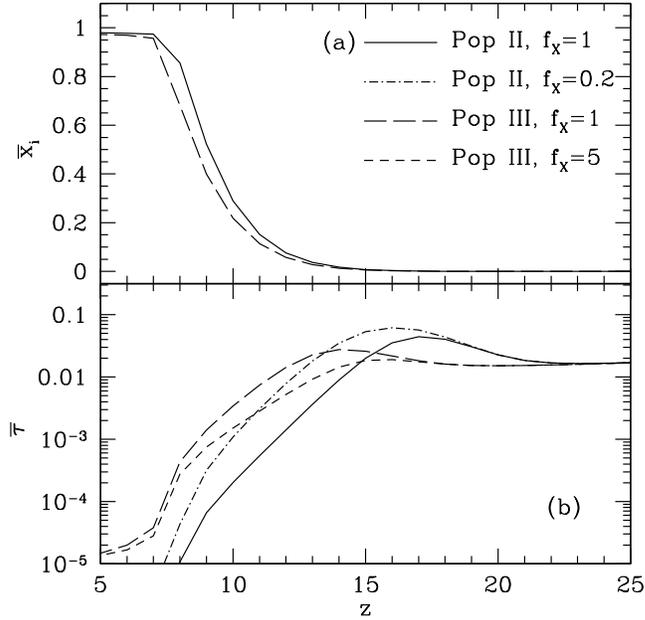,width=3.5in}}
\caption{Mean optical depth in some example reionization histories.  \emph{(a)}: Ionization history, $\bxion(z)$.  Note that $\bxion$ is independent of $f_X$, so we only show two curves here.  \emph{(b)}: Optical depth $\tau$.  The models follow those in Figs.~\ref{fig:pop2-glob} and \ref{fig:pop3-glob}.  From \cite{furl06-forest}. }
\label{fig:taumean}
\end{figure}

We will consider four aspects of the forest.  The first is the mean level of IGM absorption, which is determined by $\bxhi(z)$ and $T_S(z)$.  We have seen in \S \ref{glob} that these quantities depend on a number of unknown parameters in the star formation history, including their spectral energy distribution,  star formation efficiency, escape fraction, and stellar initial mass function.  We can use the same models as in \S \ref{glob} to study the qualitative features of the evolution.  Figure~\ref{fig:taumean} shows the  mean optical depth in several of the models from Figures~\ref{fig:pop2-glob} and \ref{fig:pop3-glob}.

%%%%%%%%%%%% FIGURE 10-2: Mean optical depth

At sufficiently high redshifts, before the Wouthuysen-Field effect has become significant, $T_S=T_\gamma$ and $\bar{\tau} \sim 0.02$.  In most of the models, it then briefly increases  by a factor of a few because \lya coupling becomes efficient while the IGM is still cold.  However, this period is short-lived, because $\bar{\tau}$ declines rapidly once X-ray heating begins.  As we saw in equations~(\ref{eq:xic}) and (\ref{eq:tx-xi}), reionization is delayed compared to \lya coupling and X-ray heating.  As a result, by the time reionization begins in earnest and large \htwo regions appear, $T_S$ is already large and $\bar{\tau} \sim 10^{-3}$.  This is a crucial characteristic of the 21 cm forest:  in the most plausible scenarios, $\bar{\tau}$ is only large long before reionization.  This is, of course, unfortunate, because powerful radio sources probably do not appear until powerful ionizing sources do.  Thus lines of sight with strong absorption are likely to be rare.  Simulations find similar results, even if X-ray heating is neglected (Fig.~\ref{fig:forest-sim} and \cite{carilli02}).  The right-hand panel of Figure~\ref{fig:forest-sim} has $\bar{\tau} \sim 0.1\%$ at $\bxion \sim 0.5$ (or $z=8$ in this particular simulation).  Once X-ray heating is included, the prospects are even worse.  

The next set of features come from sheets and filaments.  These stand out in Figure~\ref{fig:forest-sim} as $\tau \ga 2\%$ absorption spikes.  Naively, one might expect that these features would strengthen with time as the cosmic web condenses out of the IGM.  But Figure~\ref{fig:forest-sim} shows the opposite:  they actually grow weaker, because the increasing spin temperature washes them out.  Assuming the absorbers are near hydrostatic equilibrium (an excellent approximation for the analogous \lya forest \cite{schaye01}), the typical overdensity of an absorber with optical depth $\tau$ is \cite{furl06-forest}
\begin{equation}
1 + \delta \sim 94 \, \left( \frac{\tau}{0.01} \right)^2 \, \left( \frac{T_S}{100 \kel} \right)^2 \, \left( \frac{10}{1+z} \right)^3 \, \left( \frac{\Omega_m h^2}{0.15} \right) \, \left( \frac{\Omega_b h^2}{0.023} \right)^2.
\label{eq:filament-tau}
\end{equation}
Thus, if the IGM has been even moderately heated, only virialized objects host measurable absorption, and the cosmic web spikes become rare.  In the simulation of \cite{carilli02}, $T_S \sim 30 \kel$ throughout much of the IGM at $z=10$, but by $z=8$ it has risen to $T_S > 100 \kel$ everywhere.\footnote{It is worth noting again that this simulation lacked X-ray heating, so it probably underestimates the temperature.}  Over the same interval, the number density of $\tau>0.02$ lines drops from $\sim 50$ per unit redshift at $z=10$ to $\sim 4$ at $z=8$.  Again we see that strong cosmic web absorption probably requires finding bright radio sources that shine long before reionization begins.  These features are also narrow, with $\Delta \nu_{\rm obs} \sim 2 \kHz$, determined by thermal and Hubble broadening.

Minihalos also contribute to the forest \cite{furl02-21cm}.  We described 21 cm emission from these objects in \S \ref{mh}; while it can be reasonably strong, it is nearly impossible to distinguish from that of the diffuse IGM without high sensitivity on small scales.  But the 21 cm forest suffers no such confusion, because it allows us to resolve individual objects; indeed, it is the \emph{only} known method to study these objects in detail.  The excess optical depth through a minihalo depends in detail on its gas density profile and temperature distribution \cite{furl02-21cm}.  But we can estimate the typical central optical depths by assuming $N_{\rm HI} \approx n_{\rm HI} r_{\rm vir} \approx \Delta_{\rm vir} \bar{n}_H r_{\rm vir}$ (where $\Delta_{\rm vir} \approx 18 \pi^2$ is the mean overdensity of virialized objects; see \cite{bryan98} for a numerical fit) and that $T_S \approx T_K$ (because the gas is so dense).  Then, using equation (\ref{eq:optdepthcloud}), we have (at line center)\footnote{The total equivalent width of a minihalo line is proportional to $m^{-1/3}$.}
\begin{equation}
\tau_{\rm mh} \approx 0.026 \left( \frac{1+z}{10} \right)^{1/2} \, \left( \frac{m}{10^7 \Msun} \right)^{-2/3}.
\label{eq:mh-tau}
\end{equation}
Thus the optical depths can even exceed those of sheets and filaments, and they are much more robust to uncertainties about the \lya background and the IGM temperature. The typical line widths (due to thermal broadening) are $\Delta \nu_{\rm obs} \sim 2 \kHz$, comparable to those of sheets and filaments.

\begin{figure}[!t]
\centerline{\epsfig{file=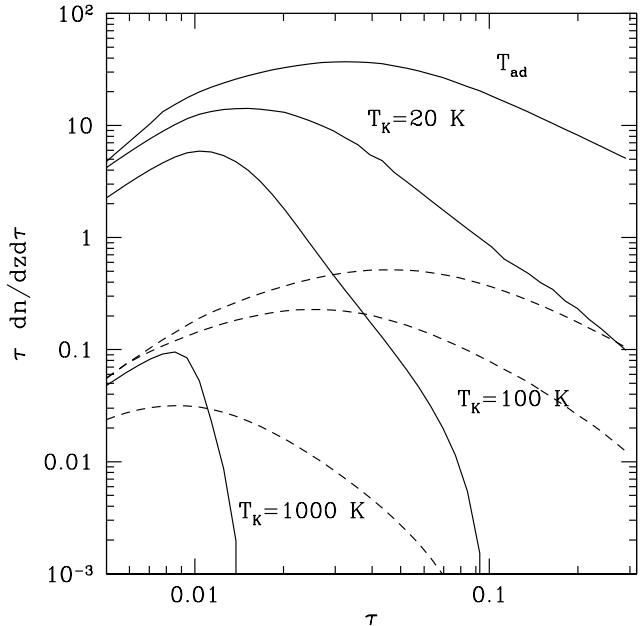,width=3.5in}}
\caption{Differential number density of minihalo absorption features at $z=10$ (solid curves) and $z=20$ (dashed curves).  From top to bottom, each set takes the minimum minihalo mass to be that of a uniform medium at $T=T_{\rm ad}(z),\,20,\, 100$ and $1000 \kel$, from top to bottom.  From \cite{furl06-forest}.}
\label{fig:forest-mh}
\end{figure}

Figure~\ref{fig:forest-mh} shows the resulting differential number densities of minihalo absorbers at $z=10$ and $20$ (solid and dashed lines, respectively).  At each redshift, the top curves assume $T_K=T_{\rm ad}$ (the temperature without X-ray heating; shown in Fig.~\ref{fig:tevol}).  The remaining curves use $T_K=20,\,100,$ and $1000 \kel$, from top to bottom, to compute the Jeans mass.  In a cold IGM, the number of minihalo features is large (with $\ga 100$ absorbers per unit redshift at $z=10$) and they can be rather strong (with $\ga 20$ absorbers with $\tau>0.1$ per unit redshift). Minihalos are of course much rarer at $z=20$, simply because structure formation is so much less advanced at that time.

%%%%%%%%%%%% FIGURE 10-3: Minihalos

Clearly the number of absorbers decreases rapidly as $T_K$ increases, especially for large optical depths.  This is because the smallest minihalos produce the strongest features; increasing the Jeans mass by even a small amount suppresses them.  The weaker absorption features with $\tau \sim 0.01$ are more robust to $T_K$ because they come from larger halos.  Comparison with simulations shows that minihalo features have comparable abundance to those of the cosmic web \cite{carilli02,furl02-21cm}, although minihalos do probably persist longer during reionization once X-ray heating is included \cite{furl06-forest}.

Measuring the frequency of minihalo features is thus a sensitive probe of the thermal history of the IGM.  More sophisticated models of minihalo formation in the presence of thermal feedback show that the lack of a minihalo forest, in combination with large-scale 21 cm emission, could identify warm regions in which an \htwo bubble had recombined \cite{oh03-entropy}.  Such regions would be invisible in most other ways.

\begin{figure}[!t]
\centerline{\epsfig{file=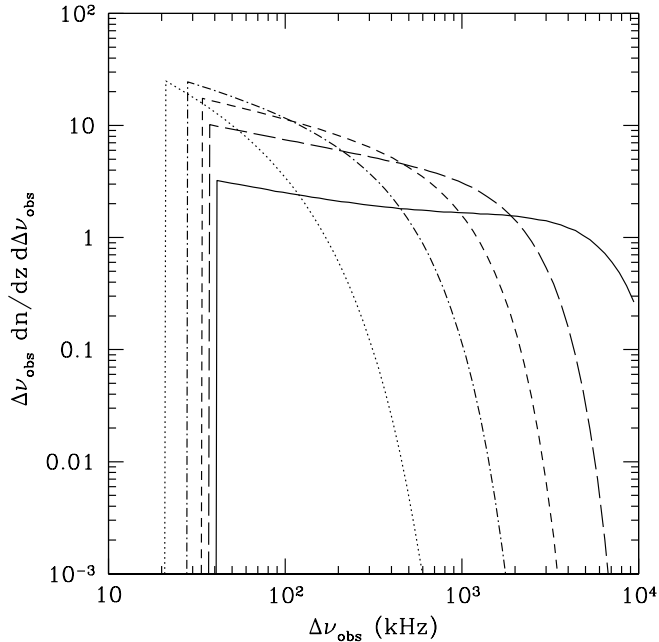,width=3.5in}}
\caption{Size distribution of \htwo region forest gaps intersected per unit redshift at $z=10$.  The dotted, dot-dashed, short-dashed, long-dashed, and solid lines take $\bxion=0.1,\,0.3,\,0.5,\,0.7,$ and $0.9$, respectively.  From \cite{furl06-forest}.}
\label{fig:forest-htwo}
\end{figure}

Ionized bubbles create gaps in the absorption, with an amplitude equal to the mean optical depth of the neutral gas.  Figure~\ref{fig:forest-htwo} shows the resulting distribution of absorption features (transformed into differential form) for $\bxion=0.1,\,0.3,\,0.5,\,0.7,$ and $0.9$ at $z=10$.  It has a number of interesting features.  First, except at $\bxion=0.1$, the distribution is flat with $\Delta \nu_{\rm obs}$.  This is because $n_b(m)$ has a well-defined characteristic size in this model.  Crucially, because the bubbles are relatively large, the spectral features are quite wide -- more than an order of magnitude larger than those of filaments or minihalos, even when $\bxion=0.1$.  Thus they may indeed be the simplest features to identify, despite their relatively small contrast (see Fig.~\ref{fig:taumean}).  Second, the total number of \htwo regions per unit redshift remains roughly constant throughout reionization, from $\sim 20$ at $\bxion=0.1$ to $\sim 6$ at $\bxion=0.9$.  This is because most ionizations occur as existing bubbles grow larger rather than by creating new bubbles \cite{furl05-rec}.  However, the overall distribution of $\Delta \nu_{\rm obs}$ does evolve rapidly as $\bxion$ increases, so the forest will certainly prove useful in constraining the reionization history.  

%%%%%%%%%%%% FIGURE 10-4: Ionized regions

Finally, for completeness, dense neutral gas inside galaxies (similar to DLAs at $z \la 4$ \cite{briggs02}) can induce rather large optical depths.  But such lines of sight are rare ($\deriv n/\deriv z \sim 0.1$ \cite{furl02-21cm}), so collecting a large sample will be difficult.

\subsection{Is It Feasible?} \label{forest-feas}

As we mentioned above, 21 cm forest observations are much simpler than tomography, because they can essentially ignore most of the challenges discussed in \S \ref{int}.  Their feasibility boils down to two questions:  raw sensitivity and the existence of background sources.  Detecting a feature with signal-to-noise ratio S/N requires a source brightness (assuming simultaneous observations of two orthogonal polarizations)
\begin{equation}
S_{\rm min} = 16 \mJy \, \left( \frac{{\rm S/N}}{5} \, \frac{0.01}{\tau} \, \frac{2500 \ {\rm m^2/K}}{A_{\rm eff}/T_{\rm sys}} \right) \, \left( \frac{1 \kHz}{\Delta \nu_{\rm ch}} \, \frac{1 {\rm \ week}}{t_{\rm int}} \right)^{1/2},
\label{eq:forest-sens}
\end{equation}
where $\Delta \nu_{\rm ch}$ is the bandwidth of each channel (assumed smaller than the feature of interest) and $t_{\rm int}$ is the total ``on-source" integration time.  These telescope parameters are similar to those expected for the SKA at $z=10$, shown in the mock spectra of Figure~\ref{fig:forest-sim}.  Clearly, observing any of the narrow features described above requires long integrations on SKA-class instruments and rather bright sources; for example, Cygnus A would have $S_\nu \la 20 \mJy$ if it existed at such redshifts.

This requirement can be alleviated in some circumstances.  First, minihalos and filaments introduce an extra fluctuating component to the spectra that could be detected statistically by, for example, computing a running rms noise statistic.  This reduces $S_{\rm min}$ by a factor of at least three \cite{carilli02}. Second, \htwo regions can have significantly larger widths.  Bubbles with $\Delta \nu_{\rm obs} \sim 100 \kHz$ reduce the required optical depth to $\tau \sim 10^{-3}$ at $S_{\rm min}=16 \mJy$, and significantly weaker sources would suffice if $\tau$ is larger.

Unfortunately, these thresholds are still relatively large.  Interestingly, when extrapolated to higher frequencies, the possible background sources are well within reach of existing surveys (such as FIRST \cite{becker95}), assuming that they have spectra typical of radio galaxies.  Thus, appropriate sources may have already been detected.  Of course, none have been identified so far, and that step will pose the real challenge.

The most promising background sources are radio-loud quasars, although their luminosity function is relatively unconstrained at $z \ga 4$ \cite{jarvis01}.  Under the reasonable assumption that their abundance declines along with the bright optical quasars, the sky would contain $\sim 1000$ $(10)$ sources at $z>8$ $(12)$ \cite{carilli02}.  Models of the radio luminosity function over a range of redshifts (matching existing data where available) predict $\sim 2000$ sources across the sky with $S>6 \mJy$ at $8<z<12$ \cite{haiman04} (although this is quite sensitive to the assumed radio-loudness parameter).  According to this model, the FIRST survey should have already found $\ga 10^3$ objects with $z>7$ -- albeit out of a total sample of $\sim 750,000$ sources!  These estimates predict reasonably large number of lines of sight to $z <12$.  Beyond that point, however, we will likely be limited to just a few extreme objects.  This is unfortunate in light of Figs.~\ref{fig:taumean}--\ref{fig:forest-htwo}, which show that the absorption rapidly weakens during reionization itself.  The rare, high-redshift sources have the potential to provide the most information about the early universe, but guessing at their abundance is nearly impossible.  It is therefore crucial to push searches to the highest possible redshifts.

Fortunately, there are no fundamental reasons to expect a cutoff in the radio source population at high redshifts.  While the evolving IGM will affect extended radio lobes, compact radio sources are driven by local jet physics and so should not depend on the large-scale environment \cite{haiman04}.  The CMB energy density does increase at higher redshifts, increasing the importance of inverse-Compton cooling relative to (radio) synchrotron emission.  The two become energetically comparable at $z \sim 6$, but this should only steepen the spectrum rather than quench the radio emission \cite{carilli02}.  Such steepening could be useful in identifying high-redshift radio sources.  Most persuasively, the fraction of radio-loud objects does not seem to evolve, even to $z \sim 6$ \cite{ivezic02,petric03}.

Another possible set of sources are GRBs and hypernovae.  Unfortunately, although they should occur at high redshifts \cite{bromm02-grb, kawai06}, afterglow models predict that only the most energetic events achieve the required flux densities, and only then if they occur in exceptionally diffuse environments \cite{ioka05}.  Thus it seems unlikely that transient sources can be used for the 21 cm forest, although detecting 21 cm absorption from their host galaxies is not at all unreasonable \cite{furl02-21cm}.

%\bibliographystyle{elsart-num}
%\bibliography{Ref_21cm}

%\end{document}

%% file: synergy-ch11.tex
%\documentclass{elsart}
%\usepackage{amssymb,cite,epsfig}

%\input{./defns.tex}

%\begin{document}

\section{Complementary Observations} \label{syn}

Before closing, we wish to consider 21 cm observations in relation to other probes of the high-redshift universe.  As we shall see, all of these other techniques require luminous sources and so focus on the eras of the first luminous objects and of reionization (\S \ref{obj} and \ref{reion}).  Between the surface of last scattering and the time that the first luminous sources appear, the 21 cm sky is the \emph{only} known probe available to us (assuming that galaxies reionize the Universe, of course).

\subsection{Galaxy Surveys} \label{galsurv}

The utility of comparing 21 cm surveys to the galaxy (and quasar) distribution is obvious:  it will identify the sources responsible for reionizing each slice of the IGM (or at least their descendants).  Of course, such galaxy surveys are intrinsically difficult, as the sources are extremely faint (because of their distance and small sizes).  Moreover, \lya absorption by the intervening IGM is essentially complete for rest wavelengths blueward of $1216$ \AA \ at $z \ga 6$, requiring near-infrared cameras that are generally less advanced than their optical counterparts. 

Nevertheless, surveys have already begun to probe the reionization epoch.  Deep integrations with the \emph{Hubble Space Telescope}, such as the GOODS survey and the UltraDeep Field, have detected a number of galaxies at $z \sim 6$ using the photometric dropout technique (e.g., \cite{stanway04, bunker04, bouwens06}) and perhaps even some objects at higher redshifts \cite{bouwens04, bouwens05}.  Population synthesis models, in conjunction with optical and infrared observations, can also be used to identify high-redshift candidates; this technique has already led to some surprising identifications of massive ($M_\star \ga 10^{11} \Msun$) galaxies at $z \sim 6$--$7$ \cite{mobasher05}.  Large-format ground-based surveys have also successfully identified photometric dropouts \cite{shimasaku05, shioya05}.  

An alternative to dropout surveys (which require deep spectroscopic followup for confirmation) is to search for objects with strong \lya emission lines.  Specific redshift windows can be searched with narrowband filters, which offer the enormous advantage of working between the bright sky lines that nearly blanket the near-infrared sky.  Observations in one such window at $z=6.56$ have already found dozens of objects \cite{hu02-lya, kodaira03, rhoads04, stern05, taniguchi05}, and windows at $z \sim 8$--$10$ are now being explored \cite{barton04, willis05, horton04, stark05}.  Unfortunately, these surveys can only detect the brightest objects (although gravitational lensing can probe further down the luminosity function in small patches of sky \cite{santos04-obs, kneib04, stark05}).  Substantial samples of less unusual galaxies must await larger telescopes more optimized for the near-infrared, such as the \emph{James Webb Space Telescope} or a ground-based 30 meter telescope.  Such surveys are interesting for any number of reasons, but drawing conclusions about reionization is difficult because of the many uncertainties in the stellar initial mass function, metallicity, escape fraction, and the IGM recombination rate \cite{stiavelli04a, stiavelli04b}.  Thus the correlation with 21 cm maps may in the end prove the best way to learn about this aspect of the sources.  Of course, such a comparison will be easiest with a well-understood galaxy population (which is relatively difficult for \lyans-selected galaxies).  

Quasar searches have also reached $z \sim 6$, although again only at the bright end of the luminosity function \cite{fan01, fan02, fan03, fan04, fan06}.  Nevertheless, these surveys have already taught us that bright quasars are much too rare to be responsible for reionization.  These luminous objects also provide an intriguing set of targets for 21 cm imaging, because their \htwo regions are likely to be much larger than average (see \S \ref{image}).

While the merits of comparing the galaxy distribution with \htwo regions detected by 21 cm observations are clear, galaxy observations can also teach us about the ionized gas distribution on their own.  Consider a galaxy embedded in the neutral IGM.  Some ionizing photons leak out into the IGM and create an \htwo region, but most are absorbed within the galaxy and reprocessed into \lya photons.  When these photons diffuse out of the galaxy, they will initially propagate through the \htwo region, suffering relatively little absorption.  However, once they encounter the neutral IGM, they will quickly be absorbed unless they have traveled far enough to redshift out of line center.  Of course, the amount of redshifting depends on the size of the host \htwo region; thus we expect strong \lya lines to gradually fade toward higher redshifts \cite{miralda98-lya, madau00, haiman02-lya}.

\begin{figure}[!t]
\centerline{\epsfig{file=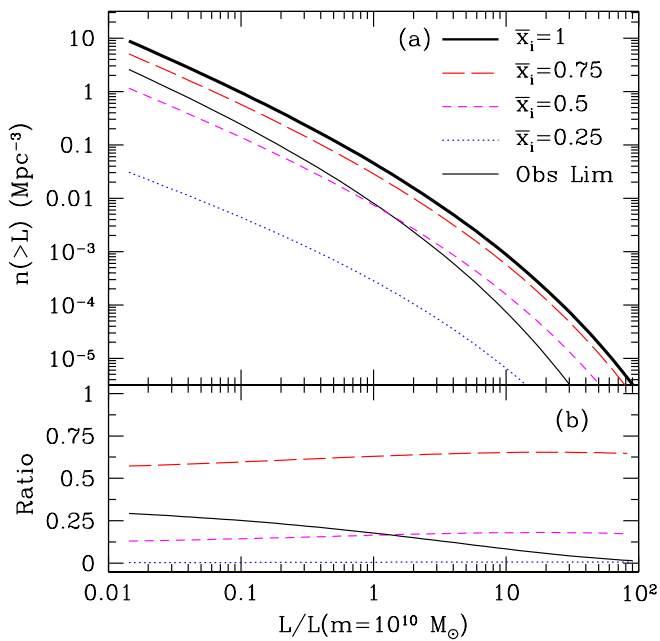,width=3.5in}}
\caption{(\emph{a}):  \lya luminosity functions at $z=6.5$.  The thick solid line shows an assumed intrinsic function, with luminosity $L \propto m_{\rm gal}$.  The observations require the measured luminosity function to lie above the thin solid line \cite{malhotra04}.  The other curves show how IGM \lya absorption affects the luminosity function at several different $\bxion$.  (\emph{b}):  Ratio of attenuated to intrinsic luminosity functions.  From \cite{furl06-lyagal}.}
\label{fig:lya}
\end{figure}

Quantitatively constraining $\bxion(z)$ with this tool is difficult for two reasons.  First, the intrinsic characteristics of the galaxies -- especially their winds -- can strongly affect their \lya line properties \cite{santos04}.  This problem can be alleviated (to some extent) by comparing galaxy populations at two different redshifts (preferably closely spaced enough in time to avoid significant cosmic evolution).  For example,  \lyans-selected samples at $z=5.7$ (which we know to be after reionization) and at $z=6.5$ (which may be \emph{in medias res}) show no significant differences between their luminosity functions, at least within the statistical errors of the observations \cite{malhotra04}.  The second ``problem" is that the bubble sizes depend on the sources of reionization \cite{gnedin04-lya, furl04-lya, wyithe05-clus, furl06-lyagal}; see Figure~\ref{fig:lya} for an illustration of the constraints that can be set. 

%%%%%%%%%%%% FIGURE 11-1: Lyman-alpha galaxies

Fortunately, this second ``bug" is actually a powerful tool for studying reionization.  Although large bubbles weaken the absorption, their existence at relatively early times also implies that \lya surveys will successfully find galaxies throughout the middle stages of reionization.  Moreover, the survey selection function will be modulated by the ionized bubbles:  lines of sight passing through large \htwo regions will contain many galaxies while other lines of sight will appear entirely empty even though their intrinsic galaxy densities could be nearly as large \cite{furl04-lya, furl06-lyagal}.  Thus \lya surveys can be used to map the bubble distribution (albeit only indirectly), making them the best known alternative to 21 cm measurements.  The major difficulties are surveying sufficiently large volumes to the required sensitivities and understanding the intrinsic \lya properties and clustering of the galaxies (which is probably easiest in conjunction with dropout surveys). Nevertheless, comparing the two kinds of maps offers exciting possibilities to learn about the sources and their interaction with the IGM.

Galaxies can also be studied indirectly through the near-infrared (NIR) background and its fluctuations \cite{kashlinsky05-rev}:  these same \lya photons travel through the IGM without being destroyed (only scattered) and eventually redshift to wavelengths $\ga 1 \, \mu{\rm m}$ at the present day.  Thus the NIR background contains information about the \emph{recombination} history (which, of course, also constrains the ionizing luminosity) \cite{santos02,salvaterra03}; provided that most recombinations occur inside or near galaxies (as is true if $\fesc \ll 1$), its fluctuations directly trace those of the galaxy population \cite{magliocchetti03, kashlinsky04, cooray04-nir} and are therefore complementary to 21 cm tomography of the IGM.  However, studies of the NIR background suffer from many of the same challenges as 21 cm observations:  the signal is strongly dominated by foregrounds (in this case the zodiacal light and moderate redshift galaxies), which must somehow be separated from the cosmological signal.  As a result, the recent claimed detection of Pop III stars via their fluctuations in deep \emph{Spitzer} fields \cite{kashlinsky05} remains controversial \cite{rozas06, salvaterra06-nir}.  There may be useful synergies between the data analysis techniques of both fluctuation searches and in cross-correlation of the two datasets (to eliminate those foregrounds that contribute to only one probe); the main challenge is the large disparity in angular scales subtended by the relevant telescopes:  the NIR searches use fields $\la 30$ arcmin across, while 21 cm observations will span hundreds of square degrees.

\subsection{Quasar and GRB Spectra} \label{quas}

Just as with the 21 cm line, Lyman series absorption in the spectra of high-redshift quasars traces neutral hydrogen in the early Universe. However, since the cross sections for such permitted transitions are $\sim 10^{7}$ times larger, optical/UV absorption spectra saturate much more rapidly than the 21 cm line (which is still optically thin even in a fully neutral universe).  Quasar spectra are therefore primarily useful for studying the tail end of reionization, when the neutral fraction is small.  This very property makes quasar absorption spectra highly complementary to 21 cm probes, because they are most powerful precisely when $\bxhi$ (and the 21 cm signal) are plummeting.  On the other hand, despite the small neutral fraction, the tail end of reionization has rich physics and any number of unanswered questions.  For example, how does the transition from the ``bubble-dominated" topology characteristic of reionization to the ``web-dominated" topology we see at lower redshifts (where Lyman-limit systems embedded in filaments consume most of the ionizing photons) occur?  How large do the ionized bubbles get?  How does the IGM clumpiness evolve as the Universe is ionized and heated?  Quasar spectra are ideally suited to answer these and other questions that arise when the 21 cm sky is fading.  However, we must keep in mind that, while quasar spectra can potentially constrain the global neutral fraction and topology of reionization, in practice these are difficult questions, because making inferences about the ionization state of the universe from a saturated absorber is fraught with uncertainty. 

As with many topics we have discussed in this review, our understanding of the $z\sim6$ universe as revealed by quasar spectra is evolving rapidly; to date, 19 quasars with redshifts $5.74 < z_{\rm em} < 6.42$ have been discovered in the SDSS \cite{fan06}. Thus far, the following spectral features have been used to deduce IGM properties: (i) the evolution of the mean Gunn-Peterson optical depth in Ly$\alpha, \,\beta$, and $\gamma$ transitions, as well as its line-of-sight scatter; (ii) the distribution of lengths of dark absorption gaps and transmission spikes; (iii) the detection of damping wings; (iv) the size of HII regions around high-redshift quasars; and (v) metal absorption lines. We will discuss each of these in turn (see Fig.~\ref{fig:obs-summ} for a summary of some of the resulting constraints). 

Because only a small fraction of the IGM need be neutral for Ly$\alpha$ Gunn-Peterson absorption to saturate (see eq.~\ref{eq:gp}), Ly$\beta$ and Ly$\gamma$ absorption provide better limits by a factor of 2--4, though they are affected by uncertainty in the foreground absorption from lower-order transitions.\footnote{In a clumpy IGM, $\tau^{\rm eff}_{\alpha}/\tau^{\rm eff}_{\beta}$ and $\tau^{\rm eff}_{\alpha}/\tau^{\rm eff}_{\gamma}$ are a factor 2--4 smaller than the ratio of oscillator strengths would suggest, reducing the leverage granted by these higher order transitions \cite{song02,oh05}.} Since most transmission arises in underdense voids (where the recombination time is longest), quasar spectra provide little information about regions of the IGM where most baryons reside \cite{oh05}; thus, estimates of the neutral fraction can at best constrain $\bxhi \ge 10^{-3}$--$10^{-2}$ and are dependent on theoretical models of the IGM density distribution. Nonetheless, striking observational patterns emerge for some reasonable models. According to \cite{fan06} (who use the IGM model of \cite{miralda00}), the effective optical depth $\tau_{\rm eff}=-{\rm ln}\langle \mathcal{T} \rangle$ (where $\mathcal{T}$ is the transmission fraction) increases rapidly at high redshift, accelerating from $\tau_{\rm eff} \propto (1+z)^{4.3}$ to $(1+z)^{\ge 11}$ at $z>5.7$. This rapid evolution has been interpreted by some to indicate sudden evolution in the radiation field and mean free path of ionizing photons, as might be expected when the ionized bubbles percolate (e.g., \cite{cen_mac02,lidz02,fan02}).  On the other hand, an empirically motivated model of the mean optical depth evolution does not show such a break and implies that no such dramatic event is occurring \cite{becker06} (see also \cite{song02}).  Better data and modeling will clearly be needed to settle this debate.

\begin{figure}[!t]
\centerline{\epsfig{file=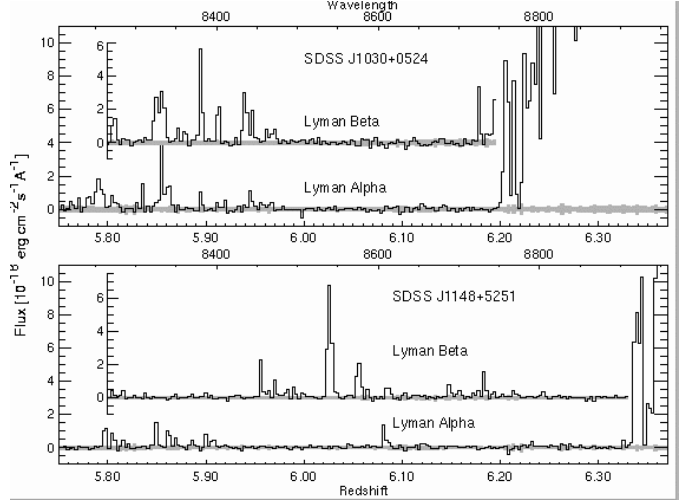,width=3.5in}}
\caption{Quasar absorption spectra.  In each panel, the lower curve shows the observed spectrum in the \lya forest region; the overlaid curve shows the corresponding Ly$\beta$ spectrum.  J1030+0524 ($z=6.28$) shows saturated absorption in both transitions from $z\approx 5.95$--$6.2$, while J1148+5251 ($z=6.42$) shows transmission throughout the spectrum over the same interval (and even to higher redshifts).  From \cite{white03}.}
\label{fig:qso}
\end{figure}

Accompanying this sudden evolution is a rapid increase in the dispersion of $\tau_{\rm eff}$, even when it is smoothed on scales $\ga 50 h^{-1} \Mpc$.  Figure~\ref{fig:qso} shows the most dramatic example, the clear detection of transmission throughout the line of sight to SDSS J1148+5251 ($z=6.42$) \cite{white03,white05,oh05} despite uniformly complete absorption in the Ly$\alpha,\,\beta$, and $\gamma$ troughs of SDSS J1030+0524 ($z=6.28$) \cite{becker01,white03}. This has been interpreted as evidence for patchy reionization, or at the very least for a strongly fluctuating radiation field, which is expected near the end of reionization \cite{wyithe06-var, fan06}.  However, such claims should be viewed with caution: numerical simulations show that the observed sightline-to-sightline variance is in fact consistent with density fluctuations in a uniform radiation field \cite{lidz06} (see also \cite{liu06}). Transmission fluctuations can be of order unity on large ($\sim 50\,h^{-1}$ Mpc) scales for two reasons: (i) transmission spectra are highly biased tracers of the underlying density fluctuations, because they are mainly sensitive to rare voids, and (ii) projected power from small-scale transverse modes is aliased to long wavelength line-of-sight modes. Thus, although it is quite likely that the ionizing background does contain substantial fluctuations at these epochs, it is extremely difficult to detect them.

%%%%%%%%%%%% FIGURE 11-2: Quasar spectra

A useful statistic related to the observed scatter in transmission is the ``dark gap distribution," or the size distribution of regions with $\tau > \tau_{\rm min}$ \cite{croft98,song02,barkana02,nusser02,furl04-lya,paschos05,kohler05-sim,gallerani05}. Because it still provides useful information when absorption is saturated, this is potentially the most useful quasar absorption probe in the nearly saturated limit. Although the relation between dark gaps and $\bxhi$ is model-dependent, the relative evolution of gap lengths is a robust statistic. Indeed, simulations find that reionization causes a dramatic decrease in the average length of dark gaps, as well as its dispersion \cite{paschos05,kohler05-sim}.  Only recently have simulations reached sizes comparable to dark gaps \cite{kohler05,gallerani05}, which is essential for making quantitative comparisons. The data show a dramatic increase in mean dark gap length (defined with $\tau_{\rm min}= 3.5$) from $R_{\rm dark}<10$ to $R_{\rm dark}>80$ comoving Mpc at $z\sim 5.7$ \cite{fan06}, with large line of sight variations: dark gaps as long as 50 Mpc appear by $z\sim 5.5$, while some lines of sight at $z>6$ are still transparent \cite{fan06}. This abrupt transition may or may not be consistent with a simple thickening of the Ly$\alpha$ forest \cite{liu06}; there is probably more information to be mined from dark gaps.  One recent study found that, with $\sim 10$ $z>6$ quasars, the dark gap distribution can sharply distinguish between different reionization histories \cite{gallerani05}. It could also constrain the topology of reionization, at least in principle \cite{nusser02}.

Note that a large dark gap does not necessarily correspond to a large neutral region, because the strong damping wings of \lya can mask ionized regions. Given an estimate for the size distribution of  HII regions, the absence or presence of transmission spikes constrains $\bxhi$ by limiting the total contribution from damping wings (see \S\ref{galsurv} for the application of this to Ly$\alpha$ emitting galaxies).  Again, large \htwo regions, created by clusters of high-redshift galaxies, increase the probability of transmission spikes\cite{furl04-lya}.  A novel test for damping wings compares the Ly$\alpha$ and Ly$\beta$ transmission near the boundary of an \htwo region \cite{mesinger04}.  The presence of Ly$\beta$ transmission coupled with saturated Ly$\alpha$ absorption over a contiguous region $\Delta z \sim 0.02$ in the proximity zone around SDSS J1030+0534 ($z=6.28$) requires strong damping wing absorption in this region. If the absorption were solely due to the fluctuating opacity of residual \hone in mostly ionized gas, there should have been a transmission spike due to an underdense region somewhere inside such a large stretch. The required damping wing implies $\bxhi \ge 0.2$. However, this feature has only been seen in one quasar, and its statistical significance is still not clear. 

Yet another probe of the high-redshift IGM is the size of proximity zones around high redshift quasars.  Provided that the region in which Lyman series transmission can be seen corresponds to the edge of an \htwo region, this should scale as $R \propto \bxhi^{-1/3} \dot{N}_{\rm ion} t_{\rm Q} (1+z)^{-1}$ in a uniform IGM, where $\dot{N}_{\rm ion}$ is the production rate of ionizing photons and $t_{\rm Q}$ is the quasar lifetime. Given appropriate ansatzen for these two parameters, it should be possible to infer $\bxhi$; several authors have suggested, from the relatively small sizes of quasar HII regions around $z\sim 6$ quasars, that $\bxhi \ge 0.2$ \cite{wyithe04-prox,wyithe05-prox,yu05}. These estimates are afflicted by uncertainties in the quasar lifetime, redshift (e.g., most quasar redshifts are determined by high-ionization lines such as CIV and SiIV, which are systematically offset from the host galaxy's redshift), spectral template (i.e., $\dot{N}_{\rm ion}$ for a given observed luminosity), IGM clumping factor, and the effects of nearby galaxies.  Nonetheless, the differential change in proximity zone sizes should be a reasonably robust indicator of evolution in the neutral fraction. The observed sizes decrease rapidly toward high redshift, as might be expected if the IGM were becoming more neutral (see Fig.~\ref{fig:obs-summ}) \cite{fan06}. 

Finally, an alternative probe of the IGM ionization state is to use metal absorption lines such as OI, SiII, and CII \cite{oh02}, which tend to trace \hone and have transitions redward of \lyans. Since metal atoms are $\sim 10^{-5}$--$10^{-7}$ times less abundant than hydrogen in the IGM, their absorption does not saturate even in a fully neutral IGM. OI is a particularly attractive tracer: its ionization potential differs from hydrogen by only $\Delta E=0.02$ eV, so it should sit in tight charge exchange equilibrium with HI. It is nevertheless imperfect, because metals do not fill all of space and tend to lie preferentially in the ionized regions around galaxies. However, if there is substantial early star formation (and hence metal pollution), and if metals lie in overdense regions (which recombine quickly once nearby ionizing sources shut off), the test may prove useful.  Recent high-resolution observations found a sharp rise in the abundance of OI absorbers at high redshifts, with abundances comparable to the predicted number of Lyman limit systems \cite{becker06}. However, there are striking line-of-sight variations; four of the six observed systems lie toward SDSS J1148+5251 ($z=6.42$), even though nine lines of sight were observed.  It is highly unlikely that this is due to variations in sensitivity or random fluctuations. The interpretation of this puzzling result is still unclear, but it likely signals inhomogeneous metal pollution and/or reionization. 

Throughout this section, we have referred exclusively to quasar spectra, but of course any other luminous, high-redshift background source will do.  GRBs are a particularly interesting alternative, because they likely occur at high redshifts (so long as they approximately trace the star formation history of the Universe \cite{bromm02-grb}).  Moreover, cosmic time dilation implies that, as the burst redshift increases, a fixed observer time corresponds to earlier and earlier times in the frame of the burst; because bursts fade rapidly, this helps counteract the usual decline of flux with luminosity distance and makes GRBs visible to higher redshifts than naively expected \cite{ciardi00, lamb00}.  They have the theoretical advantage of simpler intrinsic spectra, without large proximity zones or \lya emission lines, that should in principle allow easier extraction of the shape of the red damping wing -- which in turn offers a sensitive constraint on the neutral fraction \cite{miralda98, barkana04-grb}.  However, most GRBs have strong damped-\lya absorbers from the host galaxies, making the damping wing test much more difficult (though it can still be applied, because the line profiles differ in the two cases; see \S \ref{intro-full}).  The first (weak) constraints came from GRB 050904 at $z=6.3$ \cite{kawai06, totani06}, which was observed 3.4 days after the burst.  Faster identification and followup of high-redshift GRBs will allow even more sensitive tests of the damping wing (especially because some GRBs do appear to have weak local absorbers; see the compilation in \cite{vreeswijk04}), and of course measurements of the intervening IGM are independent of the local absorption.  

In summary, the picture painted by quasar absorption studies of the $z\sim 6$ IGM is highly complex and unclear: interpretations range from significantly neutral to highly ionized, and the topology of reionization remains quite controversial.  Prospects for improvement are good, but they require more lines of sight to be identified and better modeling of the existing data.  Ultimately, 21 cm studies have the power to strip away most of these uncertainties!

\subsection{CMB Polarization} \label{cmbpol}

While 21 cm observations are sensitive to the neutral hydrogen density, CMB polarization experiments are sensitive to the free electron density, and there is obviously a good deal of potential for synergy between the two.

CMB temperature anisotropies are relatively unaffected by reionization on large scales, but on small scales they are suppressed by a factor ${\rm exp}(-2 \tau_{\rm es})$, where $\tau_{\rm es}\equiv \sigma_T \int_0^\eta \deriv \eta' \, n_e$ is the electron scattering optical depth and $\eta$ is the conformal time. The characteristic angular scale dividing these regimes is $l_{r}=D_{A}[z(\eta_{r})]/\eta_{r}$, where $\eta_{r}$ is the visibility-weighted conformal time (essentially describing the distance to the reionization surface; see below for a precise definition of the visibility function) \cite{hu97}. Similarly, the angular power spectrum of polarization is suppressed on small scales by the same factor ${\rm exp}(-2\tau_{\rm es})$. However, on large scales, the rescattering of photons also creates a broad peak in polarization power at $l\sim l_{r}$ \cite{ng96, zalda97-reion}. This polarization anisotropy is sourced by the temperature quadrupole seen by scatterers during reionization, which is much larger than the temperature quadrupole at the surface of last scattering.  The amplitude of the bump depends most strongly on $\tau_{\rm es}$ and on the amplitude of primordial potential perturbations.  Other parameters matter much less; for instance, the signal is nearly independent of the baryon density, because the modes contributing to the reionization signal have $k \sim 2/\eta_{r}$ and enter the horizon well after reionization.  

Particularly if $\tau_{\rm es}$ is large, most of the information on reionization comes from the auto-correlation power spectrum $C^{l}_{EE}$, rather than the cross-correlation $C^{l}_{TE}$ between polarization and temperature.\footnote{Here ``E" refers to spin-free or E-mode polarization \cite{seljak97pol, zalda97, kamionkowski97}; reionization does not generate the complementary divergence-free or B-mode polarization.}  This is primarily because $C^{l}_{EE} \propto \tau_{\rm es}^{2}$, whereas $C^{l}_{TE} \propto \tau_{\rm es}$.  Also, in the cosmic variance limit, the fractional uncertainty in $C^{l}_{EE}$ is smaller than the fractional uncertainty in $C^{l}_{TE}$ unless they are perfectly correlated. Furthermore, $C^{l}_{TE}$ correlates quantities which at fixed $k$ have different angular frequencies on the sky, because the temperature anisotropies are projected from the more distant surface of last scattering.  The matching angular frequencies of $C^l_{EE}$ create well-defined secondary peaks, whereas the mismatch in $C^{l}_{TE}$ washes out fluctuation power, suppressing the peaks.

The first detection of reionization in CMB data was reported by the \emph{WMAP} team with their first year data; from $C^l_{TE}$, they found $\tau_{\rm es} = 0.17 \pm 0.08 \, (2\sigma)$ \cite{spergel03,kogut03}.  Two more years of data significantly improved the measurement, especially because they were able to detect $C^l_{EE}$.  The present best-fit value (using \emph{WMAP} data alone) is $\tau_{\rm es}=0.088^{+0.028}_{-0.034}$ \cite{page06, spergel06}.  Because $\tau_{\rm es}$ suppresses the amplitude of the CMB, other constraints on the power spectrum normalization (such as large-scale structure) are also useful; the best fit measurement including several other cosmological datasets is $\tau_{\rm es}=0.069^{+0.026}_{-0.029}$.

Is any other information available beyond a simple measure of the integrated optical depth $\tau_{\rm es}$?  Unfortunately, the dominant modes in the large scale polarization signal have $k\sim2/\eta_{r}$.  We expect little correlation between ionized bubbles on horizon scales, and CMB polarization is essentially independent of the topology of reionization. However, even if CMB polarization experiments cannot probe patchy reionization, future high-precision measurements could constrain the global reionization history $\bxion(z)$ and distinguish reionization scenarios with identical total optical depths $\tau_{\rm es}$ \cite{kapling03,holder03}. 

\begin{figure}[!t]
\centerline{\epsfig{file=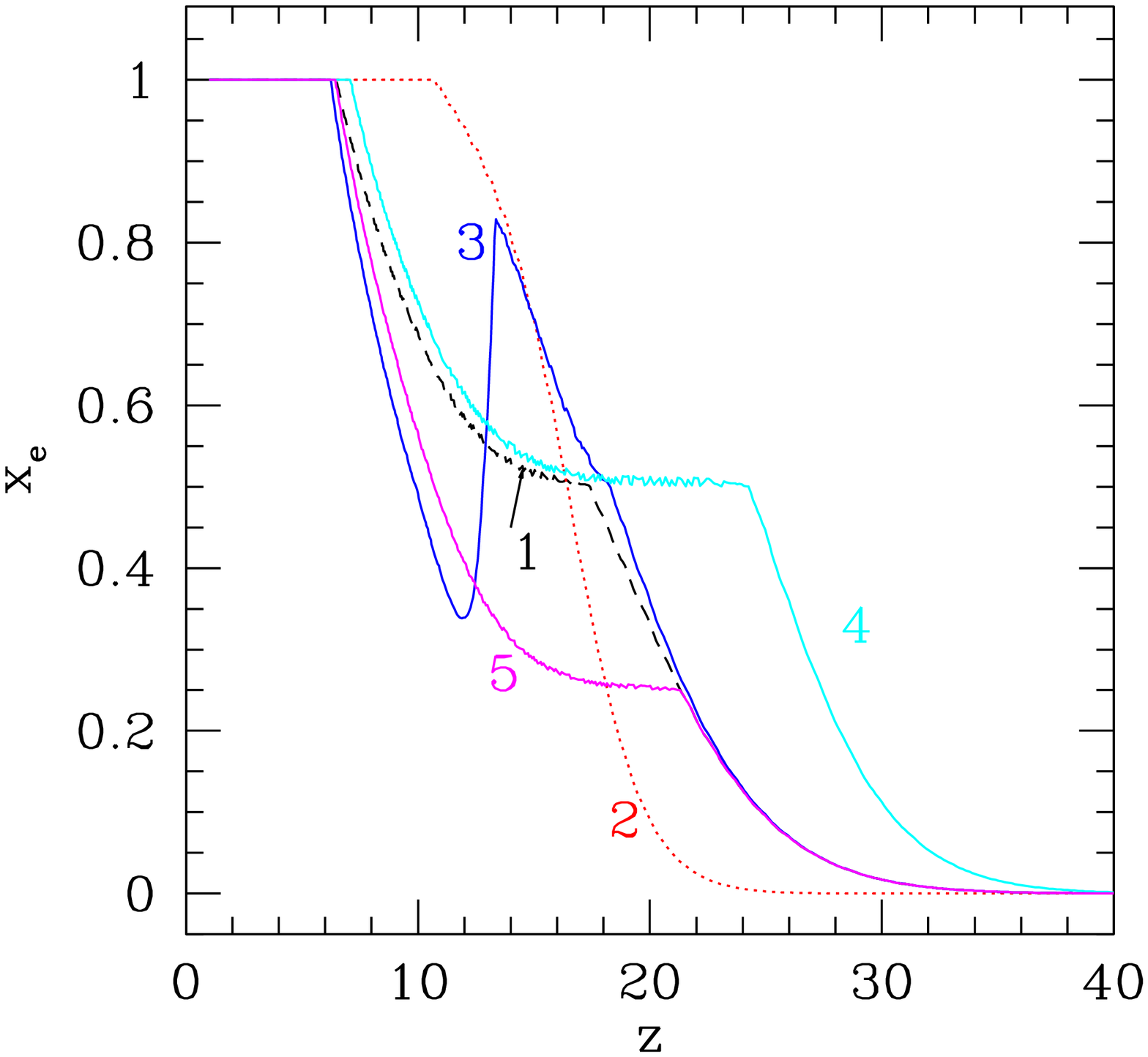,width=2.75in}
\epsfig{file=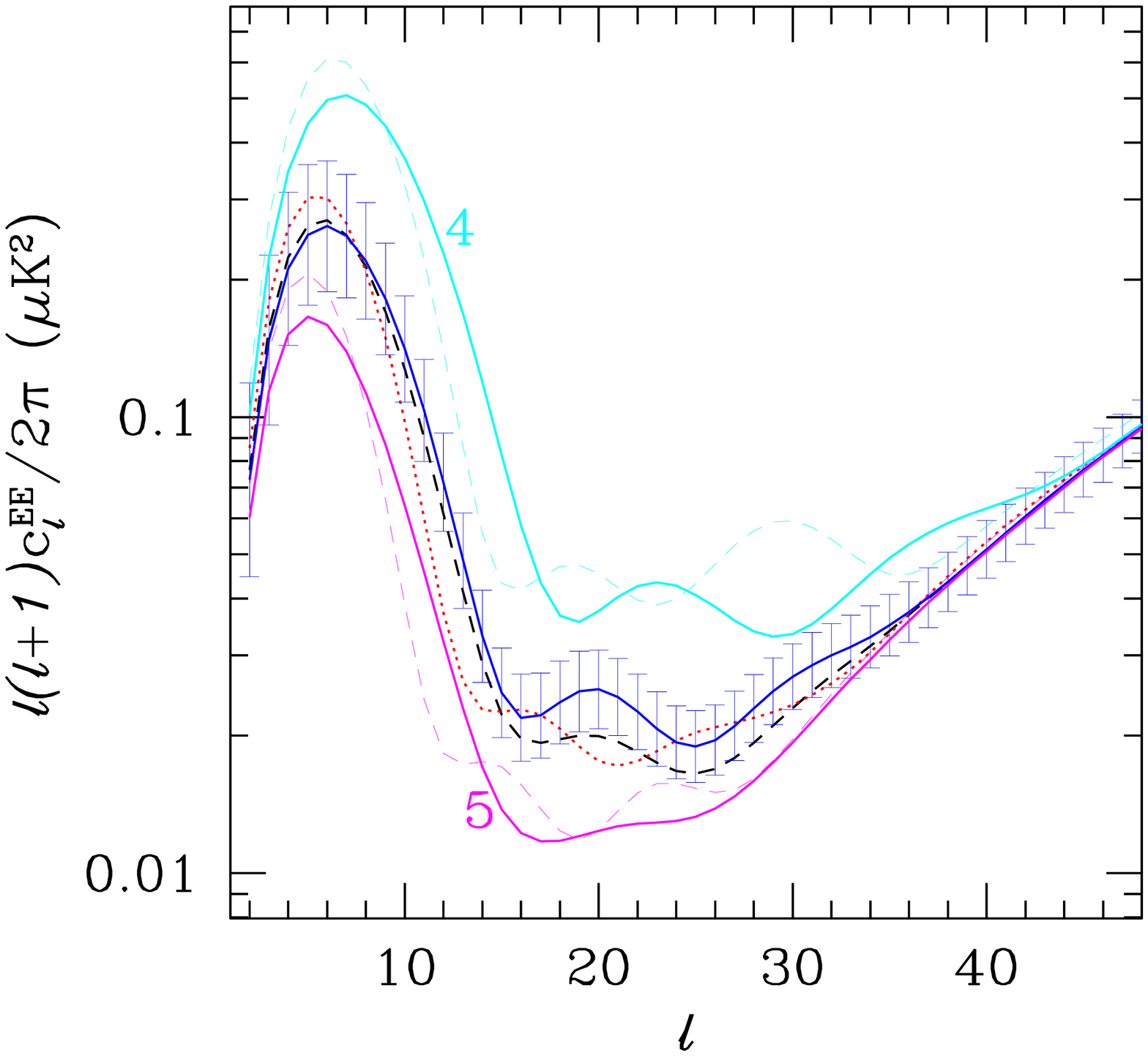,width=2.75in}}
\caption{Large-scale CMB polarization power spectra (right panel) for five different reionization histories (left panel).  All are normalized to have identical power at $l=50$. The thin dashed curves in the right panel show the best fit instantaneous reionization histories to models 4 and 5.  The error bars show the cosmic variance limit.  From \cite{holder03}.}
\label{fig:cmbpol}
\end{figure}

The leverage comes from the shape of the power spectrum; for models with substantial partial reionization at high redshifts, there is significantly more polarization power at smaller angular scales.  Qualitatively, this occurs because the horizon is smaller at high redshifts.  Again, most of the information comes from $C^l_{EE}$; we show some example power spectra, together with their corresponding reionization histories, in Figure~\ref{fig:cmbpol} (from \cite{holder03}).  Here the error bars include cosmic variance only; thus, even with these radically different reionization histories, the observable differences are relatively small and constraints will be crude.  CMB analyses typically characterize reionization as a sharp step-function at some redshift $z_{r}$. Such an unrealistic reionization history can significantly bias the estimate of $\tau_{\rm es}$ in a manner that depends on the true $\bxion(z)$ \cite{kapling03,holder03}.  This bias is smaller than the statistical errors of {\it WMAP} but could easily be $\delta \tau_{\rm es} \ge 0.01$ for the {\it Planck} satellite, compared with statistical errors $\delta \tau_{\rm es} \sim 0.005$ and the cosmic variance limit of $\delta \tau_{\rm es} \sim 0.002$--$0.003$  \cite{holder03}. It can be reduced by fitting slightly more complicated reionization histories (such as two-stage reionization), though uncertainty in this modeling may ultimately prevent optical depth measurements from reaching the true cosmic variance limit. The accuracy with which polarization can constrain the reionization history is likely similar, with information on only $\sim 2$--$3$ broad redshift bins. The accuracy of 21 cm experiments in tracking the volume filling fraction of ionized regions should be much higher (see \S \ref{reion}). 

%%%%%%%%%%%% FIGURE 11-3: CMB polarization

In practice, foregrounds are likely to be the key difficulty, and cleaning strategies are still under development (see \cite{page06} for an example).  If foregrounds do impose a limit well above cosmic variance, the advantages of all-sky surveys are less apparent, and ground-based polarization experiments could be competitive with space-based missions \cite{keating03}.  We refer the interested reader to \cite{fan06-review} for a more in-depth discussion of the future prospects of CMB polarization measurements.

\subsection{Small-Scale CMB Anisotropies} \label{ksz}

The same scattering processes that produce large-scale polarization also imprint temperature anisotropies on the CMB through the peculiar velocity of the scattering medium, which imparts a blue- or redshift to the scattered photons.  This ``Doppler" or ``kinetic Sunyaev-Zeldovich" (kSZ) effect has several observable consequences.

The Doppler anisotropy toward direction ${\bf \hat{n}}$ on the sky is
\begin{equation}
\frac{\Delta T}{T_\gamma}({\bf \hat{n}}) = \sigma_T  \int \deriv \eta \, e^{-\tau_{\rm es}} \, a \, n_e \, {\bf \hat{n} \cdot v},
\label{eq:doppler-cmb}
\end{equation}
where ${\bf v}$ is the local peculiar velocity.  We define $g_v(z) = \dot{\tau_{\rm es}} \, e^{-\tau_{\rm es}} = \sigma_T \, n_e \, a \, e^{-\tau_{\rm es}}$ as the visibility function; it is the probability that a photon last scattered at $\eta \pm \deriv \eta$.  (Throughout this section, an overdot will represent $\partial/\partial \eta$.)

From equation~(\ref{eq:doppler-cmb}), one might hope that linear order velocity fluctuations would source temperature perturbations.  Unfortunately, the kSZ effect vanishes to linear order,  because the crests and troughs of a perturbation produce equal and opposite Doppler shifts \cite{kaiser84}.  Thus anisotropy only appears if either $v$ or $g$ evolve across the perturbation, and it is a second order effect.  Obviously, $g_v(z)$ can only evolve significantly across a perturbation that spans relatively large scales.  For reasonable reionization histories, the signal has a broad peak at $l \la 100$; on larger scales, the previously described cancellation damps out the anisotropies.  However, even at $l \sim 100$, the Doppler contribution is more than an order of magnitude smaller than the primary CMB anisotropies \cite{cooray04-cmbt}:  we need some way to isolate the contribution from reionization.

The 21 cm background provides exactly such a tool \cite{alvarez06}, because it traces both large scale overdensities (which source velocity perturbations) and fluctuations in the ionized fraction.  For example, consider an overdense region during (uniform) reionization.  On the far side of the perturbation, the Universe will be mostly neutral and hence little scattering will occur; the near side, on the other hand, will be mostly ionized, imparting a net redshift to the CMB.  Thus, in the simplest model (in which overdense regions are not ionized significantly earlier than underdense regions), the 21 cm signal and the CMB will be anti-correlated.  On large scales, the cross-correlation can be written \cite{alvarez06}
\begin{equation}
\frac{l^2 \, C_l}{2 \pi} \approx - \frac{T_\gamma \, \bdtb \, D(z)}{2 \pi} \, \left[ \frac{4}{3}\bxhi \, P_{\delta \delta_L} \left( \frac{l}{r} \right) - \bxion(z) \, P_{x \delta_L} \left( \frac{l}{r} \right) \right] \frac{\partial}{\partial \eta} [\dot{D}g_v].
\label{eq:cmb-21cm}
\end{equation}
As in equation~(\ref{eq:pk_vel}), $\delta_L$ denotes the linear density field and marks those terms arising from the velocity fluctuation in equation (\ref{eq:doppler-cmb}).  The first term in square brackets comes from the correlation of this velocity with the large scale density fluctuations in the 21 cm signal; the $P_{x \delta_L}$ term arises from the evolution of the ionized fraction across the density fluctuation.  Note that these terms have opposite signs:  ionizing overdense regions first cancels part of the cross-correlation by moderating the evolution of the electron fraction across the perturbation.  In realistic models, this term can easily dominate \cite{alvarez06}, so the sign of the correlation provides information on reionization.  It obviously peaks at $\bxion \sim 0.5$ and therefore constrains $\bxion(z)$ directly as well.  Simple models predict peak signal strengths $\sim 200$--$500 \microkel^2$ \cite{alvarez06}, within reach of SKA-class observatories (although foregrounds could pose a problem on these large scales).

This linear Doppler effect peaks on large scales (hundreds of Mpc) and hence does not depend strongly on the bubble properties (except insofar as their bias affects $P_{x \delta_L}$).  It also ignores the effect of small-scale inhomogeneities in either the density or ionized fraction on $g_v(z)$.  By including these in equation (\ref{eq:doppler-cmb}), we see that the anisotropies depend on ${\bf q} \equiv {\bf v}(1 + \delta + \delta_x)$.  The angular power spectrum then has significant contributions from $P_{vv} P_{\delta \delta}$, $P_{vv} P_{xx}$, and $P_{vv} P_{\delta x}$ (as well as four-point terms) \cite{ma02, santos03}.  The first of these (the Ostriker-Vishniac or OV effect) depends only on the density field and $\bxion(z)$ \cite{ostriker86}.  At the relevant scales the OV effect is dominated by nonlinear structure formation at low redshifts \cite{hu00, baumann03} and only depends weakly on reionization \cite{santos03, mcquinn05} (though see \cite{jaffe98}).

\begin{figure}[!t]
\centerline{\epsfig{file=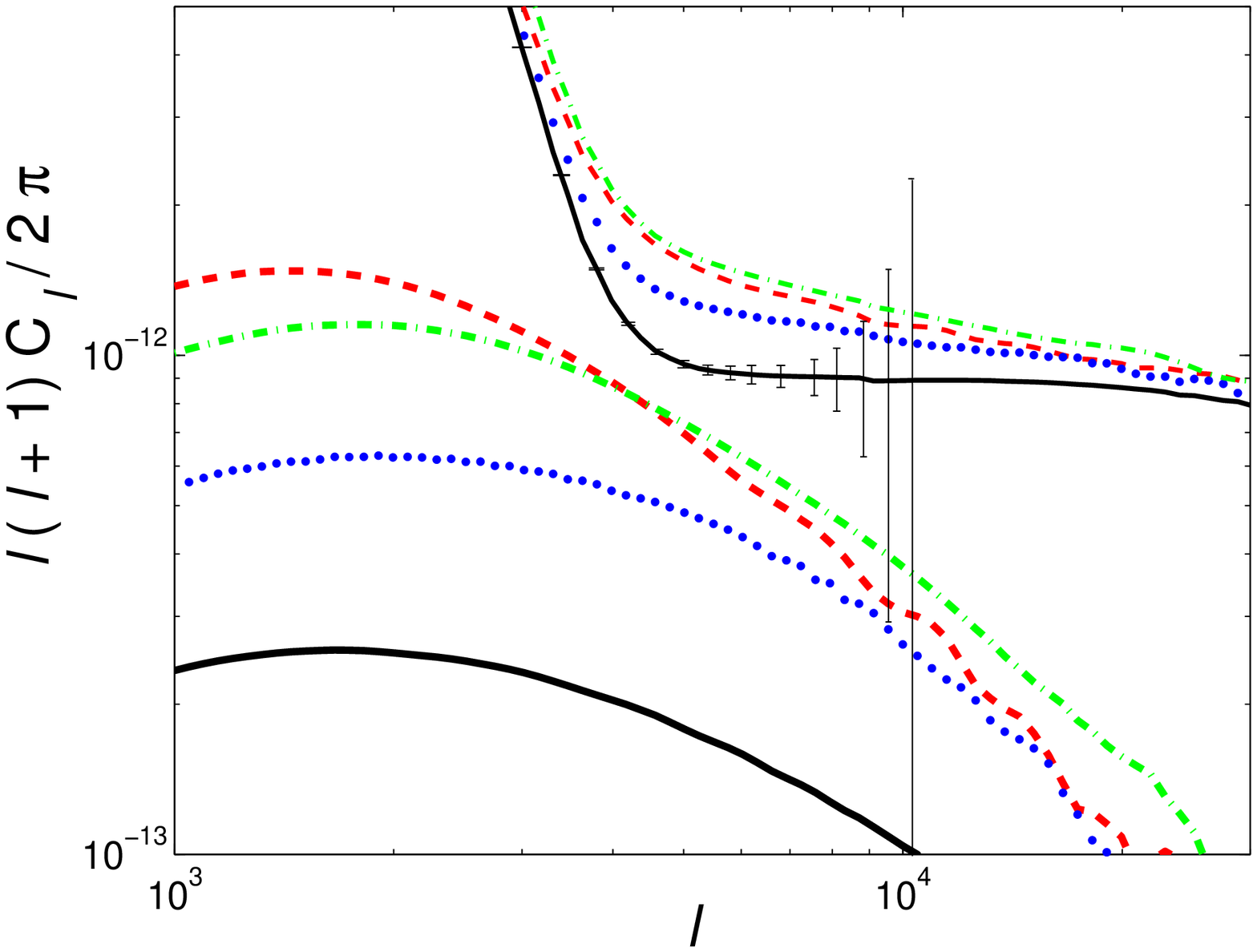,width=2.75in}\epsfig{file=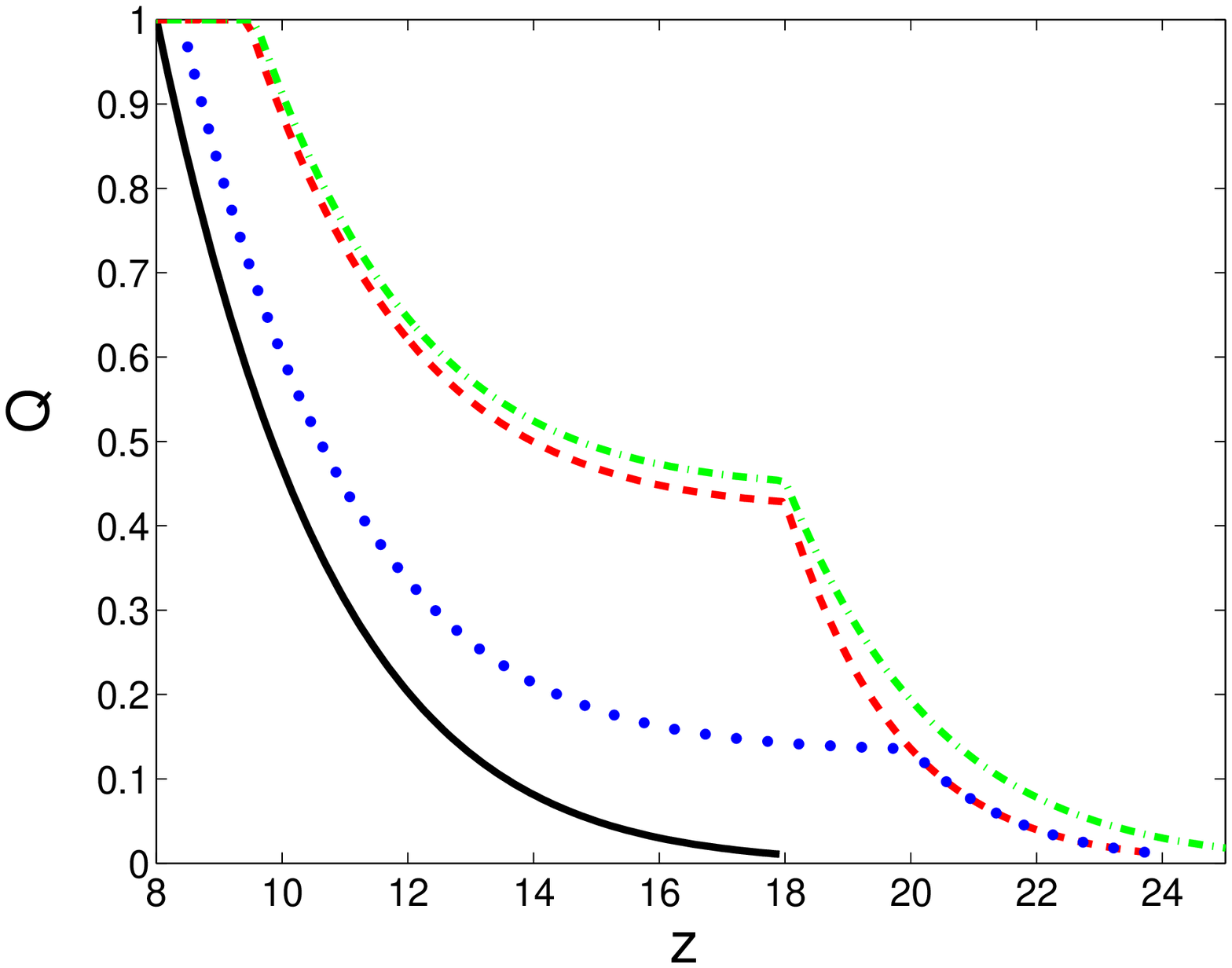,width=2.75in}}
\caption{Kinetic Sunyaev-Zeldovich anisotropies (left panel) from several different reionization histories (right panel; here $Q$ is the mean ionized fraction).  In the left panel, the lower set of curves shows the ``patchy reionization" component while the upper set includes the patchy, OV, and primary CMB anisotropies.  We also show the anticipated $1\sigma$ error bars for the Atacama Cosmology Telescope \cite{kosowsky03}, assuming perfect foreground removal.  From \cite{mcquinn05}.}
\label{fig:ksz}
\end{figure}

More interesting from our point of view are the so-called ``patchy reionization" terms $P_{vv} P_{xx}$ and $P_{vv} P_{\delta x}$ \cite{aghanim96, gruzinov98, knox98}.  Physically these are sourced by \htwo regions subtending only part of a large-scale velocity fluctuation:  the non-uniform ionized fraction prevents the usual linear-order cancellation.  The scale dependence of the patchy signal will therefore depend on the bubble size distribution, and its amplitude on the duration over which patchiness persists.  The resulting signals have been calculated analytically \cite{aghanim96, gruzinov98, knox98, santos03, mcquinn05, furl05-charsize} and via simulations \cite{bruscoli00, benson01, salvaterra05-ksz, zahn05} for a number of different reionization scenarios.  Figure~\ref{fig:ksz} shows some illustrative examples for a range of reionization histories.  We plot the patchy reionization component (the lower set of curves) along with the ``total" signals (including the patchy, OV, and primary anisotropies, but neglecting the thermal Sunyaev-Zeldovich component).  This calculation is based on the bubble model presented in \S \ref{reion-an} (see also \cite{furl04-bub}); the peak at $l \sim 3000$ corresponds to the characteristic bubble size at $\bxion \sim 0.5$ in these models. Comparison with the $\bxion(z)$ curves in the right panel clearly shows that the overall amplitude depends strongly on the duration of reionization, because persistent patchiness helps to build up the signal.  Simulations yield similar amplitudes \cite{zahn05, salvaterra05-ksz}.  

%%%%%%%%%%%% FIGURE 11-4: KSZ effect

Small-scale CMB temperature anisotropies therefore provide an integrated measurement of the duration of reionization as well as some limited information on the bubble size distribution.  (Note as well that patchy reionization will bias cosmological measurements that use the tail of the primary CMB anisotropies \cite{santos03, zahn05}.)  The expected amplitudes are well within reach of the next generation of ground-based CMB experiments (see the sample error bars in Fig.~\ref{fig:ksz}).  The main difficulty, as usual, is foreground contamination.  At most frequencies, the thermal Sunyaev-Zeldovich effect dominates by a large factor; fortunately, it has a null at $\nu=218 \MHz$ and can be cleaned efficiently through multifrequency measurements.   More worrying is point source contamination, which could present a substantial problem \cite{huffenberger05}.  Finally, Figure~\ref{fig:ksz} also shows that the patchy contribution will be smaller than the OV effect from low redshifts, which must be modeled properly to isolate the contribution from reionization.

Intuitively, one would expect these patchy anisotropies to be anti-correlated with 21 cm fluctuations, because both are produced by ionized bubbles during reionization.  However, there has been some controversy over this naive expectation.  The cross-correlation has terms proportional to $P_{v \delta} P_{x x}$ and $P_{v x} P_{\delta x}$.  In analytic models, there is a scale mismatch between the arcminute 21 cm fluctuations and the velocity fluctuations that source the CMB anisotropies \cite{cooray04-cmbt}.  The CMB signal must be integrated along the line of sight, and the highly oscillatory mode-coupling integral cancels out most of the signal.  One can avoid the cancellation by constructing a bispectrum (the Fourier-space equivalent of the three-point function) between the 21 cm signal, the small-scale kSZ anisotropy, and the large-scale Doppler anisotropy, because the last of these directly samples the velocity field \cite{cooray04-cmbt}.  Unfortunately, even with ideal experiments extracting the signal will be a challenge.

However, direct calculations with numerical simulations show that the cross-correlation between 21 cm fluctuations and small-scale CMB anisotropies does \emph{not} vanish \cite{salvaterra05-ksz}; instead they are anti-correlated on scales smaller than the characteristic bubble size at the appropriate redshift.  The evolution of the zero point therefore constrains the growth of \htwo regions during reionization and can help extract information from the 21 cm signal.  The reasons for the discrepancy with the analytic predictions is not clear, but it most likely lies in the assumption $P_{x x} \propto P_{\delta \delta}$ made in the analytic model.  As we have seen (\S \ref{reion-an}), the bubble power spectrum actually has features on much larger scales than the density field and is imperfectly correlated on smaller scales.  Thus the apparent cancellation may be a result of the simplified analytic model used by \cite{cooray04-cmbt}.  

Another CMB signature of early structure formation could take the form of the thermal rather than the kinetic SZ effect: energy from high-redshift supernovae must have been deposited in the CMB via Compton cooling, and the strong clustering bias of high-redshift halos means that strong anisotropies can develop without violating constraints on the Compton $y$-distortion parameter \cite{oh_sz}. While uncertainties are large, a thermal SZ contribution from high-redshift supernovae could in principle explain the excess anisotropy at arcminute scales seen in the CBI \cite{mason03}, BIMA \cite{dawson02,dawson06}, and ACBAR \cite{kuo04} experiments. If they are due to thermal SZ from galaxy clusters, the observed anisotropies require $\sigma_{8}\sim 1$ \cite{bond05}; since the cluster SZ signal is $c_{l} \propto \sigma_{8}^{7}$ \cite{komatsu02}, the tension increases still further given the lower value $\sigma_{8}=0.74^{+0.05}_{-0.06}$ found in the three-year \emph{WMAP} data. Because star-forming halos source HII regions an anti-correlation between the small-scale SZ and 21 cm signals may be detectable \cite{CO_SZ}. The true source of this small-scale CMB anisotropy will become clearer with more sensitive experiments such as the Atacama Cosmology Telescope, which will be able to resolve out the cluster contribution (but not the high-redshift contribution). 

Finally, inhomogeneous scattering during reionization also produces small-scale anisotropies in the CMB polarization \cite{hu00}.  These are, unfortunately, much smaller than the primary polarization anisotropies at the relevant scales, so isolating them requires a tracer field such as the 21 cm signal \cite{cooray04-cmbpol}.  The simple cross-correlation vanishes because the polarization depends on the background CMB quadrupole (which is uncorrelated with the 21 cm fluctuations), so the signal can only be extracted through the bispectrum of the large-scale CMB temperature anisotropies, the small-scale CMB polarization anisotropies, and the 21 cm (or some other tracer) field \cite{cooray04-cmbpol}.  It may be detectable with the combination of high signal-to-noise CMB polarization maps and SKA-class 21 cm telescopes.

%\bibliographystyle{elsart-num}
%\bibliography{Ref_21cm}

%\end{document}

%% file: conc-ch12.tex
%\documentclass{elsart}
%\usepackage{amssymb,cite,epsfig}

%\input{../../submission/defns.tex}

%\begin{document}

\section{Concluding Remarks} \label{conc}

We hope to have convinced the reader of the unparalleled promise of 21 cm observations for unlocking the mysteries of the high-redshift Universe  -- as well as the pitfalls that lie ahead before observations can be successful.  These challenges -- including (but most likely not limited to) terrestrial interference, ionospheric distortions, foreground contamination (especially by the polarized component), and beam-shape control -- are formidable indeed and should not be underestimated.  But, already in the past three years, enormous strides have been made both on the instrumental and data analysis sides, and we have every hope that each of these difficulties can be overcome.  In many ways, the 21 cm community today is analogous to the CMB community in the 1980s:  a clear observational goal exists, but experiments are just starting to explore the landscape.  We do not know, of course, how the potholes along this road compare to those along the path toward the CMB, but the scientific return is clearly large enough that sustained exploration is worthwhile.  On the other hand, the final destination ($\sim 20 \mkel$ fluctuations from \htwo regions and $\sim 3 \mkel$ fluctuations from the underlying density field) \emph{is} better known (or so we hope), because CMB and other observations have lifted the fog concealing the basic cosmological and structure formation parameters.  

The richness of the physics available through the 21 cm line clearly makes even a long journey down this road worthwhile.  Because it is (by definition) the last epoch to produce a strong signal, and because the observational challenges are smallest at high frequencies, the era most amenable to observations is reionization itself -- which is, fortunately, also the most interesting epoch from an astrophysical perspective.  What sources are responsible for it?  How does feedback affect them?  How do these galaxies interact with the IGM?  In conjunction with the other methods described in \S \ref{syn}, the first generation of 21 cm experiments promises to open reionization for detailed study.  But the 21 cm transition is unique in that only it can \emph{directly} probe the three-dimensional morphology and evolution of \htwo regions.  Our most nagging concern about the potential for these experiments to study reionization -- that it may have occurred at such high redshifts that it will lie beyond their reach -- seems less pressing with the latest \emph{WMAP} data \cite{page06, spergel06}.

Although foregrounds make higher redshifts more difficult to observe, those epochs also contain fascinating information -- the 21 cm line could allow us to probe structure formation from its earliest, linear phases ($z \ga 50$) through the formation of the cosmic web and the first luminous sources.  At the highest redshifts, the 21 cm line provides a rich testbed for cosmology, probing much smaller scales than the CMB and making three-dimensional measurements over a wide redshift range.  Once the first stars and galaxies form, the 21 cm line becomes a sensitive measure of their influence on the IGM -- providing perhaps our first view into the secrets of these objects (albeit an indirect one).

The first generation of 21 cm experiments, which are now under construction, will make the first strides down this road, and the next several years promise to be a truly exciting time in cosmology.

%\bibliographystyle{elsart-num}
%\bibliography{Ref_21cm}

%\end{document}